\documentclass{article}

\usepackage[english]{babel}

\usepackage[letterpaper,top=2cm,bottom=2cm,left=2.5cm,right=2.5cm,marginparwidth=1.75cm]{geometry}
\usepackage{float}
\usepackage{amsmath}
\usepackage{graphicx}
\usepackage{authblk}
\usepackage{polski}
\usepackage{caption}
\usepackage{subcaption}
\usepackage{siunitx}

\newcommand{\citeExec}[1]{\cite{#1}} % Temporary fix

\usepackage[
    colorlinks=true,
    allcolors=blue, 
    backref=page % Comment for executive summary alone
]{hyperref}
\usepackage[]{cleveref}
\usepackage{makecell}
\usepackage{multicol}
\usepackage{booktabs}
\usepackage{wrapfig}

\usepackage{titlesec}

\addto\extrasenglish{
  
}
\addto\extrasenglish{
  
}
\addto\extrasenglish{
  
}
\addto\extrasenglish{
  
}

\usepackage[dvipsnames]{xcolor}

\usepackage[shortlabels]{enumitem}
\usepackage[acronym]{glossaries}
\makeglossaries
\newglossaryentry{ECR}{
    name=ECR,
    description={Early Career Researcher}
}

\newglossaryentry{ECRs}{
    name=ECRs,
    description={Early Career Researchers}
}

\newglossaryentry{ESPPU}
{
    name=ESPPU,
    description={European Strategy for Particle Physics Update, \href{https://europeanstrategyupdate.web.cern.ch/}{https://europeanstrategyupdate.web.cern.ch/}}
}

\newglossaryentry{PPG}
{
    name=PPG,
    description={Physics Preparatory Group}
}

\newglossaryentry{ESG}
{
    name=ESG,
    description={European Strategy Group}
}

\newglossaryentry{ECFA}
{
    name=ECFA,
    description={European Committee for Future Accelerators, \href{https://ecfa.web.cern.ch/}{https://ecfa.web.cern.ch/}}
}

\newglossaryentry{RECFA}
{
    name=RECFA,
    description={Restricted ECFA Panel, \href{https://ecfa.web.cern.ch/restricted-ecfa}{https://ecfa.web.cern.ch/restricted-ecfa}}
}

\newglossaryentry{PECFA}
{
    name=PECFA,
    description={Plenary ECFA Panel, \href{https://ecfa.web.cern.ch/plenary-ecfa}{https://ecfa.web.cern.ch/plenary-ecfa}}
}

\newglossaryentry{ECFA ECR Panel}
{
    name=ECFA ECR Panel,
    description={Early-Career Researchers Panel of the European Committee for Accelerators \\\href{https://ecfa.web.cern.ch/ecfa-early-career-researchers-panel}{https://ecfa.web.cern.ch/ecfa-early-career-researchers-panel}}
}

\newglossaryentry{NuPECC}
{
    name=NuPECC,
    description={Nuclear Physics European Collaboration Committee \url{https://www.nupecc.org/}}
}

\newglossaryentry{APPEC}
{
    name=APECC,
    description={Astroparticle Physics European Consortium \url{https://www.appec.org/}}
}

\newglossaryentry{JENAA}
{
    name=JENAA,
    description={Joint ECFA-NuPECC-APPEC Activities \url{https://indico.cern.ch/category/18012/}}
}

\newglossaryentry{IPPOG}
{
    name=IPPOG,
    description={International Particle Physics Outreach Group \url{https://ippog.org/}}
}

\newglossaryentry{EPPCN}
{
    name=EPPCN,
    description={European Particle Physics Communication Network}
}

\newglossaryentry{CPL WG}
{
    name=CPL WG,
    description={Career Prospects and ECR Leadership Working Group}
}

\newglossaryentry{DIMH WG}
{
    name=DIMH WG,
    description={Diversity, Inclusion and Mental Health Working Group}
}

\newglossaryentry{Comm WG}
{
    name=Comm WG,
    description={Communicating the importance of particle physics Working Group}
}

\newglossaryentry{FC WG}
{
    name=FC WG,
    description={Future Colliders Working Group}
}

\newglossaryentry{BC WG}
{
    name=BC WG,
    description={Beyond Colliders Working Group}
}

\newglossaryentry{FCC}
{
    name=FCC,
    description={Future Circular Collider. FCC-ee: colliding electrons and positrons, FCC-hh: colliding hadrons \url{https://fcc.web.cern.ch/}}
}

\newglossaryentry{ILC}
{
    name=ILC,
    description={International Linear Collider \url{https://linearcollider.org/}}
}

\newglossaryentry{CLIC}
{
    name=CLIC,
    description={Compact Linear Collider \url{https://home.cern/science/accelerators/compact-linear-collider}}
}

\newglossaryentry{C$^3$}{
    name=C$^3$,
    description={Cool Copper Collider \url{https://web.slac.stanford.edu/c3/}}
}

\newglossaryentry{muon collider}{
    name=muon collider,
    description={muon collider \url{https://muoncollider.web.cern.ch/} and \url{https://www.muoncollider.us/}}
}

\newglossaryentry{CEPC}{
    name=CEPC,
    description={Circular Electron Positron Collider \url{http://cepc.ihep.ac.cn/}}
}

\newglossaryentry{WG}{
    name=WG,
    description={Working group}
}

\newglossaryentry{DEI}{
    name=DEI,
    description={Diversity, Equity, and Inclusion}
}

\usepackage[most]{tcolorbox}

\newtcolorbox[auto counter,number within=subsection,number freestyle={\noexpand\arabic{subsection}.\noexpand\arabic{\tcbcounter}}]{recommendationbox}[2][]{colback=blue!5!white,colframe=blue!75!black,fonttitle=\bfseries,label={rec:\arabic{subsection}.\arabic{\tcbcounter}},title=Recommendation~\thetcbcounter,#1}

\newcommand{\important}[3][]{%
    \begin{tcolorbox}[colback=green!5!white,colframe=green!75!black,fonttitle=\bfseries,title=#2,#1]
      #3
    \end{tcolorbox}
}

\newcommand{\example}[3][]{%
    \begin{tcolorbox}[colback=gray!5!white,colframe=white!75!black,fonttitle=\bfseries,title=#2,#1]
      #3
    \end{tcolorbox}
}

\newcommand{\GlsBlack}[1]{\begingroup\hypersetup{linkcolor=black}\Gls{#1}\endgroup}

\newcommand{\recommendationboxexec}[1]{{\bfseries#1}}

\newcommand{\statementboxexec}[1]{{\bfseries}}

\title{
    \vspace*{-1.9cm}
    \Large Early Career Researcher Input to the\\
    European Strategy for Particle Physics Update:
    White Paper\\
    \vspace*{0.1cm}
    \large \textit{Fifty-five recommendations for the future of our field}
}

\date{27th of March, 2025}
\author{
    \vspace*{-0.8cm}
    \parbox{0.88\textwidth}\centering
    
    \small{ \textbf{Editors}\\ \vspace{-3.5mm}
Jan-Hendrik Arling$^{1,\dagger,*}$, Alexander Burgman$^{2,*}$, Christina Dimitriadi$^{3,\dagger}$, Ulrich Einhaus$^{4}$, Axel Gallén$^{5}$, Abdelhamid Haddad$^{6}$, Laura Huhta$^{7,\dagger}$, Armin  Ilg$^{8,\S,*}$, Jan Klamka$^{9}$, Elizabeth Long$^{10}$, Thomas  Madlener$^{1}$, Arnau Morancho Tardà$^{11,\dagger}$, Emanuela Musumeci$^{12,*}$, Krzysztof Mękała$^{1,9,*}$, Elena Pompa Pacchi$^{13}$, Marvin Pfaff$^{14}$, Daniel Reichelt$^{15}$, Leonhard Reichenbach$^{15,16,\dagger}$, Birgit  Stapf$^{15}$, Francesco P. Ucci$^{17,18}$, Erik Wallin$^{19,\dagger}$ and Harriet Watson$^{20,\dagger,*}$ \\

    \vspace*{0.1cm}

    \textbf{Authors} \\
    Sagar Vidya Addepalli$^{21}$, Bruno  Alves$^{22,\dagger}$, Robert Mihai Amarinei$^{23}$, Ricardo Barrué$^{24}$, Lydia Brenner$^{25,\S}$, Giacomo Da Molin$^{24}$, Arturo de Giorgi$^{26}$, Bohdan Dudar$^{27}$, Francesco Giuli$^{28,29}$, Andrea Gurgone$^{30,31}$, César Jesús-Valls$^{32}$, Antoine Laudrain$^{1}$, Martin J. Losekamm $^{33}$, Rafał Masełek$^{34}$, Wrishik Naskar$^{35}$, Miquel Nebot-Guinot$^{20}$, Marko Pesut$^{8,\dagger}$, Thomas Pöschl$^{15}$, Efrain P. Segarra$^{36}$, Rebecca Taylor$^{14,15}$, Pavel Vana$^{10,\dagger}$, Hannah Wakeling$^{37}$ and  Aidan R. Wiederhold$^{38,\dagger}$
    \vspace*{0.2cm}

    \small{    
        $\dagger$: Member of the ECFA ECR Panel as of 31st of March 2025
        
        $\S$: Past member of the ECFA ECR Panel

        $^{*}$: Contact addresses: \href{mailto:jan-hendrik.arling@cern.ch}{jan-hendrik.arling@cern.ch}, \href{mailto:alexander.burgman@fysik.su.se}{alexander.burgman@fysik.su.se}, \href{mailto:armin.ilg@cern.ch}{armin.ilg@cern.ch}, \href{mailto:k.mekala@uw.edu.pl}{k.mekala@uw.edu.pl}, \href{mailto:emanuela.musumeci@cern.ch}{emanuela.musumeci@cern.ch}, \href{mailto:harriet.watson@cern.ch}{harriet.watson@cern.ch}, \href{mailto:eppsu-ecr-organisers@cern.ch}{eppsu-ecr-organisers@cern.ch}
    }

    \vspace*{0.3cm}
    
    {\large Endorsed by the ECFA ECR Panel} % Only once endorsed
    \vspace{-4mm}
    }
}

\begin{document}

\thispagestyle{empty} % Removing page number from titlepage

\maketitle
\thispagestyle{empty} % Removing page number from titlepage
\vspace*{-8mm}
\begin{abstract}
This document, written by early career researchers (ECRs) in particle physics, aims to represent the perspectives of the European ECR community and serves as input %supporting document
for the 2025--2026 update of the European Strategy for Particle Physics. With input from a community-wide survey, it highlights key challenges faced by ECRs --- career stability, funding access and long-term research opportunities --- while proposing policy recommendations and targeted initiatives. It underscores the importance of practices fostering diverse, equitable, inclusive and healthy workplaces, as well as of stronger ECR communities, and highlights how effective communication and interdisciplinary collaborations reinforce the societal relevance of particle physics and promote continued support for large-scale and long-term projects.
Finally, the future of both collider and beyond-collider experiments is addressed, emphasising the critical role of ECRs in shaping future projects.

The ECR contribution
is formed of two parts:
the
%this % uncomment for ES only
ten-page executive summary submitted as input to the European Strategy for Particle Physics Update and, as backup document, 
this
%an
extended white paper %which serves as a ``backup document",
providing additional context.
\vspace*{-1.5mm}
\end{abstract}
%\fontsize{9pt}{12pt}\selectfont
\noindent\rule{\textwidth}{0.1pt}
\noindent\hrulefill
\vspace{-0.05mm}
\begin{minipage}[t]{0.48\textwidth}
\raggedright
\footnotesize
$^{1}$ Deutsches Elektronen-Synchrotron DESY, Hamburg, Germany \\
$^{2}$ Stockholm University, Stockholm, Sweden \\
$^{3}$ KTH Royal Institute of Technology, Stockholm, Sweden \\
$^{4}$ Karlsruhe Institute for Technology, Karlsruhe, Germany \\
$^{5}$ Uppsala University, Uppsala, Sweden \\
$^{6}$ Laboratoire de Physique de Clermont Auvergne, CNRS/IN2P3, Université Clermont Auvergne, France \\
$^{7}$ University of Jyväskylä, Jyväskylä, Finland \\
$^{8}$ University of Zürich, Zürich, Switzerland \\
$^{9}$ University of Warsaw, Warsaw, Poland \\
$^{10}$ Charles University, Prague, Czech Republic \\
$^{11}$ Niels Bohr Institute, Copenhagen, Denmark \\
$^{12}$ Instituto de Física Corpuscular (IFIC), CSIC -- Universitat de València, Paterna (València), Spain \\
$^{13}$ The University of Oklahoma, Norman, Oklahoma, USA \\
$^{14}$ Imperial College London, London, United Kingdom \\
$^{15}$ CERN, Geneva, Switzerland \\
$^{16}$ University of Bonn, Bonn, Germany \\
$^{17}$ University of Pavia, Pavia, Italy \\
$^{18}$ INFN - Sezione di Pavia, Pavia, Italy \\
$^{19}$ Lund University, Lund, Sweden \\
$^{20}$ The University of Edinburgh, Edinburgh, United Kingdom \\
$^{21}$ SLAC National Accelerator Laboratory, Menlo Park, USA
\end{minipage}%
\hspace{0.02\textwidth}%
%\vrule%
\hspace{0.02\textwidth}%
\begin{minipage}[t]{0.48\textwidth}
\raggedright
\footnotesize
$^{22}$ Laboratoire Leprince-Ringuet, CNRS/IN2P3, Ecole Polytechnique, 
Institut Polytechnique de Paris, Palaiseau, France \\
$^{23}$ University of Geneva, Geneva, Switzerland \\
$^{24}$ Laboratório de Instrumentação e Física Experimental de Partículas (LIP), Lisbon, Portugal \\
$^{25}$ National Institute for Subatomic Physics (NIKHEF), Amsterdam, Netherlands \\
$^{26}$ Institute for Particle Physics Phenomenology, Durham University, Durham, United Kingdom  \\
$^{27}$ University of Mainz, Mainz, Germany \\
$^{28}$ Dipartimento di Fisica, Università degli Studi di Roma Tor Vergata, Rome, Italy \\
$^{29}$ INFN - Sezione di Roma 2, Rome, Italy \\
$^{30}$ Università di Pisa, Pisa, Italy \\
$^{31}$ INFN - Sezione di Pisa, Pisa, Italy \\
$^{32}$ Kavli IPMU (WPI), UTIAS, The University of Tokyo, Kashiwa, Japan \\
$^{33}$ Technical University of Munich, Munich, Germany \\
$^{34}$ Laboratoire de physique subatomique et de cosmologie de Grenoble, Grenoble, France \\
$^{35}$ University of Glasgow, Glasgow, United Kingdom \\
$^{36}$ Paul Scherrer Institute, Villigen, Switzerland \\
$^{37}$ John Adams Institute for Accelerator Science, University of Oxford, Oxford, United Kingdom \\
$^{38}$ University of Manchester, Manchester, United Kingdom 
\end{minipage}

\noindent\hrulefill
\noindent\rule{\textwidth}{0.1pt}

\newpage

% Make spacing more tight for the executive summary
\titlespacing\section{0pt}{3.5ex plus 1ex minus .2ex}{2.3ex plus .2ex}

\thispagestyle{empty} % Removing page number from TOC. Comment out for executive summary

\tableofcontents % Comment out for executive summary alone

\setcounter{page}{0} % Start at first page of executive summary. Comment for executive summary alone
\thispagestyle{empty}

\newpage % Comment out for executive summary alone

%%% Executive summary

% Use this to disable hyperlinks to glossary, for when exporting executive summary alone
% \glsdisablehyper  % Comment

\part*{Executive summary}\phantomsection\addcontentsline{toc}{section}{Executive summary}\label{executive_summary}
% \vspace*{-0.3cm}

\section*{Introduction}

% Flower text :-)
The study of the nature of matter, its constituents, and their interactions excite people all over the world. The collaborative and innovative spirit of particle physics, specifically, draws in young talents who want to contribute to this global, scientific endeavour of understanding nature.
Early career researchers (ECRs) play a crucial role in shaping the future of particle physics, yet their perspectives are often under-represented in strategic planning. The ECRs of today will be the leaders of tomorrow. Whether in accelerator science, data analysis, engineering, instrumentation, software development, or theory, they will conduct their future research based on the decisions of today.

During the last European Strategy for Particle Physics Update (\Gls{ESPPU}), which took place in 2019--2020, dedicated input from ECRs was gathered for the first time. The \Gls{ECFA} member countries were each asked to nominate up to ten ECRs who discussed together their view on the strategy and the Physics Briefing Book~\citeExec{ESU2020PBB}. The resulting document~\citeExec{ECRInput2020} was sent to the European Strategy Group (\Gls{ESG}). The final ESPPU document~\citeExec{2020_eppsu} then featured recommendation 7.B, addressing ECR issues that had been expressed in the ECR input. Emphasis was put on the importance of supervision and training and the recognition of the development and maintenance of detectors, computing infrastructure, and software. Furthermore, the importance of equality, diversity and inclusion was stressed. Beyond these recommendations, there were two major organisational outcomes from the ECR discussions. Firstly, the necessity of forming a Europe-wide representation of ECRs in particle physics was stated. On this basis, the \Gls{ECFA ECR Panel}~\citeExec{ECFAEarly-CareerResearchersECRPanel:2024trx} was created soon after by \Gls{ECFA}. Secondly, in future strategy updates, ECRs should proactively coordinate and provide regular input rather than an ad-hoc response upon request. 

In early 2023, the \Gls{ECFA ECR Panel} organised the \emph{Future Colliders for ECRs} workshop~\citeExec{ecfa_ecr_future_colliders}. With almost a hundred in-person and more than a hundred online participants, it showcased the significant interest of the ECR community in future colliders and in the future of the field in general. The main conclusion was that a decision on the next collider in Europe should be taken as soon as possible to give certainty of a bright future for the field of particle physics. 
%and help the careers of ECRs working on future colliders. 
To achieve this, the %ECFA ECR
panel sent a letter to the CERN Council (Appendix A in Ref.~\citeExec{ecfa_ecr_future_colliders}) before its March 2024 meeting, strongly urging the council to 
%proceed with as quick a pace as possible of evaluating, selecting and implementing potential future projects, thus accelerating its time frame to 
start the next strategy update as early as possible. Following that meeting, the CERN Council formally launched the ESPPU process to take place 2025--2026 --- earlier than previously anticipated.

In September 2024, the \Gls{ECFA ECR Panel} initiated the process towards the \emph{ESPPU ECR white paper} summarised in this document. The goal was to collect, discuss, aggregate, and then disseminate the views of European ECRs, building on top of the previous strategy and ECR input. It aims to provide a single, bottom-up input to the ESPPU based on the principles of openness, inclusion, and consensus building. This process has been open to all European ECRs\footnote{Studying in or employed at European institutes, with non-permanent positions or $< 10$ years after finishing the PhD} in particle physics\label{footnote:EuropeanECR}.
%After preparatory meetings of the ECFA ECR Future Colliders WG with ECR representatives from the LHC experiments and future collider communities
In October 2024, the ECR session at the \emph{3rd ECFA workshop on $e^+ e^-$ Higgs, Electroweak and Top Factories}~\citeExec{event_3rd_ecfa} brought together the ECR community to identify key topics and issues to address in the white paper. In the following, these working groups (\Glspl{WG}) were formed:
%Key outcomes included the need for stable funding, career certainty, timely decisions on future colliders, and the integration of sustainability into long-term planning. 
\vspace{0.2cm}

\begin{minipage}{0.45\textwidth}
    \begin{enumerate}[nosep,leftmargin=0.1cm,rightmargin=0.0cm]
        \item[\textbullet] Career prospects and ECR leadership,
        \item[\textbullet] Future colliders,
        \item[\textbullet] Future experiments beyond colliders,
    \end{enumerate}
\end{minipage}
\hfill
\begin{minipage}{0.51\textwidth}
    \begin{enumerate}[nosep,leftmargin=0.0cm,rightmargin=0.0cm]
        \item[\textbullet] Diversity, inclusion and mental health,
        \item[\textbullet] Communicating the importance of particle physics,
        \item[\textbullet] Interplay of particle physics with neighbouring fields.
    \end{enumerate}    
\end{minipage}
\vspace{0.0cm}

The WGs convened regularly to refine ideas and draft actionable recommendations. These efforts led to the \emph{ECR Workshop on the ESPPU} on the 14 November 2024~\citeExec{event_eppsu_ecr_cern}, adjacent to the 115th Plenary ECFA meeting, which brought the WGs together to define the next steps. To gather detailed input from the wider ECR community, a comprehensive ECR survey was conducted. The WGs then drafted the white paper document that was circulated among all ECRs interested in this effort and discussed in-depth during the open ECR Symposium in February 2025~\citeExec{event_ecr_symposium} for refinement and consolidation. The white paper gives recommendations based on the survey findings, discussions in the corresponding WGs, and the open meetings with the community. It is submitted to the ESPPU process as a backup document 
%(also available on arXiv~\cite{white_paper_arxiv}) % Comment out for complete white paper
and this executive summary of the white paper serves as the main ESPPU input document.
\clearpage
\section*{Early career researcher survey on ESPPU}

%Survey (1 page, incl. 1-2 bio plots: Country of work + Career state + maybe no. of years after PhD)
The ESPPU ECR white paper organisers decided to conduct a survey among European ECRs to collect an empirical basis for recommendations for the white paper and this executive summary.
All WGs contributed a number of questions to the survey, which was open from 18 December 2024 to 27 January 2025.
It was distributed to as many ECR and ECR-related email lists as possible, including the \Gls{ECFA ECR Panel} list, the participants in the white paper effort, national ECR communities, and (ECR) lists of large collaborations as well as future collider and beyond-collider projects.
Out of the 936 responses received, 804 were submitted by European ECRs (see footnote on \Cpageref{footnote:EuropeanECR}) and are hence considered the main data set for the survey analysis.
%In total, 936 responses were submitted. Of these, 804 fulfill the requirement to be submitted by a `European Early-Career Researcher', i.e.\ working in a European country and either having a non-permanent position or having received the PhD fewer than 10 years ago. These 804 responses are the main data set we use for the survey analysis. 
%, although some comparisons between ECRs and the ca. 50 non-ECRs as well as between European and the ca. 100 non-European ECRs were also produced.

\begin{figure}[H]
     \centering
     %\begin{subfigure}[tb]{0.38\textwidth}
     %    \centering
         \includegraphics[width=.37\textwidth,trim=0.5cm 1.2cm 0.5cm 1.4cm,clip]{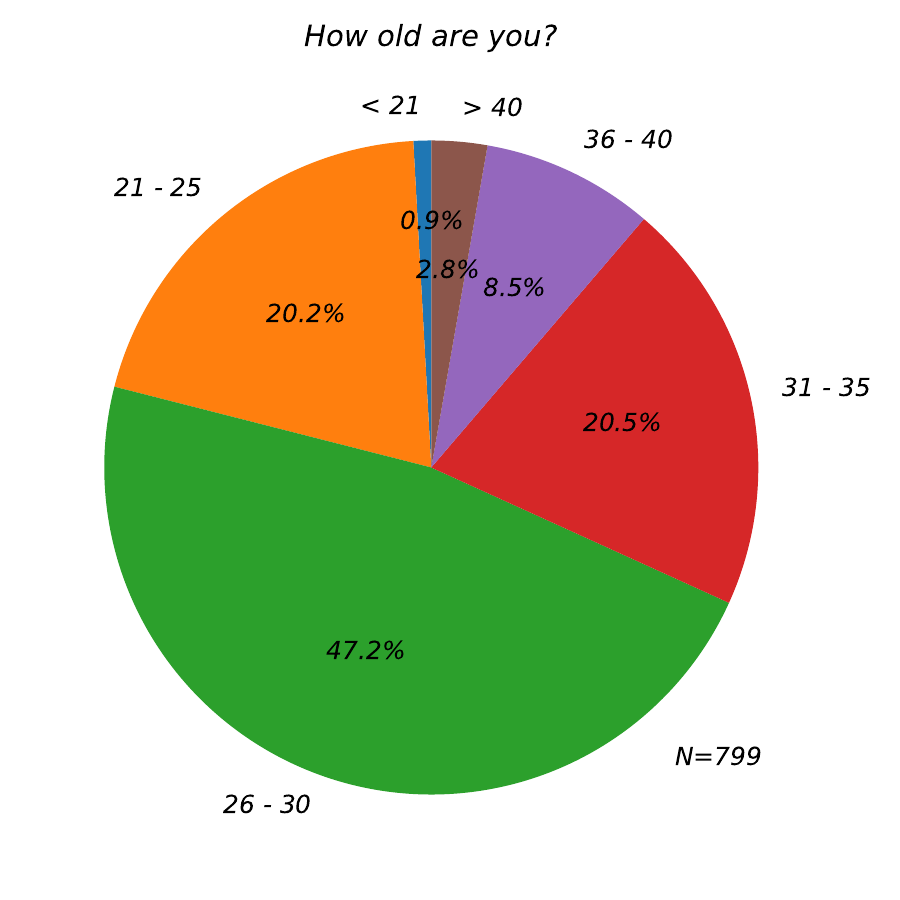}
     %    \caption{}
     %    \label{fig:experiment-exec}
     %\end{subfigure}
     \vspace*{0.2cm}
     \hfill
     %\begin{subfigure}[tb]{0.48\textwidth}
     %    \centering
         \includegraphics[width=.5\textwidth,trim=0cm 2.0cm 0.1cm 2.8cm,clip]{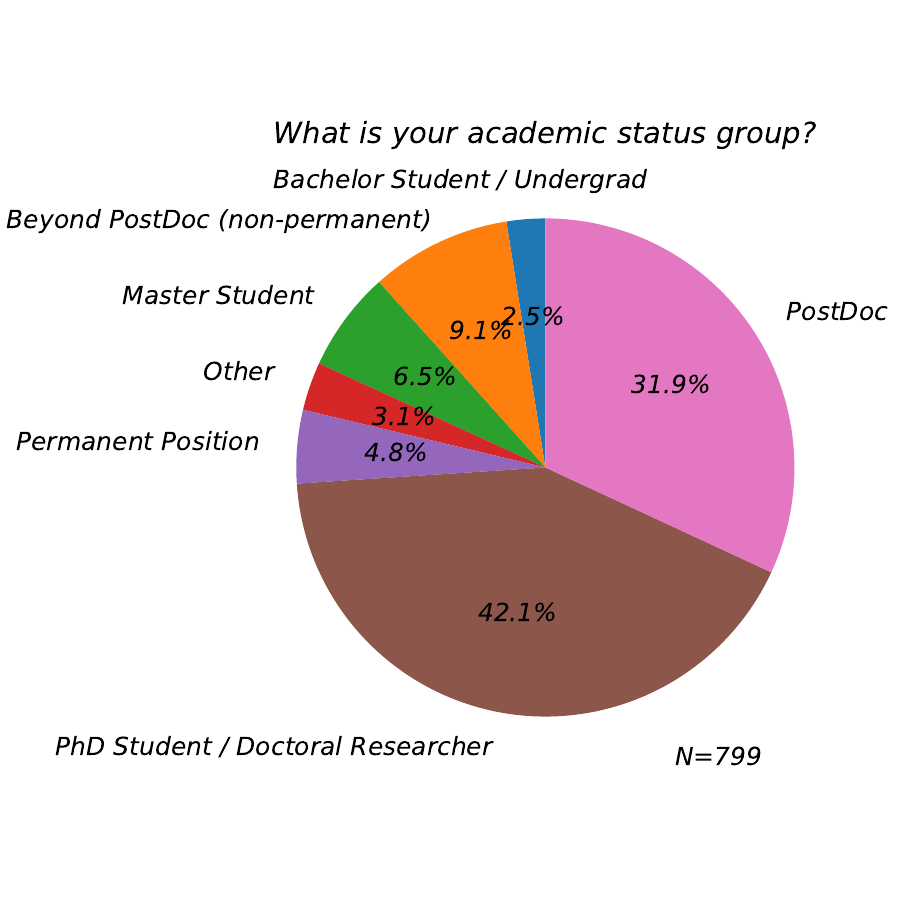}
     %    \caption{}
     %    \label{fig:future-experiment-exec}
     %\end{subfigure}
        \vspace*{-0.4cm}
        \caption{Distribution of age (left) and academic status (right) of survey respondents.}
        \label{fig:bio-age-status-exec}
\end{figure}

\begin{figure}[H]
     \centering
     %\begin{subfigure}[tb]{0.41\textwidth}
     %    \centering
         \includegraphics[width=.5\textwidth,trim=0.3cm 0.4cm 0.3cm 0.3cm,clip]{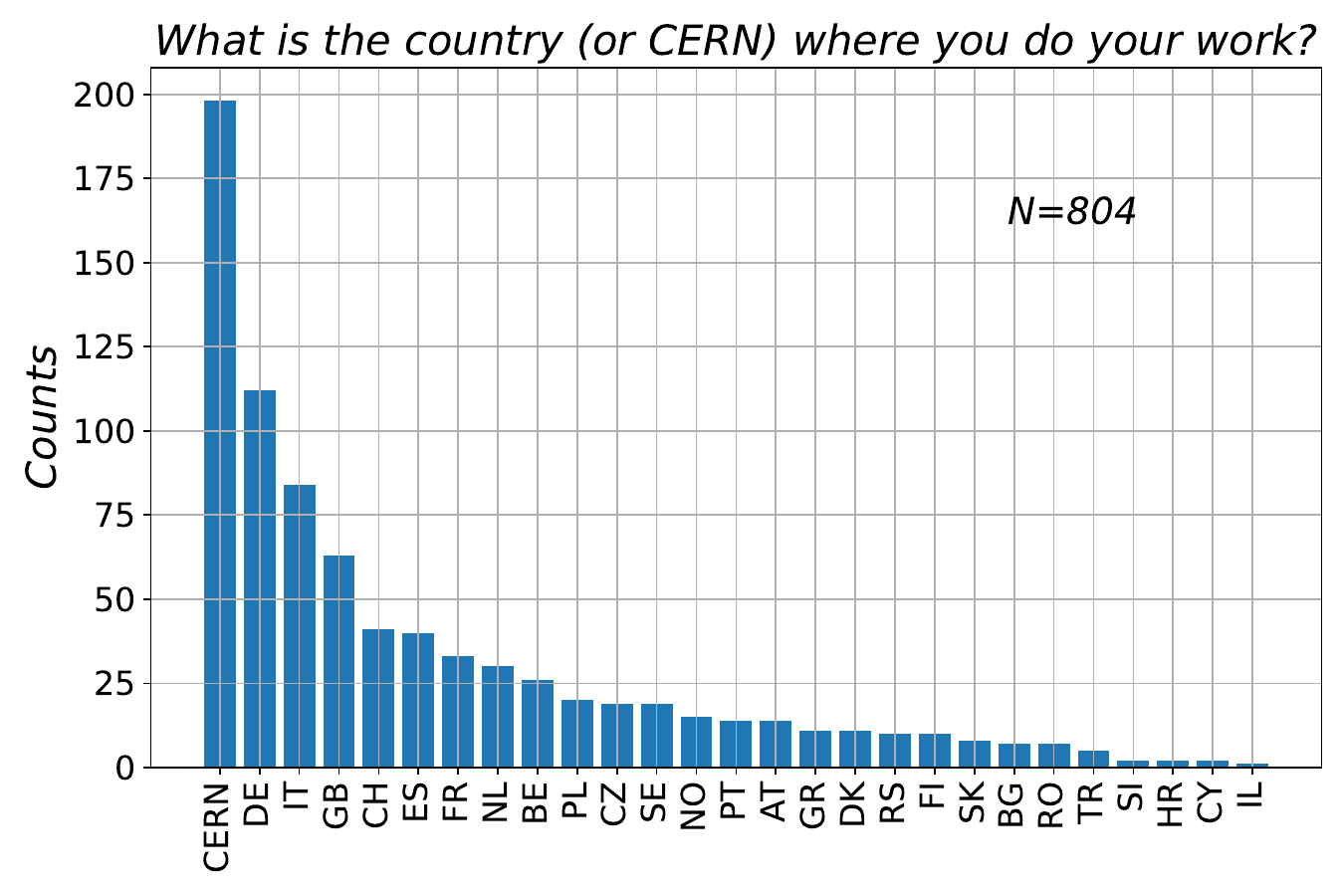}
     %    \caption{}
     %    \label{fig:region-of-origin-work-exec}
     %\end{subfigure}
     \hfill
     %\begin{subfigure}[tb]{0.35\textwidth}
     %    \centering
         \includegraphics[width=.40\textwidth,trim=0.0cm 1.4cm 0.3cm 1.0cm,clip]{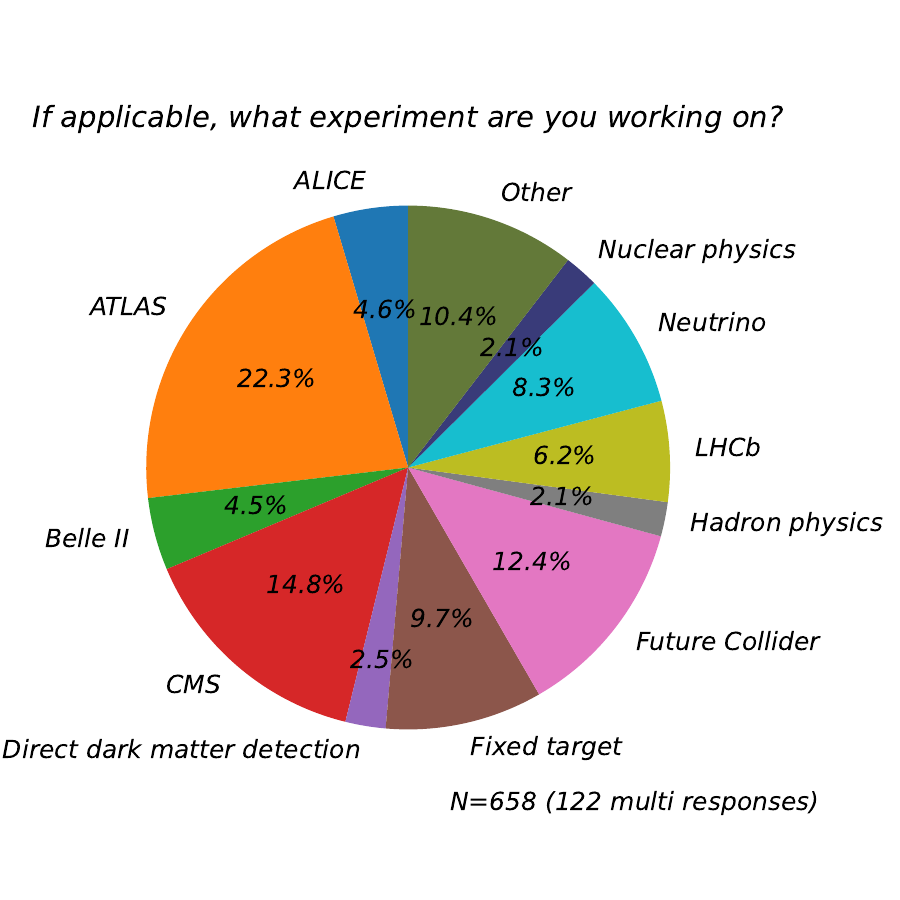}
     %    \caption{}
     %    \label{fig:country-of-work-exec}
     %\end{subfigure}
        \vspace*{-0.3cm}
        \caption{Work locations of survey respondents in European host countries or \Gls{ECFA} member states (left) and the experiments survey respondents participate in, if applicable (right).}
        \label{fig:bio-region-of-origin-experiment-exec}
\end{figure}

To limit bias, guidelines for the survey analysis were established before closing the survey and unblinding the data.
The pre-defined analysis procedure includes the statistical calculation of significance of a difference between answer options.
Each survey-based statement in this document that indicates a difference between two answers required at least a $2\sigma$ separation (for details see
%white paper~\citeExec{arxiv_whitePaper}, section 3.5).
Section~\ref{sec:analysis_procedure}).
% Comment switch for Executive Summary standalone
Additionally, specific wording with regard to majorities was also pre-defined, which is used in this document: ``relative majority'' for most votes, but below $\SI{50}{\percent}$, ``majority'' for $>\SI{50}{\percent}$, ``significant majority'' for $>\SI{70}{\percent}$ and ``the respondents'' for $>\SI{90}{\percent}$ support.
The only section of responses that was looked into before the survey closed was the biography of the respondents, which allowed to send dedicated emails to country lists and communities which had been under-represented at that time.
We believe the resulting biographical distribution of participants follows the overall community reasonably well, so we conclude the survey to be sufficiently representative to make quantitative statements and conclusions.
An overview of age, academic status, country of work and, if applicable, the experiment respondents participate in can be seen in Figures~\ref{fig:bio-age-status-exec} and \ref{fig:bio-region-of-origin-experiment-exec}.
Of the ca.~100 future collider participants, \SI{40}{\percent} are involved in \Gls{FCC}-ee, \SI{26}{\percent} in linear collider projects, \SI{15}{\percent} in the muon collider and \SI{10}{\percent} in \Gls{FCC}-hh.
The ca. \SI{20}{\percent} of respondents working in nuclear physics, neutrino physics, fixed target experiments, direct dark matter detection and strong-field QED experiments are considered to be working in ``beyond-collider'' areas for the purposes of this survey.

\vspace*{-0.2cm}

\vspace*{-0.8cm}
% \vspace*{-0.4cm}

\clearpage
\section*{Recommendations}

This section gives concrete recommendations on the topics and issues that ECRs consider critical for discussion by the ESPPU.
The recommendations are highlighted in boldface and a reference to the corresponding recommendations in the
extensive discussion% in the backup document
%white paper ~\citeExec{arxiv_whitePaper}
% Comment switch for Executive Summary standalone
is given in parentheses. Please refer to this for more details and context.
%, e.g.~on previous experiences and relevant best-practice examples of actions connected to our recommendations.
In particular, some recommendations refer to measures that have already been put in place at some research centres, as also shown by best-practice examples provided by survey respondents to free-text questions.
%(for details see the white paper~\cite{arxiv_whitePaper}).
However, there is a lack of consistency across institutes in Europe and where these measures do exist, some could be improved.
If recommendations directly address questions of the \Gls{ECFA} guidelines for national inputs~\citeExec{ecfa_national_input_guidelines}, then this is also indicated in parentheses.

First, we discuss the \textbf{careers and wellbeing} of ECRs, emphasising the importance of these topics for the future of the field. This is followed by \textbf{early career community building, leadership and recognition} which all act as catalysts for the success of ECRs. Next, we address how particle physics is \textbf{communicated}, a vital aspect for the advancement of \textbf{future particle physics projects}, from flagship collider to beyond-collider projects, and the strategy process itself is addressed thereafter.

%First, we consider the careers and wellbeing of ECRs, emphasising their importance for the future of the field. This is followed by a discussion of early career community building, leadership, and recognition, all of which are crucial for the success of ECRs. Next, we address how particle physics is communicated, a vital aspect for the advancement of future particle physics projects, and ECR opinion on those projects, ranging from flagship collider experiments to beyond-collider physics. Finally, we discuss the ESPPU process.

\paragraph{\large Careers and wellbeing of early career researchers\\}%(1.5 pages)

%This section addresses the ECR community's concerns with regards to career prospects, where the most pressing issues are related to instability and relocation. Diversity, equity and inclusion (DEI) concerns are also addressed, including mental health, supervision and mentorship, experiences of discrimination and harassment, and expected workplace conduct. 
ECRs face significant challenges, particularly instability and frequent relocation in their career paths. Concerns related to Diversity, Equity, and Inclusion (\Gls{DEI}) are also prominent and include topics of mental health, supervision and mentorship, experiences of discrimination and harassment, and expectations of workplace conduct.

The primary challenges for the ECR community associated with career progression are a lack of long-term stability and difficulty in securing permanent academic positions. Our survey found that while \SI{89}{\percent} of respondents intend to stay in particle physics, only \SI{45}{\percent}  believe they have ``okay" or ``pretty good" prospects of doing so. Approximately half reported that, before starting their academic careers, they were unaware of the percentage of PhD graduates who eventually leave academia for other paths.
To better support ECRs amid these uncertainties, providing clear and transparent information on career trajectories from the start is crucial. Therefore, \recommendationboxexec{institutions and their senior researchers should take active steps to increase awareness about employment perspectives in the field among individuals at the earliest stages of their careers~(R.~\ref{rec:1.1}).}

ECRs are concerned by the instability caused by frequent relocations and short-term contracts~\cite{Allen:2024lyp}.
\recommendationboxexec{Postdoctoral positions should not be shorter than two years. The emphasis on long-term mobility for career advancement must be reconsidered to foster a more inclusive field~(R.~\ref{rec:1.2}, \ref{rec:1.3}).} In addition, relocating for temporary positions does not guarantee long-term employment, often adding uncertainty and stress. To provide greater stability, \recommendationboxexec{a structured path to a permanent position should be provided and stated in job offers and contracts whenever possible~(R.~\ref{rec:1.4}).} Furthermore, long-term relocation is complex, requiring expertise in regulations from the departure, destination, and origin countries. \recommendationboxexec{An international body should be established to assist researchers with administrative relocation matters, potentially under CERN’s mandate, given its role as an international hub. Additionally, all academic institutions should employ experts to provide such assistance. Economic support for relocation must also be guaranteed in ECRs' contracts~(R.~\ref{rec:1.5}).}

To encourage a thriving research environment, it is crucial to support secondary projects and non-research activities, as these foster innovation, employability, and networking. \recommendationboxexec{Contracts for non-permanent positions should specifically allocate time for such projects, supported by flexible working hours. ECRs should feel able to explore secondary research interests without fear of pushback from their
supervisors. In addition to the classical criteria of academic excellence, these activities --- including public engagement, mentoring, \GlsBlack{DEI} initiatives, and institutional service --- should be formally recognised in performance evaluations and hiring decisions~(R.~\ref{rec:1.6}, \ref{rec:1.7}, \ref{rec:1.10}).} 

While larger collaborations often provide broad networking opportunities, researchers in smaller experiments can struggle with limited networks and career progression challenges. Despite producing excellent and well-rounded researchers, these smaller settings often lack sufficient networking support. Shorter timescales also often present a challenge to researchers starting their careers in a small collaboration, with ECRs potentially changing collaboration after each contract. While the LHC job matching event facilitates networking for ECRs looking for jobs in collider experiments, similar resources do not exist for small or beyond-collider collaborations. Easier access to up-to-date information regarding the field of beyond-collider physics facilitate the planning of potential career paths for ECRs. \recommendationboxexec{Career support must be provided for ECRs pursuing careers in both beyond-collider and collider physics. There should be greater clarity of potential career paths for ECRs working on small and short-timescale experiments, along with stronger support for networking opportunities to help them navigate these career trajectories~(R.~\ref{rec:1.8}).}

To broaden opportunities beyond academia,
\recommendationboxexec{ECR training programs should integrate skills that are transferable to industry. Supervisors should be required to explicitly outline these skills in their project proposals. Additionally, universities and research institutions should recognise and leverage industry experience by creating positions that attract researchers from diverse career backgrounds, including from outside academia~(R.~\ref{rec:1.9}, \ref{rec:1.11}).} 
\paragraph{}
Mental health is a concern, as the majority of those surveyed (\SI{57}{\percent}) reported struggling with it during their research career, especially among under-represented groups and those who have experienced discrimination or harassment. Mental health care is often not completely covered by public healthcare or health insurance and can be prohibitively expensive, especially for ECRs who typically earn non-competitive salaries~\cite{mental_health_europe}. 
For these reasons, \recommendationboxexec{(inter)national laboratories, research centres and institutions should allocate part of their funding to mental health support for their research community, with particular focus on ECRs who tend to have more vulnerable positions%. %Additionally, ECRs should be provided with emotional support and constructive feedback in supervision. 
~(R.~\ref{rec:1.12}).}

ECRs depend on the support of supervisors, whose interactions impact their research quality, future opportunities, and wellbeing. \recommendationboxexec{Institutes should require a mandatory supervision course for staff in supervisory roles that focuses on aligning expectations with the supervisee and understanding their needs~(R.~\ref{rec:1.13}).}
Beyond formal supervision, mentorship can also play a crucial role in career development. %Mentors provide guidance beyond research, supporting career development, work-life balance, and personal growth. 
\recommendationboxexec{ECRs should have better access to mentorship programs which offer career guidance, research skill development and personal growth opportunities, thus building confidence and advancing their careers~(R.~\ref{rec:1.14}).}

Survey respondents indicate progress is needed in inclusivity and diversity within research environments. For example, \SI{80}{\percent} of respondents feel that their research environment is not completely inclusive for people with disabilities and \SI{37}{\percent} of women respondents report experiencing gender-based discrimination or harassment. \recommendationboxexec{Universities and research centres must ensure a safe and healthy workplace, where bullying, harassment and discrimination are treated with a zero-tolerance policy and any resultant actions are applied consistently to all staff members regardless of their seniority. Robust, accessible, effective and independent support systems must be put in place for cases which violate these policies, ensuring that complaints can be made anonymously (R.~\ref{rec:1.15}).} To support this, \recommendationboxexec{\GlsBlack{DEI} Offices should be established, with a size proportionate to their research community. They should provide mandatory training on \GlsBlack{DEI} and monitor the workplace culture, taking action when needed. They should create safe spaces for marginalised groups and ensure diversity, including in age and position in academia, in hiring panels and executive roles. Finally, policies should accommodate various socio-economic backgrounds to provide equal opportunities for all, including when hosting events with external participants~(R.~\ref{rec:1.16}).} In order to prevent and solve problems before they escalate, \recommendationboxexec{\GlsBlack{DEI} Offices should establish a monitoring committee, including ECR representatives, to review findings from surveys, exit interviews and workplace assessments. Outcomes and actions taken to address identified challenges should be made public through regular \GlsBlack{DEI} progress reports~(R.~\ref{rec:1.20}).}
CERN and international research centres should support ECRs who face persecution or discrimination based on their characteristics in their home countries and cannot rely on support from their home institutes. Furthermore, as recently evidenced in the USA, \Gls{DEI} policy is not guaranteed anywhere and this is impacting ECRs~\citeExec{science-Trump-DEI}. 
\recommendationboxexec{CERN should maintain its commitment to \GlsBlack{DEI} policies that enrich its working environment and support users who originate from countries where they do not have adequate protections~(R.~\ref{rec:1.17}).}

To encourage an equitable, inclusive, diverse and healthy workplace, 
\recommendationboxexec{institutions and research centres must have an adequately-publicised code of conduct that includes zero tolerance of discrimination or harassment and that considers under-represented groups when setting out good practices and expected behaviours. A transparent and accessible complaints procedure should be put in place for when the code of conduct is violated. This procedure should be clearly signposted in the code of conduct. Anonymity and discretion should be guaranteed for those involved with complaints~(R.~\ref{rec:1.18}).}
These measures should also extend to events that are an essential part of a career in research.
%and are typically an excellent opportunity to build a network and reputation, gain a sense of community and learn. 
\recommendationboxexec{Conferences, schools and workshops should have a code of conduct that is explicitly presented to participants and that follows the same guidelines as the recommendation above. CERN and other large research centres should provide a template code of conduct that can be shared with smaller institutions and independent organising committees that lack the resources to produce them ad-hoc for events~(R.~\ref{rec:1.19}).}

\paragraph{\large Early career community building, leadership and recognition\\} %(1 page)

ECRs play a vital role in advancing particle physics, yet they encounter persistent challenges related to representation, funding, recognition and participation in decision-making.
%Addressing these challenges is essential to ensure that ECRs can fully contribute and shape the future of the field.

%ecfa ecr panel
The \Gls{ECFA ECR Panel}, established in 2020, has helped to build and engage the ECR community in Europe through events and surveys.
However, concerns remain regarding the transparency of its selection process and the extent to which it represents the broader ECR community.
National representatives are currently appointed through varying and often unclear procedures, limiting broader participation.
To address this, \recommendationboxexec{the panel should become more open to input and involvement from the community by creating national ECR forums with regular meetings and ECR mailing lists where not yet established. Additionally, a uniform, democratic and transparent selection process for national representatives must be implemented, ensuring fair representation of different research areas within each country~(R.~\ref{rec:2.1}).}
%\recommendationboxexec{The ECFA ECR Panel should become more open to input and involvement from the community by creating national forums with regular meetings and ECR mailing lists where not yet established. A uniform, democratic selection process for panel members, ensuring fair representation of different research areas within each country, must be implemented whenever possible.}

%role of ESPPU - Krzysztof - needs to be shortened
Even though the current ECR community will conduct future research shaped by today's decisions, % which makes their involvement in the ESPPU process essential.
%Therefore, it is completely natural and reasonable to demand the inclusion of ECRs in the processes of strategy updates. 
the current extent and manner of involvement are far from satisfactory.
Despite our efforts with this community white paper and in many countries with the inclusion of ECRs in the national input discussions, only \SI{20}{\percent} of respondents feel their voice is heard in the ESPPU process. At the same time, \SI{31}{\percent} state that the strategy outcome influences their decision to stay in academia. 
The appointment of ECR ``scientific secretaries" in the \Gls{PPG}~\citeExec{PPG2024} was a step forward, but a lack of transparency in the selection process and unclear responsibilities undermine their role.
Additionally, having no structured communication in any form between said secretaries and the broader ECR community limits their representativeness.
To consolidate their position, \recommendationboxexec{such appointments must be made transparently, with secretaries given an equal voice among working group conveners. Furthermore, the \GlsBlack{ECFA ECR Panel} should also be mandated to send a delegate to the ESG~(R.~\ref{rec:2.2}).}
%While the appointment of ECRs as ``scientific secretaries" in PPG~\citeExec{PPG2024} is undoubtedly a step in the right direction compared to previous strategy updates, there is a lot of room for improvement, particularly in terms of transparency. 
%The exact role and scope of responsibilities of secretaries are unclear and not well-defined, and the definition of ECR itself is also not clearly set out. 
%Moreover, their selection process is highly arbitrary, and there is no established form of communication between the secretaries and the community. 
%This undermines both their position as representatives of ECRs, and the very purpose of their appointment which \recommendationboxexec{must become transparent and their mandate should also give them an equal voice among working group conveners. Furthermore, the ECFA ECR Panel should also be mandated to also send a delegate to the ESG.}
%\recommendationboxexec{Appointment of the scientific secretaries, taking into account the broader ECR community, must be transparent. Their mandate should give them an equal voice among the working group conveners. The ECFA ECR Panel, which already provides several delegates to PECFA and one to RECFA, should be mandated to also send a delegate to the ESG.}

%acknowledgment of contributions and leadership skills
Beyond the ESPPU process, similar challenges exist within research collaborations.
While survey results show that \SI{81}{\percent} of ECRs feel heard in their local research groups, this drops to \SI{54}{\percent} for collaborations.
Some collaborations already include ECRs in executive board meetings, but this practice should be extended more broadly.
To strengthen ECR representation further, inclusion should go beyond decision-making roles to active participation in research discussions and community-building efforts.
\recommendationboxexec{ECRs should be more included in topical working groups and the organisation of events. Dedicated ECR sessions should be an integral part of conferences and similar events~(R.~\ref{rec:2.3}).}
%These measures would ensure greater visibility and provide a platform for their contributions.
%\recommendationboxexec{ECRs should be more included in the executive board of collaborations as well as in topical working groups and the organisation of events.Dedicated ECR sessions should be an integral part of conferences and similar events.}

ECRs take on significant leadership responsibilities in research, from operating experiments to driving future facility studies, yet their contributions often go unrecognised in hiring. \recommendationboxexec{Leadership roles should be formally credited~(R.~\ref{rec:2.4}),} as they demonstrate project management skills relevant both in academia and beyond.
Current evaluation metrics often fail to capture the full scope of these contributions.
\recommendationboxexec{Supporting alternative evaluation methods beyond traditional bibliometric measures in the particle physics community is necessary. Service work and R\&D for sensitivity studies of future facilities must also be recognised~(R.~\ref{rec:2.4}).}
%\recommendationboxexec{Supporting alternative evaluation methods beyond traditional bibliometric measures in the particle physics community is highly recommended. Service work and R\&D for sensitivity studies of future facilities, such as colliders, as well as leadership roles must also be recognised.}

%funding
Despite their active role in fostering collaboration and career development, ECRs struggle to secure funding for their initiatives. Most institutional resources are strictly allocated to research, leaving little support for networking and other ECR-led activities aimed at strengthening the community.
Additionally, funding is often unavailable for professional development opportunities, such as attending relevant events beyond an ECR's immediate research focus.
To address this gap, \recommendationboxexec{the \GlsBlack{ECFA ECR Panel} should receive dedicated funding to organise events and initiatives that strengthen the ECR community and support professional development~(R.~\ref{rec:2.5}).}
This should follow increased transparency by the panel itself to ensure ECRs are aware of available opportunities and how to engage with the panel~(see R.~\ref{rec:2.1}).
%\recommendationboxexec{The major institutions should establish a system of dedicating funding for ECR activities which would be a unique opportunity for young scientists to improve their professional development and enrich the research community, shaping the future of scientific research.}

% beyond-collider community
While collider research in Europe is centralised at CERN, beyond-collider experiments are more dispersed, making collaboration and networking more difficult.
\recommendationboxexec{A forum for beyond-collider particle physics researchers should be established to give coherence to this community and to facilitate collaboration, knowledge exchange and networking. Such efforts should include the whole beyond-collider particle physics community, and the establishment of additional fora for large beyond-collider subfields would be beneficial for more effective topical exchange~(R.~\ref{rec:2.6}).}

Despite its broad remit, \Gls{ECFA}
is often perceived as focused on collider-based research, partly due to its historical name and references to ``high-energy physics" in its terms of reference~\citeExec{ECFATermsOfReference}.
\recommendationboxexec{In order to create a more inclusive \GlsBlack{ECFA}, there must be fair and even representation of the different particle physics research fields, including beyond-collider particle physics, within all \GlsBlack{ECFA} bodies. Additionally, \GlsBlack{ECFA} should form a dedicated panel on future beyond-collider particle physics experiments and facilities and their physics potential, as a step towards unifying the disparate beyond-collider field and giving it a communal voice~(R.~\ref{rec:2.7}).}

\paragraph{\large Communicating the importance of particle physics\\}

The importance of effective science communication for the future of particle physics cannot be overstated. Every researcher needs to be able to convey scientific research questions and results comprehensively to fellow particle physicists, other scientists and to the general public. Only in this way the relevance of particle physics to society can be showcased and the future of the field secured, with part of the communication effort relying on ECRs.

There are many ongoing efforts to strengthen the role of science communication in particle physics, for example the International Particle Physics Outreach Group (\Gls{IPPOG})~\cite{IPPOG_website}, the European Particle Physics Communication Network (\Gls{EPPCN})~\cite{EPPCN} or CERN's Education, Communications \& Outreach Group (ECO Group)~\cite{CERN_ECO}, that have been very successful already at providing training and resources to researchers involved in outreach. However, the survey results also indicated that awareness of these initiatives is not sufficiently spread in the ECR community and more work is needed to support ECRs.

Depending on the target audience in question, different kinds of support for ECRs may be needed, and therefore the recommendations posed in this section attempt to make a distinction based on the different groups: communication to the general public (\textit{outreach}), to industry and within the field.
These recommendations aim to represent the specific voice of ECRs and they are both complementary to and aligned with the proposals made by \Gls{IPPOG} and \Gls{EPPCN} in their inputs to the previous strategy update~\cite{IPPOG,EPPCN} and their updates under preparation for this ESPPU. 

Within the ECR survey, a significant majority (\SI{84}{\percent}) of respondents expressed in general a high motivation to engage with the public on topics related to the future of particle physics. Therefore, \recommendationboxexec{the particle physics community should promote a culture where outreach and communication are integral to the practice of science, not secondary responsibilities, in order to sustain the high levels of motivation for engaging with the public amongst ECRs and supporting their efforts in ensuring public approval of future projects~(R.~\ref{rec:3.1}).}

However, more than \SI{40}{\percent} of ECR responses in the survey reported that the main factor discouraging them from pursuing outreach activities is the lack of recognition for such work. To sustain and enhance an active ECR contribution in science communication also in the future, 
\recommendationboxexec{systemic changes should be made to formally recognise and reward outreach and communication efforts and to integrate them into institutional benchmarks, performance and funding evaluations to validate their importance within the scientific profession~(R.~\ref{rec:3.2}).}

Additionally, more than \SI{40}{\percent} of survey respondents indicated that they do not feel adequately qualified for outreach activities, with the main reason being the absence of adequate professional training on outreach. Of those who have received such training, a significant majority (85\%) found it useful, showing that training programs are a valuable use of resources. Thus, \recommendationboxexec{disparities in the availability of communication training must be addressed. Institutions and collaborations should develop standardised programs, leveraging internal expertise and partnerships with professional science communicators. The particle physics community should cooperate with universities to integrate training programs into academic curricula to ensure equitable access for all ECRs~(R.~\ref{rec:3.3}).}

Apart from the lack of professional communication training, the survey responses indicated also that opportunities for outreach engagement as well as sufficient funding and resources for outreach tools are missing. Therefore,
\recommendationboxexec{the particle physics community should continue supporting outreach initiatives that bridge academia with the public, schools, and other educational institutions. A more visible and centralised platform should be established, building upon existing outreach efforts, to provide early career researchers (and beyond) with essential resources, such as materials, tools, and logistical support. Outreach materials shared with the community should follow an open-source policy~(R.~\ref{rec:3.4}).}

Furthermore, an active channel of communication with industry is highly relevant for the field and its future development. Hence,
\recommendationboxexec{the particle physics community should build upon existing collaborations with industry to further strengthen synergies. Expanding shared doctoral programs, enhancing knowledge transfer, and highlighting the societal and technological applications of research will be key. Additionally, dedicated fora should be reinforced or newly established to facilitate ongoing dialogue and foster long-term partnerships~(R.~\ref{rec:3.5}).}

Similarly, effective communication within the particle physics community itself and also cross-disciplinary approaches to extend the view of the field are important to ECRs.
\recommendationboxexec{Open communication within the scientific community must be maintained. Interdisciplinary initiatives should be supported, as they enhance knowledge sharing and foster opportunities for ECR networking, thereby maximising the impact of research~(R.~\ref{rec:3.6}).}

Overall, the understanding of nature at its most fundamental level, as well as the importance of particle physics in comprehending our universe, and the diversity of the field needs to be adequately communicated to all audiences and stakeholders. This includes highlighting the science pursued across the vast variety of experiments. However, smaller collaborations often have scarce resources, making the participation in outreach more challenging. Therefore,
\recommendationboxexec{support should be given to smaller collaborations in outreach activities. This could include finding ways for small collaborations to organise outreach together, as well as making pre-existing outreach structures more inclusive to smaller collaborations~(R.~\ref{rec:3.7}).}

Furthermore, reaching significant visibility on social media is challenging for most collaborations and institutions, irrespective of the size.
Therefore, to make this form of science communication effective for our field,
\recommendationboxexec{the European particle physics community should establish additional joint outreach channels on social media that highlight the science pursued in Europe no matter the size or location of the experiments. CERN, with its existing social media expertise and its large follower base on the different platforms, should actively support the efforts and ensure that shared content gains maximum visibility by actively promoting and distributing it~(R.~\ref{rec:3.8}).}

Of equal importance to science communication in general is the storytelling put forward by outreach activities. Future colliders are usually promoted for their (in)direct discovery potential of new hypothetical particles that might not exist in nature or not be observed due to current technological limitations. In this view,
\recommendationboxexec{the storytelling of particle physics research should emphasize the role of future experiments as particle observatories, rather than focusing solely on their potential for discovering new particles. Highlighting their ability to provide precise measurements, test fundamental theories, and explore unknown phenomena can reinforce enthusiasm within the field and enhance public support for fundamental research~(R.~\ref{rec:3.9}).}
To this end, the scientific community should promote a communication that is both accurate and engaging, avoiding misinformation or excessive simplification.

\paragraph{\large Future particle physics projects\\}

The future of particle physics depends on ECRs contributing to fundamental measurements, novel theories, and experimental techniques. The present-day ECRs will be the drivers of future particle physics experiments, both the next flagship collider and future beyond-collider experiments of various scales. Therefore, these projects should meet also the scientific interests of the ECR community.

As the operational lifetime of the HL-LHC is fixed, the question of its successor becomes pressing. The survey results indicate that a significant majority of ECRs (\SI{79}{\percent}) support the development of a next flagship collider. A strong long-term vision is necessary to maintain engagement and ensure the future of the field. Therefore, \recommendationboxexec{a future flagship collider should be developed to further the progress of particle physics~(R.~\ref{rec:4.1}).}

In recent years, CERN has been the heart of particle physics in Europe and a global leader in research, maintaining Europe's position as the main hub for particle physics and driving industrial developments. 
\SI{68}{\percent} of ECRs support building the future flagship collider at CERN, recognising its role as the natural host for the next flagship collider in Europe, while 19\% do not have a strong opinion in this regard.\recommendationboxexec{A CERN-based location of the next collider facility would be highly appreciated, but it should not be used as the justification to prioritise any of the projects proposed during the ESPPU process.  The European ECR community remains open to collaboration on any project, with a particular preference for those based in Europe~(R.~\ref{rec:4.2}).}

Whatever decision is taken in the ESPPU, today's ECRs will be affected by its results for decades. This makes it crucial that future colliders are also aligned with the aspirations of the generation that will carry them forward. 
Furthermore, if a new type of collider is selected, an entirely new community of experts will be required. Investing in skill development now will ensure a seamless transition to future machines. \recommendationboxexec{Training related to upcoming facilities should be embedded within academic programmes and ECRs should play a vital role in future-collider research~(R.~\ref{rec:4.3}).}

\paragraph{}

Survey respondents were asked to evaluate nine criteria reflecting different priorities regarding the construction of the next collider by allocating 90 points among them. 
The need to drive technological innovation scored highest in almost all communities considered, with the exception of theorists and phenomenologists, among whom it was in third place. For this group, the most important criterion was an ambitious baseline physics programme, which scored second, taking all responses into account. Openness to global collaboration secured third place. Sustainability, securing support for smaller projects, and a well-defined upgrade path garnered medium support overall. In contrast, the support for smaller projects was assessed as the most important in the community of fixed-target experiments. Minimising the time to the first collision scored third from lowest all answers considered; conversely, this answer was the third most popular among ECRs working on future colliders. The second to last place was taken by the need to maximise social acceptance (e.g. regarding cost and land use). A specific location of the next project scored significantly lowest.

All statements given above, including the ranking of the criteria, are based on the statistically significant comparison of the mean values of all responses for each criterion and subgroup of respondents (for details see
Section~\ref{sec:collider_criteria}).
%white paper~\citeExec{arxiv_whitePaper}, sec. 4.4.4).
% Comment switch for Executive Summary standalone
Thus, we conclude that \recommendationboxexec{the selection of the next collider facility should be guided by its technological innovation as the driving factor. The project should provide a comprehensive and exhaustive baseline programme; while a long-term vision for particle physics is important, the upgrade path should not be the primary motivation for building a specific collider. Regardless of the location of the next major project, it is crucial to ensure its openness to global collaboration~(R.~\ref{rec:4.4}, \ref{rec:4.5}, \ref{rec:4.6} and \GlsBlack{ECFA} question 3.b).}

It is important to highlight that R\&D is crucial for particle physics, while the same new technologies are often also synergistic with societal applications, providing an important contribution to society from academia. Despite this, ECRs are often not incentivised to take part in R\&D projects because of the potentially low-publication nature of technological developments and lack of recognition for this work. Many of the R\&D activities also rely on short-term contracts due to a lack of funds. \recommendationboxexec{European laboratories should assign greater value to innovation in preparation for the next generation of particle physics facilities. ECRs and projects driving technological innovation should be recognised and supported, to benefit both particle physics and broader societal applications~(R.~\ref{rec:4.7}).}

In the survey, a question on a specific collider preference was also asked. The selection of possible options was made after intense discussions in the WG.
Of the 782 answers, ``a circular $e^+e^-$ collider (e.g. \Gls{FCC}-ee)" received \SI{28}{\percent} of the votes, followed by ``a \lowercase{\gls{muon collider}}" at \SI{15}{\percent}, ``a hadron collider (e.g. \Gls{FCC}-hh)" at \SI{14}{\percent}, and ``a linear $e^+e^-$ collider (e.g. \Gls{CLIC}/\Gls{ILC}/\Gls{C$^3$})" at \SI{8}{\percent}. Additionally, \SI{23}{\percent} of respondents expressed that they do not have a strong opinion or do not know, while \SI{9}{\percent} supported the idea of building ``any collider, as soon as possible". Only \SI{2}{\percent} of participants opposed the concept of constructing a collider in Europe. No other options were considered in the question.
As it might be susceptible to biases resulting from the composition of ECR communities this survey reached, it is also meaningful to analyse this question in light of the respondents' specific research field. 
Notably, among all future-collider communities, the project most closely related to the participant's work typically gains around \SI{60}{\percent} support, with the second-highest priority being ``any collider, as soon as possible". Thus, we deliberately refrain from recommending any specific collider facility and conclude that \recommendationboxexec{the main collider proposals, a circular $\mathbf{e^+e^-}$ collider, a muon collider, a hadron collider and a linear $\mathbf{e^+e^-}$ collider, have received recognition from the ECR community. A relative majority prefers a circular $\mathbf{e^+e^-}$ collider, closely followed by the option ``I do not know/I do not have a strong opinion" (\GlsBlack{ECFA} question 3.a).} 

As no single collider proposal received an absolute majority mandate from the community, some people might be hesitant to join the selected project. For this reason,  \recommendationboxexec{a clear recommendation on the next flagship collider for Europe, based on the submitted proposals, should be given in this ESPPU process. The ESPPU should urge the CERN Council to make a timely decision on the next flagship collider and efforts should be made to ensure that every member of the community feels invited to contribute to the project, bringing their expertise and experience to serve the advancement of our field~(R.~\ref{rec:4.8}).}

Respondents were also asked what the next collider in Europe should be if the ``\Gls{CEPC} is approved in China and could deliver first collisions in 2035-2040'' or the ``\Gls{ILC} is approved and delivering first collisions in a timely manner". The responses did not point to a clear solution, but in total, about three-quarters of respondents answered that \recommendationboxexec{the European community should pursue a different path and start construction of a complementary collider project if any major collider is approved outside Europe~(R.~\ref{rec:4.9} and \GlsBlack{ECFA} question 3.c).}

\paragraph{}

ECRs are split among the different collider proposals for the next facility in Europe and a lot of convergence is still necessary to unite behind a single project. There is, however, an impression that this decision, which concerns the future of CERN and the whole of particle physics in Europe, is pre-determined and being driven by a limited number of people. For example, the \Gls{FCC} is increasingly often referred to as the ``baseline" project (see e.g.~\citeExec{talk_PECFA}), despite the facts that firstly, a feasibility study is still ongoing to determine \textit{whether} the project is feasible; and secondly, the last ESPPU has not concluded or determined a ``baseline". 
This impression is detrimental to the process and to the consequences of its outcome. Therefore, we advocate for \recommendationboxexec{open, transparent and democratic decision-making. After the decision has been made, a structured explanation of the criteria which led to a certain result is necessary and will help significantly in understanding the procedure and building trust for its outcome within the whole community~(R.~\ref{rec:4.10}).} For example, it is not clear to what extent securing the future of CERN and preserving its size, as well as preferences from funding agencies or national governments, impact the decision-making.

There has been significant pressure~\citeExec{ECFA_WS3_Sphicas_Talk} to converge behind one project, once the decision is taken. However, the way in which the convergence should be achieved has not been specified.
Combined with the impression of the pre-defined decision, this might create resentment. A significant majority of European ECRs are willing to support the ESPPU outcome, even if their preferred collider is not chosen as a priority, but the advocates of the prioritised project will need to convince others to join the effort.
\recommendationboxexec{A fair and inclusive selection process, along with the requested well-defined plan B, C, D, etc. is best suited to create acceptance for the decision among proponents of the alternatives. Proponents of the leading project must work to gather support for their proposal~(R.~\ref{rec:4.11}).}

\paragraph{} 

The survey participants were asked about preferred funding allocation for flagship and smaller projects, assuming the total amount of money was fixed. The results indicate that \SI{39}{\percent} of respondents supported increased funding for the flagship collider and \SI{20}{\percent} a similar increase for smaller projects, \SI{34}{\percent} prefer maintaining the current balance and the rest do not know. Therefore,  \recommendationboxexec{the budget split between flagship collider and smaller experiments should not be changed significantly~(R.~\ref{rec:4.12}).} As any future flagship collider has the potential to serve as a host facility for a variety of smaller projects beyond its primary interaction points, \recommendationboxexec{the physics potential of any future collider proposal should be maximised by considering smaller-scale specialised experiments in addition to the main interaction-point experiments~(R.~\ref{rec:4.13}).}

Further aspects of future colliders assessed in the survey concerned sustainability and more intense collaboration with CERN non-member states, which could be crucial for securing stable funding for the large-scale facility. In this regard, \recommendationboxexec{the next flagship collider should be built and run in the most sustainable way, and collaboration with institutions from countries other than CERN member states should be increased~(R.~\ref{rec:4.14}).}

\paragraph{}

A wide variety of European particle physics is conducted at dedicated experiments that are not located at the interaction points of colliders. % experiments.
%These experiments explore phenomena that often cannot be explored at collider facilities %but can be studied with beyond-collider facilities, e.g.\ 
%and cover topics ranging from neutrino physics and direct dark matter searches, to QCD, QED and fundamental symmetries. % outside of a collider environment.
Such \textit{beyond-collider} experiments appear in a great diversity of sizes and cover questions often inaccessible in collider measurements, ranging from pure particle physics to its neighbouring fields, such as astroparticle or nuclear physics.
Additionally, beyond-collider precision measurements, for example those on\ soft-QCD or QED, provide essential input to the modelling of collider interaction processes, and there are significant synergies for methodological and technological developments between the collider and beyond-collider communities.

As such, the beyond-collider field is currently an integral part of European particle physics. Going forward, \recommendationboxexec{beyond-collider experiments and activities should maintain a prominent role in the European particle physics landscape, both as groundbreaking activities in their own right and as pathfinders for collider searches.
Their diversity in scale, infrastructure and duration should be valued and sustained in order to maintain a thriving beyond-collider landscape~(R.~\ref{rec:4.15}).}
This importance is also reflected in the responses to the supplementary\footnote{As a supplementary question, it received lower participation than other questions. The fraction of respondents to this question who are working on collider experiments compared to beyond-collider experiments is similar to other questions, however.} survey question related to questions 4.a and 4.c of the \Gls{ECFA} national input guidelines~\citeExec{eppsu_national_inpus_instructions}: a significant majority of respondents replied that the extent of CERN's participation in beyond-collider activities in general should be increased, with emphasis on novel accelerator R\&D (by a significant majority) and medical physics (by a majority). For other sub- and adjacent fields (fixed target, neutrino physics, direct dark matter searches, nuclear physics, and astroparticle physics experiments), a significant majority of respondents think CERN's current level of involvement should at least be maintained.

\paragraph{}

%Funding procurement is perceived as highly challenging for beyond-collider research in Europe:
When asked what they thought were the three main challenges of beyond-collider experiments in Europe, \SI{54}{\percent} and \SI{41}{\percent} of respondents answered  ``Unclear financial situation due to lack of long-term funding commitment'' and  ``Difficulty to compete for funding with large flagship projects'', respectively, making them the two biggest challenges according to the survey.
Additionally, a significant majority (\SI{81}{\percent}) expressed that a dedicated funding scheme for beyond-collider particle physics would benefit beyond-collider experiments in Europe ``rather much'' or ``significantly''.
Therefore, 
\recommendationboxexec{a dedicated beyond-collider funding scheme should be put in place to strengthen the European beyond-collider community~(R.~\ref{rec:4.16}).}

Beyond-collider experiments are 
%not only located at major particle physics laboratories but 
often located at smaller laboratories and interdisciplinary facilities, such as neutron sources, free-electron lasers or nuclear reactors. This repurposing of infrastructure provides an efficient and cost-effective way of pursuing science.
\recommendationboxexec{The ESPPU should %acknowledge the importance of smaller laboratories for research in fundamental physics, and 
explicitly recommend the innovative usage of smaller particle physics and non-particle physics facilities for fundamental physics research. ECRs should receive training on particle physics research done at non-collider facilities to be able to use them %efficiently
in their research.
%To maximise the efficient use of such shared resources with neighbouring fields, 
Recommendations in the ESPPU should be developed in synergy with those from \GlsBlack{NuPECC} and \GlsBlack{APPEC}, and vice versa%Furthermore, 
~(R.~\ref{rec:4.17}).}

Smaller experiments also provide unique training opportunities for ECRs, where they can participate in all stages and aspects of an experiment. Easier career transitions between the collider and beyond-collider fields, and therefore a further diversification of ECRs' skills, would benefit the whole particle physics community. This could be particularly beneficial during the transition period between the current and next flagship collider project, when the personnel needs of the collider community might be different.
%With the often shorter timespan of beyond-collider experiments, the beyond-collider field is more versatile in terms of workforce than the collider field.
Therefore,
\recommendationboxexec{a concrete plan for particle physics in Europe between the end of the HL-LHC and the next major flagship project should be laid out at the latest in 
%within the scope of the next
the next ESPPU in around 5 years' time, %. Such a plan 
and should contain a dedicated beyond-collider programme %for this period 
that allows for the retention of knowledge and talent~(R.~\ref{rec:4.18}).}

\paragraph{}

As an international strategy document, the ESPPU will carry weight with national and international funding agencies. %.Additionally, the ESPPU is also expected to
It is expected to present a concrete, united European strategy for the next flagship collider project, however, it is unclear whether it will present a similar strategy for beyond-collider projects.
When asked in the survey, a majority of the participants (\SI{68}{\percent}), and a significant majority of beyond-collider respondents (\SI{79}{\percent}), agree that beyond-collider particle physics experiments in Europe would benefit from concrete recommendations for the beyond-collider fields in the ESPPU either rather much or significantly.
With this in mind, \recommendationboxexec{the ESPPU should include concrete recommendations for beyond-collider particle physics research, in order to support and sustain the full spectrum of European particle physics~(R.~\ref{rec:4.19}).}
These recommendations should define European priorities within each subfield, % (neutrino physics, dark matter physics, the intensity and precision frontiers, etc.),
including general experimental efforts and the most pressing measurements, and the theoretical advances required to support them.

\section*{Conclusions}

This document summarises the views of the European ECR community in particle physics and related fields. The ECRs of today will be the ones conducting the future research of tomorrow. Thus, the ESPPU must consider their views, ideas, and concerns. We appreciate the opportunity to have our voices heard. Implementing the recommendations and taking into account the further statements for the ESPPU further strengthens the ECR community --- ensuring a bright future for particle physics in Europe and beyond. 

The white paper has been endorsed by the \Gls{ECFA ECR Panel}.
The full list of supporters is available at \href{https://indico.cern.ch/e/espp-ecr-white-paper-support}{https://indico.cern.ch/e/espp-ecr-white-paper-support} where everyone, including other panels or organisations, can show their support of this document by signing it. The editors and authors will disseminate the messages of this document through presentations to the particle physics community and will follow and participate in the ESPPU process.

We hope that this strategy input is useful to the \Gls{ESG} when drafting their summary document and beyond to showcase the opinions, concerns, and ideas of the enthusiastic community of European early career researchers in the field of particle physics.

%%% Extensive document

\begingroup
\let\clearpage\relax
\glsenablehyper

\renewcommand{\thesection}{\arabic{section}}
\setcounter{section}{0}

% Normal spacing for extensive document
\titlespacing\section{0pt}{3.5ex plus 1ex minus .2ex}{2.3ex plus .2ex}

\newpage

% \part*{Extensive discussion}\phantomsection\addcontentsline{toc}{section}{Extensive discussion}\label{extensive_discussion}

\part*{Extensive discussion}\phantomsection\addcontentsline{toc}{section}{Extensive discussion}\label{extensive_discussion}

\section{Introduction} % {\color{red}Frozen, contact Uli before changes!}}

The last European Strategy for Particle Physics Update (\Gls{ESPPU}) took place in 2019--2020. The Physics Briefing Book~\cite{ESU2020PBB}, written by the Physics Preparatory Group (\Gls{PPG}), summarised the discussions from the Granada Open Symposium~\cite{eppsusymposium2019}. The European Strategy Group (\Gls{ESG}) then reached convergence and drafted scientific recommendations, which the CERN Council adopted in June 2020~\cite{2020_eppsu}. In the resulting strategy document, twenty recommendations were given concerning the future of particle physics in Europe, with Recommendation 7.B about Early Career Researchers (ECRs). Emphasis was put on the importance of supervision and training, the recognition of detector development and maintenance, computing and software. Furthermore, the importance of equality, diversity and inclusion was stressed.
    
During the last ESPPU, for the first time, there was dedicated input from ECRs.
The then \Gls{ECFA} chair invited ECRs to a meeting at CERN to discuss their view on the strategy and specifically on the Physics Briefing Book~\cite{ESU2020PBB} and produce a document to be brought to the ESG's attention. Since, at that time, no Europe-wide ECR representation in particle physics existed, the \Gls{ECFA} member countries were asked to each nominate up to ten ECRs who were invited to contribute at the end of September 2019.
Details on this procedure can be found in Ref.~\cite{ECFA-NL3}. The report on the ECR debate on the ESPPU, including the results of a survey among delegates, was published on arXiv in February 2020~\cite{ECRInput2020} and given to the ESG by the \Gls{ECFA} chair.

In addition to the above content, there were two major organisational outcomes of this process. Firstly, the desire to form a Europe-wide representation of ECRs in particle physics. On this basis, the \Gls{ECFA ECR Panel}~\cite{ECFAEarly-CareerResearchersECRPanel:2024trx} was created soon after and officially mandated by Plenary ECFA (\Gls{PECFA}) in July 2020. Five ECR panel members represent the ECRs in PECFA, with one of them permanently invited as an observer to \Gls{RECFA} meetings.
Secondly, in future strategy update processes, ECRs should proactively coordinate and provide regular input rather than an ad hoc response upon request. The result is this document for the ongoing strategy update.

\subsection*{Current ESPPU update}
In early 2023, the \Gls{ECFA ECR Panel} created the Future Colliders working group (WG) to bring the topic of future colliders closer to the ECR community. It organised a one-day hybrid workshop at CERN and online \cite{event_future_colliders_ecrs} with almost a hundred in-person participants and more than a hundred online, showcasing the significant interest of the ECR community in future colliders and, in general, the future of particle physics. The workshop is summarised in Ref.~\cite{ecfa_ecr_future_colliders}. One of the main conclusions was that a decision on the next collider in Europe should be taken as soon as possible to give certainty to a bright future of particle physics and help the careers of ECRs working on future colliders. To achieve this, the \Gls{ECFA ECR Panel} sent a letter to the CERN Council (Appendix A in Ref.~\cite{ecfa_ecr_future_colliders}) before the meeting of March 2024, with the following statement:
\begin{quotation}
\textit{We, the ECFA Early-Career Researchers Panel, on behalf of the ECR community, would like to strongly urge the Council to make every effort to ensure that the process of evaluating, selecting and implementing potential future projects, which will define this century of particle physics for Europe and the World, proceed with as quick a pace as possible, accelerating its time frame to start the European strategy process as early as possible and conclude by early 2026.}
\end{quotation}
Following the March 2024 CERN Council meeting, the CERN Council formally launched the ESPPU process, which will take place in 2025--2026 --- earlier than previously anticipated.

In September 2024, the \Gls{ECFA ECR Panel} initiated the process towards an ECR white paper input to the ESPPU. The drafting process has been open to all European ECRs, defined as students and researchers employed at European institutes who hold non-permanent positions or who completed their PhD less than 10 years ago. After preparatory meetings of the ECFA ECR Future Colliders WG with ECR representatives from the LHC experiments and future collider communities, on 10 October 2024, the ECR session at the 3rd ECFA workshop on $e^+ e^-$ Higgs, Electroweak and Top Factories in Paris~\cite{event_3rd_ecfa} brought together approximately 55 participants, comprising both in-person and remote attendees. The session aimed to engage the ECR community, encourage contributions, and identify key topics to address in the white paper. Key outcomes included the need for stable funding, career certainty, timely decisions on future colliders, and the integration of sustainability into long-term planning. WGs, introduced in Section~\ref{sec:topics}, were established to focus on the most critical issues identified during the session.

From October to November 2024, each ESPPU ECR white paper WG convened regularly to refine ideas, draft statements, and develop actionable recommendations. These efforts culminated in the ECR Workshop on the ESPPU held at CERN on 14 November 2024~\cite{event_eppsu_ecr_cern}, adjacent to the 115th Plenary ECFA meeting. More than thirty participants attended in person, with around sixty joining via Zoom. The workshop featured presentations from the WGs, parallel sessions for deeper discussions and informal interactions to solidify the scope of each WG's work and outline the next steps.

To gather extensive input from the ECR community, a comprehensive survey was created and widely circulated between 18 December 2024 and 27 January 2025, with detailed results and analysis presented in Section~\ref{sec:survey}. The WGs carefully reviewed the survey responses and drafted concrete recommendations in Section~\ref{sec:recommendations}, which were discussed in-depth during the ECR Symposium in February 2025~\cite{event_ecr_symposium}. Lastly, Section~\ref{sec:conclusions} lists the next steps throughout the ESPPU process.

\section{Topical working groups } %{\color{red}Frozen, contact Armin before changes!}} 
\label{sec:topics}

This section describes the six WGs established following the main discussion points raised at the first ESPPU ECR white paper event, hosted at the 3rd ECFA Workshop on $e^+ e^-$ Higgs, Electroweak, Top factories~\cite{event_3rd_ecfa}.

\subsection{Career prospects and ECR leadership } %{\color{red}Frozen, contact Armin before changes!}}
\label{subsec:intro-careers}

Many discussions about the well-being of ECRs are centred around career-related issues, particularly the difficulties in pursuing long-term careers in academia. An extensive survey conducted in 2022 reaching 759 particle physics ECRs~\cite{Allen:2024lyp} revealed multiple pressing concerns that ECRs have regarding their future, such as job insecurity due to the prevalence of short-term contracts, the scarcity of permanent academic positions, the necessity for frequent international relocations, inadequate career development support, challenges in achieving a healthy work-life balance, and limited recognition of transferable skills outside of academia. These concerns are reflected in the significant fraction of ECRs who transition out of academia, often not by choice but due to structural constraints.
   
For ECRs who stay in the field, progressing in their career usually requires taking multiple back-to-back short-term positions, usually in different cities or countries, with no ultimate guarantee of a permanent position. Constant relocation can cause challenges, especially in one's personal life. With no guarantee of a long-term position, many ECRs cite the lack of stability as a reason to consider leaving the field. The short-term nature of these contracts often means significant pressure to finish a project in the time available, discouraging ECRs from pursuing their own research topics and interests in addition to the main projects they are required to work on by their contracts.  

As a result, many ECRs find themselves at a crossroads, forced to weigh their passion for research against the reality of an unstable career path. If these systemic issues persist, particle physics risks losing a substantial number of skilled and motivated researchers, further exacerbating concerns about the long-term sustainability of the field.

In Section~\ref{theme:careers_wellbeing}, the \emph{Career Prospects and ECR Leadership} Working Group (\Gls{CPL WG})  covers common challenges and offers some solutions which can enhance our field's appeal and support ECRs in their career paths. Moreover, recommendations for enhancing the work of ECRs in the field in terms of community building, leadership and recognition are addressed in Section~\ref{theme:community_leadership}.

\subsection{Diversity, inclusion and mental health } %{\color{red}Frozen, contact Armin before changes!}}

The well-being of researchers affects the whole community and is an important factor ECRs may consider when deciding on their future careers in or out of the field. Keeping ECRs engaged in particle physics requires that they feel welcome and secure in their work environment at all levels, have adequate support and supervision, and are heard in their community. 

One key aspect of feeling welcomed is having fair and equal job opportunities without discrimination. Even though the early stages of an academic career path have traditionally included many years of international mobility, valuing diverse ways of achieving mobility and networking could ensure better compatibility with family and other personal aspects of life, making the field more attractive in the long term.

In the case of researchers in the early stage of their careers, the role of their supervisory team is crucial for developing their research and identity in the field. Consistent and understanding communication between the supervisor and ECR can help create a supervisory relationship which benefits both the researcher and their research. ECRs can also act in supervisory roles, so it is essential to train them in healthy supervision. 

In addition, support beyond academic supervision is an important aspect of ECR well-being and productivity. For instance, dedicated HR support or a group secretary can help with matters like salary, working conditions, time off, organisation of teaching responsibilities, and workplace issues. Access to mental health education, resources and support is vital to a healthy research environment.

A healthy research career includes evaluating the importance of both personal life and research, as well as understanding how these two aspects affect each other. Without a balance between the two, a researcher may feel unsupported either by their research centre or by the community. This, in turn, may lead to negative effects on productivity and/or mental health and may influence the individual's career choice. More features of a healthy career from the point-of-view of ECRs, as well as aspects that are counterproductive to the well-being and success of ECRs, are discussed in Section~\ref{theme:careers_wellbeing} by the \emph{Diversity, Inclusion and Mental Health} Working Group (\Gls{DIMH WG}), along with recommendations to mitigate potential problems. 
\subsection{Communicating the importance of particle physics}
Communicating results and challenges of particle physics research keeps the field dynamic and engaging to the general public. It inspires new generations of scientists, ensures continued support and funding, helps build interdisciplinary collaborations, and gives back to society. Particle physics research institutions have recognised the importance of effective science communication and have put a lot of effort into improving their communication. 

Rising levels of disinformation about particle physics and other fields of science and the decreasing public and even internal support for large-scale future projects, however, remain a cause for concern. 
% These issues can be addressed by reassessing how outreach is conducted, what key resources are committed, and ensuring that the general public and other scientific communities and stakeholders understand particle physics research's societal impact. 
In this context, it is particularly important to assess the storytelling of particle physics communication. Since today's ECRs are the future leaders of any particle physics project to be realised, it is crucial to involve ECRs from the ground up and take their voices into account in how future projects are promoted to the public. 

Additionally, ECRs face specific challenges in balancing scientific work and outreach efforts due to a lack of formal training and recognition for their contributions to public engagement. 

The \emph{Communicating the importance of particle physics} Working Group (\Gls{Comm WG}) sought to identify and address these barriers and challenges ECRs face regarding science communication. 
The aim of the survey questions was to understand the level of involvement of ECRs in outreach activities and the level of training they are receiving. 

Based on these results, as well as reviewing personal experiences with outreach in the working group meetings, recommendations for and good practices in science communication are summarised in Section~\ref{subsec:comm_recs}. To improve the quality and effectiveness of scientific communication and enhance the visibility and impact of particle physics in diverse contexts, the community should consider these aspects. 
\subsection{Future colliders } %{\color{red}Frozen, contact Armin before changes}}

CERN is a global leader in particle physics, driving fundamental research and technological advancements and maintaining Europe's position at the forefront of particle physics. Securing its future with a next-generation flagship collider is essential not only for continued scientific discovery but also for retaining talent and innovation within Europe and reaffirming Europe's commitment to scientific excellence. The absence of such a vision risks losing both the innovation that has defined Europe's particle physics success and the momentum that this has created.

In this context, it is essential to consider how we continue scientific exploration beyond the High Luminosity (HL)-LHC era. For ECRs, this continuity is crucial to sustaining long-term growth and engagement in fundamental research, technological innovation, and particle physics community-building.
The uncertainty surrounding the approval of a future collider at CERN makes the prospects of particle physics in Europe beyond the HL-LHC less clear. This reduces the motivation of talented ECRs to commit to a career in particle physics, particularly for those working on future colliders directly. 
This was one of the main conclusions from discussions at the \textit{Future Colliders for Early Career Researchers} workshop in September 2023 \cite{event_future_colliders_ecrs}. The attendance of 200 people at this workshop showcased the great interest of ECRs in future colliders, many of whom are already involved in future collider studies, driving the development of computing facilities, accelerator R\&D, reconstruction algorithms, theoretical calculations, and much more. The active involvement of the ECR community in this sector demonstrates that the vision of ECRs on future colliders is crucial to the ESPPU and should firmly guide the discussions and decisions taken.

The main goal of the \emph{Future Colliders} Working Group (\Gls{FC WG}) was to understand how ECRs perceive different collider options. In this context, the WG aimed to identify which criteria ECRs believe should be prioritised when evaluating proposals for future colliders. Essential factors include not only the technological aspects of the collider facilities but also the broader goals of the projects in both physics and social impact. Ultimately, the next flagship project will only be successful if it aligns with the needs and aspirations of the upcoming generation of physicists who will execute today's vision. 

The recommendations laid out in Section~\ref{theme:future_particle_physics} are based on results from the survey as well as discussions in the regular meetings of the WG. The intention is not only to ask which future collider ECRs prefer but rather to focus on the points common to as many people as possible.
\subsection{Future experiments beyond colliders } %{\color{red}Frozen, contact Armin before changes}}
\label{subsec:WG-BC}

Experiments that are not situated at the interaction point of a collider\footnote{Beyond-collider in this white paper also refers to smaller collider-based experiments such as, for example, FASER and SND@LHC} are an integral part of the particle physics landscape, offering complementary approaches and sensitivities to searches at energy- and intensity-frontier collider experiments. Such projects, denoted here as \emph{beyond-collider} experiments, vary in scope and scale from small ``table-top" experiments with $\mathcal{O}(30)$ collaborators to medium-sized collaborations of $\mathcal{O}(200)$ members, such as Muon $g-2$, right up to large-scale infrastructure projects with $\mathcal{O}(1,000)$ collaborators worldwide, such as the Deep Underground Neutrino Experiment (DUNE). Although they vary greatly in size, this sector is primarily dominated by collaborations between $\mathcal{O}(50)$ and $\mathcal{O}(300)$ collaborators. Because of their smaller sizes compared to collider experiments, these experiments offer a dynamic environment that many ECRs value, as they enable a diverse research experience and the acquisition of varied skills. Furthermore, through the generally shorter timescales of these projects, ECRs can see through large parts of the experiment's life-cycle during their PhD and subsequent post-doctoral positions, enabling involvement from construction or even conception to physics results.

On the other hand, the same traits that bring these opportunities also pose challenges within the community. For example, shorter timescales also mean career paths that are less well defined, and a diverse community of smaller experiments makes it harder to form a network, making knowledge transfer between experiments less straightforward and complicating the process of finding career opportunities in other experiments.

The \emph{Beyond Colliders} Working Group (\Gls{BC WG}) identifies several challenges specific to the beyond-collider community and recommends tailored solutions to these problems in Section~\ref{theme:future_particle_physics}.

\subsection{Interplay of particle physics with neighbouring fields } %{\color{red}Frozen, contact Armin before changes}}

The interplay between a research field and its adjacent fields is integral to any research area. This is also true for particle physics and its relationship to fields such as astroparticle and nuclear physics, as well as accelerator physics, computer science, medical physics and many others.
Particle physics interacts extensively with each of these fields, and interactions with neighbouring disciplines often influence a particle physicist's career path. This interplay can be classified into the following categories:
\begin{itemize}
    \item \textbf{Knowledge exchange:} The exchange of ideas, technologies and results with neighbouring fields, facilitated by academic papers, conferences and direct communication.
    \item \textbf{Direct cross-disciplinary collaboration:} Direct collaboration with researchers from neighbouring fields with aligned scientific goals on topics requiring both areas of expertise.
    \item \textbf{Exchange of personnel:} Expertise-exchange through changing career paths between fields, temporarily or permanently, either to or from particle physics.
\end{itemize}

Many ECRs are involved in activities that fall into at least one of these categories. Knowledge exchange and cross-disciplinary collaboration between adjacent research fields can build efficiencies, preventing different groups from making the same effort twice. A contemporary example of this kind of collaboration is the Joint \Gls{ECFA}-\Gls{NuPECC}-\Gls{APPEC} Activities (\Gls{JENAA}), where the particle, nuclear and astroparticle physics communities work together on common challenges and scientific goals.

The exchange of people and expertise across disciplinary boundaries is an extension of direct collaboration. In this case, the experts who swap scientific fields bring their unique perspectives to their new context. 

These themes of communication, collaboration, and exchange of personnel also apply to the relationship between academia and industry. This is an integral part of ECR career paths since, as discussed in Section~\ref{subsec:intro-careers}, most ECRs will leave the academic path due to the deficit of senior opportunities. Closer communication and collaboration with industry partners could facilitate the transition from academia to industry and mean that greater value is placed on experience within industry. This may, therefore, facilitate similar transfers in the other direction: from industry into the research domain.

This WG had common interests and highly overlapping discussions with the working groups on \emph{Career prospects and ECR leadership}, \emph{Future experiments beyond colliders}, and \emph{Communicating the importance of particle physics}. It thus contributed to recommendations throughout all themes presented in Section~\ref{sec:recommendations}.

\newpage
\section{Early career researcher survey on ESPPU } %{\color{red}\textbf{Frozen, contact Harriet if you want to change this section}}}
\label{sec:survey}

Following the formation of topical WGs, discussions took place online to decide the relevant issues for each WG. The WGs met on the 14th of November for the \textit{ECR Workshop on the European Particle Physics Strategy} \cite{event_eppsu_ecr_cern}. The community discussed various ideas for the content of the white paper and identified overlaps and synergies between the WGs. These discussions indicated that more detailed input was needed from the ECR community in order to give recommendations on behalf of not only the editors and authors of this white paper but the entire European particle physics ECR community. For this reason, a survey was created with input from each WG to collect quantitative input on the stance of ECRs on the topics outlined in Section~\ref{sec:topics}. The full set of survey plots is available in Ref.~\cite{zenodo_plots}. The ones relevant to statements and recommendations in this document are listed in appendix~\ref{sec:survey-plots} with some of particular importance for the corresponding discussions distributed across the thematic recommendations.
\subsection{Methodology} %{\color{red}\textbf{Frozen, contact Harriet if you want to change this section}}
\label{subsec:Methodology}

The survey aimed to gather wider community input among ECRs, using the aggregate of anonymous answers to identify common opinions and areas of concern shared by the respondents. A considerable effort was made to engage European ECRs from as many different sub-fields as possible, recruiting participants across collider and beyond-collider experiments, as well as theory and related fields. %In this way, the survey findings illuminate the collective voice of ECRs across the field. 
The results from this survey were used to substantiate  thematic recommendations in Section~\ref{sec:recommendations} to address the identified challenges.

An introductory text preceded the survey questions, informing participants of its purpose and goals: 

\begin{quote}
   \textit{ %\textbf{Introduction and motivation:}
This survey aims to shape the input of Early Career Researchers (ECRs) to the Update of the European Strategy for Particle Physics. It was designed by a broad group of ECRs that gathered following open calls for participation. [...] Our objective is to understand how young researchers perceive the future of particle physics, including both its scientific programs and social aspects, and to explore potential challenges that may arise in this context and how they might be addressed. Participation in any question or section is optional, but your feedback is invaluable in helping to develop a comprehensive statement that reflects the perspectives of the broader ECR community.}
\end{quote}
The complete survey consisted of six topical sections: A set of general questions to record the demography and background of respondents, described in Section~\ref{subsec:survey_bio}, followed by one set of questions each from the WGs on \emph{Future colliders}; \emph{Future experiments beyond colliders}; \emph{Career prospects and ECR leadership}; \emph{Communicating the importance of particle physics}, and \emph{Diversity, inclusion and mental health} --- in this order. 
%The questions from the \emph{Interplay of particle physics with neighbouring fields} WG were inserted into the section best fitting to the individual question.
Section~\ref{subsec:survey_topics} describes the question topics and the motivation behind them in detail. In order to maximise the number of responses, the survey was kept short in length: each topical WG question section was divided into a set of a few main questions and a supplementary section. The main questions were shown to every participant and focussed explicitly on topics relating to the future of the field, both in terms of future particle physics projects and the societal aspects of the field, as this poses the main topic this document aims to address. Supplementary questions from each WG were only shown to participants who actively ticked an ``opt-in" checkbox. The exact number of questions answered by each participant and, thus also, the time spent to finish the survey varied between participants. The average time spent on the survey is estimated to be between 15 to 25 minutes. 

Various types of questions were employed throughout the survey: single-choice questions with simple yes/no answers, multiple-choice questions allowing for only a single answer, multiple-choice questions allowing for multiple answers,  questions requesting answers in agree/disagree matrices, and priority ranking by point allocation. Multiple-choice questions typically provided an ``Other" option as a free text field for participants to share views deviating from the pre-defined answers. 
None of the questions were mandatory, and a default ``No answer" option was always pre-selected, allowing participants to skip any question they did not wish to answer. In the survey analysis, ``No answer" data points are not considered. For this reason, and because the questions were split into the main body and supplementary questions, the total number of answers varies across the questions. 

The survey was hosted on CERN's instance of Limesurvey~\cite{limesurvey}. 
All collected answers were anonymous and the acquired dataset was only analysed in aggregate form, ensuring that individual responses could not be identified and protecting the privacy of participants. The survey data was treated in a GDPR-compliant manner, ensuring that only a select set of survey analysers from each WG, who actively produced analysis results, had access to the raw data. The raw data will be permanently deleted after the ESPPU Open Symposium in June 2025. Participants were informed of the data protection strategy by the disclaimer reported in \Cref{app:data-prot}, which was shown to them before starting the survey.

%Overall, 936 complete responses were collected, with 804 of the respondents fulfilling the ECR definition set out for the purpose of this white paper effort\footnote{non-permanent researchers or researchers that finished their PhD less than 10 years ago, and are working or studying in a European country or ECFA member state}, thus comprising the main dataset analysed and presented in Section~\ref{sec:recommendations}. Incomplete survey responses, where the participants did not submit the survey, are not considered. 
\subsection{Topics and questions } %{\color{red}\textbf{Frozen, contact Harriet if you want to change this section}}}
\label{subsec:survey_topics} 

\subsubsection{Career prospects and ECR leadership}

Thanks to the extensive survey on career prospects and diversity conducted in 2022 by the \Gls{ECFA ECR Panel}~\cite{Allen:2024lyp}, many issues raised in discussions on this topic have already been addressed. The members of the \Gls{CPL WG}) concluded that asking the same questions just two years later would be redundant, as a significant shift in ECRs' opinions is not expected over such a short period. Instead, efforts focused on identifying complementary matters, which were discussed in the \Gls{CPL WG} meetings and not covered in the previous survey.

One such aspect was the awareness of job availability in academia among the ECR community during the earliest stages of their careers. Discussions within the WG suggested that this awareness is limited, implying that career decisions may have been different if ECRs had known how many researchers leave academia shortly after obtaining their PhD. For this reason, the \Gls{CPL WG} decided to include questions both on the level of awareness and, for those who were not fully aware, whether this knowledge would have altered their career choices.
    
The question of whether to leave academia is very significant for many ECRs. It was directly asked to the participants in the 2022 \Gls{ECFA ECR Panel} survey, revealing that while most respondents wished to remain in HEP\footnote{The previous survey focused on ECRs in HEP, while this white paper is concerned with the field of particle physics more generally.}, only \SI{12}{\percent} considered their chances to be ``pretty good''. Given the importance of this issue at the early stages of an academic career and the interesting findings from that survey, it was decided to make an exception and revisit this question.
    
ECR leadership was not covered in the 2022 survey and is considered extremely important for this round of the ESPPU. Since the extent of ECR involvement in decision-making processes was unclear, the \Gls{CPL WG} decided to assess ECRs' perceived level of influence at three levels: their research group, their collaboration, and the ESPPU process. A similar question was asked in the 2022 survey, however, it concerned only recognition and visibility of one's work in a group or collaboration, rather than the extent to which one's opinions are heard. Also, the 2022 question did not study the group and collaboration separately, while the survey for this white paper did.

\Gls{CPL WG} members also decided to ask about the extent to which ECRs consider the ESPPU outcome when deciding whether to pursue a long-term career in academia. If these decisions significantly influence their decision to remain in academia, this would indicate that directly engaging ECRs in the ESPPU process is vital for the future of the field.

\subsubsection{Diversity, inclusion and mental health} 

Similarly to the \Gls{CPL WG}, the \Gls{DIMH WG} chose to address topics not covered in the previous study by the 2022 \Gls{ECFA ECR Panel}~\cite{Allen:2024lyp}. The choice of questions in this survey results from discussions within the \Gls{DIMH WG} and with other WGs. Four major topics were identified:
\begin{itemize}
    \item \textbf{Supervision from ``official" supervisors and mentors:} The 2022 survey identified `More guidelines and accountability on supervisors to fulfil their role well' as a factor that could improve their personal situation. Good or bad supervision can determine the success of research careers and impact the mental health of ECRs. To expand on the findings of the previous survey, one question aimed to assess the level of satisfaction of ECRs towards senior supervision they have experienced, while another captured the need for official training for supervisors.  
    \item \textbf{Inclusivity and diversity in the workplace:} Equal opportunities should be given to people from all backgrounds, including to those who belong to different marginalised groups. Furthermore, science progresses only through the effort of a diverse community. Therefore, it was extremely important to capture ECRs' perceptions of the inclusivity and diversity of their research environment while assessing how often ECRs have experienced discrimination or harassment.  
    \item \textbf{Mental health and support:} Poor mental health in research is a concern, as many challenges come with a research career, and treatment is often unaffordable on an ECR salary. For these reasons, this survey asked whether ECRs have struggled with their mental health and whether support structures are in place in their home institutions and/or laboratories.
    \item \textbf{Code of conduct:} The absence of clear behavioural guidelines or the lack of publicity around existing guidelines means that ECRs are often left without guidance on their conduct or knowledge of where to seek help if they feel that they have experienced or witnessed unacceptable behaviour. Therefore, the \Gls{DIMH WG} asked whether ECRs' home institutions have written their own codes of conduct and, if they have, whether they are adequately publicised and accessible.
\end{itemize}

\subsubsection{Communicating the importance of particle physics}
\label{sub:survey_topics_comm}
The \Gls{Comm WG} focused on identifying good practices for effective science communication about particle physics and its strategic goals, as implemented by different institutions. It also reviewed the key factors that contribute to the success of these practices. The \Gls{Comm WG} aimed to take into account all aspects of communication that ECRs face in their career path. For this reason, the thematic recommendations are given separately for three different target groups: the general public, representatives of industry, and fellow scientists, as further defined in Section~\ref{subsec:comm_recs}. Ensuring that ECRs can successfully communicate the importance of particle physics to each of these audiences is crucial for the future of the field.  

In the ECR survey, the \Gls{Comm WG} focussed its questions mainly on communication to the general public, as the key aspects of effective communication between scientists and in interactions with industry representatives are covered by the other WGs.

The \Gls{Comm WG} section of the survey included four main questions shown to every participant, with a supplementary set of more in-depth questions (see Section~\ref{subsec:Methodology} for details). The exact set of questions shown to each respondent depended on the specific answers of the respondent, with some questions only shown as follow-up questions to specific answers to a previous question. The total number of answers therefore varies between each question.

The survey questions addressed two major themes:
\begin{itemize}
    \item \textbf{Motivation and interest:} Respondents were asked whether they are motivated to engage in outreach activities relating to future projects and their main reasons for this (lack of) motivation. Testing the reasons behind a lack of interest in pursuing outreach activities is key to proposing strategies to overcome such limitations. Moreover, the supplementary questions served to understand how strong ECRs' present commitment to outreach is and what the limiting factors are. In particular, respondents were asked whether they feel sufficiently valued and rewarded for any outreach they do.
    \item \textbf{Training and preparedness:} Investigating and reviewing the experience and opinion of ECRs on communication training they may have received at their institutions is crucial to deriving concrete recommendations and guidelines for the kind of training programs most useful to ECRs. Respondents were therefore asked if they felt sufficiently prepared to engage with the public and, if not, what they felt was missing. Respondents were also asked to share what (additional) training on science communication they would like to receive. Additionally, optional questions were posed to investigate the type and amount of training ECRs have received during their career to define practices considered most beneficial.
\end{itemize}

Since this document focuses on the ESPPU and, therefore, the future of the field, the main questions reflected this by specifically asking about the motivation for outreach regarding future projects. The supplementary questions on interest, training and preparedness were phrased generically, without referring specifically to future projects.

Understanding how to best train and acknowledge ECRs for outreach work is critical for the success of any future project. This training is necessary to maintain public support and attract the next generation(s) of physicists, who will help to make the future project strategy plans a reality. Without ECRs and younger generations confident in promoting future projects to a general audience, this future cannot happen. Therefore, the general survey results on science communication training are relevant to the strategy discussion.

\subsubsection{Future colliders}
\label{sub:survey_topics_FC}
 One of the challenges of the current ESPPU process is establishing a clear-cut path for European particle physics research conducted at colliders for the upcoming decades. This task is impeded by the abundance of collider proposals and their multiple variants, which differ in achievable energies, the particles used in collisions, luminosities, and beam polarisation. Consequently, using labels based on collider names is often misleading; two people using the same acronym may refer to two distinguishable experimental facilities colliding different particles at energy scales that differ by orders of magnitude. Thus, we have focused on distinctive features characterising the favoured collider proposals (circular and linear $e^+ e^-$, hadron and muon colliders) to detach from the naming conventions. 
 % Furthermore, also ``any collider, as soon as possible", ``I do not know/do not have a strong opinion", and ``I do not support the idea of building a collider in Europe" were given as options.

The \Gls{FC WG} discussed possible evaluation criteria to compile a list of independent aspects to be rated by the respondents. The following factors were considered in the survey: ambitious baseline physics programme, well-defined long-term upgrade path, project timeline, specific location, technological innovativeness, sustainability, worldwide collaboration, public acceptance, and stability of funding for smaller projects. Respondents were asked to allocate 90 points over these nine aspects. This ``budget-allocation" format was chosen even though it is more complicated than typical ranking or multiple-choice formats since it allows respondents to assign points deliberately to emphasise or diminish given aspects according to personal preference. Furthermore, it forces participants to choose between priorities rather than assigning high importance to each one.

The survey included further questions to complete the picture. Firstly, the respondents were asked to indicate their level of agreement with the proposal to build a new collider. Further, they were also asked about their willingness to relocate to specific places to work on a future collider, namely CERN, China, Japan and the US --- the main alternative hosts for the next collider.

Since national inputs were requested to provide a prioritised list of preferred options, the \Gls{FC WG} decided to take a step in a similar direction. The question of choosing a specific collider was particularly challenging, given that not all the proposals can be realised at the same time. Anticipating a diverse range of competing options, it is important to gather informed opinions on which direction will best serve the field. In preparing the survey, a significant point of discussion was the value of an answer to this question, given that it will be derived from a potentially highly biased sample with limited control over the participants' demographics. Despite these concerns, it was decided to include the question and analyse the results thoroughly. The \Gls{FC WG} is confident that, by gathering supplementary information, it is possible to derive a meaningful statement while remaining aware of the limitations of an openly distributed survey.

One question asked whether respondents would be willing to support the outcome of the strategy process, even if their preferred collider option was not selected as a priority. This helped to assess how strongly the community is attached to their preferred options. Respondents were also asked to assess their knowledge of the future collider landscape.

In the supplementary section, several other questions were posed. Respondents were asked how far into the future the ESPPU should plan, aiming to gauge how broad the community perceives the scope of the process. Respondents were also asked to agree or disagree with statements regarding a sustainable path towards the next collider, the inclusion of more than two interaction points, and the potential for broader collaboration with non-member states. While these questions were relatively technical, they served to provide a more nuanced vision for the implementation of the next flagship project. Finally, recognising the significant cost of some proposals relative to the current CERN budget, the \Gls{FC WG} inquired about possible budget balancing scenarios between the next flagship project on the one and smaller projects on the other hand. The answer possibilities ranged from a significant budget reduction for smaller projects in favour of an increased budget for the next major flagship project, to the opposite scenario where the money available for the next major flagship project was significantly reduced in favour of smaller projects.

\subsubsection{Future experiments beyond colliders}

The \Gls{BC WG} focused on the concerns of those working in experiments outside the major collider projects, in particular the unique challenges that the diversity and, often times, smaller scale of these experiments pose and the possible actions that may help to alleviate these. Therefore, survey questions on future experiments beyond colliders were written to gauge the respondents' views on the state of this broad and diverse field of physics.
The first question assessed the main challenges that researchers in fields other than  particle physics with colliders face today.
The respondents were prompted to choose up to three answers, where the alternatives concerned e.g.\ recognition in the wider particle-physics landscape, inter-experimental networking and methods-exchange, career paths and funding situations, as well as the effect on beyond-collider projects of the ESPPU.
Correspondingly, the second question suggested a number of actions to mitigate the listed challenges, now prompting the respondent to grade the effect of each action on a five-degree scale from ``Not at all" to ``Significantly".
Alternatives here included the establishment of additional dedicated inter-disciplinary conferences, funding schemes, and a beyond-collider forum, as well as facilitated transitions between collider and beyond-collider experiments and more concrete recommendations for beyond-collider physics in the ESPPU. Additionally, there were options to specify other challenges and solutions in form of free-text answers.

The two first questions were thereby devoted to understanding which challenges the respondents view as most pressing in these fields, as well as what the most important actions would be to mitigate these challenges.
These questions were written to be considered in conjunction with the demographic information of the respondents, in particular their research field (beyond-collider experiment or not). 

An additional question was included in the supplementary section to reflect the respondents' prioritisation of activities CERN should engage with.
This was included as a direct parallel to questions 4.a and 4.c from the \emph{ECFA guidelines for inputs from national HEP communities to the European Strategy for Particle Physics}~\cite{eppsu_national_inpus_instructions}.
The respondents were asked to report their preferred engagement level of CERN in beyond-collider physics generally, fixed-target experiments, neutrino physics, direct dark-matter experiments, nuclear physics, astroparticle physics, medical physics, and novel accelerator R\&D beyond next-generation colliders.
\subsection{Survey circulation } %{\color{red}\textbf{Frozen, contact Harriet if you want to change this section}}}
\label{sec:survey-circulation}

The survey was circulated between mid-December 2024 and late January 2025, together with an invitation to the ECR Symposium in February 2025 at CERN to discuss the white paper draft. The following channels were used to advertise the survey:
\begin{itemize}
    \item All participants in the ECR white paper effort
    \item \Gls{ECFA ECR Panel} members and announcement channel (ecfa-ecr-announcements@cern.ch)
    \item National ECR communities through \Gls{ECFA ECR Panel} contacts (circulation confirmed in all but two \Gls{ECFA} countries)
    \item ECRs from the large LHC experiments, through LHC Early-Career Scientist Fora contacts
    \item Mailing lists of other particle physics experiments (AMBER, NA62, DUNE, ICARUS, MicroBooNE, SBND, HyperK, SuperK, T2K, Darkside-20k, LZ, and XLZD) % ICECUBE, P-ONE, and LEGEND)
    \item Future collider mailing lists (for FCC, ILC/linear colliders, muon colliders) and participants of the 2023 CERN event on future colliders \cite{ecfa_ecr_future_colliders}
    \item CERN Theory (TH) and Experimental Physics (EP) ECRs through TH and EP secretariats
    \item young-at-cern@cern.ch mailing list
\end{itemize}

The recipients were encouraged to share the survey with colleagues and communities that may not have received it through the mailing lists. In addition to the experiments listed above, efforts were made to distribute the survey to many other particle physics experiments, however, due to a lack of response, it is not clear whether the message reached the respective ECRs. The survey was also advertised in all presentations given on the white paper effort: specifically at the FCC Early-Career Forum \cite{Krzysztof_FCC_early_career}, the Linear Collider Vision ECR session \cite{Krzysztof_LCvision}, the neutrinos@CERN workshop \cite{Marvin_neutrinosAtCern}, and at the ECR discussion at the US Higgs Factory Planning \cite{Abdelhamid_US_Higgs_Factory_planning}.

The survey closed on January 27th, 12:00 CET, with 936 complete responses recorded.
\subsection{Demography of respondents } %{\color{red}\textbf{Frozen, contact Harriet if you want to change this section}}}
\label{subsec:survey_bio}

Of the 936 responses to the survey, 804 respondents are \Gls{ECFA} ECRs according to the definition of an \Gls{ECFA} ECR as ``\textit{non-permanent or less than 10 years after PhD, studying in or employed in \Gls{ECFA} member states or CERN}" as circulated with the survey. These 804 responses comprise our base data set, to which we refer from here on, unless stated otherwise.
The respondents originate from 59 countries, with the two most frequent answers being Italy and Germany. As shown in Figure~\ref{fig:country-of-work}, the respondents work at institutes across 26 European countries or \Gls{ECFA} member states and at CERN, the latter accounting for almost a quarter of total respondents.
When the countries of origin are grouped into geographical regions, as shown in  Figure~\ref{fig:region-of-origin-work}, it shows that just over a third of respondents originate from the Mediterranean, another third from Northern Europe and around \SI{17}{\percent} from Central and Eastern Europe, with the rest originating from North or South America, Africa or Oceania. The grouping of the geographical regions is described in Appendix~\ref{app:geo-regions-def}. Grouped by the same geographical regions, just over \SI{65}{\percent} of respondents work in Northern Europe, over \SI{20}{\percent} in the Mediterranean, just under \SI{10}{\percent} in Central and Eastern Europe and the remainder in Asia. 

Figure~\ref{fig:status} shows the career stage of survey respondents. Most respondents are PhD students (\SI{42}{\percent}), closely followed by PostDocs (\SI{32}{\percent}). Other respondents hold other non-permanent positions (\SI{9}{\percent}), are Masters students (\SI{7}{\percent}) or hold a permanent position (\SI{5}{\percent}). Most of the respondents are aged between 26 and 30 (\SI{47}{\percent}), followed by those aged 31 to 35 (\SI{21}{\percent}) and 21 to 25 (\SI{20}{\percent}). A minority were over 36, and very few were under 21 years old. For respondents that have completed their PhD, Figure~\ref{fig:years-since-PhD} shows the number of years since PhD completion. The majority (\SI{69.2}{\percent}) of respondents completed their PhD less than 5 years ago, while \SI{27.2}{\percent} completed their PhD between 5 and 10 years ago. A small percentage (\SI{3.6}{\percent}) of respondents completed their PhD more than 10 years ago and do not have a permanent position.  Where the total number of responses is lower than 804 this means that some respondents did not answer the question, and no questions were mandatory.

\begin{figure}[H]
     \centering
     \begin{subfigure}[b]{0.45\textwidth}
         \centering
    \includegraphics[width=\textwidth]{figures/biography/country-of-work.pdf}
    \vspace*{0.6cm}
         \caption{}
         \label{fig:country-of-work}
     \end{subfigure}
     \hfill
     \begin{subfigure}[b]{0.51\textwidth}
         \centering
         \includegraphics[width=\textwidth]{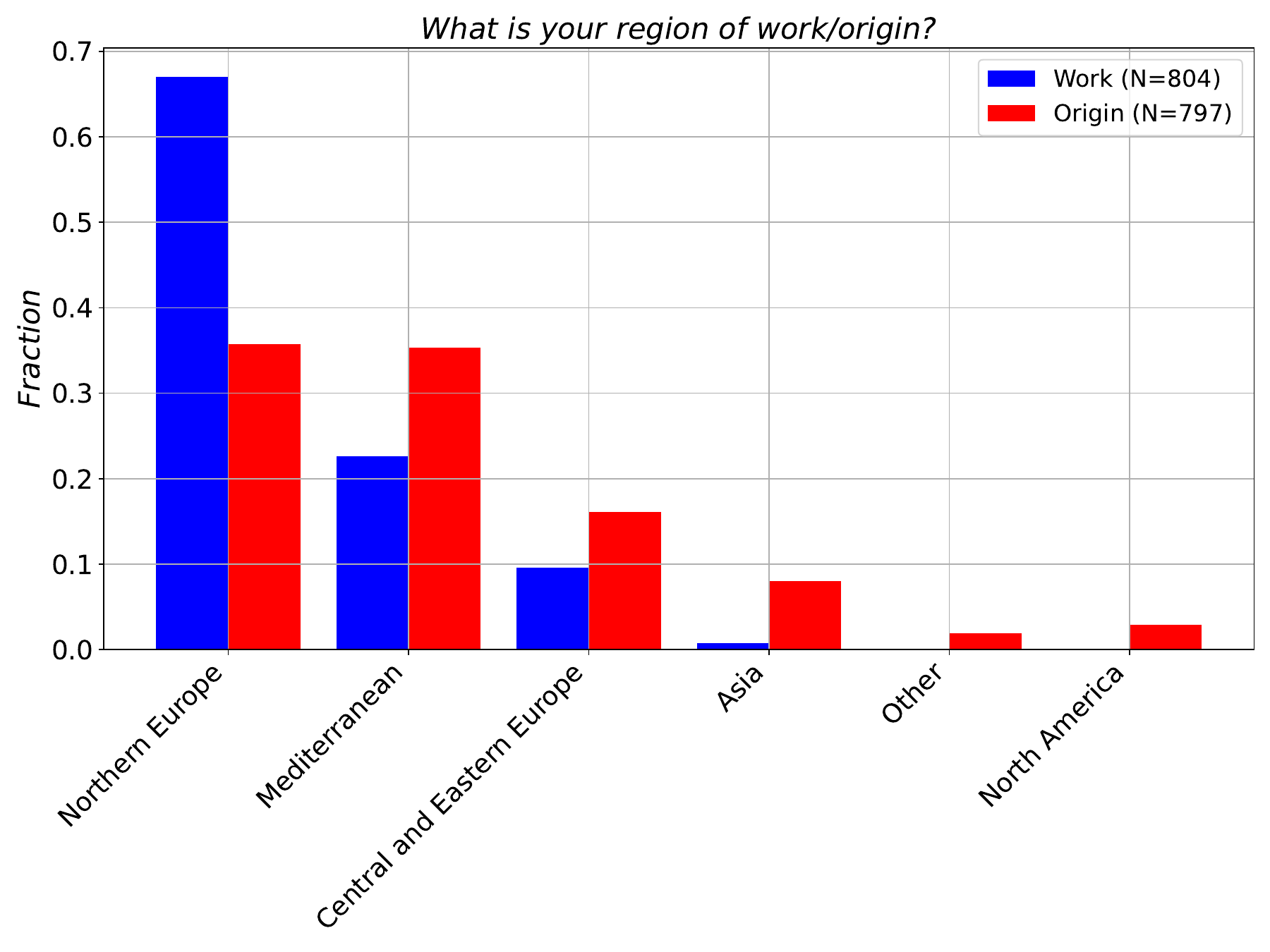}
         \vspace*{-0.8cm}
         \caption{}
         \label{fig:region-of-origin-work}
     \end{subfigure}
        \caption{(a) Work locations of survey respondents in European host countries or \Gls{ECFA} member states. CERN is listed as an independent option to either of its geographical host countries. (b) Geographic distribution of survey respondents by region of origin and work location. The region grouping is described in Appendix~\ref{app:geo-regions-def}.}
        \label{fig:bio-country-of-work-region-of-origin}
\end{figure}

\begin{figure}[H]
     \centering
     \begin{subfigure}[b]{0.53\textwidth}
         \centering
         \includegraphics[width=\textwidth]{figures/biography/status.pdf}
        \vspace*{-1.4cm}
         \caption{}
         \label{fig:status}
     \end{subfigure}
     \hfill
     \begin{subfigure}[b]{0.41\textwidth}
         \centering
         \includegraphics[width=\textwidth]{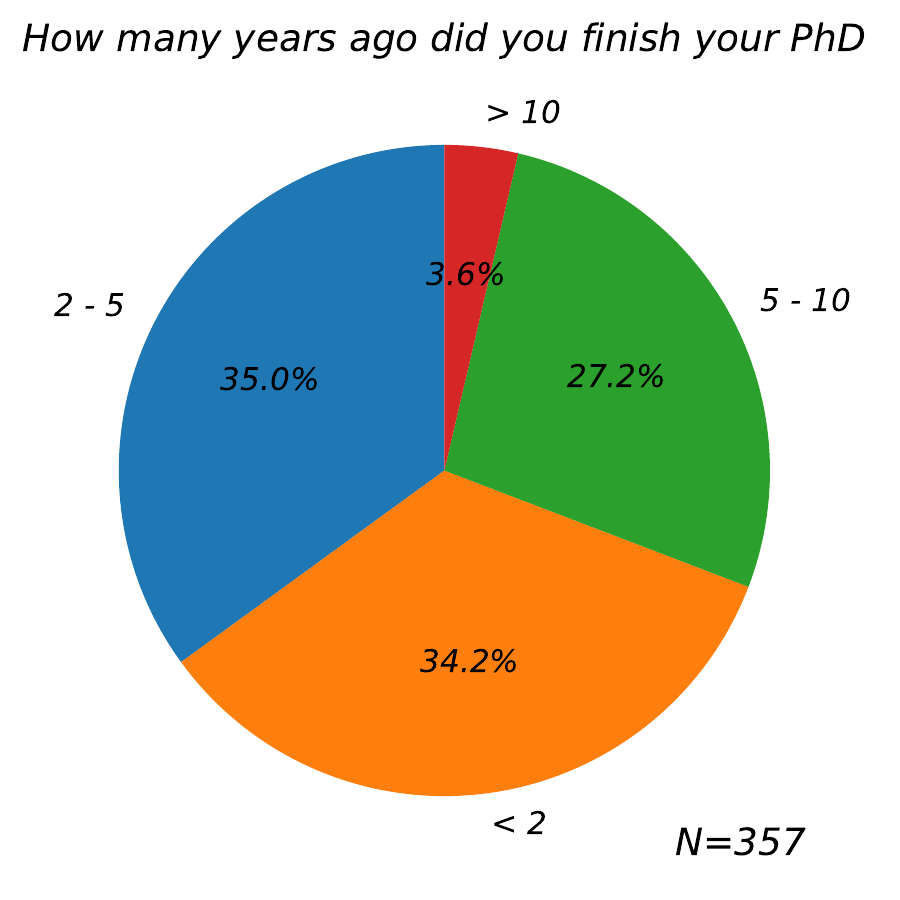}
         \caption{}
         \label{fig:years-since-PhD}
     \end{subfigure}
        \caption{Distributions of the (a) career stages of survey respondents and (b) for those that have completed their PhD, years since its completion.}
        \label{fig:bio-age-and-status}
\end{figure}

A significant majority of respondents identify as male (\SI{72}{\percent}), \SI{26}{\percent} as female and \SI{2}{\percent} as non-binary or gender-diverse. In addition, a small but not insignificant percentage (\SI{5}{\percent}) of respondents identify as having a disability or chronic condition. 

Figure~\ref{fig:bio-field-and-experiment} shows the field of work, experiments and future experiments that survey respondents work on. The majority of respondents (\SI{57}{\percent}) work in Experiment, followed by Phenomenology (\SI{13}{\percent}), Engineering (\SI{11}{\percent}), Theory (\SI{8}{\percent}), Accelerator (\SI{8}{\percent}), Software \& Computing (\SI{1}{\percent}) and Other (\SI{2}{\percent}), where multiple responses were allowed. For those that answered Other, the most common clarification was detector R\&D. For respondents that work in experimental particle physics, Figure~\ref{fig:experiment} shows the various experiments that they work on. The five most frequent answers are the ATLAS collaboration (\SI{22}{\percent}), followed by the CMS collaboration (\SI{15}{\percent}) and Future Collider (\SI{12}{\percent}), before the beyond-collider subfields of fixed target (\SI{10}{\percent}) and neutrino physics (\SI{8}{\percent}) experiments. For those that answered Other (\SI{10}{\percent}), most respondents work on accelerator/beamline physics relating to the LHC or CERN accelerator complex, cosmic/gamma-ray and telescope experiments or general detector R\&D.  The large majority of survey respondents in Experiment work on collider experiments.

Survey respondents are considered to be working in beyond-collider physics if they are involved with at least one beyond-collider sub- or adjacent field which seems to have been reached by the survey, namely: fixed target experiments, nuclear physics, neutrino physics, direct dark matter detection or strong-field QED experiments. For survey respondents involved with future colliders, Figure~\ref{fig:future-experiment} shows the projects that they work on. 

\begin{figure}[H]
    \centering
    \begin{subfigure}[tb]{0.48\textwidth}
        \centering
         \includegraphics[width=\textwidth]{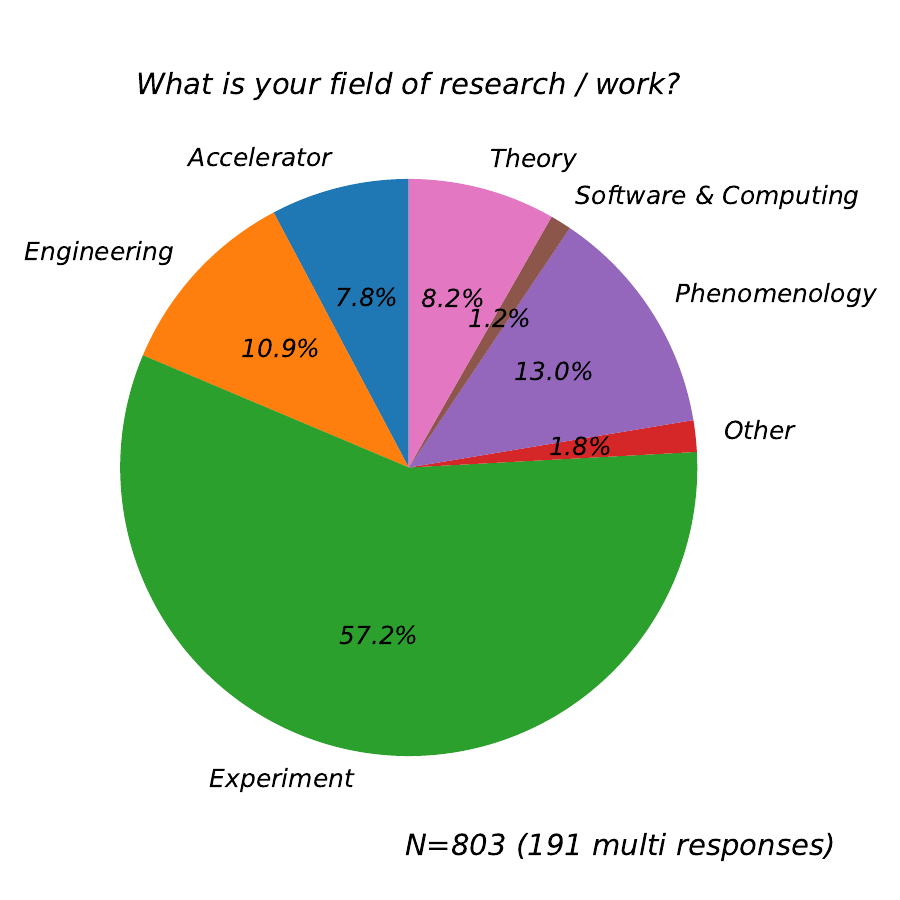}
         \caption{}
         \label{fig:field-of-work}
    \end{subfigure}
    %\hspace{4mm}
    \begin{subfigure}[tb]{0.48\textwidth}
        \centering
         \includegraphics[width=\textwidth]{figures/biography/experiment.pdf}
         \caption{}
         \label{fig:experiment}
    \end{subfigure}
    \hspace{3cm}
    \begin{subfigure}[tb]{0.65\textwidth}
        % \centering
         \includegraphics[width=\textwidth]{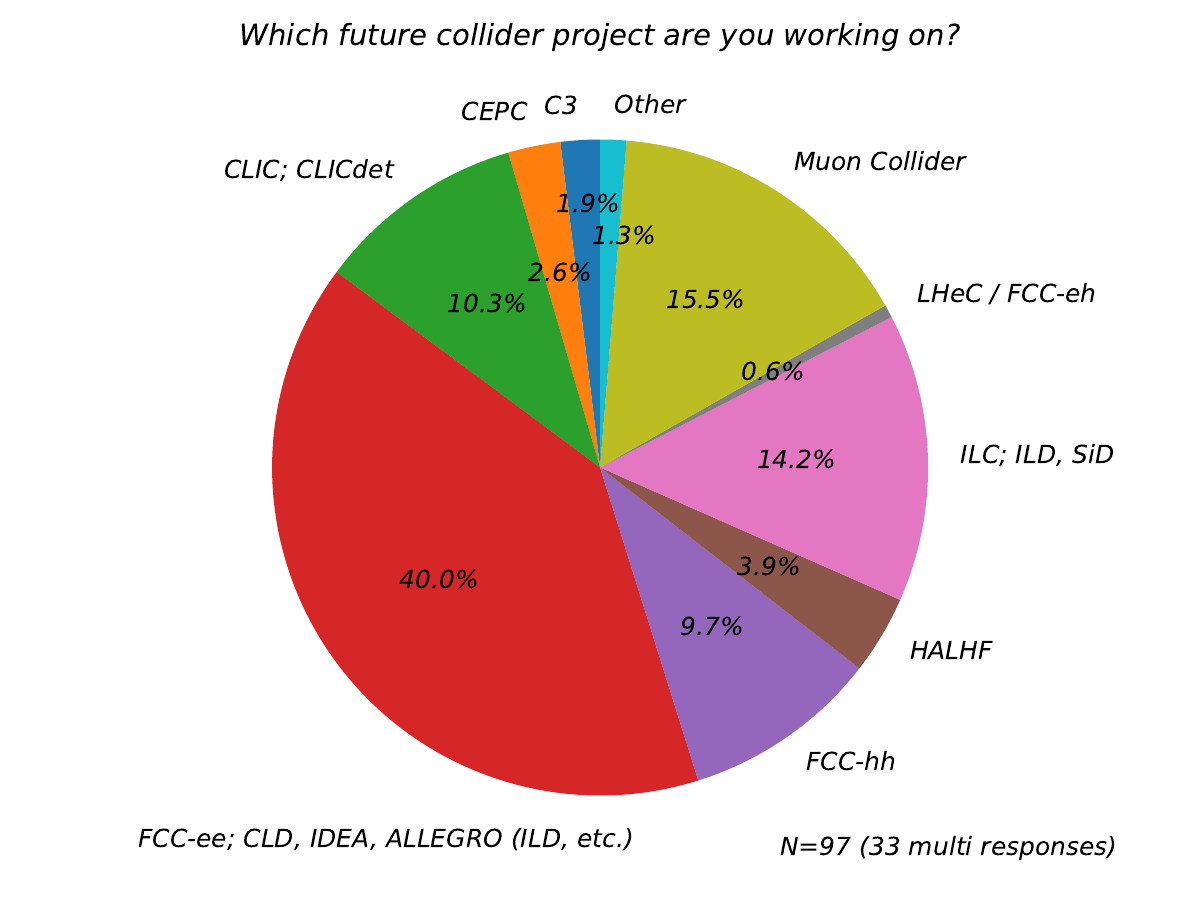}
         \caption{}
         \label{fig:future-experiment}
    \end{subfigure}
    \hspace{2.5cm}
    \caption{(a) Fields of work, (b) experiments and (c) future collider projects that survey respondents are active in.}
    \label{fig:bio-field-and-experiment}
\end{figure}

% \begin{figure}
%     \centering
%     \includegraphics[width=0.45\linewidth]{figures/biography/gender-identity.png}
%     \includegraphics[width=0.45\linewidth]{figures/biography/disability.png}
%     \includegraphics[width=0.45\linewidth]{figures/biography/age.png}
%     \caption{Gender identity, disability identitiy and age of survey respondents.}
%     \label{fig:bio-gender-disability-age}
% \end{figure}
\subsection{Analysis procedure and wording } %{\color{red}\textbf{Frozen, contact Harriet if you want to change this section}}} 
\label{sec:analysis_procedure}

This section explains the methodology employed for the statistical evaluation of the survey responses. 
The statistical significance of the difference between two distinct response counts, $a$ and $b$, assuming Poisson statistics, is calculated as:
\begin{equation*}
    S = \frac{| a - b |}{\sqrt{a + b}}.
\end{equation*}
%Statistical significances are then interpreted using standard thresholds, e.g. $S > 2$ for greater than $2\sigma$ significance and $S > 3.5$ for $3.5\sigma$ significance.
The typical threshold used is then $S > 2$, which is colloquially known as a $2\sigma$ separation.
For more complex comparisons, namely determining the difference between multiple answers at once or determining correlations in multiple dimensions, $\chi^2$-tests are used where applicable. Here, the criterion $S > 2$ corresponds to $p < 0.05$ with a standard p-value.

In order to limit selection bias, the majority of plots to be shown and the exact meaning of quoted `majorities', depending on the exact result, were defined prior to the unblinding of the survey data. The wording used in this document is as follows:
\begin{itemize}
    \item``relative majority'': more votes than any other option but less than $\SI{50}{\percent}$ and $S > 2$  to second highest option,
    \item ``majority'': more than $\SI{50}{\percent}$ of the votes and $S> 2$ to second highest option,
    \item ``significant majority'': more than $\SI{70}{\percent}$ of the votes and $S> 2$ to second highest option,
    \item ``the respondents'' or  ``the ECRs support'': making a recommendation on behalf of the ECRs: more than $\SI{90}{\percent}$ of the votes and $S> 2$ to second highest option,
    % \item ``all respondents'' or ``unanimously'': $\SI{100}{\percent}$ and at least 50 votes.
\end{itemize}

\subsection{Biases and limitations of the survey } %{\color{red}\textbf{Frozen, contact Harriet if you want to change this section}}}

While this survey provides valuable insights into the perspectives of ECRs across Europe, several biases and limitations must be acknowledged when interpreting the results:

\begin{itemize}
    \item \textbf{Selection bias:} The survey was primarily distributed through experiment mailing lists, ECR-specific mailing lists, where available, and shared among ECRs involved in this initiative and their colleagues. This may have led to over-representation of researchers already engaged in these communities and under-representation of researchers in hard-to-reach communities, such as those not belonging to large collaborations.

    \item \textbf{Response bias:} ECRs with strong opinions on career challenges and scientific policy may have been more likely to respond, potentially skewing the results toward more engaged individuals. The \Gls{ECFA} national input guidelines~\cite{ecfa_national_input_guidelines} can be seen to focus on the choice of the next collider, and individuals from the beyond-collider community may have been discouraged from participating in the process. In contrast, researchers from the collider community may feel this process is instrumental to their future in the field and therefore feel more compelled to participate.

    \item \textbf{Geographical imbalance:} Despite efforts to reach ECRs across Europe, responses may be unevenly distributed, favouring countries or institutions with higher survey dissemination. As mentioned in Section~\ref{sec:survey-circulation}, mailing lists for all but two \Gls{ECFA} countries were reached. %Additionally, the challenges faced by ECRs can vary significantly between countries.

    \item \textbf{Limited sample size:} With approximately 800 ECR responses, the survey provides meaningful insights but may not fully capture the diversity of ECR experiences and challenges across all institutions and career stages.

    \item \textbf{Survey design constraints:} The wording of questions and predefined answer choices may have influenced responses, limiting the ability to fully capture more nuanced perspectives. Considerable effort was put into the phrasing of  questions in order to limit these constraints as much as possible.
\end{itemize}

Despite these limitations, the survey offers a reliable and representative overview of common concerns and aspirations shared by ECRs in European particle physics. While it may not capture every challenge, its findings provide a solid foundation for informed recommendations in this white paper.

\newpage
\section{Thematic recommendations}\label{sec:recommendations}

The survey described in Section~\ref{sec:survey} and the discussions at the workshops and WG meetings were used to draft the thematic recommendations described in this section. These recommendations are not grouped by WGs, but rather by topic to showcase the connection between all the aspects investigated by the WGs.
\subsection{Careers and well-being of early career researchers } %{\color{red}Frozen, contact Emanuela/Elena before changes!}}
\label{theme:careers_wellbeing}

The most pressing issues raised by ECRs are ``getting a permanent job in academia" and ``lack of long-term planning and stability" which far outweigh concerns such as ``completing the PhD" or ``continuing high-quality research in an area of interest", as shown by the 2022 \Gls{ECFA ECR Panel}’s survey on career prospects and diversity~\cite{Allen:2024lyp}. 
Addressing these challenges, along with other career-related aspects highlighted in the survey, would significantly improve the appeal of careers in particle physics. %While many of these challenges may not be fully resolved in the short-term or by the ESPPU, steps must still be taken to enhance the field’s attractiveness. 

%\important{ECR career struggles}{The most pressing issues for ECRs' careers revolve around the general lack of long-term planning and stability, including the difficulty of getting a permanent position in academia.}

In the survey conducted for this white paper, respondents were explicitly asked whether they want to remain in academia. While a significant majority (\SI{89}{\percent}) want to stay in the field, only \SI{45}{\percent} consider their chances to be good, including \SI{15}{\percent} who think that their chances are ``pretty good'' (see Figure~\ref{fig:career_future-in-academia}).
These results are visibly more positive compared to the corresponding question in the 2022 survey. No significant demographic shifts explain this change, and a similar trend appears across age groups and work regions.  Variation across fields is minimal, except in engineering and small-sample cases (see Figure~\ref{fig:career_future-in-academia_correlations}). However, different field categorisations in 2022 prevent direct comparisons. Considering that there is only a two-year gap between the surveys, a simple change in ECR opinion seems unlikely;  external factors like the Covid-19 pandemic may have played a role. Additionally, the launch of the Next Generation EU recovery plan, which opened numerous positions and funded several PhD programs, could have significantly influenced ECR perspectives, highlighting the broader socioeconomic impact during this period~\cite{EC_NextGenEU}.

\important{Statement: ECR career goals}{The most pressing issues for ECRs' careers concern the general lack of long-term planning and stability, including the difficulty of getting a permanent position in academia. A significant majority of respondents to our survey wish to remain in the field, but only \SI{45}{\percent} of them consider their chances to be okay or pretty good. }

Beyond career uncertainties, ECRs face challenges that impact their productivity and well-being. While a supportive research environment is essential, it is not always guaranteed. Seeking guidance from mentors, peers, or institutional resources can help navigate frustration, isolation and unrealistic expectations. 
ECRs often face these issues without having been given tools to tackle them, and their mentors/supervisors are not always prepared or trained to support them through this journey. 

The researcher's life is famously flexible with irregular working hours and frequent remote work. This can be of great help in some cases, such as scheduling appointments and childcare, however it can also be a source of difficulty in setting work-life boundaries. Flexibility often leads to unpredictability due to unclear schedules, frequent travel and relocation.

International mobility is essential for knowledge exchange and cultural enrichment, but it also presents significant challenges. 
Relocating can be very stressful, particularly when moving with or far away from one's family or partner.
Belonging to marginalised groups can be particularly challenging in flexible and unclearly structured working environments, where supportive measures or support centres are often not in place.

%\important{ECR mobility}{In a recent survey~\cite{Allen:2024lyp}, the most significant measure to improve the ECR's personal situation included more job security and location-stability.}

Researchers' mental health is an additional concern. According to studies \cite{busch2024behind}, the severity of a student's anxiety and depression is associated with them contemplating departure from graduate school. 
The relationship between graduate student and mentor correlates with levels of anxiety and depression, and women, transgender and/or gender-nonconforming populations face increased risk. %Results from our survey here? Also could cite a recent LHC survey on scientists' mental health (https://cds.cern.ch/record/2922676/files/).

\important{Statement: ECR mental health and well-being}{The mental health of ECRs is a topic of concern, as rates of depression and anxiety among ECRs consistently exceed those of the general public. In a recent survey~\cite{Allen:2024lyp}, the most significant measures to improve ECRs' personal situation included more job security and location-stability.}

%\begin{recommendationbox}{}
%This subsection is full of important ideas which may get overlooked if they're not turned into one or more recommendations
%\end{recommendationbox}

It is essential to maintain control throughout these challenges. When seeking advice or exploring options, understanding institutional procedures ensures clarity and confidence in the process. Researchers should feel empowered to gather information, carefully consider their options, and make well-informed decisions that support their long-term goals. 
Sections~\ref{subsubsec:awareness}-\ref{subsubsec:monitoring} further explore the most important challenges and propose some solutions.

%\textcolor{green}{CareerWG: Emanuela, Jan}

\subsubsection{Awareness of job availability}
\label{subsubsec:awareness}
Recent trends indicate that more than \SI{50}{\percent} of ECRs will transition out of academia into other sectors~\cite{kwiek2024quantifying, Woolston:2022,Woolston2022Narrowing}. This phenomenon reflects the significant challenges faced by ECRs in maintaining long-term careers in academia, underscoring the importance of raising awareness about career trajectories among students at the bachelor's and master's levels. Providing accurate statistics and transparent information about post-PhD career outcomes is crucial to ensuring that prospective PhD candidates make informed decisions. By understanding the broader career landscape, students can enter doctoral programs with realistic expectations and a clearer understanding of the diverse career paths available.

According to the survey, approximately half of ECRs reported that, before starting their academic careers, they were unaware of the percentage of PhD graduates who eventually leave academia for other career paths. This shows the need for improvement, as informed decision-making should be actively supported. With that in mind, increasing awareness should not be seen as a barrier to attracting young researchers to the field. Among respondents who were not ``well aware'' of the current statistics, about \SI{80}{\percent} claimed they would still have pursued this career path anyway, had they known about the job availability.  The results are shown in Figure~\ref{fig:career_awareness}.

Universities and research institutions should take an active role in the effort of informing students by integrating career development resources into their programs. A more transparent dialogue about career prospects can also mitigate feelings of disillusionment that many ECRs face when transitioning out of academia.

The current employment trends in academia are unlikely to incentivise anyone to enter this field and pursue a long-term career. However, it is ECRs who provide the backbone of the workforce in particle physics, carrying out most hands-on tasks and delivering key results, while senior researchers primarily focus on supervision, management and coordination. This is another reason why increasing the attractiveness of our field to ECRs is crucial, and this includes students and others just beginning their careers in particle physics. This effort should not depend on their limited awareness regarding job availability or on their willingness to pursue a career driven by passion regardless of the personal or professional costs. Instead, the field should aim to remain appealing even when ECRs are fully informed about the challenges of the job market.

\begin{recommendationbox}{}The academic community, particularly senior researchers and institutions, should take active steps to increase awareness about employment perspectives in the field among individuals at the earliest stages of their careers.
\end{recommendationbox}

\subsubsection{Job security and mobility challenges}\label{subsubsec:job-security-mobility} %{\textcolor{green}{Emanuela, Jan}} \\
The unpredictability of academic careers often deters individuals from staying, particularly due to the scarcity of permanent roles, limited job opportunities, and the instability caused by frequent relocations.
A structured pathway towards permanent positions, accompanied by a clearer timeline, would provide much-needed encouragement for ECRs to remain in the field. This may also make them more willing to undertake multiple postdoctoral positions. 
% make recommendation about long-term contracts

The 2022 \Gls{ECFA ECR Panel}'s survey~\cite{Allen:2024lyp} has revealed that lack of long-term planning, stability, or job security are the biggest concerns to ECRs. Short-term postdoctoral contracts lay at the core of these issues, since they force a person to move abroad often with very little time for preparation in advance. In the case of two-year contracts, a large fraction of this period is spent not only on settling into a foreign country and handling administrative matters, but also on searching for another job before the current one ends, leaving limited time for performing actual work and research. This situation is even worse for those hired only for one year or less, as they are forced to start job hunting almost immediately upon arrival, leaving little to no time for meaningful research contributions. Such short contracts exacerbate stress and instability, making it extremely difficult to build a sustainable academic career. Therefore, the community should move towards  post-doctoral positions lasting three years or more, and furthermore, to move away from contracts shorter than two years.

\begin{recommendationbox}{}Postdoctoral positions should not be shorter than two years. Contracts for three years and longer should be actively promoted.
\end{recommendationbox}

In particle physics, international mobility is often considered a necessary component of career progression, with many postdoctoral positions requiring relocation to different countries or even continents. While international mobility broadens research experiences, it also presents significant challenges, especially for those with family or other personal commitments. Furthermore, this field already involves extensive international collaboration which does not always necessitate permanent relocation. Therefore, an inability to relocate should not disadvantage researchers or limit their career prospects. It is unclear whether, in these circumstances, conducting work on a single topic in multiple institutions increases one's scientific potential more than working on several different topics in a single institution. Currently, the former scenario appears to be more favourable for a successful career.

Creating alternative pathways that allow researchers to remain competitive while minimising relocation requirements can promote a more inclusive and equitable academic environment. By implementing these measures, academia can become more stable, appealing, and supportive, encouraging researchers to pursue long-term careers in the field while respecting individual choices. 

\begin{recommendationbox}{}The emphasis on long-term mobility for career advancement must be reconsidered in order to create a more inclusive field.
\end{recommendationbox}

Even for those who relocate, there is no guarantee that such a move will result in a permanent position, creating further uncertainty and stress. One possible improvement is creating a clear pathway towards permanent positions. Currently, the requirements needed to get a tenure track position in most cases are extremely vague and perceived as a combination of luck (``being in the right place at the right time") and connections. Job offers containing concrete demands, e.g. on the number of postdocs at foreign institutions necessary to obtain the position, would at least present unambiguous information about a timescale. A much more transparent solution would be to present an explicit path after which the job is guaranteed.
%\example{Example}{
For example, a permanent position at institution A could require a long-term relocation to institution B, then to institution C, etc., assuming some form of collaboration between those institutions. After fulfilling the demands, the position at institution A would have to be granted.%}

\begin{recommendationbox}{}
    A structured path to a permanent position should be stated in job offers and contracts whenever possible. 
\end{recommendationbox}

To support those who need to relocate, institutions should increase financial, psychological and logistical support for mobility. Financially, covering relocation costs and offering higher salaries that account for the expenses of moving to regions with different living standards can ease the burden. Family support, including assistance with childcare, partner employment, or schooling for children, is another critical aspect that would make international mobility less discouraging.

Moreover, relocating often entails navigating complex administrative processes, such as visa applications, fulfilling tax obligations, health insurance registration, and housing arrangements. These issues can be time-consuming, stressful, and expensive, particularly when lacking organised support. Institutions and funding bodies should ease these processes by providing dedicated administrative assistance with an adequate expertise to researchers~\cite{EdinburghRelocation}. Centralised support offices specialised in handling international relocations could significantly reduce the administrative burden, allowing researchers to focus on their work. 

\example{Example}{A good example of an office put in place to ease the moving to a new country is the International Office at DESY~\cite{DesyIntOffice}.}

\begin{recommendationbox}{}
 An international body should be established to assist with administrative relocation matters, potentially under CERN's mandate, given its role as an international hub. All academic institutions should also employ experts to provide such assistance. Economic support for relocation must also be guaranteed in ECRs' contracts. 
\end{recommendationbox}

\subsubsection{Dedicated time for ECRs to pursue activities beyond main duties}\label{sec:time_beyond_duties}
The current funding model of early career contracts means, in most cases, that ECRs are employed to work full-time on a single project and have no flexibility to pursue scientific interests outside of their main focus. The standardisation of ``side project time" would give researchers the flexibility to finish projects left over from previous contracts or broaden their horizons and skill sets. Researchers would also have the opportunity to expand their network and experience different cultures and workflows from different projects. Allowing researchers to spend up to \SI{20}{\percent} of their time on a secondary project would foster greater scientific innovation, while also allowing researchers to increase their future employability. 

\example{Example}{Side project time is already standard in many companies in industry, most famously Google and Atlassian, as well as LinkedIn \cite{SideProjectTime}.}

While the transfer of knowledge and ways of working is usually unidirectional from academia to industry, this is an excellent example of an opportunity to learn from industry and to import best practices into academia. It should be noted that this is also already a possibility with some ECR contracts, for example as a CERN research fellow.

\begin{recommendationbox}{}Contracts for non-permanent positions should explicitly include a fraction of researchers' time which can be spent on projects outside their main focus, the topic of which is decided by the ECR themself.
\end{recommendationbox}

ECRs often find themselves immersed in the rigorous demands of research and academia. However, dedicating time to activities beyond their primary research duties is crucial for career development. Non-research-related activities such as organising events, coordinating journal clubs, even contributing to this white paper, are essential for a thriving academic community and to enhance networking opportunities. 
Engaging in these activities also allows ECRs to develop essential soft skills such as leadership and project management.

The time invested in these projects must be supported and recognised as valuable. 
Institutions' support to ECRs in pursuing activities beyond their main duties is fundamental to enriching the academic community and fostering the development of well-rounded researchers.

\begin{recommendationbox}{}The time investment in non research-related projects should be supported with flexible working hours and formally recognised, e.g. in performance evaluations and hiring decisions. Such projects may include public engagement, mentoring, \GlsBlack{DEI} initiatives, or institutional service.
\end{recommendationbox}

\subsubsection{Early career researchers in smaller collaborations}\label{subsubsec:careers:ECRs_in_small_collab}

Beyond-collider and smaller-scale projects offer particular opportunities for ECRs, however this career path also comes with a unique set of challenges. While the relatively short time-scale of smaller experiments allows ECRs to participate in different stages of an experiment, from planning to construction to commissioning to data analysis, it also means that ECRs cannot plan to spend their entire career working on only one project. This can also provide the advantage of being involved in a variety of different tasks early in one’s career, giving ECRs the opportunity to become well-rounded researchers with a diversity of experiences from the very beginning of their professional lives. However, the lack of continuity leads to challenges in career planning, and by the nature of small collaborations, without strong support for external networking, the network of ECRs working in these groups can be relatively limited. While LHC researchers have access to events such as the LHC Job Matching Event \cite{LHCJMev}, this kind of event for beyond-collider researchers currently does not exist. The ESPPU should highlight the importance of career support also for beyond-collider researchers. The creation of a beyond-collider physics forum discussed in \Cref{subsubsec:Forum_and_non_collider_community_building} would provide a structure within which to organise networking events and publicise lists of open job postings. This would be very beneficial to the career development of ECRs working in beyond-collider physics.

The importance of this topic to ECRs is demonstrated by the majority of respondents to the survey (\SI{69}{\percent}) who said that easier career transitions between beyond-collider experiments would benefit beyond-collider physics in Europe and the significant majority (\SI{76}{\percent}) who said the same about easier transitions between collider and beyond-collider experiments (see Figure \ref{fig:bc_career_paths}). It should be noted that neither of these figures changes significantly when considering survey respondents as a whole or respondents working on beyond-collider experiments.

\begin{recommendationbox}{}Career support must be provided for ECRs pursuing careers in both beyond-collider and collider physics. There should be greater clarity of potential career paths for ECRs working on small and short-timescale experiments. Smaller experiments produce excellent and well-rounded researchers but with small networks, meaning that greater support for networking is needed for researchers in these collaborations to help them navigate these career trajectories.
\end{recommendationbox}

\subsubsection{Career shifts across research fields and industry}\label{subsubsec:CareerShifts}
Particle physics is a highly collaborative discipline that thrives on exchanging ideas with other scientific fields and industries. 
These interactions often lead to transitions between different research fields or, more commonly, from academia to industry. 
In today's multidisciplinary research landscape, fostering synergy across fields is increasingly encouraged, and ECRs should be well prepared and supported for such transitions.  
The dynamics of these exchanges can vary significantly across Europe, influenced by differences in national funding models and policy frameworks.
Encouraging a European exchange of best practices and developing standardised pathways for career transitions could help ensure more equitable opportunities and smoother transitions for ECRs.

As discussed in \Cref{subsubsec:job-security-mobility}, the shift from academia to industry occurs for various reasons, sometimes as a deliberate choice but often driven by low salaries, job instability, limited academic career prospects, or personal challenges related to family responsibilities and the enforced requirement of frequent relocation. 
Regardless of the motivation, leaving academia is often accompanied by stigma, perceived as academic failure, a pursuit of financial gain at the expense of ideals, or even a betrayal of mentors and colleagues. 
While the success of research groups and the broader field is vital, the particle physics community must not lose sight of the foundational principles of academia: fostering critical thinking and preparing individuals to contribute meaningfully to society. 
Career transitions should be viewed as opportunities for personal growth and a step towards self-realisation, rather than shortcomings. 
Given that many ECRs will eventually transition to industry it is essential to acknowledge this openly and prepare researchers for this possibility.

\begin{recommendationbox}{}ECR training programs should integrate skills transferable to industry. Supervisors should be required to outline these skills explicitly in their project proposals.
\end{recommendationbox}

Transitions are not limited to moving from academia to industry. 
ECRs often also develop interests in secondary research fields, which may include topics having potential societal impact, synergistic with their main area of work. 
However, these interests are sometimes kept hidden, due to concerns that supervisors may view them as unrelated to their own scientific priorities. 
As discussed above, the interdisciplinary nature of the field makes such compartmentalisation increasingly impractical. For instance, physics today is deeply intertwined with computational techniques and engineering methods.
Conversely, many fields benefit from technologies and methodologies developed within particle and accelerator physics. 
Notable examples are the widespread use of particle detection devices in medical physics and the modelling of multivariate systems in climate research. 
Encouraging ECRs to engage in interdisciplinary research exchanges can foster innovative synergistic approaches to address long-standing challenges.

Mentors and supervisors have a responsibility to acknowledge the broader scientific interests of ECRs and equip them with a versatile skill set that supports both academic and non-academic careers. 
This includes cultivating soft skills, teamwork, time and resource management, and technical expertise. 
By embracing a holistic approach to training, the particle physics community can empower ECRs to thrive, regardless of their career trajectory.

\begin{recommendationbox}{}ECRs should feel able to explore secondary research interests without fear of pushback from their supervisors.
\end{recommendationbox}

Lastly, but no less importantly, transitions can also occur from industry to particle physics, as specialised experts from industry may be attracted to the cutting-edge technologies developed by European particle physics laboratories. Such cases are currently rare, as individuals in industry are often reluctant to give up financial stability and job security to ``simply'' pursue a greater cause. 
Recent advancements in particle physics demonstrate the field's growing dependence on cutting-edge technologies. 
While physicists ultimately make use of these technologies, their development necessitates attracting specialised expertise from industry.
%Modern particle physics cannot thrive without expertise in software, mechanics, electronics, and other domains.
However, this requires abandoning stigmatisation of people coming from industry -- both those educated in particle physics and those not previously involved in academic studies. Their additional experience should instead be seen as a strength which can be used to benefit research excellence in our field.

One effective way to draw talent from outside particle physics is by creating positions that do not require a PhD in particle physics. 
For example, CERN has successfully implemented programs offering applied physicist roles that prioritise strong expertise in areas such as computing or engineering, instead of requiring a formal academic background in particle physics.
These positions appeal to industry professionals and highlight how CERN's competitive salaries and benefits can bridge the gap between industry and research. 
Another improvement in the same direction would be an intensified collaboration between existing engineering and computing teams at laboratories involving temporary placements and training within particle physics and vice-versa.
While CERN provides an excellent model, similar initiatives must be adopted by other institutions and universities to attract specialised talent. 

Although research institutions may not match industry funding levels, other incentives can make such positions appealing.
For instance, offering additional paid leave, comprehensive insurance plans, or flexible work-from-home policies can significantly enhance job attractiveness. 
Policymakers need to consider these factors and work to optimise conditions for non-physicists. 
This approach enables particle physics to build the diverse and specialised workforce essential for its advancement.
Moreover, promoting diversity in career pathways can bring a wealth of perspectives and experiences to particle physics, enriching the field and driving innovation across disciplines.

\begin{recommendationbox}{}Universities and research institutions should recognise and leverage industry experience by creating positions that attract researchers from diverse career backgrounds, including from outside academia.
\end{recommendationbox}

\subsubsection{Mental health}

A recent study on the mental health of LHC scientists~\cite{LHC-mentalhealth-2025} found that \SI{56}{\percent} of respondents struggled with mental health issues on a daily or monthly basis. This is consistent with data collected by the survey for this white paper: \SI{57}{\percent} of respondents agreed that they had struggled with their mental health during their research career, compared to \SI{28}{\percent} that disagreed with the same statement, as shown in Figure~\ref{fig:mental_health}. While the ECRs targeted for the survey work in areas beyond the LHC, there is a large overlap in working environment between these two surveys. 

Furthermore, it was found that women are more likely to have struggled their mental health with respect to men: \SI{66}{\percent} of female respondents said they struggled with their mental health, while \SI{53}{\percent} of male respondents said the same, as shown in Figures~\ref{fig:mental_health_gender_disability}a and \ref{fig:mental_health_gender_disability}b. Additionally, 10 gender-diverse/non-binary people out of 13 said they struggled with their mental health. People with a disability or chronic condition are also more likely to have experienced mental health struggles, with \SI{79}{\percent} of respondents agreeing, as shown in Figures~\ref{fig:mental_health_gender_disability}c and \ref{fig:mental_health_gender_disability}d. In addition, \SI{80}{\percent} of respondents that experienced some level of discrimination (see Section~\ref{subsec:codeofconduct}) struggled with their mental health (Figure~\ref{fig:mental_health_discrimination}).
These findings show how mental health support would be particularly crucial, even more for people who belong to marginalised groups and who have suffered discrimination or harassment.

Despite the number of researchers who have struggled with their mental health, mental health care is often prohibitively expensive. ECRs tend to have relatively low salaries and mental health counselling is frequently costly and perceived as a luxury, often not being covered by health insurance. For these reasons, institutions, research centres and (inter)national laboratories interested in the quality of their research as well as of their researchers' lives should provide forms of mental health support or funds to be used by researchers struggling with poor mental health.

Finally, a study performed on the mental wellbeing of ECRs \cite{mentalhealth}, advocates as well for greater financial support and improved access to mental health care. Additionally, it was found that the greatest alleviating impact on ECRs mental wellbeing came from making progress and receiving emotional support. Feeling confident about their teaching abilities also had a large impact on the wellbeing of graduate students. The study encourages supervisors and mentors to focus on providing students with emotional support and constructive feedback, rather than harsh criticism. The study also advocates for adequate teacher training (see Section~\ref{subsubsec:supervision-training}).

\begin{recommendationbox}{}Institutions, research centres and (inter)national laboratories should allocate part of their funds to mental health services for the research community, with specific focus on ECRs for their more vulnerable position. Additionally, focus should be put on providing ECRs with emotional support and constructive feedback in supervision.
\end{recommendationbox}

\subsubsection{Supervision and training}
\label{subsubsec:supervision-training}

ECRs rely heavily on the support and mentoring of their supervisors and other senior researchers. 
The quality of these interactions significantly impacts academic performance, psychological wellbeing, and future career choices. 
Mutual respect between both parties should be the guiding principle, with supervisors clearly communicating their expectations while remaining open to feedback from ECRs, fostering a culture of transparency and trust.
Supervisors should actively respect an ECR's right to a healthy work-life balance, acknowledging this as an integral part of productivity and well-being.

The survey indicates that \SI{76}{\percent} of respondents are satisfied with their supervisor, while \SI{73}{\percent} feel the same about their mentor (Figure~\ref{fig:supervisor_mentor_satisfaction}). However, \SI{14}{\percent} of respondents are dissatisfied with their supervisor. This situation could be improved through supervision training provided by institutes. According to the survey, \SI{59}{\percent} of respondents state that their institute does not offer such training, while \SI{80}{\percent} believe it should be mandatory, as shown in Figure~\ref{fig:supervisor_training}.
 
\begin{recommendationbox}{}Institutes should offer a mandatory supervision course for staff in supervisory roles focused on aligning expectations with the supervisee and understanding their needs.
\end{recommendationbox}
 
Beyond formal supervision, mentorship also plays a crucial role. A mentor, such as a more experienced ECR or a permanent staff member, can offer guidance beyond research, supporting career development, work-life balance, and personal growth. Building mentor-mentee relationships through networking, collaboration, and shared research interests can provide invaluable support throughout an academic career.
Such mentoring can be particularly important for people belonging to marginalised groups, where role-models are often lacking.
\example{Example}{Survey respondents identified the following good examples of structured mentoring schemes: \begin{itemize}
    \item Mentoring@CERN programme~\cite{MentoringCERN2024},
    \item INFN gender mentoring programme~\cite{INFN_gender_mentoring},
    \item LHC ECRs mentoring programme~\cite{LHC_ECRS_mentoring},
    \item ATLAS Analysis Contact training~\cite{ATLAS_Contact_training},
    \item UMO Mental Health mentoring programme~\cite{UCL_MentalHealth_mentoring}, provided at University College London;
    \item University of Hamburg gender mentoring programme~\cite{Hamburg_Gender_mentoring}.
\end{itemize}}

While ECRs often contribute to predetermined research topics due to their early career stage or restrictive employment conditions, they should still be recognised as independent researchers with valuable ideas and perspectives.
Effective mentorship encourages the growth of individual talents 
and nurtures creativity, rather than imposing the supervisor’s perspective. 
Empowering ECRs to explore their own ideas promotes innovation and builds confidence in their abilities.

%For those lacking supervisory support, a second supervisor may provide fresh perspectives and collaboration, with guidance from a graduate coordinator or department head. If challenges persist, changing institutions, or roles, though difficult, maybe a necessary step towards long-term career and well-being goals.

\begin{recommendationbox}{}ECRs should have better access to mentorship programs which offer career guidance, help develop research skills and provide opportunities for personal growth, helping them build confidence and advance in their careers.
\end{recommendationbox}

\subsubsection{Workplace environment}
A thriving scientific community requires an environment that fosters critical thinking and academic inquiry. 
Senior researchers play a vital role in creating a space where ECRs feel comfortable asking questions, regardless of how naive they may seem. 
%The training of ECRs should aim to equip them with a robust foundation of theoretical knowledge, practical skills, and an ethical mindset essential for conducting valuable and innovative research.
Universities and research institutes must therefore commit to maintaining a safe and respectful workplace. Institutions must offer clear policies, accessible reporting mechanisms, and robust support systems for those affected by inappropriate behaviour. 

Large scientific collaborations have both the responsibility and the resources to uphold high ethical standards among their members. To safeguard ECRs, who are particularly vulnerable, internal regulations must be put in place. Measures could include assigning 
each ECR an external mentor from a different institution, requiring that supervisors provide clear project plans for new PhD students, and conducting mid-term evaluations of project progress. In severe 
cases, non-compliance should lead to the removal of groups from the collaboration.

%Figure~\ref{fig:harassment-self} shows that harassment and discriminatory behaviors related to disability, family status, gender, nationality, sexual orientation and wealth have been experienced in varying degrees, being gender, nationality and ethnicity the most common. Also concerning is that \SI{49}{\percent} of the respondents who had experienced themselves harassment or discrimination think they were not given adequate support by their main institution as seen in Figure~\ref{fig:support_harassment}.

\begin{recommendationbox}{}
Universities and research centres must ensure a safe and healthy workplace, where bullying, harassment and discrimination are treated with a zero-tolerance policy and any resultant actions are applied consistently to all staff members regardless of their seniority. Robust, accessible, effective and independent support systems must be put in place for cases which violate these policies, ensuring that complaints can be made anonymously. 
\end{recommendationbox}

\subsubsection{Diversity and Inclusion}

\emph{Diversity} in an environment refers to its varied composition in personal, physical, and social characteristics, such as ethnicity or gender.

The demographic information of the survey respondents, as shown in Section~\ref{subsec:survey_bio}, reflects the reduced diversity of the physics research fields: only \SI{26}{\percent} of respondents identify as women and \SI{2}{\percent} as non-binary or gender-diverse. This data is in line with the percentage of people not identifying as male in the field of physics more generally \cite{UK_physics_data,UKUS_PhysicsData,US_physics_data}. Furthermore, \SI{5}{\percent} of respondents identify as people with disabilities or chronic diseases, where this percentage in Europe for people aged between 25 and 45 ranges between 12--\SI{15}{\percent} \cite{disabilities_data}. 

In order to ensure a more diverse environment, strategies and procedures are to be put in place to integrate everyone and provide equal possibilities to all. Such strategies and procedure are referred to as \emph{inclusion} policies.

Among the survey questions, one was aimed at identifying the respondents' perception\footnote{Respondents were not obliged to reveal that they belong to one or more marginalised groups. Therefore, the perception of inclusivity and diversity may be different from the reality.} of inclusiveness and diversity of the research environment. 

Interestingly, respondents on average feel that their research environment is more diverse than it is inclusive, as shown in Figure~\ref{fig:diverse} and Figure~\ref{fig:inclusion} and summarised in Table~\ref{tab:frac_inclusiveDiverse}.%: while \SI{81}{\percent} of respondents feel that their environment is not totally inclusive towards people with disabilities, \SI{70}{\percent} of respondents feel that the environment is not totally diverse considering this group. For family status, the numbers are \SI{72}{\percent} and \SI{47}{\percent} respectively, for gender representation \SI{54}{\percent} and \SI{56}{\percent} respectively, for nationality/ethnicity the figures are \SI{54}{\percent} and \SI{44}{\percent}, for religion the numbers are \SI{65}{\percent} and \SI{34}{\percent}, for sexual orientation the figures are \SI{75}{\percent} and \SI{40}{\percent} respectively and finally for wealth the numbers are \SI{73}{\percent} and \SI{58}{\percent} respectively. 

\begin{table}[!h]
\centering
\begin{tabular}{|c|c c|}
\hline
                      & \multicolumn{2}{c|}{Respondents who feel their research environment is:} \\ 
                      & \textbf{Not fully inclusive} &  \textbf{Not fully diverse}\\ \hline
Disability            &  \SI{88}{\percent}                                                                 & \SI{70}{\percent}                                                               \\ \hline
Family status         & \SI{70}{\percent}                                                                 & \SI{45}{\percent}                                                               \\ \hline
Gender                & \SI{52}{\percent}                                                                 & \SI{45}{\percent}                                                               \\ \hline
Nationality/Ethnicity & \SI{54}{\percent}                                                                 & \SI{43}{\percent}                                                               \\ \hline
Religion              & \SI{65}{\percent}                                                                 & \SI{34}{\percent}                                                               \\ \hline
Sexual orientation    & \SI{67}{\percent}                                                                 & \SI{41}{\percent}                                                               \\ \hline
Wealth                & \SI{72}{\percent}                                                                 & \SI{55}{\percent}                                                               \\ \hline
\end{tabular}
\caption{Fraction of respondents feeling that inclusiveness (second column) and diversity (third column) of their research environment can be improved, towards different marginalised groups (first column). Family status refers to ``an individual's relationship with their family members, including dependents such as children, parents, or any other family responsibility" \cite{familystatus}, while wealth refers to an individual's abundance of valuable possessions or money. These percentages are obtained by considering respondents who felt that their research environment was ``not very" diverse/inclusive or ``somewhat" diverse/inclusive. The percentages found in the text are obtained in the same way.}
\label{tab:frac_inclusiveDiverse}
\end{table}

This evidence suggests how the research environment, though being perceived as somewhat diverse, still presents inequalities which make it not very inclusive. 

Another finding from the survey is that people belonging to a marginalised/under-represented group feel that their research environment is less inclusive and diverse than people who do not belong to such groups. For example, as shown in Figure~\ref{fig:inclusion-gender}, \SI{55}{\percent} of male respondents find the research environment to be very inclusive in terms of gender representation, while the same opinion is shared only by \SI{31}{\percent} of female respondents. Out of 13 gender-diverse/non-binary people, 6 believe the research environment to be very inclusive with respect to gender representation. 

Additionally, as shown in Figure~\ref{fig:inclusion-disability}, \SI{74}{\percent} of people with disabilities find the research environment to be not very inclusive with respect to disabilities, while only \SI{54}{\percent} of people without disabilities share the same opinion.
This finding shows how there is still work to be done to change the culture around the concept of inclusivity and diversity. However, a small amount of the open answers collected in the survey presented a counter-view to this, expressing the wish that hiring processes should be based on merit and not take into account \Gls{DEI} characteristics.
%\footnote{For example, ``DEI politics are destroying meritocracy and affecting very negatively the productivity and results" or ``Too much diversity making it hard to find its place as a 'standard' white male" or also ``Most of the previous questions are non sense for me".}

Researchers should hence be provided with tools to foster, encourage and understand the importance of diversity, equity and inclusion. The need of a real change of culture around the concepts of \Gls{DEI} is deeply felt. Measures which are put in place are sometimes perceived as a box-ticking exercise, rather than actions that affect real change. For these reasons, training on these topics should be provided both to seniors and to ECRs to promote a change of culture\footnote{A few survey respondents shared in the open answers how they felt that they are not educated enough to tackle these problems.}.

Additionally, people belonging to under-represented groups must be ensured safe places~\cite{safeplaces} and they must be more represented in senior positions, hiring panels and in the research environment generally. Furthermore, the opinions of ECRs should be heard in executive decisions.

In open-form comment boxes, survey respondents expressed that a background in poverty and socio-economic disadvantages is often neglected, e.g. for talk invitations or visa requirements. Such conditions should be taken into account and properly addressed by research centres to help and ensure equal possibilities for all their researchers. 

Such measures should be put in place and overseen when institutes are hosting conferences, workshops or schools, ensuring provisions for accessibility and caring requirements so that everyone is able to attend, where such provisions should be made clear on registration for the events.

For all these reasons, large institutions, research centres and laboratories should always have diversity, equity and inclusion support centres\footnote{It has been noted how often these duties are given to a single person for very large institutions, thus making it unfeasible for their impact to be relevant.}, to provide \Gls{DEI} trainings to researchers,  monitor researchers' behaviour and act to ensure an inclusive working environment.
Smaller institutions can use single counsellors/officers if a support centre cannot be put in place.

\begin{recommendationbox}{} \GlsBlack{DEI} offices should be present in laboratories, institutions and research centres. The size of such offices should reflect the dimension of the community they cater. These support centres should provide mandatory training around the concept of \GlsBlack{DEI} and monitor the workplace culture, taking action if someone's rights have been violated. Additionally, they should put in place safe spaces for people belonging to marginalised groups and they should ensure diversity, including in age and position in academia, in hiring panels and executive positions. Finally, they must enforce measures to take into account different socio-economic backgrounds and provide equal possibilities to everybody. Such measures should be enforced as well when institutes are hosting conferences, workshops, schools and events foreseeing external participants in general. 
\end{recommendationbox}

\example{Example}{A good example of a Diversity and Inclusion programme is put in place at CERN~\cite{CERN_D&I}.}

When \Gls{DEI} policies are not enforced, under-represented groups are left wondering about their safety and position in their career. As stated above, diversity and inclusion are required to unlock the full potential of a workforce. European research facilities have a responsibility towards their users, perhaps most crucially for ECRs, when it comes to creating an inclusive and equitable working environment. CERN and international laboratories/research centres should support ECRs who cannot rely on support from their home institutes, when they face persecution or discrimination based on their characteristics in the home countries. As recently evidenced in the USA, \Gls{DEI} policy is not guaranteed and this is impacting ECRs~\cite{science-Trump-DEI}. It is imperative that Europe does not follow the same path if it wants to ensure it remains a centre for excellent particle physics research.

\begin{recommendationbox}{}{CERN should maintain its commitment to \GlsBlack{DEI} policies that enrich its working environment.}
\end{recommendationbox}

\subsubsection{Code of conduct, harassment and discrimination}
\label{subsec:codeofconduct}
Home institutions must hold their own code of conduct and adequately publicise it. Survey data showed that \SI{19}{\percent} believe that existing codes of conduct are not publicised enough, as seen in Figure~\ref{fig:code-of-conduct_publicised}. Furthermore, \SI{9}{\percent} of respondents don't agree that the code of conduct of their home institution represents them, as seen in Figure~\ref{fig:code-of-conduct_representation}. From the survey responses in Figures~\ref{fig:code-of-conduct_representation-gender} and \ref{fig:code-of-conduct_representation-disability}, it can be seen however that people of gender minorities and people with a disability or chronic condition feel less represented than men or able-bodied people respectively. 

The following findings are strictly linked to the discrimination experienced by people belonging to under-represented groups.
From the white paper survey, as shown in %Figures~\ref{fig:harassment-self},~\ref{fig:harassment-others},~\ref{fig:harassment-self-gender},~\ref{fig:harassment-others-gender},~\ref{fig:harassment-self-disability},~\ref{fig:harassment-others-disability}
Figures~\labelcref{fig:harassment-self,fig:harassment-others,fig:harassment-self-gender,fig:harassment-others-gender,fig:harassment-self-disability,fig:harassment-others-disability}, \SI{37}{\percent} of female respondents and 3 out of 13 of gender-diverse respondents have experienced harassment or discrimination on the basis of their gender and \SI{18}{\percent} of people with disabilities have experienced harassment or discrimination on the basis of their disability. Around \SI{6}{\percent} of survey respondents have experienced discrimination or harassment due to their nationality or ethnicity, independent of their gender or their disability. Around \SI{2}{\percent} of respondents have been discriminated against or harassed for their economic background. Women disproportionately experience discrimination on the basis of their family status: \SI{7}{\percent} vs. \SI{3}{\percent} for other groups. Non-binary and gender-diverse people disproportionately experience discrimination on the basis of their sexual orientation: 3 out of 13 gender-diverse or non-binary people vs. roughly \SI{1}{\percent} for the other groups.
The latter finding is part of a more general trend, where different types of discrimination are correlated, meaning that people belonging to more than one under-represented group are more likely to experience discrimination on the basis of their different characteristics. 

Finally, people belonging to under-represented groups are more likely to report having observed harassment or discriminatory behaviours towards others. As an example, while \SI{2}{\percent} of male respondents have observed discrimination towards people with disabilities, \SI{11}{\percent} of both female respondents and respondents with disabilities have observed such discrimination. The same pattern is observed for all of the different kinds of discrimination investigated in the survey. Focusing on the support given, \SI{47}{\percent} of the respondents who had experienced harassment or discrimination themselves consider they were not given adequate support by their main institution, as seen in Figure~\ref{fig:support_harassment}. 

For these reasons, a code of conduct should be put in place and adequately publicised. In the case that the code of conduct is breached, the steps that are taken to report such incidents should be transparent. An appropriate code of conduct should set out good practices that allow people belonging to under-represented groups to feel comfortable in their workplace: professional and inclusive language \cite{inclusive-language-guide} and approaching conversations with colleagues of all levels with respect and dignity.

\begin{recommendationbox}{}Home institutions and research centres must have an adequately publicised code of conduct that includes zero tolerance of discrimination or harassment and that considers under-represented groups when setting out good practices and expected behaviours. Procedures should be put in place to act when such code of conducts are violated. Said procedures should be clearly described in the code of conducts and they should be easily accessible. Anonymity and discretion should be guaranteed for those involved with complaints.
\end{recommendationbox}

Conferences, schools and other events are an essential part of a career in research, perhaps particularly so for those in the early stages. These events are typically an excellent opportunity to build a network and reputation, gain a sense of community and learn. It is vital, then, that ECRs feel comfortable in these environments. 

From the very first stage in the organisation of conferences and schools, diversity, equality and inclusion should be considered --- not as a checklist item, but as an aspect that will genuinely improve the quality and experience of the event itself. 
Conference environments can be intimidating for ECRs and a symptom of this can be a reluctance to ask questions. This is compounded when responses from senior staff are not sensitive to the experience of ECR members. 
The code of conduct of a conference is used to set the expected acceptable behaviour of its participants. It is important to explicitly point out the code of conduct to all conference participants, as well as how to report any misconduct they encounter during the event. If an issue occurs, ECRs may not feel comfortable reporting it for fear of how the sensitive information may be shared. The process of management of complaints should be transparent and easily accessible. The local organising committee should be prepared for potential complaints, not because they are expected, but so that they can be dealt with discreetly and professionally if they do occur.

\begin{recommendationbox}{}Conferences, schools and similar events should have a code of conduct which is explicitly presented to participants, with clear information on how to make a complaint and the process of complaint management. CERN and other large research centres should provide templates of such code of conducts that can be shared with smaller institutions and independent organising committees which don't have enough resources to produce them ad-hoc for events. 
\end{recommendationbox}

\subsubsection{Monitoring practices}
\label{subsubsec:monitoring}
Effective monitoring practices are essential for assessing the implementation of \Gls{DEI} efforts and ensuring that policies translate into meaningful change. 
Regular assessment allows research institutions and collaborations to track progress, identify gaps in support systems and address problems before they become systemic issues. 
Establishing clear monitoring frameworks ensures that measures such as codes of conduct, inclusive hiring policies, and mentorship programs lead to concrete actions and continuous improvements rather than remaining symbolic commitments.

One key approach is the systematic collection of data to evaluate workplace conditions. 
Institutions should conduct anonymous surveys at regular intervals to assess researchers’ experiences related to inclusion, supervision, and access to key resources.
Additionally, tracking participation in mentoring programs, funding applications, leadership roles and professional development opportunities can help identify disparities in institutional support. 
Exit interviews with researchers leaving their positions, whether after a PhD or postdoc, can provide valuable insights into underlying issues such as workplace culture and challenges faced during their time in that position. 
These interviews can also help identify factors influencing decisions to leave academia.
Institutions should publish regular \Gls{DEI} progress reports, summarising key findings from surveys and workplace assessments. 
These reports should outline specific actions taken in response to identified problems, ensuring that \Gls{DEI} efforts are not just tracked but actively improved.
\Gls{DEI} Offices should also establish a dedicated \Gls{DEI} monitoring committee, responsible for reviewing findings, recommending corrective measures and ensuring their implementation. 
In large research collaborations, independent reviews by external advisory boards can provide an objective assessment of mentorship structure and inclusion efforts.
To ensure diverse perspectives in decision-making, institutions should also include ECR representatives in \Gls{DEI} oversight committees, allowing them to contribute directly to the evaluation and improvement of policies that impact their careers.

\begin{recommendationbox}{}\GlsBlack{DEI} Offices should establish a monitoring committee, including ECR representatives, to review findings from surveys, exit interviews and workplace assessments. Outcomes and actions taken to address identified challenges should be made public through regular \GlsBlack{DEI} progress reports.
\end{recommendationbox}
\subsection{Early career community building, leadership and recognition } %{\color{red}Frozen, contact Christina before changes!}}
\label{theme:community_leadership}
ECRs are essential contributors to the advancement of particle physics, yet they face persistent challenges that limit their full participation in the scientific community. 
Representation structures often lack transparency and inclusiveness, making it difficult for ECRs to have a clear voice in decision-making. The \Gls{ECFA ECR Panel}, which aims to support ECRs, plays an important role in connecting the community. However, concerns remain regarding the selection process of national representatives for the panel and the extent to which decisions reflect the broader ECR community. 

Funding opportunities for ECR-led initiatives are also limited. While ECRs actively organise collaborative efforts, financial support for such activities is often unavailable or difficult to secure. This creates barriers to participation and professional development, particularly for those without access to institutional resources.

Another major challenge is the recognition of ECR contributions. Much of their work, including service tasks and technical efforts, is essential but not always acknowledged. Traditional evaluation metrics, such as publication records, often fail to capture the full extent of their impact. This can slow career progression, as these contributions are undervalued in hiring and funding decisions. 

Beyond-collider physics also faces unique structural difficulties. Unlike collider-based research, which has well-established networks and institutions, beyond-collider experiments are more dispersed, making it harder for researchers to collaborate and engage with the broader community. This lack of cohesion limits opportunities for knowledge exchange and career development within the field.

This section discusses these challenges in detail and proposes concrete recommendations to improve transparency, funding opportunities, recognition and community-building for ECRs.

\subsubsection{ECFA ECR Panel} \label{sec:ecfa_ecr_panel}
The \Gls{ECFA ECR Panel}~\cite{ECFA-ECRs}, formed after receiving the mandate from \Gls{ECFA} in July 2020, plays a crucial role in building the ECR community in Europe. It enables the coordination of ECR activities beyond physics studies, aiming to support the interests of young scientists and to publicly express their needs. So far, the panel has been successful in efforts to study and better understand the community itself, by creating and distributing surveys on training in instrumentation~\cite{ECFAEarly-CareerResearcherPanel:2021flc} and on career prospects and diversity~\cite{Allen:2024lyp}. It also organised a \textit{Future Colliders for Early-Career Researchers} event at CERN~\cite{ecfa_ecr_future_colliders}, initiated many national ECR events discussing future colliders or the ESPPU~\cite{Ilg:2024,14thECFAnewsletter}, as well as launching the process for preparing this document. A more comprehensive overview of the \Gls{ECFA ECR Panel}'s recent activities can be found in Refs.~\cite{ AlexandruGeanta:2022wbm, ECFAEarly-CareerResearchersECRPanel:2024trx}.

Despite the notable achievements of the panel, there is room for improvement in some aspects of its operation, particularly regarding transparency and representativeness. Currently, the national ECR representatives are selected by the corresponding \Gls{RECFA} members, but the exact selection process depends on the country and is often vague and unclear. Furthermore, all \Gls{ECFA ECR Panel} activities are decided in closed meetings by a relatively small group, limiting input from the broader community which the panel is meant to represent. With ECR representatives from each \Gls{ECFA} member state, there is a great opportunity to form national ECR structures and communities, as some countries have already demonstrated. This would not only engage more people in the panel's activities and create an open forum for discussions but also enhance transparency, inclusiveness and visibility of its initiatives. In addition, implementing, whenever possible, a \textit{uniform}, open, and democratic selection process for panel members, not depending on senior researchers, would help ensure their commitment to the role of national representatives. 

\begin{recommendationbox}{}The \GlsBlack{ECFA ECR Panel} should become more open to input and involvement from the community by creating national forums with regular meetings and ECR mailing lists where not yet established. 
A uniform, democratic and transparent selection process for panel members, ensuring fair representation of different research areas within each country, must be implemented whenever possible.
\end{recommendationbox}

\subsubsection{The role of ECRs in the ESPPU process \label{fc:decisionmaking} } %{\color{red}Frozen, contact Krzysztof before changes!}} 

The current ECR community will be responsible for conducting research based on today's decisions, whether in accelerator science, data analysis, detector R\&D, engineering, software development, or theory. 
Therefore, it is completely natural and reasonable to demand the inclusion of ECRs in the processes of strategy updates. However, the current extent and manner of involvement are far from satisfactory. According to the survey, only \SI{20}{\percent} of respondents feel their voice is heard in the ESPPU process, while \SI{31}{\percent} state their decision about pursuing a future in academia depends on the strategy outcome, as shown in Figure~\ref{fig:career_opinions-heard}. Among ECRs involved in future collider studies, these results are even more worrying, rising to \SI{24}{\percent} and \SI{44}{\percent}, respectively (see Figure~\ref{fig:career_opinions-heard_correlations}).

While the appointment of ECRs as ``scientific secretaries'' in \Gls{PPG}~\cite{PPG2024} is undoubtedly a step in the right direction compared to previous strategy updates, there is a lot of room for improvement, particularly in terms of transparency. The exact role and scope of responsibilities of secretaries are unclear and not well-defined, and the definition of ECR itself is also not clearly set out. Moreover, their selection process is highly arbitrary, and there is no established form of communication between the secretaries and the community. This undermines both their position as representatives of ECRs, and the very purpose of their appointment.
Beyond the appointment of scientific secretaries in the working groups, the participation of ECRs in the \Gls{ESG} itself can be seen as the next logical step.
\begin{recommendationbox}{}
Appointment of the scientific secretaries, taking into account the broader ECR community, must be transparent. Their mandate should give them an equal voice among the working group conveners.
The \GlsBlack{ECFA ECR Panel}, which already provides several delegates to \GlsBlack{PECFA} and one to \GlsBlack{RECFA}, should be mandated to also send a delegate to the \GlsBlack{ESG}.
\end{recommendationbox}

\subsubsection{Acknowledgement of ECR contributions and leadership skills}

According to the survey, a significant majority of ECRs (\SI{81}{\percent}) agree that their voice is heard in their local research groups. However, when asked the same question about their collaboration, the percentage of agreement drops to \SI{54}{\percent}. These results are presented in Figure~\ref{fig:career_opinions-heard}. ECRs tend to feel less heard in large LHC collaborations (\SI{42}{\percent}) and more heard in neutrino (\SI{60}{\percent}) or fixed target (\SI{69}{\percent}) experiments (see Figure~\ref{fig:career_opinions-heard_correlations}). 

ECRs are already represented in some of the larger international collaborations, for example by being included in executive board meetings of LHCb~\cite{Hilton:2021osy} and ATLAS, a situation which should be extended to more collaborations. We also advocate for including ECRs in the organisation of events, in topical working groups, and having ECR-dedicated sessions as an integral part of events, conferences and other meetings. Dedicated sessions just for ECRs can play a major role in making ECRs' voices heard, as the threshold to speak up can be significantly lower for many.

\begin{recommendationbox}{}ECRs should be more included in the executive board of collaborations as well as in topical working groups and the organisation of events.
Dedicated ECR sessions should be an integral part of conferences and similar events.
\end{recommendationbox}

Outside of these dedicated sessions, feedback from ECRs should be taken into account in decision-making. The opinions and voices of ECRs should be actively included in discussions about our field. An example of this is including one ECR as a scientific secretary in each working group of the \Gls{PPG}~\cite{PPG2024}. However, the responsibilities of this type of tasks should be communicated clearly, and the selection process should be transparent, as discussed in Section~\ref{fc:decisionmaking}.

Leadership skills are a valuable asset both within our field, as well as in careers beyond particle physics. ECRs already take on significant responsibilities, and this work needs to be visible and formally credited.  This includes key roles in high-impact projects, such as extensive work on running experiments and detectors, as well as future collider studies. Recognising ECR contributions can enhance their professional growth, whether they continue in the field or pursue opportunities elsewhere. Formal recognition shows the appreciation of the work of ECRs, offering positive feedback and further motivating ECRs in their work. 
Leadership skills should be properly considered in the hiring process, as they reflect the ability to manage projects and contribute to the wider scientific community. However, one should note that concerns exist about potential conflicts of interest when ECRs - who are often on short-term contracts - take on leadership roles, as their decisions may affect senior researchers who could later review their job or grant applications or become their future employers. %One solution to this problem could be a selection process in the job application which is blinded until the very last moment, which is the interview; the review committees can also differ at subsequent stages of the process, etc.

According to the 2022 survey~\cite{Allen:2024lyp}, most of the ECRs surveyed are uncertain or do not believe that traditional bibliometric indices, such as the h-index, fairly reflect their work and achievements. In fact, much of the work conducted by researchers leads to outcomes other than publishable results. This needs to be reflected and assessed when evaluating careers of scientists. As the authors of Ref.~\cite{Allen:2024lyp} propose based on the Declaration of Research Assessment~\cite{ResearchAssessment}, ``using narrative CVs for job applications and improving collaboration or group policies could enhance the recognition of service work." 

Future collider studies often offer publication opportunities that other sub-fields of particle physics do not, due to the many unresolved questions surrounding potential future projects. However, establishing a solid foundation for the future requires significant amounts of R\&D and technical service work, much of which is undertaken by ECRs. These tasks demand full-time commitment and cannot be effectively managed alongside primary involvement in ongoing experiments. Yet a lack of experience with real data and current experiments can be a disadvantage in the job market. Therefore, we also advocate for greater recognition of the crucial service work and R\&D performed in future collider studies to address this imbalance.

\begin{recommendationbox}{}Supporting alternative evaluation methods beyond traditional bibliometric measures in the particle physics community is highly recommended. Service work and R\&D for sensitivity studies of future facilities, such as colliders, as well as leadership roles must also be recognised.
\end{recommendationbox}

\subsubsection{Funding for ECR activities}
\label{sec:ECR_funding}
In recent years, ECRs have shown enthusiasm and capability by establishing organisations and events focused on their scientific interests. These activities foster collaboration, skill-development, and networking, which are essential for career growth. However, most resources in large research institutions are allocated strictly to research, leaving little funding for ECR-led initiatives in the same way as contracts of ECRs do not allocate any time for such activities (see Section~\ref{sec:time_beyond_duties}). To address this gap, national funding agencies should establish dedicated funding lines to support ECR-driven activities. In parallel, the \Gls{ECFA ECR Panel} should receive dedicated funding to organise events and initiatives aimed at strengthening the ECR community and supporting professional development, ensuring centralised support at the European level. 

Such funding would enable ECRs to organise (or simply attend) conferences and workshops, providing platforms for ECRs to present their work and create spaces for networking with peers and senior researchers.
A positive example is the CMS event ``Coffee with Senior Scientists" where the CMS Young Scientists Committee secured modest funding for coffee and sandwiches, facilitating informal yet valuable interactions.
Dedicated support for similar initiatives would further encourage building collaborative projects, sharing ideas, and discussing career development. Besides research topics, ECRs could establish peer mentoring groups to assist with research strategies, work-life balance, and career advice, all fundamental elements often not discussed within the ECRs' institutions. 

%A dedicated source of funding could also help ECRs engage in science communication and public outreach. This could include public lectures, writing articles, participating in podcasts, or creating social media content to explain scientific concepts to a broader audience. These initiatives increase public understanding of research and help ECRs improve their communication skills, which are essential for a successful career.

\begin{recommendationbox}{}
The \GlsBlack{ECFA ECR Panel} should receive dedicated funding to organise events and initiatives that strengthen the ECR community and support professional development.
%The major institutions should establish a system of dedicated funding for ECR activities which would be a unique opportunity for young scientists to improve their professional development and enrich the research community, shaping the future of scientific research.
\end{recommendationbox}

Related topics include \Cref{sec:time_beyond_duties} on recognition for time spent on activities outside their main duties and %\Cref{subsubsec:supervision-training} on mentoring schemes, 
\Cref{subsec:comm_recs} on communication.

\subsubsection{Strengthening the beyond-collider community}\label{subsubsec:Forum_and_non_collider_community_building}

While collider physics in Europe has a central focal point geographically at CERN and scientifically with the LHC experiments, beyond-collider physics is much more disparate, with relatively small experiments being carried out at labs across the continent and beyond. This means that connections and collaboration between these experiments are more difficult to develop. The creation of a forum specifically for beyond-collider physics, and potentially related fora for its subfields, would allow the great physics potential of small and medium-sized beyond-collider collaborations to be fully exploited by facilitating collaboration, knowledge exchange and networking between researchers in these experiments. This would also enhance the coherence of, and give a communal voice to, the beyond-collider community. 

Additionally, as discussed in Section~\ref{subsubsec:importance_beyond_collider_physics}, interdisciplinary collaboration between particle physics, astroparticle physics and nuclear physics (facilitated through \Gls{JENAA}) must be strengthened. In this sense, a forum for beyond-collider particle physics would be a vehicle for collaboration and information exchange between these interconnected communities.

The importance of such a beyond-collider forum is supported by the fact that a majority of survey respondents (\SI{56}{\percent}) agree that ``A dedicated beyond-collider particle physics forum for enhanced networking, community formation, and career opportunity sharing'' would benefit beyond-collider physics significantly or rather much, see Figure \ref{fig:bc_forum_funding_scheme}.

\begin{recommendationbox}{}A forum for beyond-collider particle physics researchers should be established to give coherence to this community and to facilitate collaboration, knowledge exchange and networking. Such efforts should include the whole beyond-collider particle physics community, and the establishment of additional fora for large beyond-collider subfields would be beneficial for more effective topical exchange.
\end{recommendationbox}

\Gls{ECFA} should, as its core mission, represent the full spectrum of particle physics within Europe. %, which includes beyond-collider particle physics.
Currently, it is not always well understood that the whole field of beyond-collider particle physics also falls under \Gls{ECFA}'s core remit, partly due to its historical name and the usage of the ambiguous term ``high-energy physics'' in its terms of reference~\cite{ECFATermsOfReference}.
\Gls{ECFA} should move away from such terminology in favour of ``particle physics'' in order to be more inclusive to the wider field and in particular to low-energy particle physics.
Its mission should furthermore incorporate a fair and even representation of the different research fields within \Gls{ECFA}, \Gls{PECFA} and the \Gls{ECFA ECR Panel} (see also Section~\ref{sec:ecfa_ecr_panel}) for each country across the whole domain of particle physics, including the beyond-collider field, to strengthen its participation within the larger community.
Additionally, a clear statement that \Gls{ECFA} represents beyond-collider particle physics and a corresponding amendment of \Gls{ECFA}'s terms of reference %, along with a call for fair representation 
will directly create a more inclusive \Gls{ECFA} by giving the beyond-collider community a more grounded position within the field.

As a step towards connecting and unifying the field of beyond-collider particle physics, we propose the formation of a dedicated \Gls{ECFA} panel focusing on future beyond-collider facilities and their physics potential.
This would be analogous to the present \emph{ECFA Detector Panel} and \emph{ECFA Study on Higgs / EW / Top factories}, and aligns with our recommendations in Sections \ref{sec:futurepp:maximizing_yield} and \ref{sec:ESPPU_importance_bc}. 

\begin{recommendationbox}{}
In order to create a more inclusive \GlsBlack{ECFA}, there must be fair and even representation of the different particle physics research fields, including beyond-collider particle physics, within all \GlsBlack{ECFA} bodies. Additionally, \GlsBlack{ECFA} should form a dedicated panel on future beyond-collider particle physics experiments and facilities and their physics potential, as a step towards unifying the disparate beyond-collider field and giving it a communal voice.
\end{recommendationbox}
\subsection{Communicating the importance of particle physics}
\label{subsec:comm_recs}

The importance of effective science communication for the future of particle physics cannot be overstated. Every researcher needs to be able to convey scientific research questions and results comprehensively to fellow particle physicists, other scientists and non-experts from the general public.
This poses an additional challenge on top of researchers' professional work, where they are typically expected to learn communication skills on the job. However, there is little to no formal training on how to address different target audiences with varying levels of pre-existing understanding of complex subject matters. 

ECRs in particular may have had fewer opportunities to develop their communication skills, not necessarily due to a lack of involvement but because they are still at the beginning of their careers. Support in this area would benefit the entire community in the future.

Depending on the audience, different kinds of support for ECRs may be needed to enhance communication. Therefore, the recommendations in this section distinguish between various audience types:
\begin{itemize}
    \item  Efforts to engage with the \textbf{general public} are referred to as \textit{outreach}.  Sections~\ref{sub:comm_recs_outreach_motivation} to~\ref{sub:comm_outreach_opportunities} consider ECRs' motivation and involvement in outreach, as well as the  main limiting factors.

    \item Specific recommendations to ensure efficient exchange with \textbf{industry} representatives in order to foster innovation are made in Section~\ref{sub:comm_industry}. 

    \item The need for clear guidelines on communication \textbf{within the field}, and in particular between ECR colleagues is addressed in section~\ref{sub:comm_interfield}. Section~\ref{sec:comm:small_collab} discusses the particular challenges small collaborations face. 
\end{itemize}

To widely and effectively reach the different audiences, communication and information via social media channels play an increasing role, which is covered in Section~\ref{sub:comm:social_media}.
Finally, Section~\ref{sub:comm_narrative} provides recommendations on how to refine the way future collider projects are framed and communicated (\textit{storytelling}) to better promote them.

As described in Section~\ref{sub:survey_topics_comm}, the survey questions relating to topics of science communication focussed on outreach to the general public. They investigated the motivation and preparedness of ECRs for contributing to outreach beyond academia. The survey results are therefore mostly used to substantiate the recommendations in Sections~\ref{sub:comm_recs_outreach_motivation} to~\ref{sub:comm_outreach_opportunities}, with cross-correlations drawn to questions from other sections of the survey in order to gain more detailed insights. The set of supplementary questions on communication was answered by almost half of the respondents (382 out of 804). All data gathered from simple yes/no questions are listed in Table~\ref{tab:comm_wg_raw_data}.

%--------------- OUTREACH MOTIVATION -----------------------------------------------------------------------------------------------------------------------
\subsubsection{Sustaining motivation and involvement in outreach}
\label{sub:comm_recs_outreach_motivation}

As a first step to understand how best to foster ECRs' commitment to outreach work, the levels of motivation for and involvement in such work ECRs currently report were reviewed.

\important{Statement: ECR motivation on outreach}
{A significant majority (\SI{84}{\percent}) of survey respondents expressed motivation to engage with the public on topics related to the future of particle physics, with this high level of motivation being stable across several sub-groups, for example respondents' career stage or preferred option for a future collider project in Europe.} 

Levels of motivation for outreach are highest among respondents for whom it is important to have a future flagship project (\SI{88}{\percent}) and those who plan to remain in the field after their current position (\SI{89}{\percent}). These are very encouraging findings as they demonstrate that ECRs are highly driven regardless of their current position, and that they are aware that experience and skills in outreach are important not only for the future of the field but also for their individual career paths. 

An exception to the above is the subgroup of respondents who reported that they do not support the idea of a future collider project in Europe, where the fraction of respondents motivated to engage with the public drops to just below \SI{60}{\percent} (see Figure~\ref{fig:comm_motivation_splits}). Although this group is too small to draw statistically significant conclusions, it is worth drawing attention to the fact that there is a minority amongst ECRs who are clearly not convinced of the current particle physics communication strategy. This issue is addressed and discussed more extensively in Section~\ref{sub:comm_narrative}.

\begin{figure}[H]
    \raggedright
    \hspace{0pt}
    \includegraphics[width=0.7\textwidth]{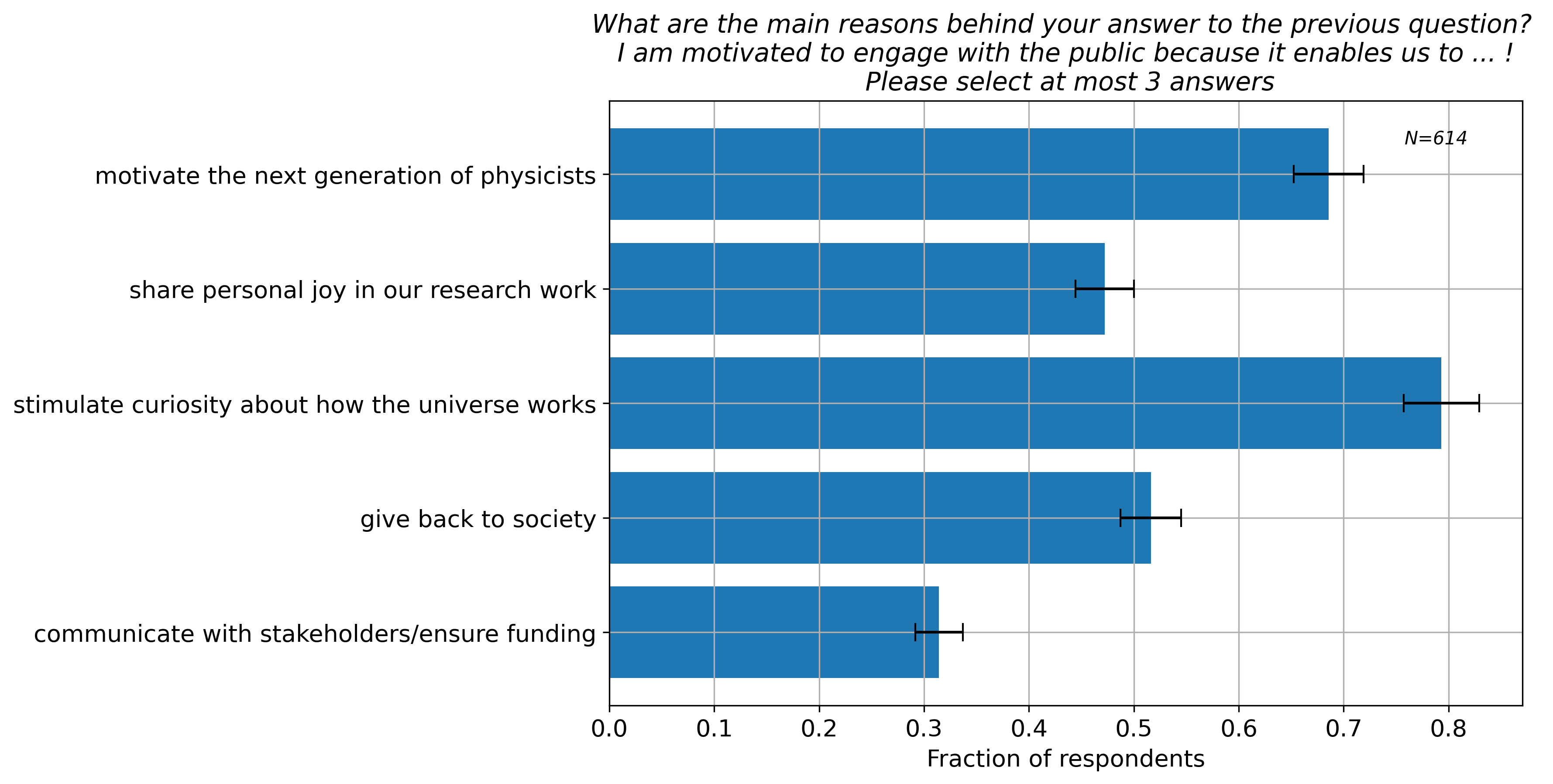}
    \caption{Fraction of respondents indicating the reasons why they are motivated to engage with the public. Respondents were able to choose up to three answers.}
    \label{fig:comm_reas_mot}
\end{figure}

The most frequent reason behind the commitment to outreach work is the wish to stimulate curiosity about how the universe works, as shown in Figure~\ref{fig:comm_reas_mot}, with the desire to motivate the next generation of physicists coming second and giving back to society third. In contrast, the need to communicate with stakeholders to ensure funding is the least frequently cited reason, showing that ECRs do not see science communication with non-experts just as an obligation or even a burden, but as an important social responsibility of a scientist.

These findings are further supported by the data on ECRs' current active involvement in outreach activities: \SI{59}{\percent} of respondents reported that they are involved in outreach, with this fraction increasing for those who reported that they are motivated to engage with the public (\SI{65}{\percent}) and those who additionally reported that they plan to stay in the field after their current position (\SI{66}{\percent}). 

In the supplementary set of questions, answered by approximately half of respondents, ongoing outreach efforts in which ECRs are engaged were investigated. Here, the significant majority of respondents who are active in outreach reported that they do it out of their own motivation, rather than for any additional benefits they receive or due to it being mandatory for their current position (see Figure~\ref{fig:comm_reas_participation}). Responses using the free text ``Other" answer option highlight that outreach is considered ``fun and rewarding" and of ``moral value" as well as that there is ``joy in explaining".

Respondents who reported being active in outreach were further asked whether their outreach work takes place during regular working hours or during their free time. A simple majority of respondents reported using both of these times for outreach activities (\SI{59}{\percent}), while fewer than one third of respondents (\SI{27}{\percent}) only spend regular working hours on their outreach efforts (see Figure~\ref{fig:comm_time}). 

While this may suggest that ECRs feel motivated to go above and beyond to do something they consider important, it may also be a warning sign that outreach activities are not adequately valued as a regular part of scientific work, becoming instead an additional burden, thereby increasing the risk of burnout and loss of motivation. Clear community guidelines on how to best acknowledge contributions to outreach efforts are therefore needed, as discussed in the following section.

\begin{recommendationbox}{}
The particle physics community should promote a culture where outreach and communication are integral to the practice of science, not secondary responsibilities, in order to sustain the high levels of motivation for engaging with the public amongst ECRs and support their efforts in ensuring public approval of future projects.
\end{recommendationbox}

%--------------- OUTREACH RECOGNITION ----------------------------------------------------------------------------------------------------------------------
\subsubsection{Recognition for outreach work}
\label{sub:comm_recognition}

A primary factor discouraging ECRs from engaging in outreach activities is the lack of recognition for such efforts in career advancement. As shown in Figure~\ref{fig:comm_no_motiv}, more than \SI{40}{\percent} of respondents who reported that they are not motivated to engage with the public express this reason as the main source of discouragement, with the other factors being significantly less important. Apart from the pre-defined answers shown in the figure, responses in the free-text ``Other" field showed a clear prevalence for factors related to the general situation of ECRs, such as lack of job stability, or a lack of confidence in one's self and one's own future in the field. On top of that, a few respondents cited their personal disagreement with the current strategy for future projects as the main reason behind a lack of motivation to engage with the public. This issue may be addressed by the recommendations on storytelling for future projects given in Section~\ref{sub:comm_narrative}. 

\begin{figure}[H]
    \raggedright
    \hspace{0pt}
    \includegraphics[width=0.8\textwidth]{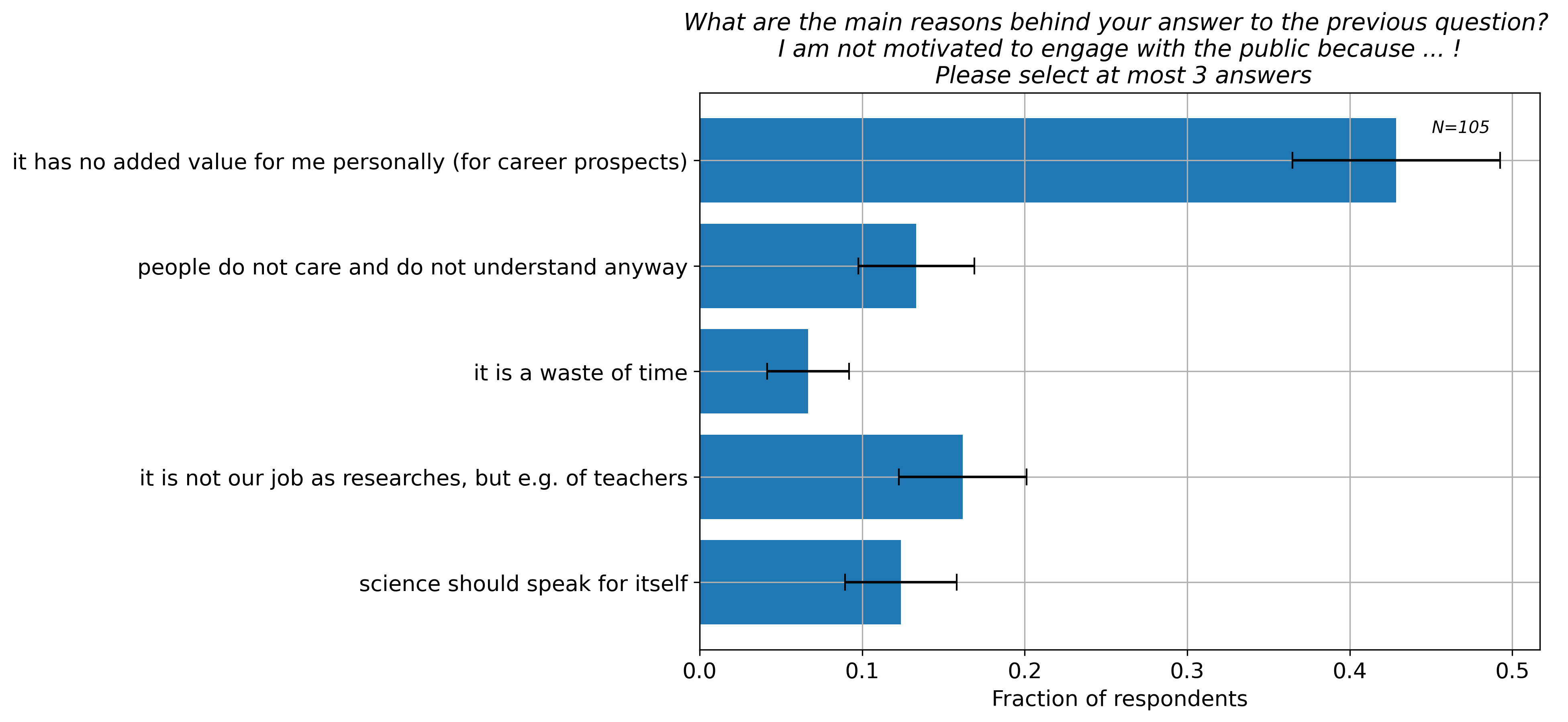}
    \caption{Fraction of respondents indicating the reasons preventing them from engaging in outreach. Respondents were able to choose up to three answers.}
    \label{fig:comm_no_motiv}
\end{figure}

Additionally, of the subset of respondents who answered the supplementary questions and are actively engaged in outreach, roughly \SI{40}{\percent} report not feeling adequately valued for the science communication work they do. In a follow-up question, those respondents were asked to describe what would need to change for them to feel valued in free text form. The answers here echo the above considerations, pointing to a direct lack of support and even personal discouragement from supervisors and other senior scientists, the general perception that outreach is ``not considered work" nor valued by hiring committees and even the concern that participating in outreach comes at the cost of ``being taken less seriously" and thus harming one's career. A few respondents also cite a lack of specific funding for outreach projects, and consider the disparity in the economic recognition of outreach among different institutions unfair. These findings are a major cause for concern, as in both academia and the broader professional landscape, communication skills are indispensable.

To learn by positive example, those respondents who reported feeling adequately valued for their outreach work were asked what the main factor behind their answer is. Although a majority of respondents (\SI{59}{\percent}) reported that they feel the work in itself is rewarding enough, personal feedback from supervisors or other senior colleagues was important to nearly \SI{30}{\percent} of respondents (see Figure~\ref{fig:comm_reas_valued}). 

Advocacy is needed to ensure that outreach and science communication efforts receive formal recognition within the field. The general recommendations given in \ref{sec:time_beyond_duties} on supporting ECRs' involvement in projects other than research apply to this issue as well, however the recommendations here specifically aim to encourage engagement in science communication activities. 

\begin{recommendationbox}{}
Systemic changes should be made to formally recognise and reward outreach and communication efforts and to integrate them into institutional benchmarks, performance and funding evaluations to validate their importance within the scientific profession.
\end{recommendationbox}

%--------------- OUTREACH TRAINING -------------------------------------------------------------------------------------------------------------------------
\subsubsection{Training and resources for science communication}
\label{sub:comm_training}

The availability and quality of training opportunities vary greatly between institutes and countries, often leaving researchers to rely on their personal commitment to seek training from (private) institutions. 
This inconsistency creates a gap in the preparedness of researchers to effectively engage with the public and communicate the value of particle physics. Survey results indicate that more than \SI{40}{\percent} of ECRs do not feel adequately qualified for outreach activities, with the main reason being the absence of adequate professional training, as shown in Figure~\ref{fig:comm_missing}. Although the fraction of respondents who feel adequately prepared for outreach activities rises with seniority of the current position, even among ECRs who hold a beyond post-doc position (permanent or non-permanent), more than \SI{30}{\percent} feel underprepared for the science communication challenges they face (see Figure~\ref{fig:comm_preparedness_splits}). 

\begin{figure}[H]
    \raggedright
    \hspace{0pt}
    \includegraphics[width=0.7\linewidth]{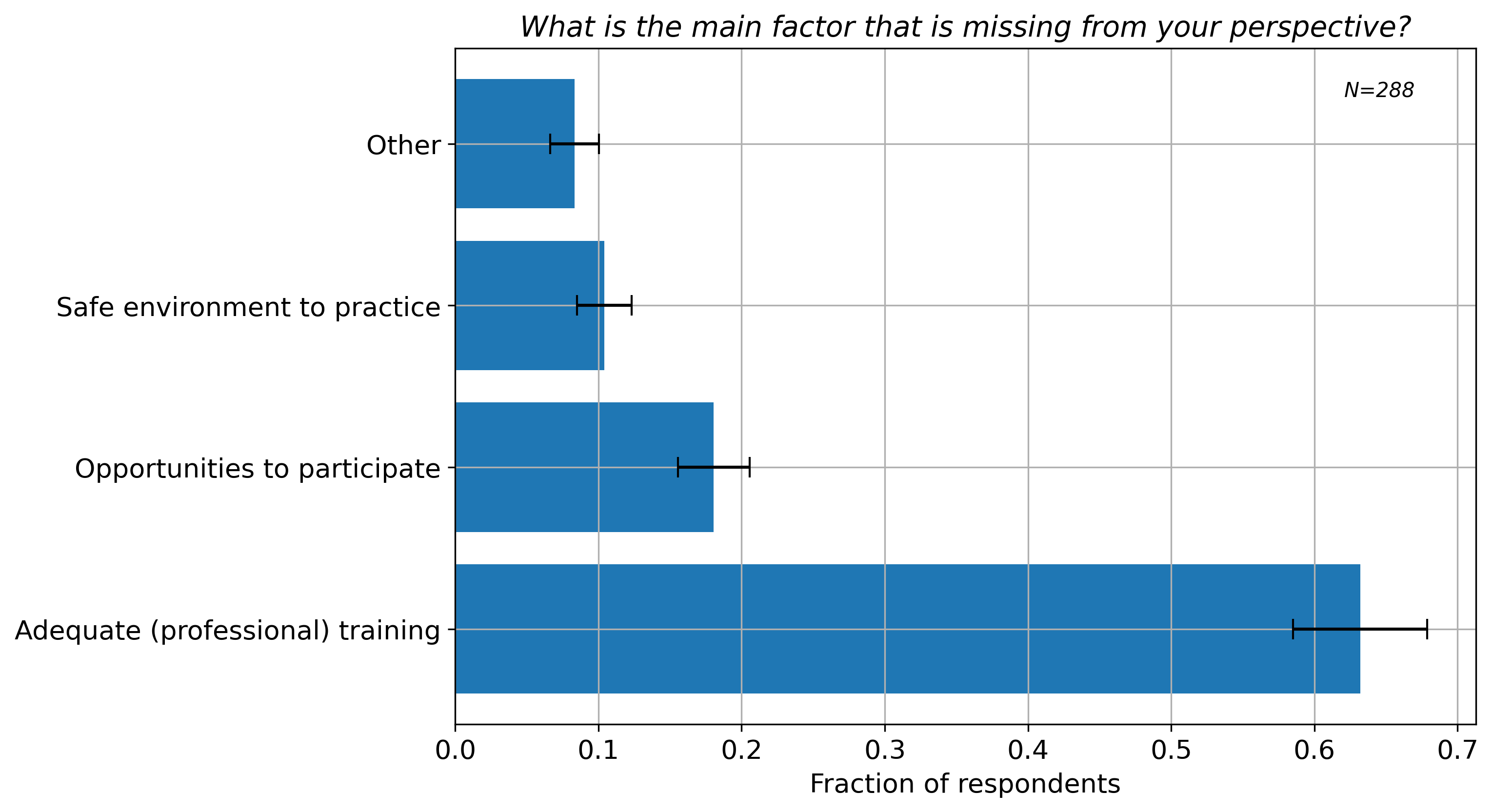}
    \caption{Proportional responses regarding missing factor in motivation to engage with the public. The question was shown to the respondents that do not feel prepared for science communication.}
    \label{fig:comm_missing}
\end{figure}

Formal training by professional science communicators is requested by the majority of respondents (\SI{64}{\percent}) as (additional) training they would like to receive, followed by practice sessions and training for university teaching and social media, as shown in \Cref{fig:comm_training}.

\important{Statement: Training on science communication}
{Over half (\SI{55}{\percent}) of the survey respondents who chose to answer the supplementary questions on science communication indicated that they have not received any kind of formal training on communication matters throughout their career.}

Individual efforts by institutes, such as CERN’s noteworthy initiatives~\cite{cern_comm}, are clearly not wide-spread enough to address this issue. A community strategy is therefore overdue to ensure broad and equitable access to adequate training on science communication for ECRs across Europe. 

\begin{figure}[!htb]
    \raggedright
    \hspace{0pt}
    \includegraphics[width=0.7\linewidth]{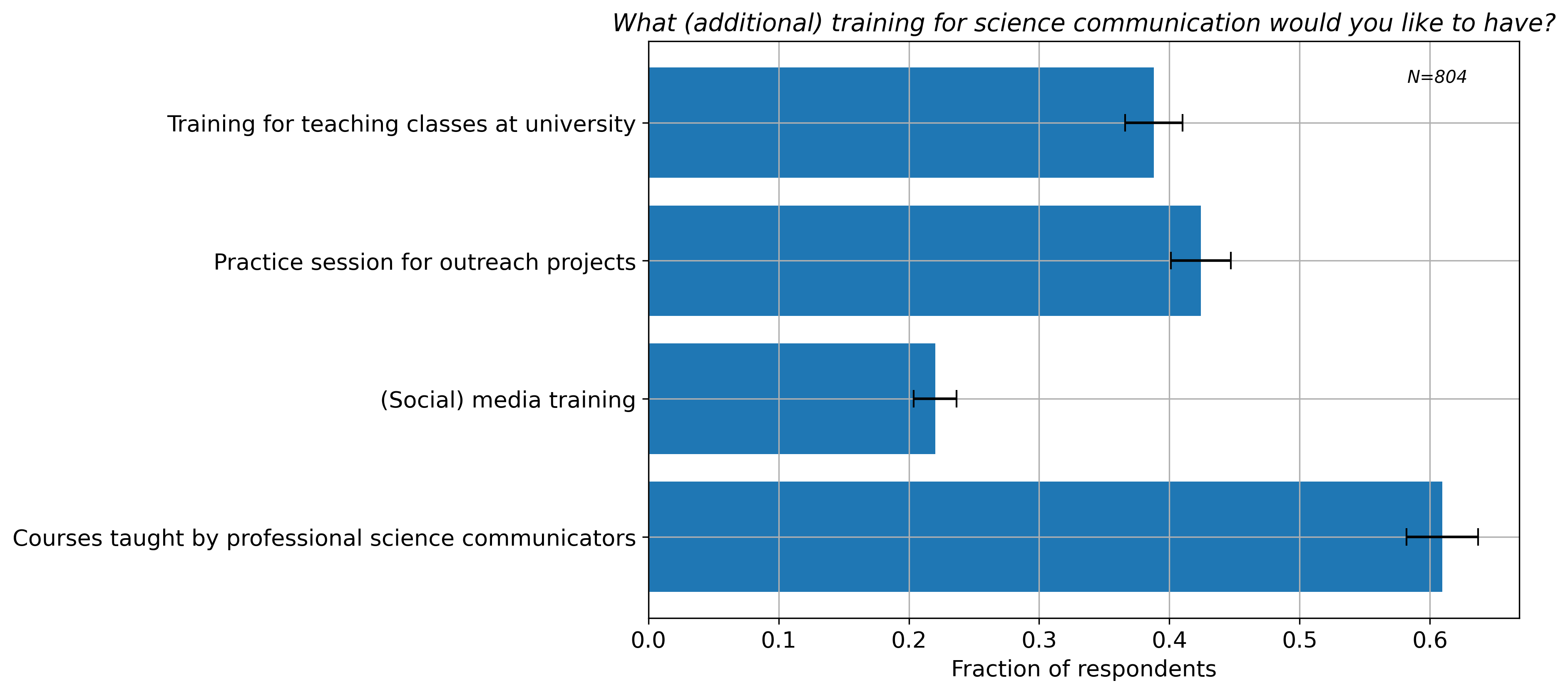}
    \caption{Fraction of respondents requesting additional training. Survey respondents could select multiple options.}
    \label{fig:comm_training}
\end{figure}

A systematic review of existing science communication training programmes and ECRs' experiences and satisfaction with these programmes would be a good starting point for the development of such a common strategy. The supplementary survey questions included a section asking respondents who have received formal training in the past on the specifics of this training, but this subset of answers is too small to be truly representative (N = 167). Nonetheless, it should be noted that a significant majority (\SI{85}{\percent}) of this subgroup of respondents found the training they received useful, showing that such programs are a valuable use of resources. Understanding how to adapt to the level of the audience was reported most frequently as the most useful learning experience of formal communication training among those respondents, closely followed by learning how to develop a storyline for a science topic (see Figure~\ref{fig:comm_training_useful}).

\begin{recommendationbox}{}
Disparities in the availability of communication training must be addressed. Institutions and collaborations should develop standardised programs, leveraging internal expertise and partnerships with professional science communicators. The particle physics community should cooperate with universities to integrate training programs into academic curricula to ensure equitable access for all ECRs.
\end{recommendationbox}

%--------------- OUTREACH OPPORTUNITIES --------------------------------------------------------------------------------------------------------------------
\subsubsection{Creating outreach opportunities for ECRs}
\label{sub:comm_outreach_opportunities}

ECRs who do not participate in outreach activities also face a lack of opportunities to practice and get involved (see Figure~\ref{fig:comm_reason_no_inv}): more than \SI{60}{\percent} of respondents said that the absence of a clear starting point prevents them from participating in communication activities.
Moreover, \SI{30}{\percent} also express concerns about the availability of interesting activities to join. These two answers strongly suggest that the absence of structured outreach programs is an issue. For many motivated ECRs, there is no obvious path that brings them into the outreach world by providing not only training but also sufficient opportunity to practice and organise events.
This is a particular problem for researchers in small collaborations, as discussed in Section~\ref{sec:comm:small_collab}. To address this problem, institutions should broaden and diversify their offer of outreach opportunities to involve early career researchers.

One of the main limiting factors here appears to be a lack of awareness about existing outreach efforts and opportunities. As became clear in WG discussions as well as from feedback received from the broader ECR community during the drafting process, many commendable initiatives and networks have been formed, which supply researchers with inspiring materials for outreach projects.

\example{Example }{
Science communication initiatives and networks:
\begin{itemize}
    \item CERN's Education, Communications \& Outreach Group~\cite{CERN_ECO}
    \item European Particle Physics Communication Network (\Gls{EPPCN})~\cite{EPPCN}
    \item International Particle Physics Outreach Group (\Gls{IPPOG})~\cite{IPPOG}
\end{itemize}
}

The lack of awareness of such initiatives among ECRs is likely due to their advertisement being limited to the participating institutes and groups of researchers and not communicated sufficiently to the whole community.  
A common platform across countries, institutes and experiments to gather information and resources on existing outreach projects is needed. Awareness of such a platform should be spread broadly, ensuring specifically that ECRs are sufficiently informed and involved. 
ECR engagement could be specifically highlighted by regular updates, such as newsletters showcasing ECR contributions to outreach, in order to inspire broader participation.

Such a common pooling of resources would facilitate knowledge transfer from more senior colleagues experienced in science communication to ECRs and allow institutes in different areas to equalise their outreach efforts by replicating common projects. 
As science itself is an asset of the public domain, all outreach projects should follow the same open-source policies as research data. In particular, the development of tools and software employed to engage with the public, such as virtual reality experiences, should not be outsourced to private companies but kept open-source within the field.

\important{Statement: Tools for outreach}
{ECRs can actively contribute to the development of outreach tools and even pursue a career in science communication and its development only if adequate funding and resources are secured.} 

\example{Example }{
Resources and tools for outreach projects:
\begin{itemize}
    \item CERN MediaLab~\cite{cern_medialab}
    \item LHC@Home~\cite{lhcathome,lhcathome_paper}
    \item LHC and ATLAS initiatives~\cite{10.3389/fphy.2024.1393355}
    \item Extreme Energy Events project~\cite{Doser_2022} 
    \item Lab2Go~\cite{lab2go}
    \item Physics Masterclass~\cite{masterclass}
\end{itemize}
}

ECRs working outside of their country of origin often also encounter a language barrier when trying to participate in outreach. Typically, information about and encouragement to join local outreach efforts is circulated amongst colleagues of the same institute, which means that colleagues who are not proficient in the respective language are excluded. This also could be addressed by the common platform proposed above: for example, it could include a database to match interested researchers with outreach projects across countries based on their native language. A positive example of how this can work is the method employed to staff ATLAS visits with guides, where different language groups are contacted separately depending on the immediate needs. However, this is an experiment-specific strategy, that needs to be reproduced across experiments and even disciplines to allow equal participation of ECRs across the whole field. 

\begin{recommendationbox}{}
The particle physics community should continue supporting outreach initiatives that bridge academia with the public, schools, and other educational institutions. A more visible and centralised platform should be established, building upon existing outreach efforts, to provide early career researchers (and beyond) with essential resources, such as materials, tools, and logistical support.
Outreach materials shared with the community should follow an open-source policy.
\end{recommendationbox}

%--------------- INDUSTRY ----------------------------------------------------------------------------------------------------------------------------------
\subsubsection{Communication with Industry}
\label{sub:comm_industry}

Engaging with industry is a critical component of promoting innovation, ensuring knowledge transfer, and maintaining long-term partnerships that benefit both the particle physics community and industrial stakeholders. 
In dedicated channels for dialogue and collaboration, institutes can ensure that industry partners are aware of potential contributions of particle physics to industrial innovation and the other way around. This involves organising regular meetings, workshops, and networking events where ideas can be exchanged and opportunities for collaboration can be identified.

Shared doctoral programs which allow researchers to gain experience in industry settings are an important way of fostering collaboration. These programs can be designed to align academic research goals with industry needs, providing ECRs with valuable skills while offering industry partners access to cutting-edge expertise. Such initiatives also help to bridge the gap between academia and industry, strengthening long-term ties. 

\example{Example}{
As one example, Centres for Doctoral Training (CDT) programs in the UK include a 6-month industry placement during the course of the PhD plus networking events with industry in general.
}

Identifying and promoting synergetic links between particle physics and industry can also unlock new opportunities for innovation. Technological advancements in particle physics, such as detector technology, computing, and data analysis, often have applications in fields like medicine, energy, and aerospace. Highlighting these connections can attract industry interest and investment, paving the way for impactful collaborations and as mentioned in Section~\ref{subsubsec:CareerShifts} these connections can also provide opportunities for ECRs looking for a career shift.

Dedicated forums for collaboration between particle physics and industry serve as a platform for the exchange of ideas, showcasing research, and exploring future initiatives. These forums can take the form of annual conferences, joint research symposia, or virtual platforms where scientists and industry representatives can share resources and expertise. By fostering an environment of open collaboration, such forums can lead to groundbreaking partnerships and innovative projects.

Targeted science communication to companies can help raise awareness of the potential applications of particle physics research. It is essential to craft messaging carefully to inspire continued engagement in particle physics and highlight the exciting opportunities within the field. Outreach efforts should focus on promoting partnerships and collaboration, emphasising the complementary roles of academia and industry rather than framing industry as an alternative to a career in particle physics.

\begin{recommendationbox}{}
The particle physics community should build upon existing collaborations with industry to further strengthen synergies. Expanding shared doctoral programs, enhancing knowledge transfer, and highlighting the societal and technological applications of research will be key. Additionally, dedicated fora should be reinforced or newly established to facilitate ongoing dialogue and foster long-term partnerships.
\end{recommendationbox}

%--------------- INTER-COMMUNITY ---------------------------------------------------------------------------------------------------------------------------
\subsubsection{Communication within the field and among ECRs}
\label{sub:comm_interfield}

Open communication within hierarchical structures in institutes and collaborations is crucial for ensuring that knowledge flows effectively across these different levels.
Initiatives should aim to address these challenges, which often act as barriers for ECRs.
Communication between experimentalists and theorists, as well as among experimentalists in different collaborations, should also be improved. Strengthening these links can lead to a more unified understanding of the field and help bridge gaps between theoretical predictions and experimental outcomes. As discussed in \Cref{subsubsec:Forum_and_non_collider_community_building}, missing links are a particular problem for beyond-collider collaborations, and a forum for exchanging knowledge and ideas in that sector would go a long way to improving the situation. Conferences addressing common challenges and progress across different experimental collaborations should be sustained, gathering together expertise from diverse experimental setups and promoting horizontal knowledge transfer. Similarly, theorists and experimentalists should communicate effectively in order to push research in the same direction.

\example{Example}
{A good example of this synergy is the Muon $g-2$ Theory Initiative~\cite{gminus2}}

Facilitating peer-to-peer exchanges and networking opportunities among ECRs is essential for building a supportive and collaborative community. Institutes and collaborations should organise dedicated forums, mailing lists and virtual discussion spaces to provide a platform for ECRs to share experiences, discuss challenges, and explore collaboration opportunities.
These channels would serve as hubs for information sharing, problem-solving, and building connections across experiments and institutions.

Cross-disciplinary communication and collaboration can unlock new perspectives and foster innovation by drawing on diverse expertise. To facilitate this, dedicated initiatives such as cross-disciplinary conferences and joint publications can be organised. These platforms would allow scientists from different fields to exchange methods, ideas and tools, ultimately increasing the efficiency and impact of their research.

\example{Example}
{A promising example of cross-disciplinary collaboration is the Joint \Gls{ECFA}--\Gls{NuPECC}--\Gls{APPEC} Activities (\Gls{JENAA}) committee, which acts as a bridge between existing organisations in accelerator and particle physics, nuclear physics, and astroparticle physics.} 

Moreover, this formula could inspire similar initiatives involving consortia for materials research, high-performance computing, engineering, and other relevant disciplines. Beyond such community-organised activities, funding agencies should encourage cross-disciplinary collaboration and the exploitation of synergies wherever reasonably possible.

\begin{recommendationbox}{}
Open communication within the scientific community must be maintained. Interdisciplinary initiatives should be supported, as they enhance knowledge sharing and foster opportunities for ECR networking, thereby maximising the impact of research.
\end{recommendationbox}

%--------------- SMALL-COLLABORATION -----------------------------------------------------------------------------------------------------------------------
\subsubsection{Small-collaboration outreach}
\label{sec:comm:small_collab}

Although science communication and outreach should generally reflect the broad diversity of the field of particle physics, it can be challenging for
%Building ECR communication infrastructure would also be beneficial to 
smaller collaborations to engage in outreach, as they may struggle to find the time and resources to promote themselves and their science objectives to the wider community and the general public. Therefore, many smaller experiments and theory collaborations have disproportionately fewer opportunities for such activities compared to large collaborations and should receive support to widen their reach and enable the public to learn about the interesting physics pursued within the whole field. % as well as its broad diversity, tailored programs should be established by funding agencies, national physical societies, and laboratories, to provide dedicated outreach specialists and training programmes for interested researchers. 

\example{Example}
{The Physics Beyond Colliders Initiative at CERN provides a dedicated outreach professional to help its member collaborations in outreach activities.}

\begin{recommendationbox}{}
Support should be given to smaller collaborations in outreach activities. This could include finding ways for small collaborations to organise outreach together, as well as making pre-existing outreach structures more inclusive to smaller collaborations.
\end{recommendationbox}

%--------------- SOCIAL MEDIA ------------------------------------------------------------------------------------------------------------------------------
\subsubsection{Social media outreach}
\label{sub:comm:social_media}

Over the last two decades, the influence of social media in our society has increased significantly. In particular for the young generation it has become a cornerstone of everyday life, shaping their communication and representing their main way of sharing and gathering information. Therefore, science outreach has also expanded into this realm, to appeal to and inform this young generation.
%Communication and information is nowadays increasingly happening via social media channels and reaching there also different parts of audiences then with more traditional media ways. This is especially true for the young generation, which could be attracted to the field of particle physics by effective communication via these means.
Achieving significant visibility for outreach platforms on social media, however, is a challenge for both experiments and institutions. Smaller experiments are particularly affected by this, often not having enough resources to create dedicated social media accounts for their outreach, therefore relying on the accounts of their host institutions to share their science. However, even larger collaborations and most institutions struggle to reach wide audiences on these platforms.

As a demonstration of the scale, CERN's Instagram profile has more than one million followers, while the profile of DESY, a large national laboratory in Europe, has only about eight thousand followers on the same platform. Further examples, as listed below, highlight that large collaborations such as the LHC experiments are also far from reaching a level of visibility similar to that of CERN.

\example{Example}
{Social media followers of selected European and international institutions (Instagram):
\begin{itemize}
    \item CERN (\texttt{@cern}) - 1M 
    \item Fermilab (\texttt{@fermilab}) - 46.3k
    \item DESY (\texttt{@follow.desy}) - 8.0k
    \item PSI (\texttt{@paul.scherrer.institut}) - 2.1k
\end{itemize}
and experiments of different sizes:
\begin{itemize}
    \item ATLAS (\texttt{@atlasexperiment}) - 69.2k
    \item CMS (\texttt{@cmsexperiment}) - 31.5k
    \item ALICE (\texttt{@alice\_experiment}) - 23.8k
    \item LHCb (\texttt{@lhcbexperiment}) - 17.9k 
    \item DUNE (\texttt{@dunescience}) - 2.7k
    \item LZ (\texttt{@lzdarkmatter}) - 1.2k
    \item NA62 (\texttt{@na62experiment}) - 1.0k
    \item NA61/SHINE (\texttt{@shine.experiment}) - 0.7k
    \item Belle II (\texttt{@belle2collab}) - 0.5k
    \item Mu2e (\texttt{@mu2eexperiment}) - 0.3k 
\end{itemize}
}

As a European community coming together to form a joint strategy for our future, we should also maintain joint outreach platforms with significant visibility to share European progress in particle physics. Such platforms, together with the right support, could enhance visibility and excitement for the fascinating research being performed, no matter how large the collaboration may be, or where an experiment might be based, in particular among younger generations.
%Effective science communication via these channels can allow the public and in particular younger generations to get excited about and inspired by the diversity within the community of particle physics.

The community should especially exploit the potential among ECRs who are naturally closer to social media and -- with appropriate training -- can take the role of ambassadors and role models for particle physics research.
As the heart of European particle physics and a centre equipped with a well-functioning outreach department, in particular with respect to social media, and a large follower base on its different channels, CERN should actively support this endeavour.

\begin{recommendationbox}{}
The European particle physics community should establish additional joint outreach channels on social media that highlight the science pursued in Europe no matter the size or location of the experiments. CERN, with its existing social media expertise and its large follower base on the different platforms, should actively support the efforts and ensure that shared content gains maximum visibility by actively promoting and distributing it.
\end{recommendationbox}

%--------------- STORYTELLING ------------------------------------------------------------------------------------------------------------------------------
\subsubsection{Changing the storytelling of future projects in particle physics}
\label{sub:comm_narrative}

As motivated before, science communication in general is extremely important to the future of our field. Of equal importance, however, is the storytelling put forward by outreach activities, particularly for activities aimed at non-specialists. The impact of the communication around the discovery of the Higgs boson is often underestimated. Inspired by the history of scientific success and knowledge advancements, future particle physics experiments are frequently framed around their potential for direct or indirect discoveries. However, in the short term, such breakthroughs are constrained by current technological limitations.

The previous generation of physicists were fortunate that advances in accelerator technology enabled the discovery of all the fundamental particles in the Standard Model in a relatively short time frame. In contrast, modern fundamental research increasingly focuses on the \textit{observation} of the universe at microscopic scales with unprecedented precision.

This approach to communication has been very successfully employed by the astrophysics community which has been very successful in framing its research mission as an exploration of the universe on a large scale.

\example{Example}
{In recent years, the James Webb Space Telescope \cite{webb} has generated a lot of enthusiasm in the general public, with its communication strategy being focused on the key concept of \textbf{observation}.}

Moreover, accelerators are often described as ``time machines" taking us back to the Big Bang conditions. Such statements, although very effective at attracting public interest, should be used with caution. The scientific community should promote a communication that is both accurate and engaging, avoiding misinformation or excessive simplification. Public interest should not be pursued at any cost, nor should the field and its research be presented in a unrealistic way. Communication should highlight the strengths of research in particle physics--such as its contributions to fundamental knowledge, technological innovation, and international collaboration--while also acknowledging its challenges.

The particle physics community must rethink not only the means of communication but also its substance. 
From leadership to public outreach we must emphasise the value of observational research. By rethinking our communication, we must ensure that future generations are involved in particle physics research both as protagonists and supporters. Free text answers by survey respondents, listed as ``other" reasons to not engage with the public, show that some ($<10$) ECRs are not highly motivated to promote future projects specifically, since they personally disagree with the current plans or ideas for the future of the field. These answers may suggest that the way in which the future of particle physics has been communicated from senior researchers to ECRs could lack vision and enthusiasm, propagating a sense of dissatisfaction through generations. This trend must be addressed, with the timeline of future projects offering the opportunity to implement a more inspiring storytelling.

\begin{recommendationbox}{}
The storytelling of particle physics research should emphasize the role of future experiments as particle observatories, rather than focusing solely on their potential for discovering new particles. Highlighting their ability to provide precise measurements, test fundamental theories, and explore unknown phenomena can reinforce enthusiasm within the field and enhance public support for fundamental research.
\end{recommendationbox}
\subsection{Future particle physics projects} %{\color{red}Frozen, contact Krzysztof before changes!}}
\label{theme:future_particle_physics}

The identification of the 125-GeV particle discovered in 2012 at the Large Hadron Collider as the Higgs boson presents an enticing possibility for marking the completion of the Standard Model of particle physics. However, many questions about this particle and, in general, our Universe at large remain unresolved.
These include, for example, the matter-antimatter asymmetry of the Universe, the nature of dark matter and the origin of dark energy.

Further investigation into the properties of the Higgs boson can offer answers to several of these subjects. Achieving this requires the construction of a next-generation particle collider, a goal underscored in both the previous ESPPU~\cite{2020_eppsu} and the recent P5 report from the United States~\cite{P5:2023wyd}. This new facility will also yield important insights into the top, electroweak, and flavour sectors and enable exploration of theories that extend beyond the Standard Model. Furthermore, essential breakthroughs can be made by the exploration of the precision or intensity frontier at dedicated beyond-collider facilities.

Early career researchers are the backbone of the future of particle physics. The current generation of ECRs is heavily involved in the ongoing LHC data taking and will play a prominent role in the upgrade and operation of the HL-LHC. After the HL-LHC program ($\sim 2040$), a new large-scale facility is expected which will inevitably be led by the current ECRs. Correspondingly, ECRs are deeply engaged in small- and large-scale beyond-collider experiments, both within pure particle physics and on the boundaries with nuclear, astroparticle and atomic physics. The current ECRs will also take leadership over future beyond-collider experiments, and direct the future development of the field. Therefore, we believe that the opinions and perspectives of ECRs are important components of the ESPPU and essential for the successful future of the field.

This section presents the recommendations from ECRs on several key aspects of the ESPPU, namely the significance of the future collider, particularly in Europe/at CERN, the importance of beyond-collider particle physics, the need to value technological innovation, and the vision of the ESPPU process and its decision-making. 
%The recommendations are based on discussions within the corresponding working groups and are supported by the results of the survey. 

\subsubsection{Importance of future colliders for early career researchers } %{\color{red}Frozen, contact Krzysztof before changes!}}

The future of particle physics depends on ECRs having long-term opportunities to contribute to fundamental discoveries, development of state-of-the-art technologies and advancements in data analysis techniques. As the operational lifetime of the HL-LHC is fixed, the question of what comes next becomes increasingly pressing. Without a clear roadmap for the post-HL-LHC era, the field risks losing talent due to limited career prospects and research opportunities.

Historically, major accelerators such as the Large Electron-Positron Collider (LEP) and its four major experiments provided a clear trajectory towards the LHC, ensuring that young researchers saw a future in the field, and a similar transition is crucial today. The LHC experiments have demonstrated the immense value of long-term projects, but without a next-generation collider, the knowledge and expertise built over decades will be lost. Beyond fundamental physics, a future collider would drive advancements in R\&D, applied physics, and hardware innovation. These technological spin-offs benefit society at large while maintaining a strong demand for skilled physicists.

The survey results clearly indicate that most ECRs strongly support the development of a next flagship collider, with \SI{50}{\percent} of 787 respondents strongly agreeing that it is important to have a future flagship collider in general and \SI{29}{\percent} rather agreeing with this statement. While \SI{12}{\percent} remain neutral, only \SI{10}{\percent} actively oppose it (see Figure~\ref{fig:fc-priority-next-flagship}), reinforcing the idea that a strong long-term vision is necessary to maintain engagement and ensure the future of the field. Ultimately, ECRs will be the ones working on the projects resulting from these decisions, making ECR involvement and support crucial for the success of any future collider. 

\begin{recommendationbox}{}A future flagship collider should be developed to further the progress of particle physics.\end{recommendationbox}

\subsubsection{Importance of a future collider at CERN } %{\color{red}Frozen, contact Krzysztof before changes!}} 

In recent years, CERN has been the heart of particle physics in Europe and a globally leading organisation in fundamental research. Securing its future with a next-generation flagship collider is essential, not only for continued scientific discovery but also for maintaining Europe's position as the leading hub for particle physics.
CERN's impact extends far beyond those who work directly with its facilities. As a centre of knowledge, it drives technological advancements that influence industries worldwide, from medical applications to computing and materials science. The transfer of expertise from CERN to industry has been instrumental in fostering innovation, demonstrating the far-reaching benefits of investment in fundamental research. Furthermore, CERN is a cornerstone of scientific collaboration across its member states and beyond, providing universities and research institutions with access to cutting-edge infrastructure and resources. 

Beyond its technical and economic contributions, a new collider at CERN would reaffirm Europe's commitment to fundamental science and international collaboration. It would serve as a beacon of scientific ambition, attracting top talent and reinforcing CERN's role as the global hub of particle physics -- as highlighted in the Draghi report~\cite{Draghi2024EU}. Without such a vision, there is a real danger of losing not only scientific momentum but also the expertise and innovation that have long defined Europe's leadership in the field. 

In addition to the previous recommendation, a majority of ECRs support locating the future flagship collider at CERN, with \SI{37}{\percent} strongly agreeing and \SI{32}{\percent} rather agreeing with this statement. While \SI{19}{\percent} remain neutral on the issue, only \SI{12}{\percent} oppose it (see Figure~\ref{fig:fc-priority-next-flagship}). This indicates a broad recognition of CERN's role as the natural host for the next flagship collider in Europe, ensuring stability and continuity for the field. It is also important to emphasise that this support comes not only from ECRs working on collider projects but from all communities concerned in the survey. However, the location of the next facility should not serve as a decisive factor in the decision-making process, as detailed in Section~\ref{sec:collider_criteria}.

\begin{recommendationbox}{}
A CERN-based location of the next collider facility would be highly appreciated, but it should not be used as the justification to prioritise any of the projects proposed during the ESPPU process. The European ECR community remains open to collaboration on any project, with a particular preference for those based in Europe.
\end{recommendationbox}

\subsubsection{Importance of early career researchers for future colliders } %{\color{red}Frozen, contact Krzysztof before changes!}} 
\label{subsubsec:importance_ecrs_for_fc}
ECRs are not just the future workforce of particle physics; they will dedicate their careers to designing, building, and analysing data from the next-generation collider. Whatever decision is taken as a result of this ESPPU, today's ECRs will live with its results for decades. This makes it crucial that future colliders are not only scientifically ambitious but also aligned with the interests and aspirations of the generation that will carry them forward. Otherwise, ECRs may disengage from the field.

A significant challenge in particle physics is knowledge retention. Many of today's experts in collider physics, accelerator design, and detector technology were trained in the early days of the LHC or even earlier during the LEP era. Without a clear next step, we risk losing this expertise as senior expert physicists retire, leaving a generational gap in critical skills. Ensuring a smooth transition means actively engaging ECRs in the planning and development phases of future colliders, giving them hands-on experience early on. Furthermore, if a lepton collider of any topology is chosen as the next flagship collider, an entirely new community of experts will be needed to secure its successful operation. Starting from scratch decades down the line is inefficient and an unnecessary risk. Instead, investment in training and skill development now will ensure a seamless transition to future machines.
Furthermore, ECRs also bring a fresh perspective to the field. While senior researchers make the key strategic decisions today, younger physicists must have a voice in shaping the projects for which they will ultimately be responsible. 

\begin{recommendationbox}{}Training related to upcoming facilities should be embedded within academic programmes and ECRs should play a vital role in future-collider research.
\end{recommendationbox}

\subsubsection{Criteria for a future collider } %{\color{red}Frozen, contact Krzysztof before changes!}}  
\label{sec:collider_criteria}

Survey respondents were asked to evaluate nine statements reflecting different priorities regarding the construction of the next collider facility by allocating 90 points among the following nine phrases: 

\textit{The next collider facility should...
\begin{itemize}
\item have an ambitious baseline physics programme (without upgrades). \hfill \textnormal{\textbf{[Physics Baseline]}}
\item have a well-defined long-term upgrade path. \hfill \textnormal{\textbf{[Upgrade Path]}}
\item minimise the time to first collision (timeline). \hfill \textnormal{\textbf{[Timeline]}}
\item be built at a specific location. \hfill \textnormal{\textbf{[Location]}}
\item drive technology R\&D and innovation. \hfill \textnormal{\textbf{[Innovation]}}
\item allow stable support for smaller projects. \hfill \textnormal{\textbf{[Smaller Project Support]}}
\item minimise the environmental impact (sustainability). \hfill \textnormal{\textbf{[Sustainability]}}
\item be open to world-wide collaboration. \hfill \textnormal{\textbf{[Collaboration]}}
\item maximise social/public acceptance (e.g. regarding cost and land use). \hfill \textnormal{\textbf{[Social Acceptance]}}
\end{itemize}}

The criteria were presented in a randomised order to each survey participant to avoid any ordering bias. To simplify the analysis, each statement has been short-labelled, as indicated in square brackets. The general results are given in Figure~\ref{fig:fc_priorities_all} and Table~\ref{tab:fc_priorities_all_ecrs}. They were cross-correlated with other answers in the Future Collider section and participants' biographies and several specific observations are highlighted in Figure~\ref{fig:fc_priorities_highlights}. Before drawing conclusions and giving our final recommendations, each criterion is discussed separately, in the order corresponding to participants' preferences, starting from the most important factors. All statements given in the following, including the ranking of the criteria, are based on the mean values of all responses for each criterion and subgroup of respondents. The $2\sigma$ condition, as defined in Section \ref{sec:analysis_procedure}, allows for an unambiguous ranking of all the criteria from the highest to the lowest priority. This holds for the complete set of responses as well as for the specific subgroup selections mentioned in the text.

\begin{itemize}
\item \textbf{Innovation} scored the highest in almost all considered communities, especially among engineers, computer scientists and accelerator physicists. One meaningful exception is the group of theorists and phenomenologists, among whom it was in third place. It is clear that this subgroup is not directly concerned with R\&D activities and thus, does not prioritise this criterion. However, the general vote shows that technical innovativeness should be a driving factor for the next collider facility, allowing for an exciting physics programme, offering many workplaces in neighbouring fields at the same time and providing links between science, industry and society at large.
\end{itemize}
\begin{recommendationbox}{}The selection of the next collider facility should be guided by its technological innovation as the driving factor. 
\end{recommendationbox}
\begin{itemize}
\item \textbf{Physics Baseline} is considered particularly important among theorists and phenomenologists. All answers considered, this criterion scored second and should be regarded as highly important for the decision-making process. The fact that this factor is also particularly supported by respondents who consider themselves well-informed might indicate that this group is more aware of differences between particular collider proposals.

\item \textbf{Collaboration}
scored among the most important criteria for all groups of participants. The respondents perceive the openness to worldwide cooperation at the future facility as a factor which should already be taken into consideration at the planning stage. 

\item \textbf{Sustainability} scored fourth among all categories of respondents, with slightly lower support from older participants. Minimising the environmental impact should be carefully considered when planning for the next facility, but not as the decisive factor. This may be connected to the fact that proponents of the major collider proposals have already been working on this issue at the current stage.

\item \textbf{Smaller Project Support} was typically perceived essential by people working on beyond-collider experiments (as defined in \Cref{subsec:survey_bio}), for example in the community of fixed-target experiments this criterion scored highest, 
however others did not give it a high priority. This factor is a prominent example of the importance of collecting data from different groups of people working in particle physics. 

\item \textbf{Upgrade Path} does not seem to be a crucial factor for most respondents, being significantly less important than the baseline programme. It is worth noting that in some cases, it is not clear how to distinguish between ``baseline" and ``extended" scenarios, which may be a source of confusion in this question. For this reason, we suggest focusing on the initial, clearly-stated programme when planning for the next facility.
\end{itemize}
\begin{recommendationbox}{}The next large-scale project should provide a comprehensive and exhaustive baseline programme. While a long-term vision for particle physics is important, the upgrade path should not be the primary motivation for building a specific collider.
\end{recommendationbox}
\begin{itemize}
\item \textbf{Timeline} scored third from lowest. The comparison between responses for scientists involved in future collider activities and those working in different fields reveals, however, that this factor is the third most important priority for the former. For the future collider community, minimising the time to the first collision is necessary to keep people involved and enable them to fully engage in the excitement of operating the collider for which they have spent years preparing. This aspect is of less importance for ECRs working on different projects.

\item \textbf{Social Acceptance} is necessary for any large-scale project to be successful, but this criterion did not receive much attention from any voting subgroup. Even though maximising support for the next collider from the outside is important, the survey shows it should not be a driving factor for the decision.

\item \textbf{Location} scored significantly lowest in the general voting, as well as among all the considered subgroups. Thus, we conclude that the European ECR community does not perceive this criterion as an important factor to be considered in the decision-making process.
\end{itemize}

\begin{recommendationbox}{}
Regardless of the location of the next major project, it is crucial to ensure its openness to global collaboration.\end{recommendationbox}

\begin{figure}[H]
    \centering
    \includegraphics[width=0.7\linewidth]{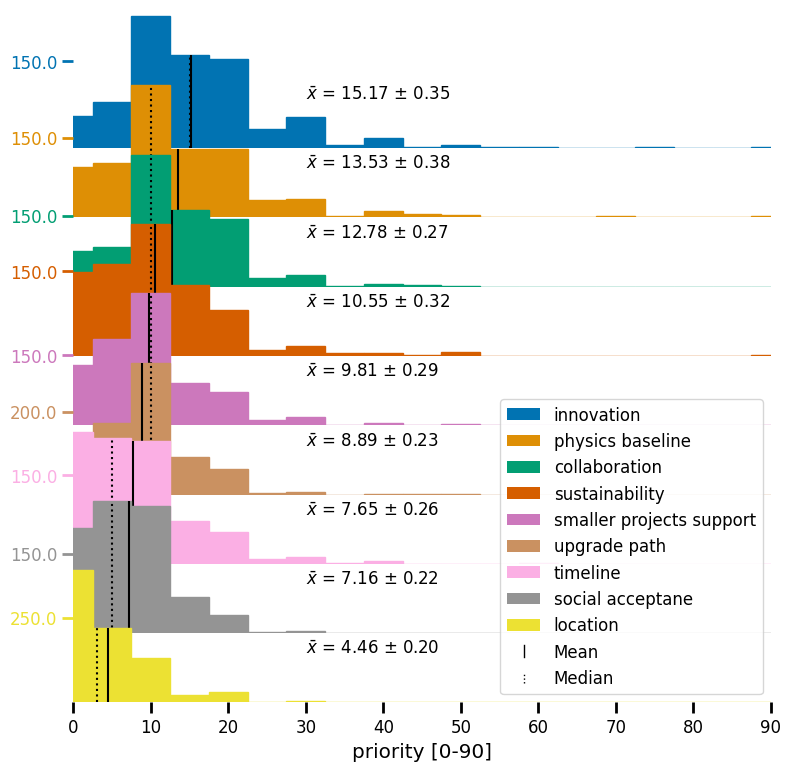}
    \caption{Ridgeline plot of the response distributions for all criteria ordered by the mean value of responses. The vertical solid line shows the mean value of all responses, while the vertical dotted line depicts the median of all responses. Note the differences in normalisation for the different criteria.}
    \label{fig:fc_priorities_all}
\end{figure}

\begin{table}[H]
\begin{tabular}{lrrrrrr}

 & mean & SEM & median & N & non-zero & max \\
\hline
innovation & 15.17 & 0.35 & 15.00 & 799 & 747 & 90.00 \\
physics baseline & 13.53 & 0.38 & 10.00 & 799 & 717 & 90.00 \\
collaboration & 12.78 & 0.27 & 10.00 & 799 & 735 & 50.00 \\
sustainability & 10.55 & 0.32 & 10.00 & 799 & 672 & 90.00 \\
smaller projects support & 9.81 & 0.29 & 10.00 & 799 & 686 & 75.00 \\
upgrade path & 8.89 & 0.23 & 10.00 & 799 & 672 & 50.00 \\
timeline & 7.65 & 0.26 & 5.00 & 799 & 598 & 60.00 \\
social acceptance & 7.16 & 0.22 & 5.00 & 799 & 627 & 55.00 \\
location & 4.46 & 0.20 & 3.00 & 799 & 455 & 30.00 \\

\hline\hline
physics baseline (Theory \& Pheno) & 15.79 & 0.91 & 13.00 & 169 & 154 & 70.00 \\
smaller projects support (Beyond collider exp.) & 14.67 & 0.94 & 13.00 & 137 & 123 & 70.00 \\
innovation (Theory \& Pheno) & 11.82 & 0.65 & 10.00 & 169 & 150 & 55.00 \\
timeline (Future Collider) & 11.41 & 0.93 & 10.00 & 99 & 85 & 40.00 \\
sustainability (age $>$ 40) & 8.41 & 0.92 & 10.00 & 70 & 54 & 40.00 \\

\end{tabular}

\caption{Table of statistics derived from the responses of all survey participants. The \emph{SEM} column reports the standard error on the mean. The \emph{non-zero} column reports the number of responses that were not zero. The rows below the double lines are the respective statistics for the data selections shown in Figure~\ref{fig:fc_priorities_highlights}. Note that the selection of \emph{age $>$ 40} is done on the set of all answers and includes non-ECR and non-\Gls{ECFA} participants.}
\label{tab:fc_priorities_all_ecrs}
\end{table}

\subsubsection{Innovation in particle physics research } %{\color{red}Frozen, contact Krzysztof before changes!}}
R\&D is crucial for the development of particle physics. Whichever new flagship project is built in Europe, its construction and operation will rely on technical advancements in fields such as detector and accelerator technologies, including magnets and RF cavities.
As \Cref{fig:fc_priorities_all} 
shows, ECRs involved in experimental research 
prioritise technological innovation as the driving factor for the next generation of particle physics experiments. 
The new technologies developed for pure particle physics research are often also synergistic with societal applications such as in imaging and medical technology, providing an important contribution to society from academia.

Despite this, ECRs are often not incentivised to take part in R\&D projects. This is for reasons such as the low-publication nature of technological developments, as well as the lack of recognition for this work, often driven by the lack of immediate benefit to physics projects since much of this work will require decades to reach maturity.

Small collaborations within European laboratories working on specific technological developments often experience similar challenges. Many of these collaborations rely on early career contracts, since a lack of funds means it can be difficult to hire and retain more senior researchers. ECRs in these collaborations must therefore be supported in this work, otherwise the field risks losing a large fraction of the workforce in this area.

\begin{recommendationbox}{}
European laboratories should assign greater value to innovation in preparation for the next generation of particle physics facilities. ECRs and projects driving technological innovation should be recognised and supported, to benefit both particle physics and broader societal applications.
\end{recommendationbox}

\subsubsection{Recommendation on specific collider } %{\color{red}Frozen, contact Krzysztof before changes!}} 

As explained in Section \ref{sub:survey_topics_FC}, the question on the collider choice was particularly challenging due to possible biases resulting from the composition of ECR communities this survey reached. For completeness, we report the result, asking the reader to take this into account. Before analysing the survey data, the WG collectively decided that the results would be reported exclusively in numerical form to avoid the possible simplification of all the considerations listed below by presenting only one plot.

After elaborated discussions in the WG, the following options were identified: ``a circular $e^+e^-$ collider (e.g. FCC-ee)", ``a hadron collider (e.g. \Gls{FCC}-hh)", ``a linear $e^+e^-$ collider (e.g. \Gls{CLIC}/\Gls{ILC}/\Gls{C$^3$})", "a \lowercase{\gls{muon collider}}", "any collider, as soon as possible", "I do not support the idea of building a collider in Europe", "I do not know / I do not have a strong opinion". Of all 782 answers considered, a circular $e^+e^-$ collider received \SI{28}{\percent} of the votes, followed by a \lowercase{\gls{muon collider}} at \SI{15}{\percent}, a hadron collider at \SI{14}{\percent}, and a linear $e^+e^-$ collider at \SI{8}{\percent}. Additionally, \SI{23}{\percent} of respondents expressed that they do not have a strong opinion or do not know, while \SI{9}{\percent} supported the idea of any collider being built as soon as possible. Only \SI{2}{\percent} of participants opposed the concept of constructing a collider in Europe. 

Notably, no single collider proposal received an absolute majority mandate from the community, which might potentially pose a challenge of forming a vivid collaboration, as discussed in \Cref{fc:transparencyinESPPU}. Some parts of the community might be hesitant to join the chosen collider project for various reasons. The community has to be aware of that. Most of the collective knowledge in the field of future collider research is transferable, as proved by projects crossing boundaries of particular proposals, e.g. Key4hep.
In order not to lose any valuable expertise when choosing a specific collider, efforts should be made to engage people who have already been working on HL-LHC or other future projects.

\important{Statement: Collaboration on the future flagship project}{Regardless of which collider is selected by the Strategy Group as the European priority, its advocates will need to convince the other communities to join the effort and foster a dynamic, collaborative group of scientists committed to advancing the project together.}

It is also meaningful to analyse this question in comparison with other answers of the respondents. 
Participants who report lower knowledge of the future collider landscape tend not to have strong opinions about the choice; out of 159 participants who identify as ``not well informed" or ``not informed at all" about \SI{50}{\percent} chose the option ``I do not know/I do not have a strong opinion". It is also evident that projects not promoted as a ``baseline" scenario for CERN\footnote{The term ``baseline" is used, for example, by the Strategy Secretary to refer to the \Gls{FCC} project~\cite{talk_PECFA}.} receive greater support from better-informed respondents. Interestingly, the ``very well-informed" participants (74 respondents) show similar levels of support -- around \SI{30}{\percent} -- for both the circular $e^+e^-$ and \lowercase{\gls{muon collider}}s, while those 159 participants who identify as ``not well informed" or ``not informed at all" tend to prioritise a hadron collider slightly.

Further cross-referencing responses with respondents' respective experiments reveals interesting patterns. For example, respondents working on heavy-ion collisions favour the idea of building a hadron collider over other proposals. Similarly, responses such as ``I do not know / I do not have a strong opinion" or ``I do not support the idea of building a collider in Europe" are significantly more common within beyond-collider communities, reaching up to about \SI{36}{\percent} and \SI{6}{\percent} among 141 participants working on fixed-target experiments, neutrino physics, EIC or direct dark matter detection. Notably, among all future-collider communities, the project most closely related to a participant's work typically gains around \SI{60}{\percent} support, with the second-highest priority being ``any collider, as soon as possible".

\important{Statement: Support for specific future collider}{The main collider proposals, a circular $e^+e^-$ collider, a muon collider, a hadron collider and a linear $e^+e^-$ collider, have received recognition from the ECR community. A relative majority prefers a circular $e^+e^-$ collider, closely followed by the option ``I do not know/I do not have a strong opinion".}

On the basis of these results, we deliberately refrain from recommending any specific collider facility and urge the Strategy Group to decide this matter based on different proposals submitted during the process, considering thoroughly the aspects important for the ECR community, as described in Section~\ref{sec:collider_criteria}.

\begin{recommendationbox}{} 
A clear recommendation on the next flagship collider for Europe, based on the submitted proposals, should be given in this ESPPU process. The ESPPU should urge the CERN Council to make a timely decision on the next flagship collider and efforts should be made to ensure that every member of the community feels invited to contribute to the project, bringing their expertise and experience to serve the advancement of our field.
\end{recommendationbox}

We also asked participants what the next flagship project in Europe should be if \Gls{CEPC} is approved in China. The responses did not point to a clear solution. Most of the 784 respondents indicated that any collider project complementary to \Gls{CEPC} would be a good option, with a closely followed preference for ``a hadron collider" and ``a \lowercase{\gls{muon collider}}", each receiving around \SI{20}{\percent}. \SI{8}{\percent} of participants selected either a linear $e^+e^-$ collider or a circular $e^+e^-$ collider, with the numbers being similar for both. In total, about three-quarters of respondents stated that Europe should pursue a different path and refrain from building a similar facility. Interestingly, the idea of building a collider in Europe under these circumstances was rejected by a similar proportion of \SI{2}{\percent} of respondents, as in the general question on the collider choice. \SI{11}{\percent} of participants chose ``I do not know / do not have a strong opinion", which represents only half of the responses in the previous question. Only \SI{2}{\percent} of respondents favoured a project other than those mentioned above, for example, a suite of beyond-collider experiments.
Similarly, we asked what the next priority should be if \Gls{ILC} is approved in Japan. The 778 responses did not differ significantly in this case. The idea of building another linear collider in Europe was supported by only  \SI{2}{\percent} of participants. Options such as ``any collider project complementary to \Gls{ILC}," a \lowercase{\gls{muon collider}}, a hadron collider, and a circular $e^+e^-$ collider all received similar support, with about \SI{20}{\percent} of participants endorsing each. The other categories garnered comparable levels of response as in the previous question. Due to the overall length of the survey, we refrained from asking further questions, e.g. regarding the \lowercase{\gls{muon collider}} in the US.

\begin{recommendationbox}{}
    If a major collider project is approved outside Europe, the European community should start the construction of a complementary collider project.
\end{recommendationbox}

Participants were also asked whether they would support the outcome of the ESPPU, even if their preferred collider was not chosen as a priority. \SI{84}{\percent} agree or rather agree with this statement, even ECRs engaged in future collider studies themselves, \SI{83}{\percent} of whom agreed. Overall, only \SI{5}{\percent} of respondents were against supporting an unfavourable decision (see Figure~\ref{fig:fc-support-non-favorite}). This indicates strong cohesion in the ECR community. 
At the same time, no member of the community should be overlooked, and efforts should be made to provide them with alternative opportunities that best serve the advancement of particle physics.

\important{Statement: Potential support for unfavourable decisions}{A significant majority of European ECRs are willing to support the ESPPU outcome, even if their preferred collider option is not prioritised.}

\subsubsection{Transparency and consensus building in the ESPPU \label{fc:transparencyinESPPU}  } %{\color{red}Frozen, contact Krzysztof before changes!}}

A number of different options are being discussed for the next flagship collider at CERN.
This discussion is not new, but it is evident that in this strategy update the community needs to come to a decision.
ECRs are split among the different proposals, and no single project exceeds a preference of \SI{28}{\percent}, or \SI{38}{\percent} if ``any collider, as soon as possible'' is added.
Therefore, we advocate for a transparent and fair decision-making process.
A lot of convincing and convergence is still necessary to unite behind a single proposal.

There is, however, an impression that this decision, which concerns the future of CERN and the whole of particle physics in Europe, is pre-determined and being driven by a limited number of people rather than the entire community. For example, the \Gls{FCC} is increasingly often referred to as the ``baseline" project (see e.g.~\cite{talk_PECFA}), despite the facts that firstly, it is still undergoing a \textit{feasibility study}, i.e.\ determining \textit{whether} the project is feasible; and secondly, the last ESPPU has not concluded or determined a ``baseline". 
This impression is detrimental to the process and to the consequences of its outcome.

We advocate for democratic participation of the whole community in the decision-making process. It was for this reason that we conducted a survey among ECRs to collect their opinions --- one scientist, one voice. This of course implicates the duty of representativeness, which we took as a major task in distributing the survey, albeit a feasible one. We recommend implementing democratic methods in the same spirit for the whole ESPPU.

An important part of transparency is laying out the criteria that contribute to the decision and their relative weights.
This includes criteria such as physics reach, upgrade path, cost and sustainability.
Moreover, considerations of technical maturity not only apply to the accelerator, but also to detector technology and simulation and analysis software used for physics studies.
A similar recommendation has already been voiced in the report of the European ECRs in the last ESPPU (\cite{ECRInput2020}, 3.6.1).
Real transparency also includes less obvious criteria, like socio-political considerations.
It is much less clear to what extent e.g.\ securing the future of CERN and preserving its size, as well as preferences from funding agencies or national governments, impact the decision-making.
\begin{recommendationbox}{}The process towards defining the European strategy for particle physics must be more transparent and democratic. After the decision has been made, a structured explanation of the criteria which led to a certain result is necessary.
%This will significantly help in understanding the procedure and building trust for the outcome within the whole community.
\end{recommendationbox}

Another aspect of democratic participation appears when considering the following:\\
``This time, we [...] absolutely must converge on [...] the next collider at CERN [...].
There is no room for disagreements after we converge.
The fastest way to getting no new collider is non-convergence of the community [...].'' \cite{ECFA_WS3_Sphicas_Talk}

While this may be true, it creates significant pressure to converge behind one project, since otherwise there may not be any, while it does not specify the way that convergence is achieved or how to ensure that every voice is heard. Combined with the impression of the pre-defined decision, this has the potential to create resentment.
A fair and inclusive selection process, along with the requested well-defined plan B, C, D, etc. is best suited to create acceptance for the decision among proponents of the alternatives.
It is paramount to act in the decision-making process to unify the community independently of the result.
An inversion of the ``pressure rhetoric'' may also be worthwhile to consider: rather than the pressure on plan-B proponents to get behind the leading project, the pressure should be on leading-project proponents to get everyone behind theirs.

\begin{recommendationbox}{}A fair and inclusive decision process is essential to create acceptance for the decision among proponents of the future collider alternatives and is vital for trust in the procedure.
Proponents of the leading project must work to gather support for their project.
\end{recommendationbox}

\subsubsection{Other aspects of future colliders } %{\color{red}Frozen, contact Krzysztof before changes!}} 

When asked how far ahead the particle physics community should plan in this strategy update, a significant majority of the survey participants are neither in favour of short-term, nor very long-term planning. A relative majority of \SI{43}{\percent} supports planning until 2045–2065, while \SI{29}{\percent} prefer 2035–2045. Fewer respondents favour either shorter (\SI{10}{\percent}) or longer horizons (\SI{9}{\percent}), with \SI{8}{\percent} being uncertain. Full results are presented in Figure~\ref{fig:fc-strategy-reach}.

Beyond these findings, the debate over funding allocation remains a crucial issue. The survey participants were asked about preferred funding allocation for flagship and smaller projects, assuming the total amount of money was fixed. The results indicate that \SI{39}{\percent} of respondents supported increased funding for the flagship collider, and \SI{20}{\percent} favour a similar increase for smaller projects, as shown in Figure~\ref{fig:fc-funding-split}. Meanwhile, \SI{34}{\percent} prefer maintaining the current balance and \SI{7}{\percent} were uncertain.
Achieving the right balance between these perspectives will be essential for optimising the use of limited resources and ensuring a dynamic, future-oriented research environment.

\begin{recommendationbox}{}The budget split between flagship collider and smaller experiments should not be changed significantly.
\end{recommendationbox}

Additionally, any future flagship collider has the potential to serve as a host facility for a variety of smaller projects, much like the dedicated LHC experiments such as FASER, LHCf, MoEDAL-MAPP, SND@LHC, and TOTEM, which have successfully extended the experimental programme of the LHC. It is important to note that several proposed future collider facilities present unique opportunities for experiments beyond their primary interaction points.
For instance, they could enable the study of neutrino properties from muon decays, the exploration of strong-field QED by dumping high-intensity beams, or the search for exotic particles.
The potential for hosting complementary experiments should be emphasised in the proposals and considered by the ESG.

\begin{recommendationbox}{}The physics potential of any future collider option should be maximised by considering smaller-scale specialised experiments in addition to the main interaction-point experiments.
\end{recommendationbox}

Another key aspect to consider is the willingness of respondents to relocate for work on future collider projects. A significant majority of survey participants would be willing to move to CERN (\SI{87}{\percent}). Moving to the U.S.\ received moderate support (\SI{26}{\percent} in favour and \SI{47}{\percent} against), with an overall tendency against it, while attitudes toward relocating to Japan (\SI{25}{\percent} in favour and \SI{49}{\percent} against) were very similar to those for the U.S. In contrast, a significant majority (\SI{82}{\percent}) is reluctant to relocate to China (see Figure~\ref{fig:fc-mobility})\footnote{Note that the survey ended on the 27th of January and may not reflect any political developments since then.}.

Further aspects of future colliders assessed in the survey concerned sustainability, more intense collaboration with CERN non-member states and number of interaction points (IPs) in the next flagship project. These aspects could be crucial for securing stable funding for the large-scale facility. A significant majority of participants (\SI{74}{\percent}) agree with the statement ``the next collider should be built and run in the most sustainable way'', while only \SI{9}{\percent} disagree. A majority (\SI{58}{\percent}) supports strengthening the collaboration with non-member states, such as sharing collider data for analysis, with only \SI{14}{\percent} against it. Finally, to nearly \SI{50}{\percent} of the respondents, it is important that the next flagship project has more than two IPs, while \SI{15}{\percent} does not consider it vital.  For more details, refer to Figure~\ref{fig:fc-specific-prios}.

\begin{recommendationbox}{}The next flagship collider should be built and run in the most sustainable way. Collaboration with institutions from countries other than CERN member states should be increased. 
\end{recommendationbox}

\subsubsection{Importance of beyond-collider particle physics in the European landscape } %{\color{red}Frozen, contact Krzysztof before changes!}}
\label{subsubsec:importance_beyond_collider_physics}

A significant part of the particle physics community investigates phenomena that are explored through means other than high-energy collider experiments. This includes the field of neutrino physics, as well as several aspects of direct and indirect dark matter searches, studies of QCD and extreme baryonic matter, or precision tests of QED and fundamental symmetries which may be better probed at smaller, dedicated experiments.

A defining feature of particle physics beyond colliders is the remarkable diversity of experiments in scale, required infrastructure, and project duration. As discussed in \Cref{subsec:WG-BC}, experiments range from compact ``tabletop'' setups conducted in university laboratories to large-scale research facilities requiring dedicated underground sites, extensive detector arrays, accelerators, or research reactors. 
These experiments and the associated infrastructure are often highly interdisciplinary, providing valuable results not only for particle physics, but also for nuclear physics and astroparticle physics.

Furthermore, results from beyond-collider experiments can provide valuable guidance for collider research by providing complementary studies. Specific fundamental questions cannot be exclusively addressed by general-purpose experiments at a future flagship collider. 
Since the exact nature and energy scale of beyond-standard-model physics remains unknown, dedicated and specialised beyond-collider experiments are critical in broadening the search. They enable the exploration of models and parameter spaces that may be inaccessible to even the most powerful future colliders, ensuring a more comprehensive approach to new physics discovery.
Moreover, beyond-collider results are essential for the interpretation of high-energy collider data. For example, precision measurements of soft-QCD and QED processes provide crucial input for modelling collision dynamics and particle interactions within detectors, which is required to enhance the accuracy of theoretical predictions and analysis of collider data.

While collider and beyond-collider experiments typically address different physics questions, they share fundamental detector technologies and expertise.
When one field develops new software tools or analysis methods, these innovations often lead to important advances in the other. Supporting both experimental approaches in parallel yields benefits more significant than the sum of their scientific programs, creating a rich ecosystem where advances in methodology, technology, and expertise naturally flow between communities and accelerate progress across the field of particle physics.

Therefore, maintaining and potentially broadening the efforts within the beyond-collider domains should continue to be a focus of the European particle physics community. When asked in the survey, a significant majority of respondents to the supplementary question said that the participation in beyond-collider activities at CERN 
should be increased (see Figure~\ref{fig:bc_q3_beyond_colliders_generally}). In particular, a majority would like to see an increase in research in medical physics and a significant majority thinks an increase in novel accelerator R\&D beyond next-generation colliders is needed (see Figure~\ref{fig:bc_q3_subfields_part2}). For all other sub- and adjacent fields, namely fixed target, neutrino physics, direct dark matter searches, nuclear physics, and astroparticle physics experiments, a significant majority is of the opinion that the current level of involvement should be at least maintained (see Figures~\ref{fig:bc_q3_subfields_part1}-\ref{fig:bc_q3_subfields_part2}).

\begin{recommendationbox}{}Beyond-collider experiments and activities should maintain a prominent role in the European particle physics landscape, both as groundbreaking activities in their own right and as pathfinders for collider searches.
Their diversity in scale, infrastructure and duration should be valued and sustained in order to maintain a thriving beyond-collider landscape.
\end{recommendationbox}

\subsubsection{Maximising the yield of European beyond-collider particle physics } %{\color{red}Frozen, contact Krzysztof before changes!}}
\label{sec:futurepp:maximizing_yield}

A common problem that many beyond-collider experiments face is the difficulty to procure funding, especially through schemes where they have to directly compete with major flagship projects. This issue is underlined by the survey conducted within the scope of the white paper, where \SI{55}{\percent} of respondents answered that the ``Unclear financial situation due to lack of long-term funding commitment'' is one of the three most important challenges for beyond-collider experiments in Europe (see top plot in Figure~\ref{fig:bc_question1}). This was the single biggest challenge according to the survey. Furthermore, \SI{41}{\percent} responded that ``Difficulty to compete for funding with large flagship projects'' was one of the main challenges for these experiments, making it tied for the place of the second biggest challenge. Interestingly, when analysing only the answers of ECRs working in beyond-collider experiments, the percentage for the first answer stayed roughly the same at \SI{54}{\percent} but the second answer received more weight with \SI{54}{\percent} as well (see bottom plot in Figure~\ref{fig:bc_question1}), showing that funding procurement is perceived as a major challenge for beyond-collider experiments.

The survey also highlighted that this is a topic which ECRs consider to be very important, with a significant majority (\SI{81}{\percent}) expressing that ``A dedicated funding scheme for beyond-collider particle physics would benefit beyond-collider particle physics experiments in Europe'' rather much or significantly (see Figure~\ref{fig:bc_forum_funding_scheme}). This would give ECRs working in this field more certainty of continuity within their domains and raise their career prospects, thereby elevating all of beyond-collider particle physics.
\begin{recommendationbox}{}A dedicated beyond-collider funding scheme should be put in place to strengthen the European beyond-collider community by keeping talent in the field through ensured continuity and therefore greater career prospects.
\end{recommendationbox}

Apart from the diverse range of collider and beyond-collider particle physics conducted at CERN, a wide variety of particle physics is studied at smaller European laboratories.
Some of these experiments take place at facilities where particle physics is not the primary focus. Instead, these institutions may specialise in fields such as astronomy, nuclear physics, materials science, or life sciences.
Such institutions include, but are certainly not limited to, neutron sources, free-electron laser and electron-synchrotron facilities, as well as nuclear reactors. Neutrino telescopes fall between the two categories, with a dual focus on astrophysics and neutrino physics.

These laboratories often provide unique experimental opportunities focused on a particular question in particle physics. 
Additionally, smaller-laboratory research environments are particularly important for the ECR community as places to grow and distinguish themselves as independent researchers.
In order to enable the continuation and growth of these environments, it is imperative that the European particle physics community acknowledges the importance of smaller laboratories in the ESPPU.

A significant fraction of the particle physics experiments that are conducted at other laboratories are interdisciplinary, spanning into e.g.\ astroparticle or nuclear physics, and are crucial for knowledge exchange across scientific disciplines.
Given these experiments’ interconnected nature, the ESPPU should emphasise a well-integrated program that leverages shared infrastructure with neighbouring disciplines. To maximise the efficient use of shared resources, recommendations in the ESPPU should be developed in synergy with those from NuPECC and APPEC, and vice versa. 

Similarly to the recommendation made in \Cref{subsubsec:importance_ecrs_for_fc}, training should also be given to ECRs on facilities other than the major colliders where particle physics research is performed. Well-rounded ECRs with a strong knowledge of the field are fundamental for the strong interdisciplinary collaborations needed for the future of particle physics.

\begin{recommendationbox}{}The ESPPU should acknowledge the importance of smaller laboratories for research in fundamental physics, and explicitly recommend the innovative usage of non-particle physics facilities for this purpose. 
ECRs should receive training on particle physics research done at non-collider facilities to be able to use them efficiently in their research. Furthermore, to maximise the efficient use of such shared resources with neighbouring fields, recommendations in the ESPPU should be developed in synergy with those from \GlsBlack{NuPECC} and \GlsBlack{APPEC}, and vice versa. 
\end{recommendationbox}

Furthermore, regardless of the location, investment in smaller, beyond-collider experiments can ensure the realisation of meaningful and challenging projects during the transition period between the HL-LHC and the next flagship collider. Often only requiring a small fraction of CERN's budget, such experiments have the potential to deliver groundbreaking science while also ensuring the retention of ECRs in the field. At the same time, additional experience in smaller beyond-collider experiments during this transition phase will allow collider ECRs to grow and to diversify their knowledge.
These ECRs will then later have the choice to continue in the beyond-collider community, or to return to collider projects, once they are finalised. This flow of talent between the collider and beyond-collider communities will strengthen the bond between these communities and allow for more organic exchange of knowledge, methods and talent, no matter where they originate from, making it easier also for other beyond-collider ECRs to move to the collider domain. As shown in the survey results (see Figure~\ref{fig:bc_career_paths}), a significant majority believes that ``Easier career transitions between collider and beyond-collider experiments" would benefit beyond-collider experiments in Europe.

Since these efforts beyond colliders will be vital during the interim period between collider programmes to retain talent and knowledge, a clear plan should be laid out. Although many of these experiments are comparatively small in scale and therefore faster to realise, they still require significant preparation and lead time through the development of any new infrastructure or novel instrumentation which may be needed, as well as the time necessary for project approval and the procurement of funding.

\begin{recommendationbox}{}A concrete plan for particle physics in Europe between the end of the HL-LHC and the next major flagship project should be laid out within the scope of the next ESPPU.
Such a plan should contain a dedicated beyond-collider programme for this period that allows for the retention of knowledge and talent. 
\end{recommendationbox}

\subsubsection{Importance of the ESPPU for ECRs in beyond-collider particle physics } %{\color{red}Frozen, contact Krzysztof before changes!}}
\label{sec:ESPPU_importance_bc}

The ESPPU is expected to present a concrete, united European strategy for future collider activities. 
Correspondingly, the ESPPU should pose concrete strategies for experiments and activities in the beyond-collider domains.
This would ensure that European universities, institutes and laboratories (including CERN), have clear recommendations about which activities to prioritise, along with a basis for measuring progress in adopting the recommendations.
Additionally, it is regarded that European-level strategic documents, such as the ESPPU, carry weight with national and international funding agencies for individual projects, which is of great importance for the often small-scale beyond-collider experiments.
It is imperative to support a strong and sustainable field of beyond-collider physics, regardless of decisions about the next flagship project. This can only be achieved by clearly defining an ambitious roadmap for the future.

The dedicated strategies should define European priorities within each subfield (neutrino physics, astroparticle physics,
the intensity and precision frontiers, etc.), including the most pressing measurements, approaches and general experimental efforts, as well as required theoretical progress.
These recommendations should be thoughtfully produced by the \Gls{PPG} and \Gls{ESG}. 

Such recommendations are particularly important to the ECR community, as they will help guide their future career choices and provide concrete motivations to present to funding agencies. A majority of all survey participants (\SI{68}{\percent}), and a significant majority (\SI{79}{\percent}) of beyond-collider physics respondents, agree that beyond-collider particle physics experiments in Europe would benefit from concrete recommendations in the ESPPU for the beyond-collider fields either significantly or rather much (see Figure \ref{fig:bc_recommendations}).

\begin{recommendationbox}{}
The ESPPU should include concrete recommendations for beyond-collider particle physics research, in order to facilitate the growth and advancement of these fields in Europe irrespective of the choice of the future flagship collider project.
\end{recommendationbox}

\newpage
\section{Conclusions and next steps throughout the ESPPU} % {\color{red}Frozen, contact Armin before changes!}}
\label{sec:conclusions}

This white paper summarises the views of the European ECR community in particle physics and related fields. The ECRs of today will be the ones conducting the future research of tomorrow. Thus, the ESPPU must consider their views, ideas, and concerns. We appreciate the opportunity to have our voices heard. Implementing the recommendations and taking into account the further statements for the ESPPU further strengthens the ECR community --- ensuring a bright future for particle physics in Europe and beyond. 

This white paper has been endorsed by the \Gls{ECFA ECR Panel}. The full list of supporters is available at \href{https://indico.cern.ch/e/espp-ecr-white-paper-support}{https://indico.cern.ch/e/espp-ecr-white-paper-support} where everyone, including other panels or organisations, can show their support of this document by signing it. The editors and authors will disseminate the messages of this document through presentations to the particle physics community and will follow and participate in the ESPPU process.

We hope that this strategy input is useful to the \Gls{ESG} when drafting their summary document and beyond to showcase the opinions, concerns, and ideas of the enthusiastic community of European early career researchers in the field of particle physics.

\newpage
\phantomsection\addcontentsline{toc}{section}{Acknowledgements}\section*{Acknowledgements}

The authors thank \Gls{ECFA} for their help in the organisation of the ESPPU ECR discussions adjacent to the ECFA workshop in Paris and at the 115th Open Plenary ECFA meeting at CERN. Further thanks go to the \Gls{ECFA ECR Panel}, the LHC Early-Career Scientist Fora, the CERN EP and Theory departments, and everyone else who helped circulate the survey and advertise our efforts. 
We also greatly appreciated the feedback we received from the ECR community as well as the external reviewers. 
We also thank all the ECRs from other regions of the world for sharing their thoughts with us as well.
Finally, we want to thank everyone who gave input throughout the process, may it be over coffee, by text message, during meetings, or by filling out the survey.

\newpage
\phantomsection
\addcontentsline{toc}{section}{References}
% \setcounter{page}{10} % Comment out for extensive document
%%% References
{
    \footnotesize
    \bibliographystyle{JHEP}
    \bibliography{biblio} % Comment out if printing executive summary only
    % \bibliography{biblio_exec} % Use this to print executive summary only

}
\endgroup

%%% Appendix
\renewcommand\thefigure{Appendix.\arabic{figure}}
\renewcommand\thetable{Appendix.\arabic{table}}
\appendix
\section{Definition of geographical regions}
\label{app:geo-regions-def}

For survey questions relating to countries, the following groups are used to define regions with a larger sample size:
\begin{itemize}
    \item \textbf{Northern Europe}: Austria, Belgium, Denmark, Finland, Germany, Ireland, Netherlands, Norway, Sweden, Switzerland, United Kingdom  
    \item \textbf{Mediterranean}: France, Greece, Italy, Portugal, Spain  
    \item \textbf{Central and Eastern Europe}: Belarus, Bosnia and Herzegovina, Bulgaria, Croatia, Czech Republic, Georgia, Hungary, Lithuania, Moldova, Poland, Romania, Russia, Serbia, Slovakia, Slovenia, Ukraine  
    \item \textbf{North America}: Canada, Mexico, United States of America  
    \item \textbf{Asia}: China, India, Iran, Israel, Japan, Kazakhstan, Korea, Malaysia, Pakistan, Taiwan, Thailand, Turkey, Vietnam  
    \item \textbf{South America}: Argentina, Brazil, Chile, Colombia, Guatemala, Peru, Venezuela  
    \item \textbf{Africa}: Algeria, Egypt, Morocco, Sudan, Zambia  
    \item \textbf{Oceania}: Australia, New Zealand
\end{itemize}
Countries not included here had no responses. Where CERN was listed as a host it was grouped with Northern Europe.

\section{Survey anonymity and data protection}
\label{app:data-prot}

The following disclaimer was displayed to survey respondents before starting the survey:

\begin{quote}
\textit{\textbf{Anonymity and data protection}:
The questions are asked to enable comparisons of responses across individuals and groups in order to understand the views of the community as a whole, but also the different opinions and experiences of people with different identities. All of the responses are fully anonymous, and results will only ever be shared in aggregate form, thus ensuring your privacy. There are some text boxes, but in such cases we would encourage you to provide answers while keeping the replies anonymous (e.g. not mentioning names). The selection of question in this survey ensures the anonymity of all participants, which is maintained all the time. The data are collected on CERN hosted servers running the LimeSurvey tool. The data will be only used to accomplish the purpose of this survey, and will be removed after performing a statistical analysis. The final results will make it impossible to track down individual answers, and will be released in a form of publicly available report. All information should be provided on a voluntary basis. By taking this survey, you agree to this privacy policy.  }
\end{quote}

\section{Further survey plots}
\label{sec:survey-plots}

\subsection{Career prospects and ECR leadership}

\begin{figure}[H]
    \centering
    \includegraphics[angle=0,width=.9\textwidth]{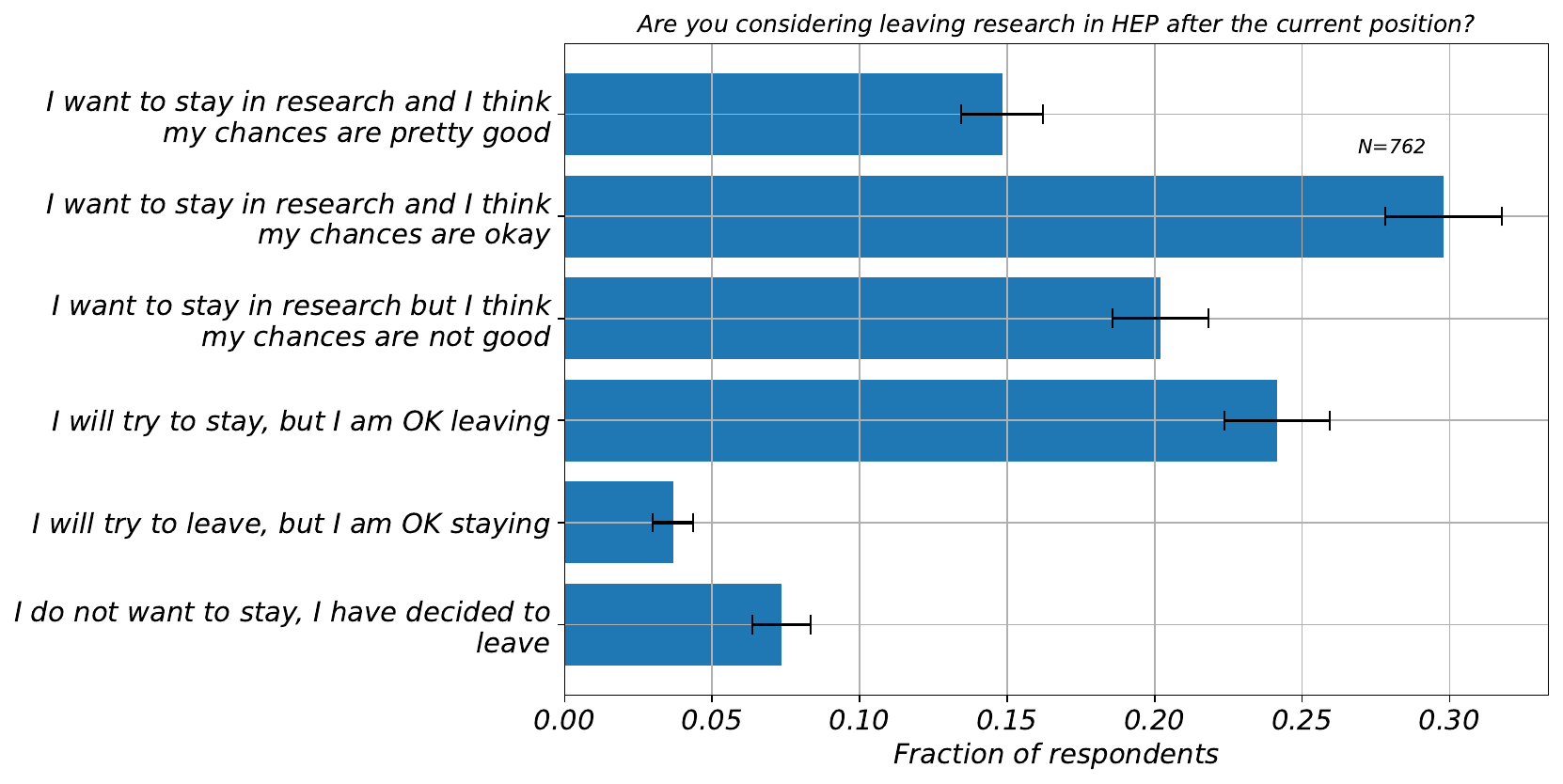}
\caption{Proportional responses to the question: ``Are you considering leaving research in HEP after the current position?'' \textit{Note:} The term `HEP' has been retained for consistency with the previous survey~\cite{Allen:2024lyp} but is intended to refer to the broader particle physics community. Based on the collected data, there is no indication that participants misunderstood this intent. } % \color{blue}(Plot needs larger labels for the ticks, as well as the axes. A size a bit larger than the plot suptitle should be okay.)\color{black}} 
    \label{fig:career_future-in-academia}
\end{figure}

\begin{figure}[h]
	    \centering
	 	 \begin{subfigure}{.51\textwidth}
	 	 	\centering
	 	 	\includegraphics[width=\linewidth]{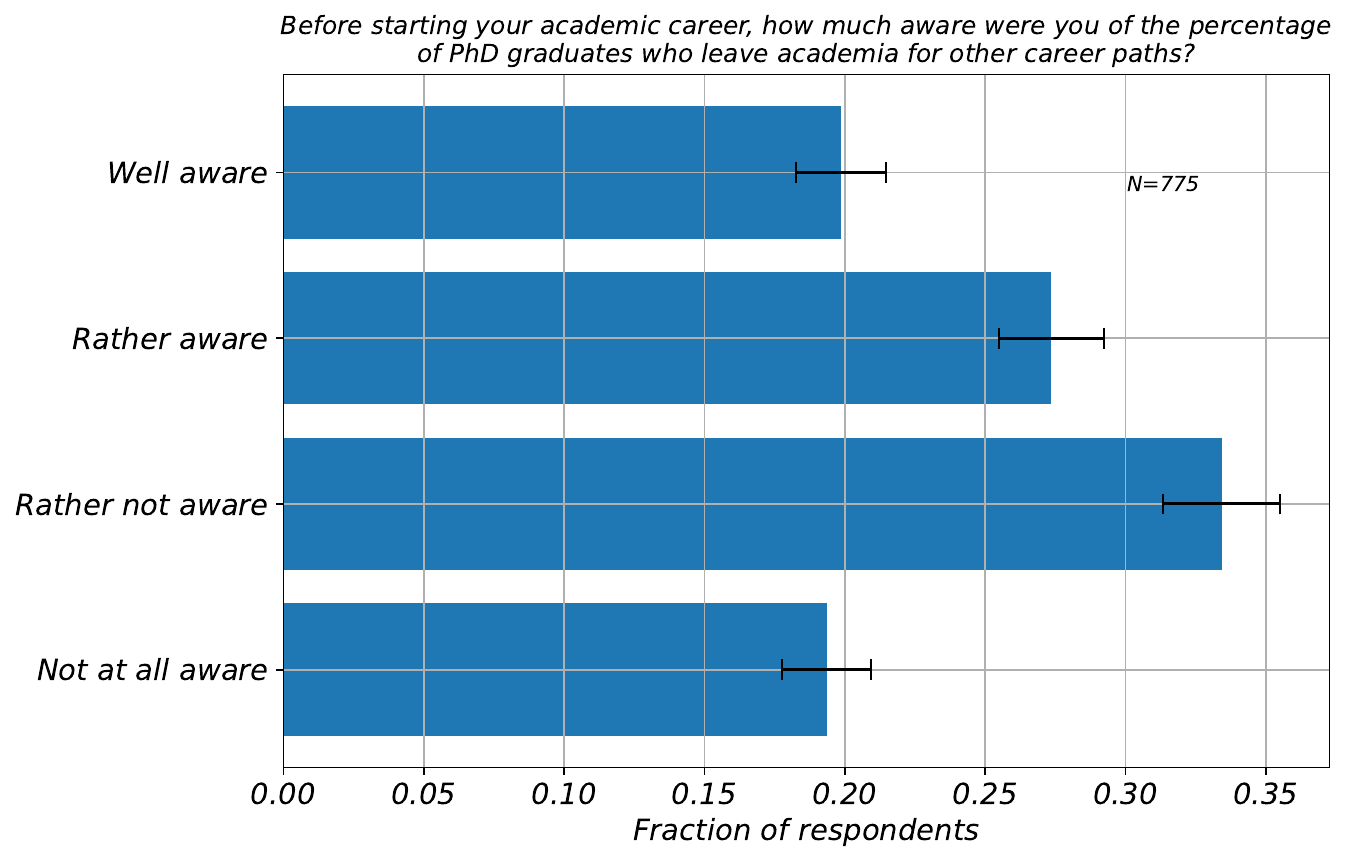}
	 	 \end{subfigure}
	 	 \begin{subfigure}{.45\textwidth}
	 	 	\centering
	 	 	\includegraphics[width=\linewidth]{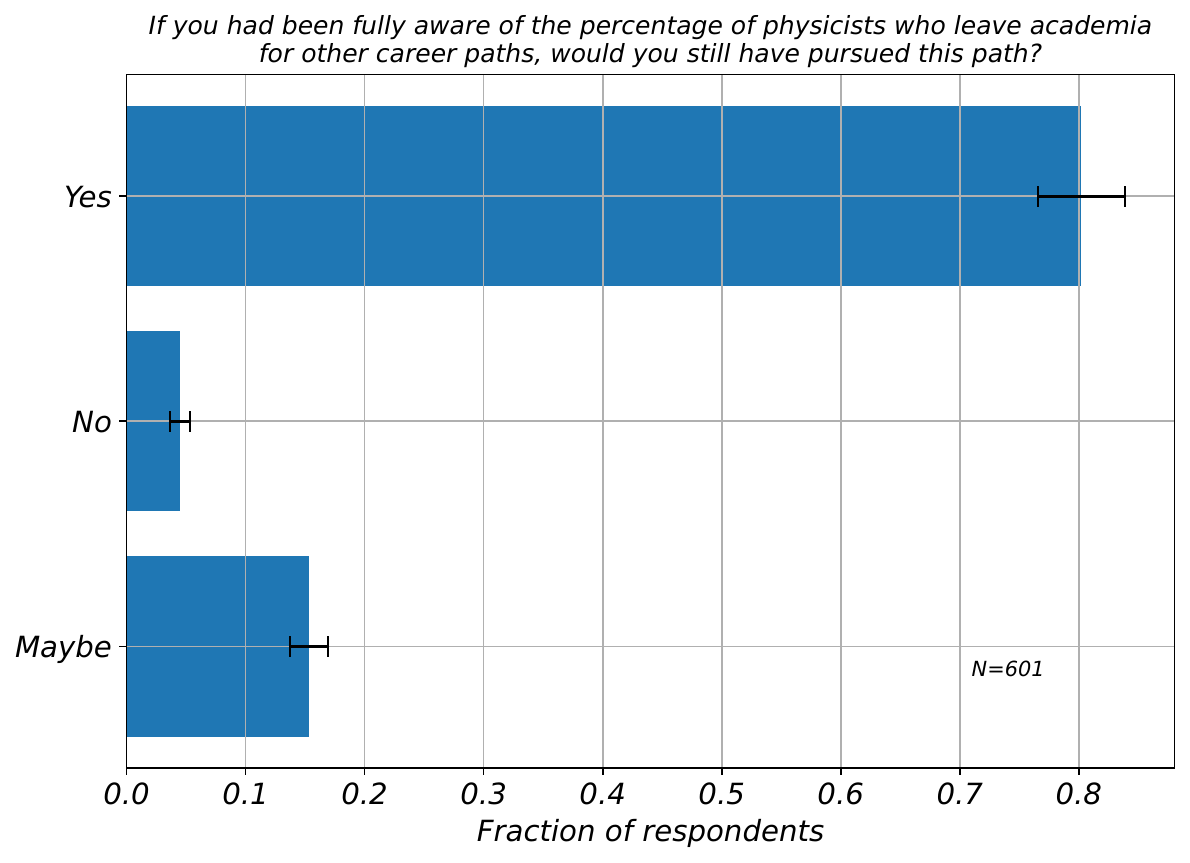}
	 	 \end{subfigure}
\caption{Proportional responses to the second (left) and the third (right) questions in the \textit{Career prospects and ECR leadership} section. The latter was asked only to the participants who did not answer ``Well aware'' to the former. } % \color{blue}(Both plots need larger labels for the ticks, axes and the titles. A suggestion would be to make the axes,tick-labels the same sizes as the ``Yes, No, Maybe" on the RHS plot, the title can have the same size as well.)\color{black}} 
	 	 \label{fig:career_awareness}
\end{figure}

\begin{figure}[H]
    	\centering
    	\includegraphics[angle=0,width=.7\textwidth]{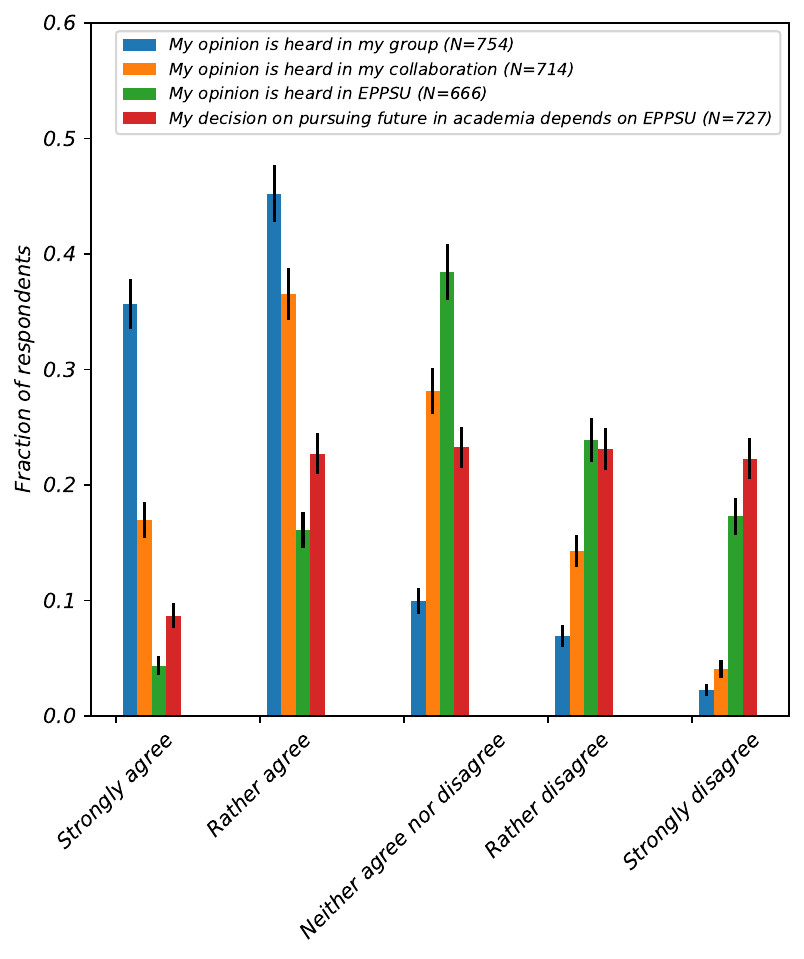}
\caption{Proportional responses to the set of questions on the perceived reception of ECRs' opinions at three different levels---research group, collaboration, EPPSU process---and on their decision about pursuing a future in academia based on the EPPSU outcome. } 
    	\label{fig:career_opinions-heard}
    \end{figure}

\begin{figure}[H]
	    \centering
	 	 \begin{subfigure}{.9\textwidth}
	 	 	\centering
	 	 	\includegraphics[width=\linewidth]{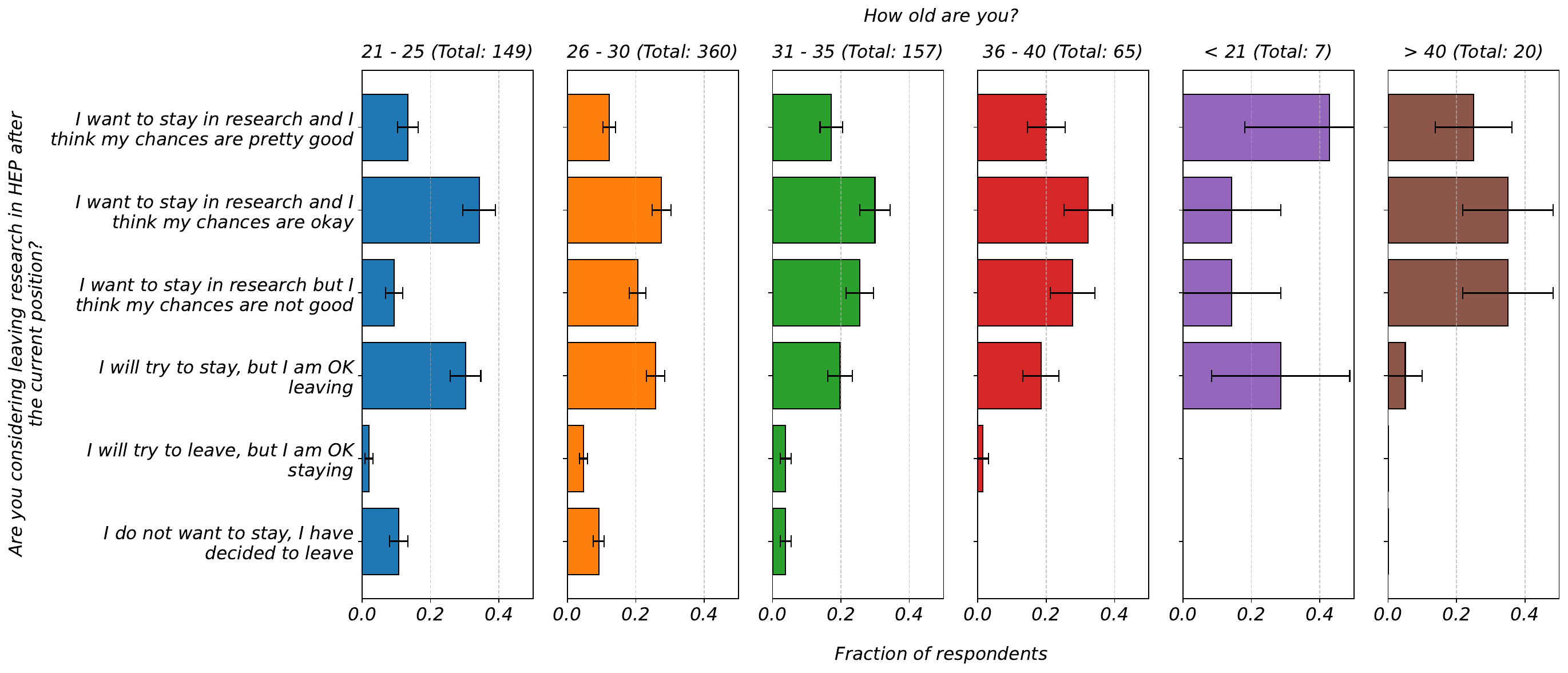}
	 	 \end{subfigure} \\
	 	 \begin{subfigure}{.9\textwidth}
	 	 	\centering
	 	 	\includegraphics[width=\linewidth]{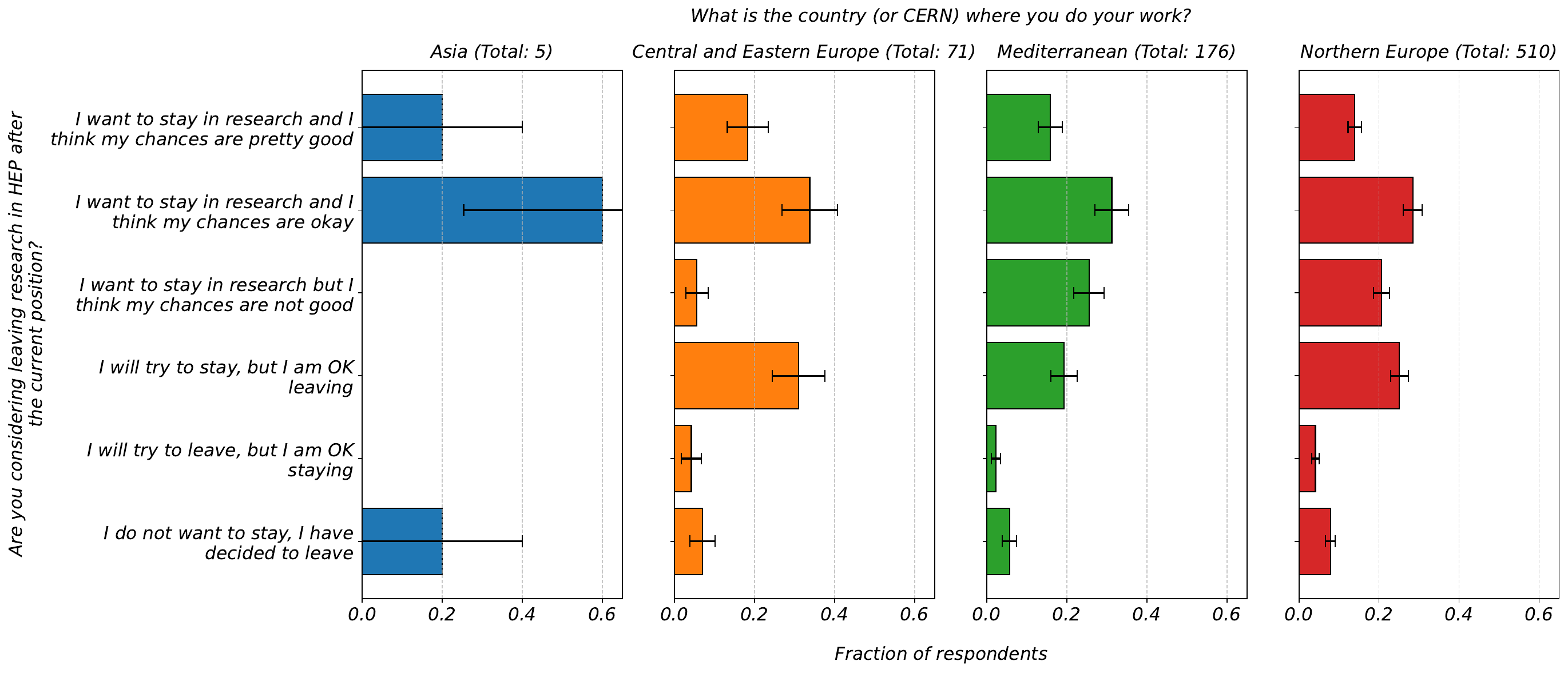}
	 	 \end{subfigure} \\
              \begin{subfigure}{.9\textwidth}
	 	 	\centering
	 	 	\includegraphics[width=\linewidth]{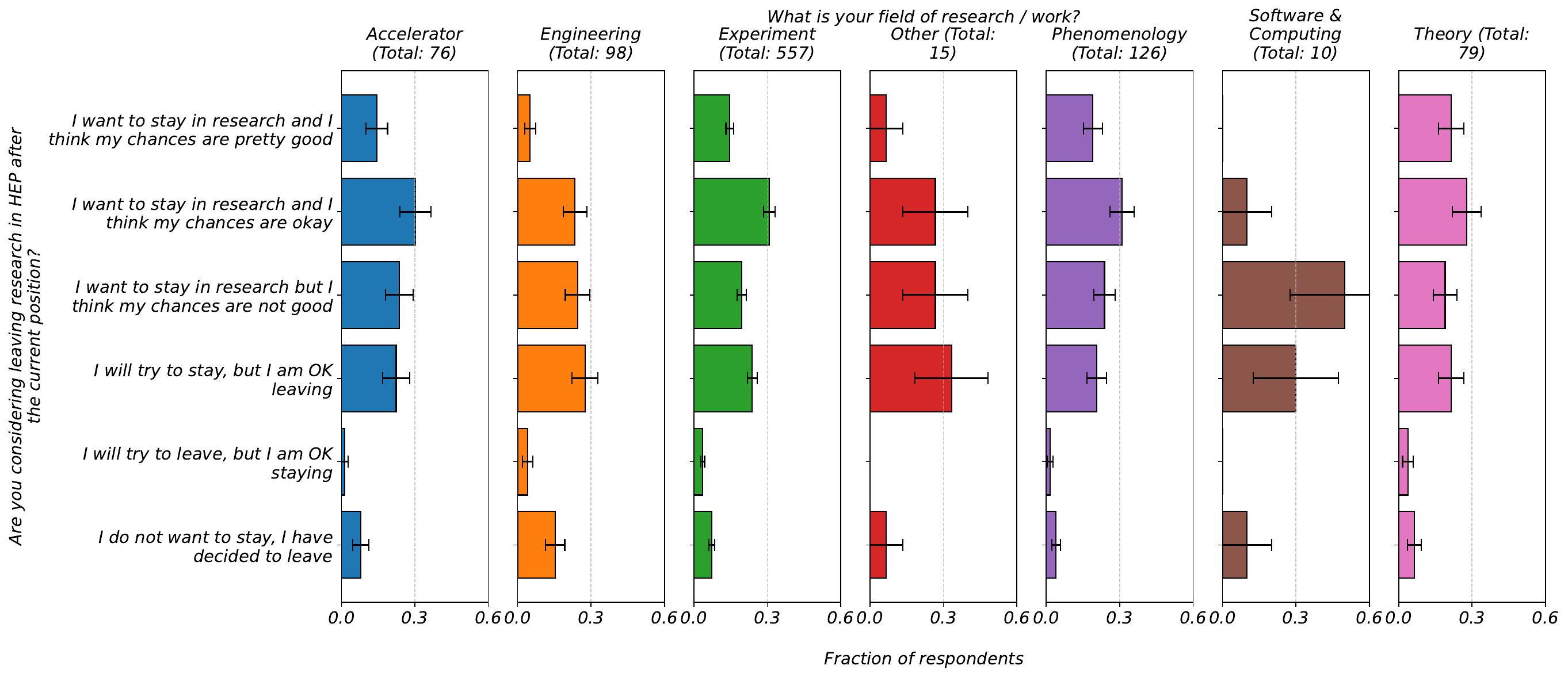}
	 	 \end{subfigure}
\caption{Proportional responses to the question: ``Are you considering leaving research in HEP after the current position?'' presented for different groups of age (upper panel), work region (middle panel) and field of work (bottom panel). Assignment of countries to regions is given in Appendix~\ref{app:geo-regions-def}. } % \color{blue}(All plots need larger axes labels, tick labels and subplot titles. The same size as the suptitle should work.)\color{black}} 
	 	 \label{fig:career_future-in-academia_correlations}
\end{figure}

\begin{figure}[H]
	    \centering
	 	 \begin{subfigure}{.9\textwidth}
	 	 	\centering
	 	 	\includegraphics[width=\linewidth]{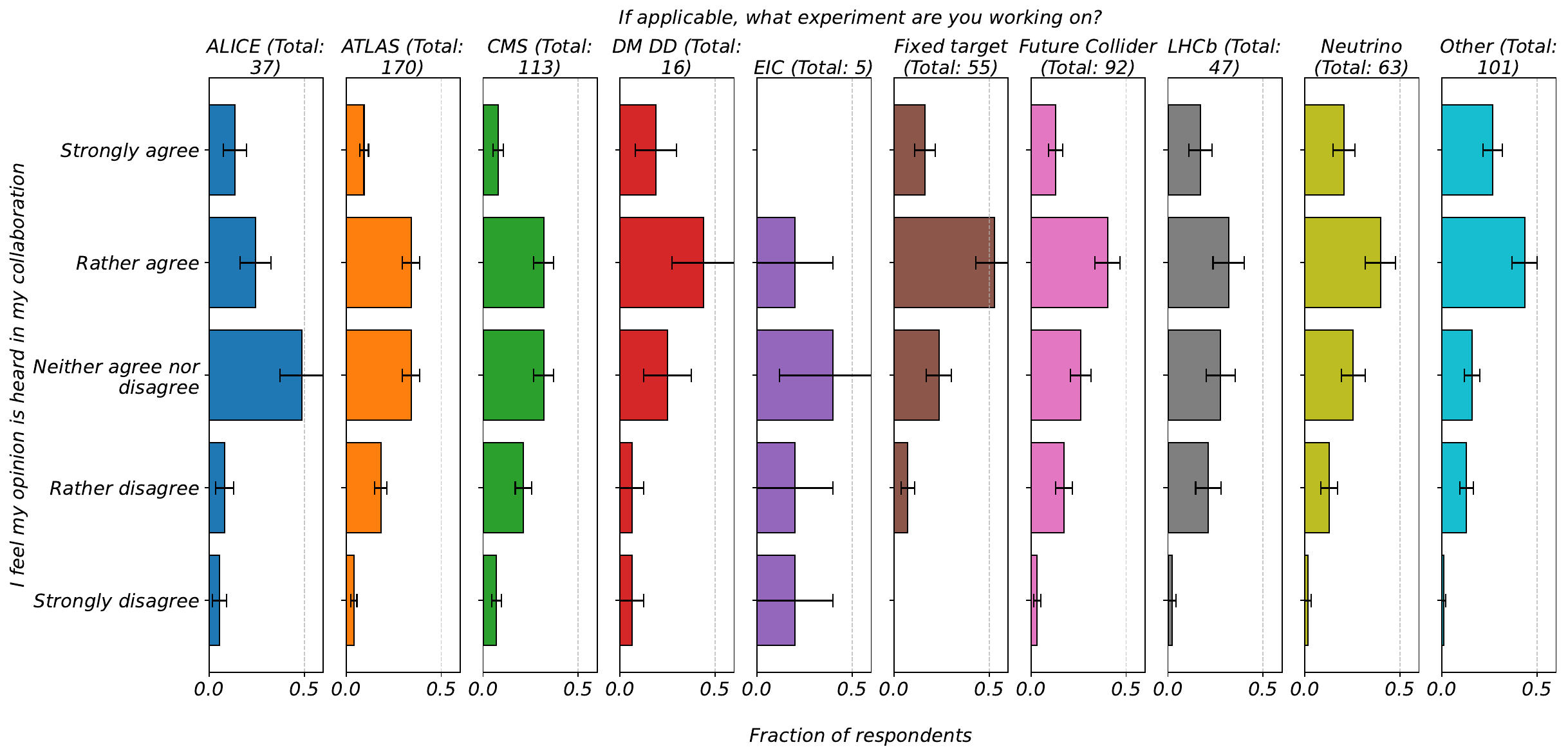}
	 	 \end{subfigure} \\
	 	 \begin{subfigure}{.9\textwidth}
	 	 	\centering
	 	 	\includegraphics[width=\linewidth]{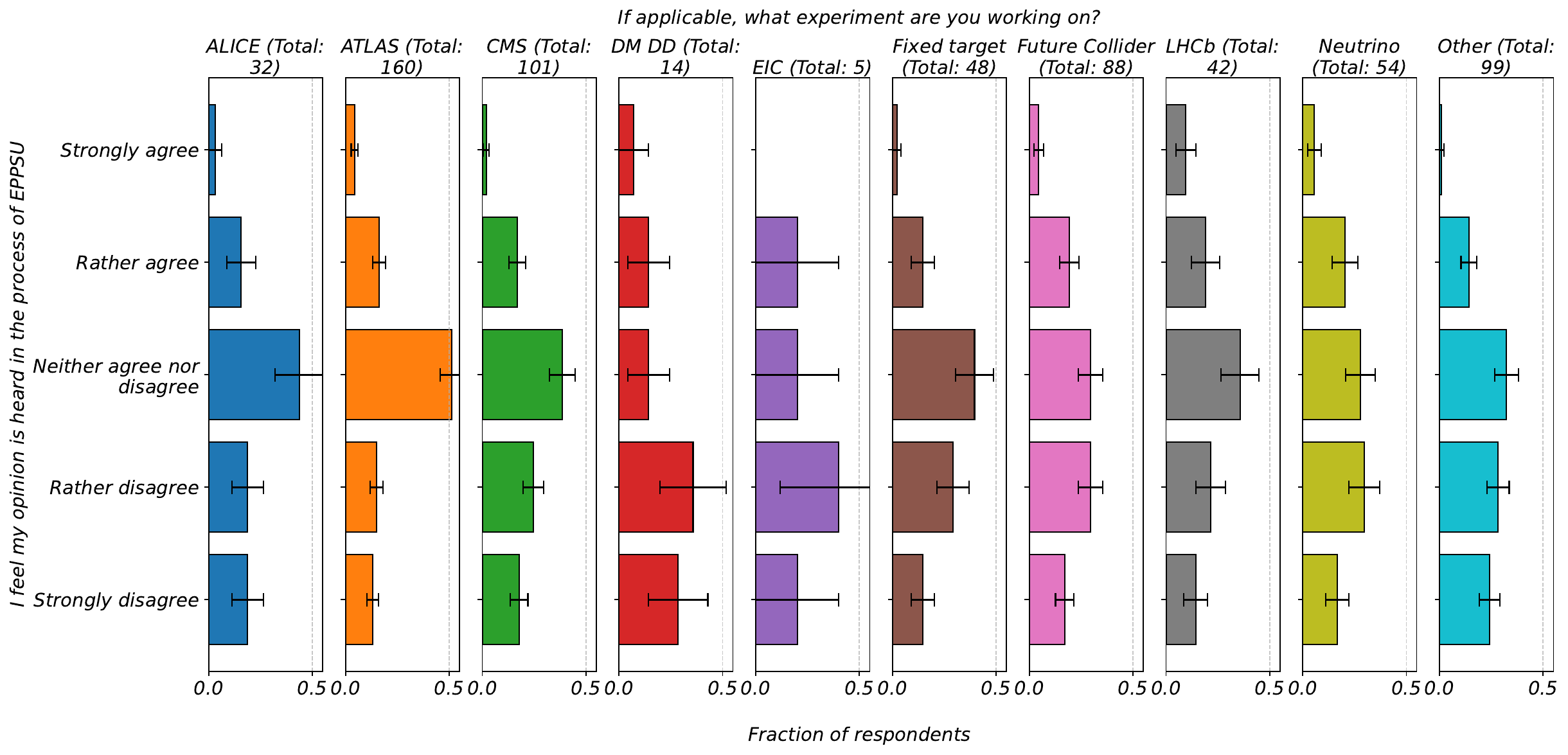}
	 	 \end{subfigure} \\
              \begin{subfigure}{.9\textwidth}
	 	 	\centering
	 	 	\includegraphics[width=\linewidth]{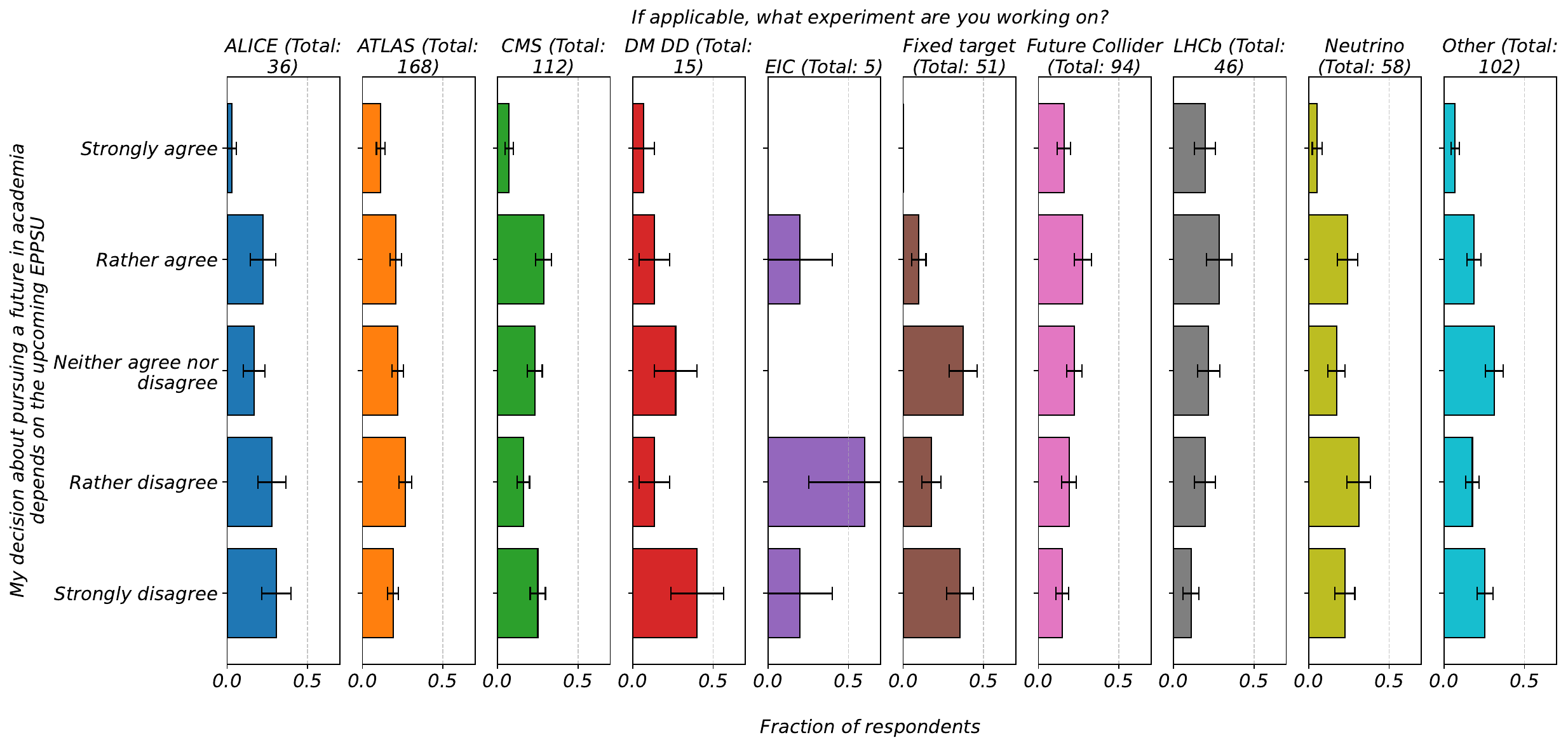}
	 	 \end{subfigure}
\caption{Proportional responses to the questions on the perceived reception of ECRs' opinions in their collaborations (upper panel) and in the EPPSU process (middle panel), as well as on their decision about pursuing future in academia based on the EPPSU outcome. The responses are shown for different experiments the respondents work on. ``DM DD'' stands for ``Dark matter direct detection''. } % \color{blue}(All plots need larger axes labels, tick labels and subplot titles. The same size as the suptitle should work.)\color{black}} 
	 	 \label{fig:career_opinions-heard_correlations}
\end{figure}

\subsection{Diversity, inclusion and mental health}
\begin{figure}[H]
    \centering
    \includegraphics[width=0.7\linewidth]{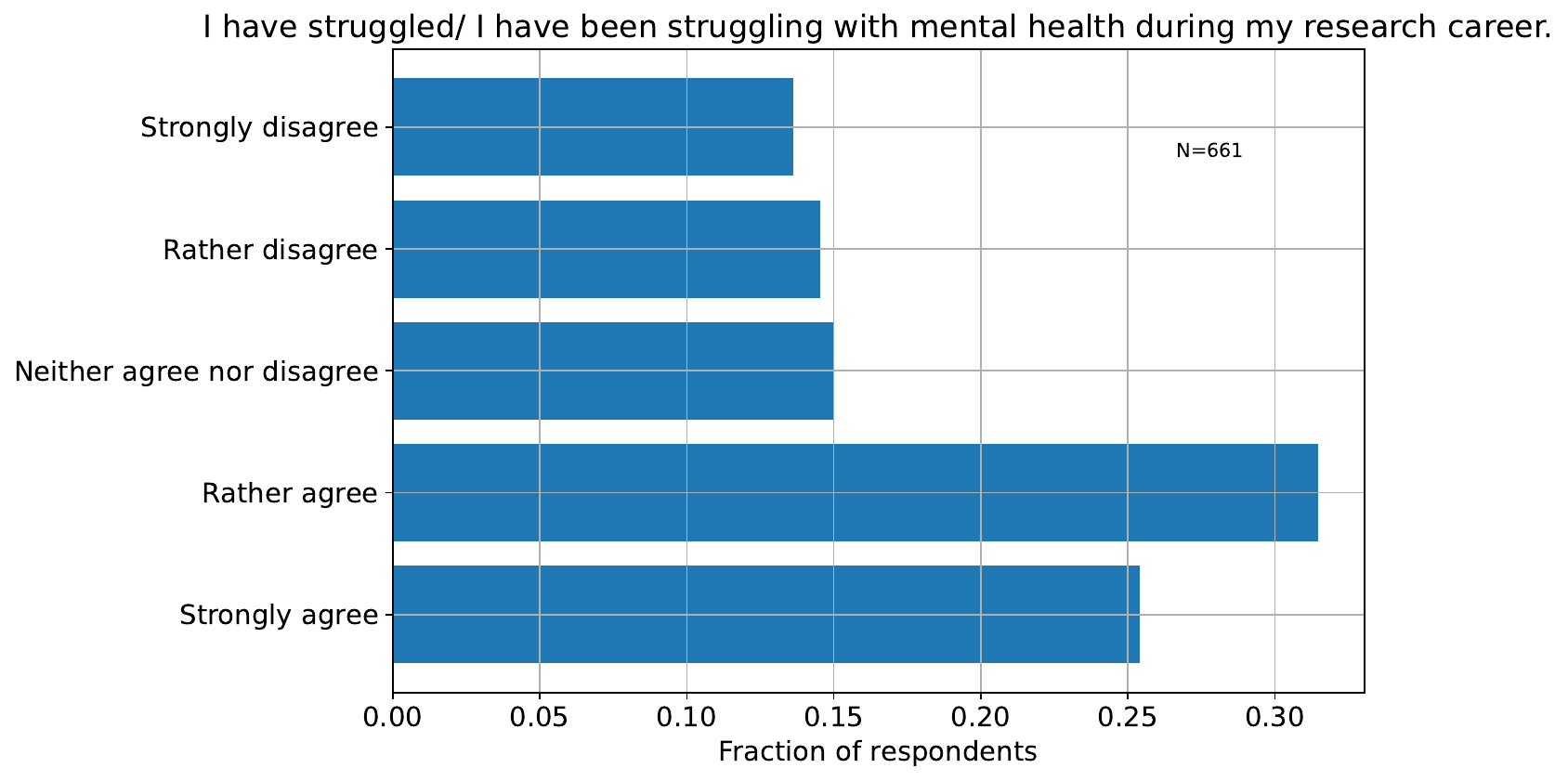}
    \caption{Survey responses on mental health struggles during research career. } % \color{blue}(The plot is a bit smeared, needs either better resolution/vector graphics. Needs error bars.)\color{black}}
    \label{fig:mental_health}
\end{figure}

\begin{figure}[H]
    \centering
    \begin{subfigure}{.49\textwidth}
        \centering
        \includegraphics[width=\linewidth]{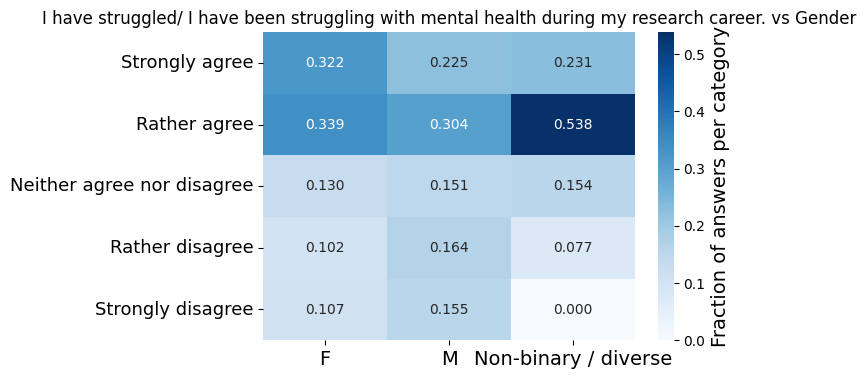}
        \label{fig:mental-health-gender-frac}
        \caption{}
    \end{subfigure}
    \hfill
    \begin{subfigure}{.49\textwidth}
        \centering
        \includegraphics[width=\linewidth]{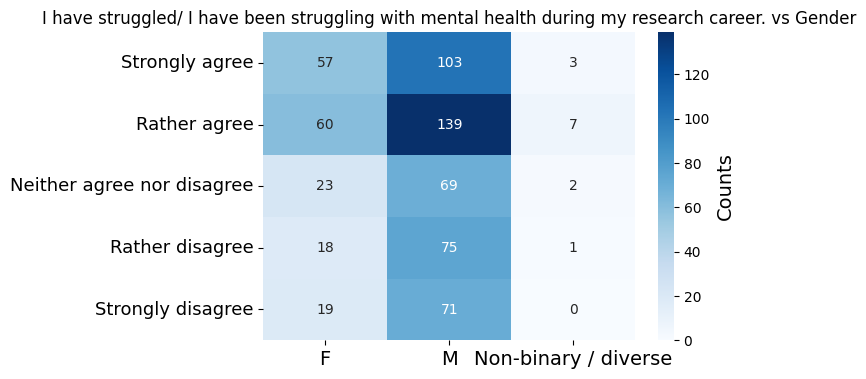}
        \label{fig:mental-health-gender-abs}
        \caption{}
    \end{subfigure}
    \centering
    \begin{subfigure}{.49\textwidth}
        \centering
        \includegraphics[width=\linewidth]{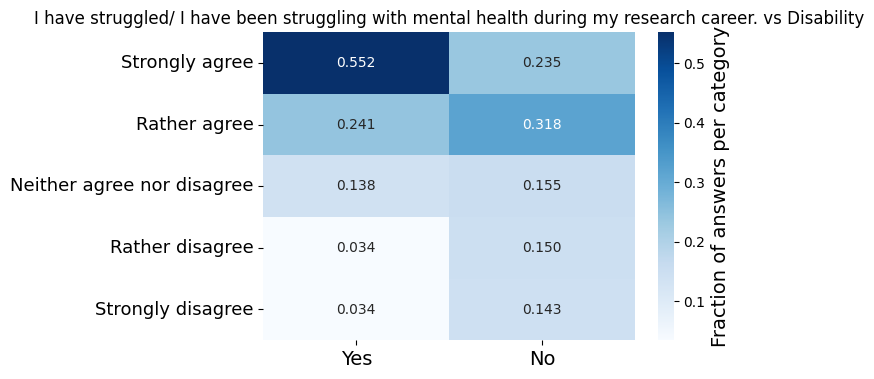}
        \label{fig:mental-health-disability-frac}
        \caption{}
    \end{subfigure}
    \hfill
    \begin{subfigure}{.49\textwidth}
        \centering
        \includegraphics[width=\linewidth]{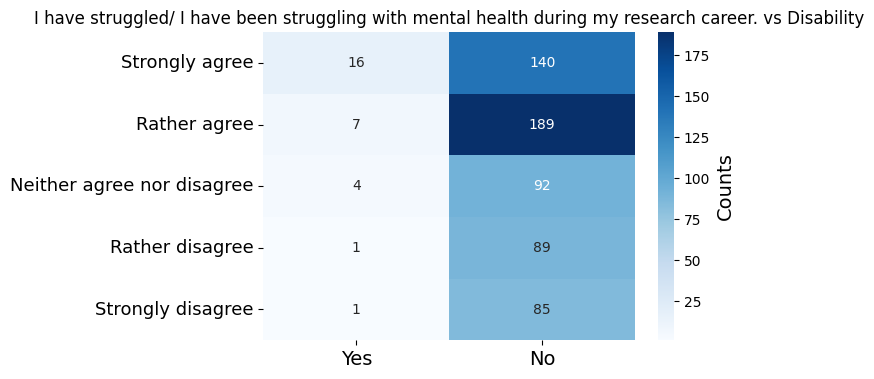}
        \label{fig:mental-health-disability-abs}
        \caption{}
    \end{subfigure}
    \caption{Survey responses on mental health struggles correlated with gender (top row) and ability (bottom row) presented as fractions (left) and counts (right). } % \color{blue}(Plots require larger labels)\color{black}}
    \label{fig:mental_health_gender_disability}
\end{figure}

\begin{figure}[H]
    \centering
    \includegraphics[width=0.9\linewidth]{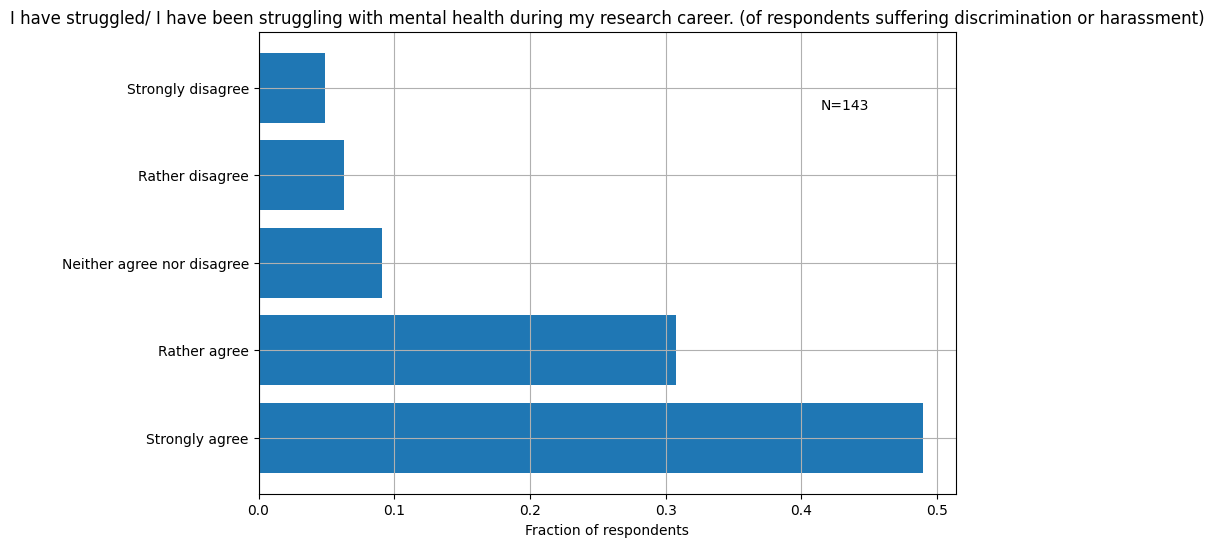}
    \caption{Survey responses on mental health struggles during research career for people that experienced discrimination or harassment. } % \color{blue}(Needs error bars.) \color{black}}
    \label{fig:mental_health_discrimination}
\end{figure}

\begin{figure}[H]
    \centering
    \begin{subfigure}{.49\textwidth}
        \centering
        \includegraphics[width=\linewidth]{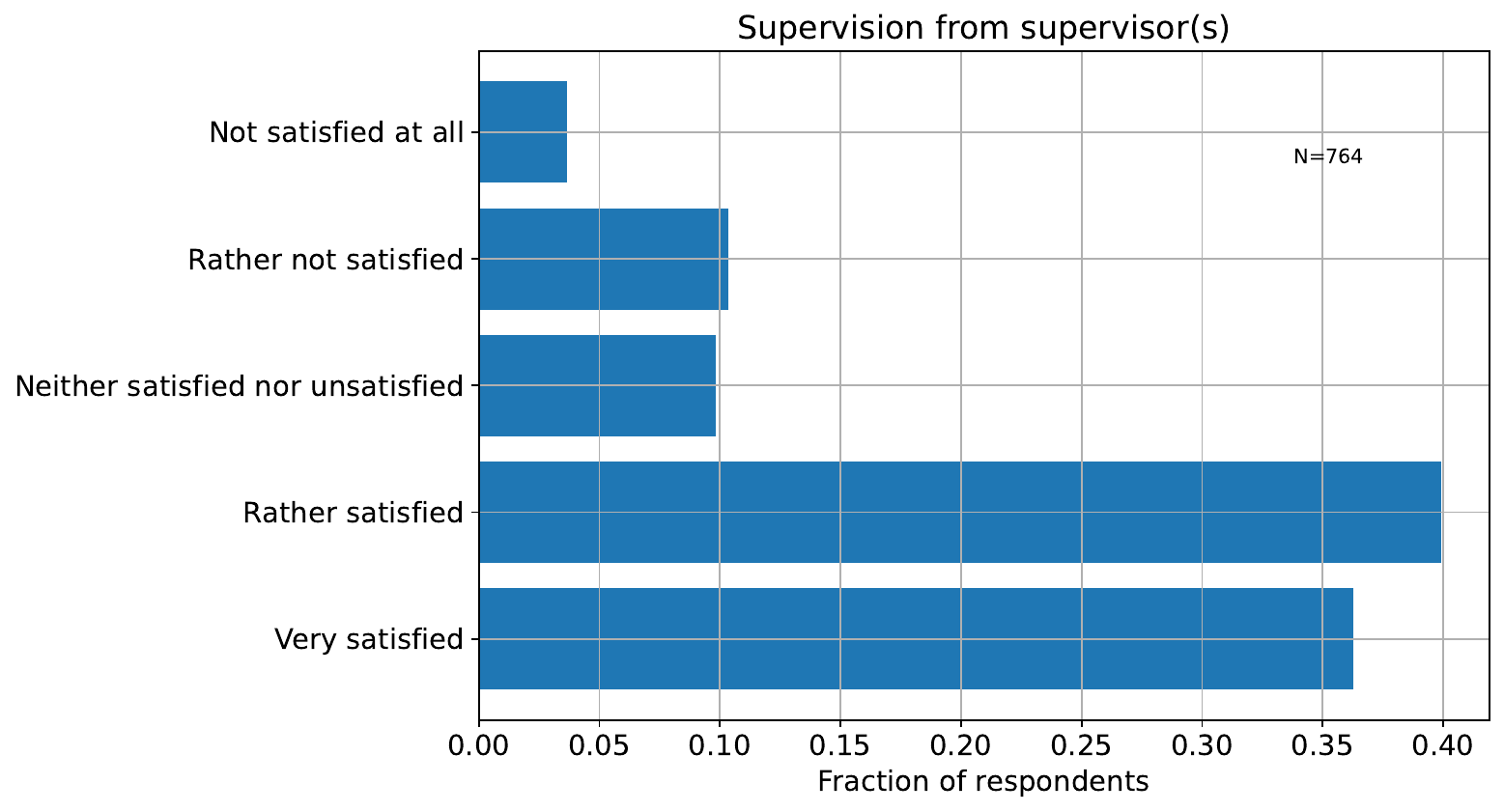}
        \label{fig:supervisor-satisfaction}
        \caption{}
    \end{subfigure}
    \hfill
    \begin{subfigure}{.49\textwidth}
        \centering
        \includegraphics[width=\linewidth]{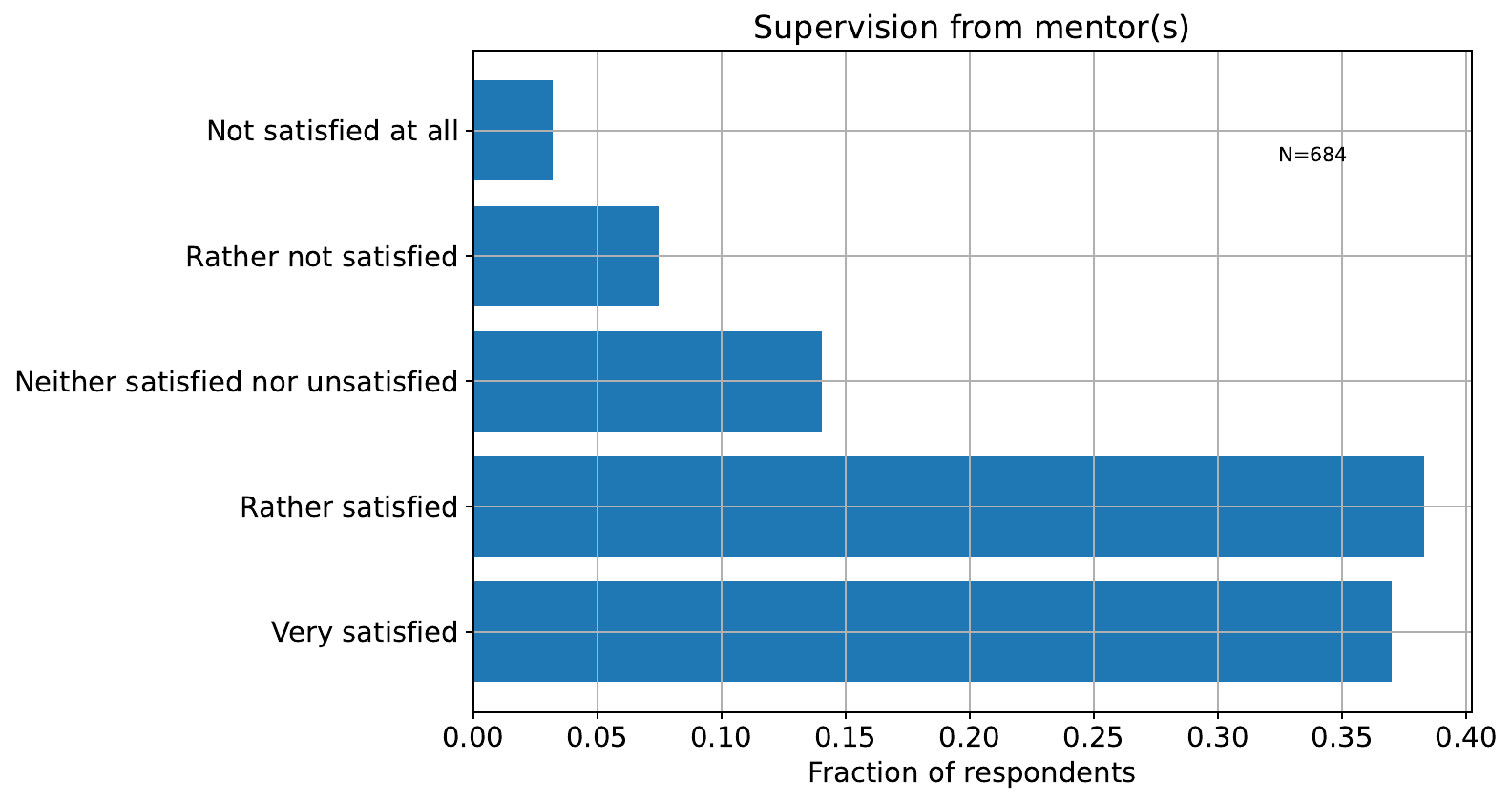}
        \label{fig:mentor-satisfaction}
        \caption{}
    \end{subfigure}
    \caption{Survey responses on satisfaction with a) supervision and b) mentorship. } % \color{blue}(Plots require slightly larger axes labels. Needs error bars.)\color{black}}
    \label{fig:supervisor_mentor_satisfaction}
\end{figure}

\begin{figure}[H]
    \centering
    \includegraphics[width=\linewidth]{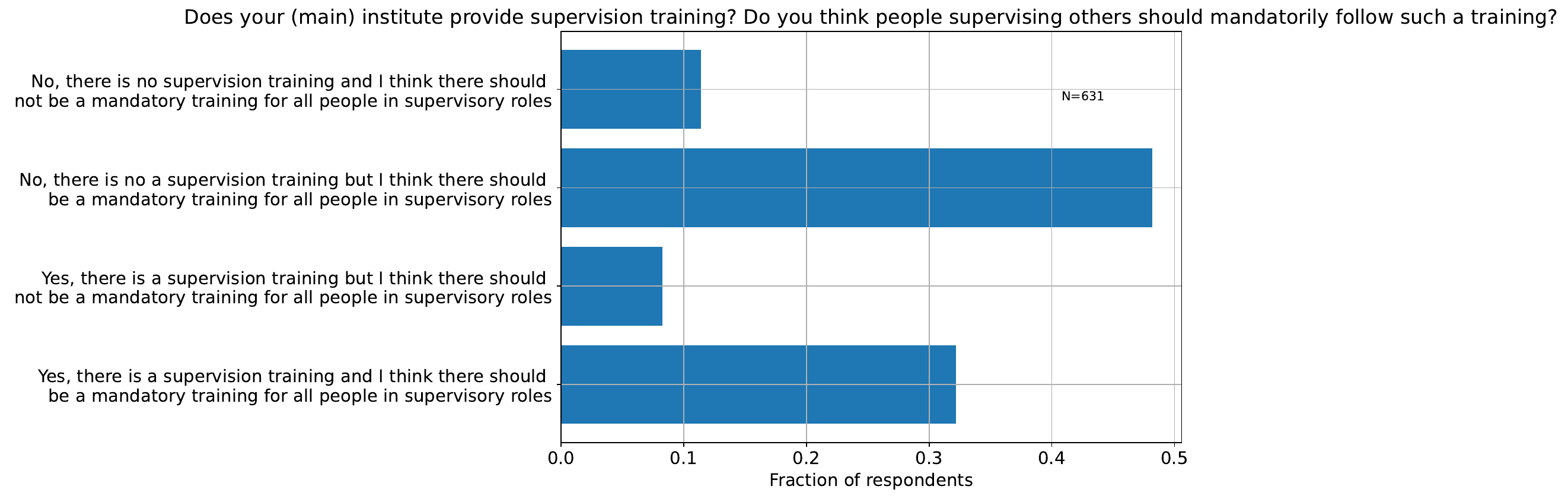}
    \caption{Survey responses on supervisor training.} % \color{blue}(Plots require slightly larger axes, tick labels. Needs error bars.)\color{black}}
    \label{fig:supervisor_training}
\end{figure}

\begin{figure}[H]
    \centering
    \includegraphics[width=0.75\linewidth]{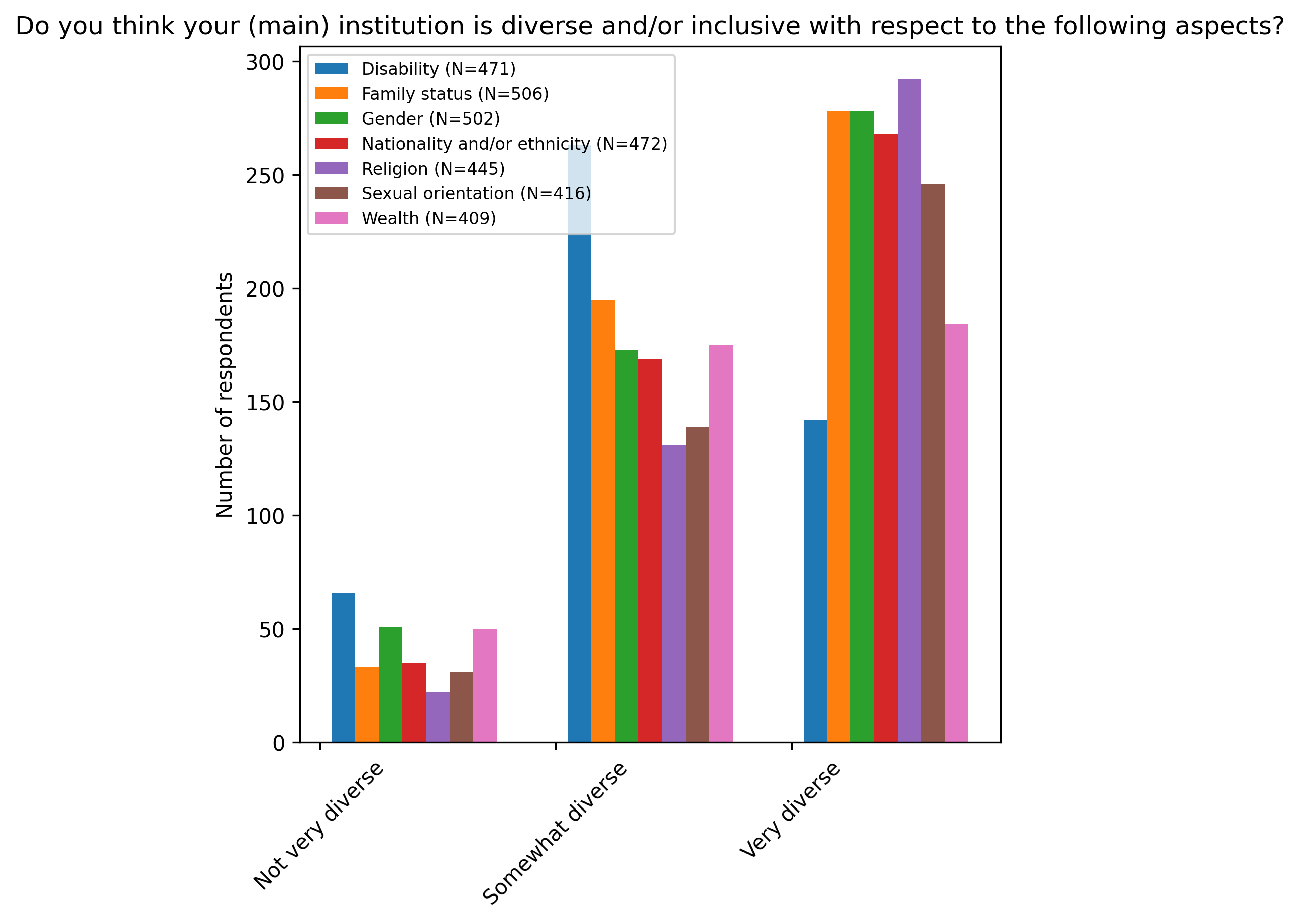}
    \caption{Survey perceptions of diversity at home institutes across various characteristics. } % \color{blue}(Needs error bars.)\color{black}}
    \label{fig:diverse}
\end{figure}

\begin{figure}[H]
    \centering
    \includegraphics[width=0.7\linewidth]{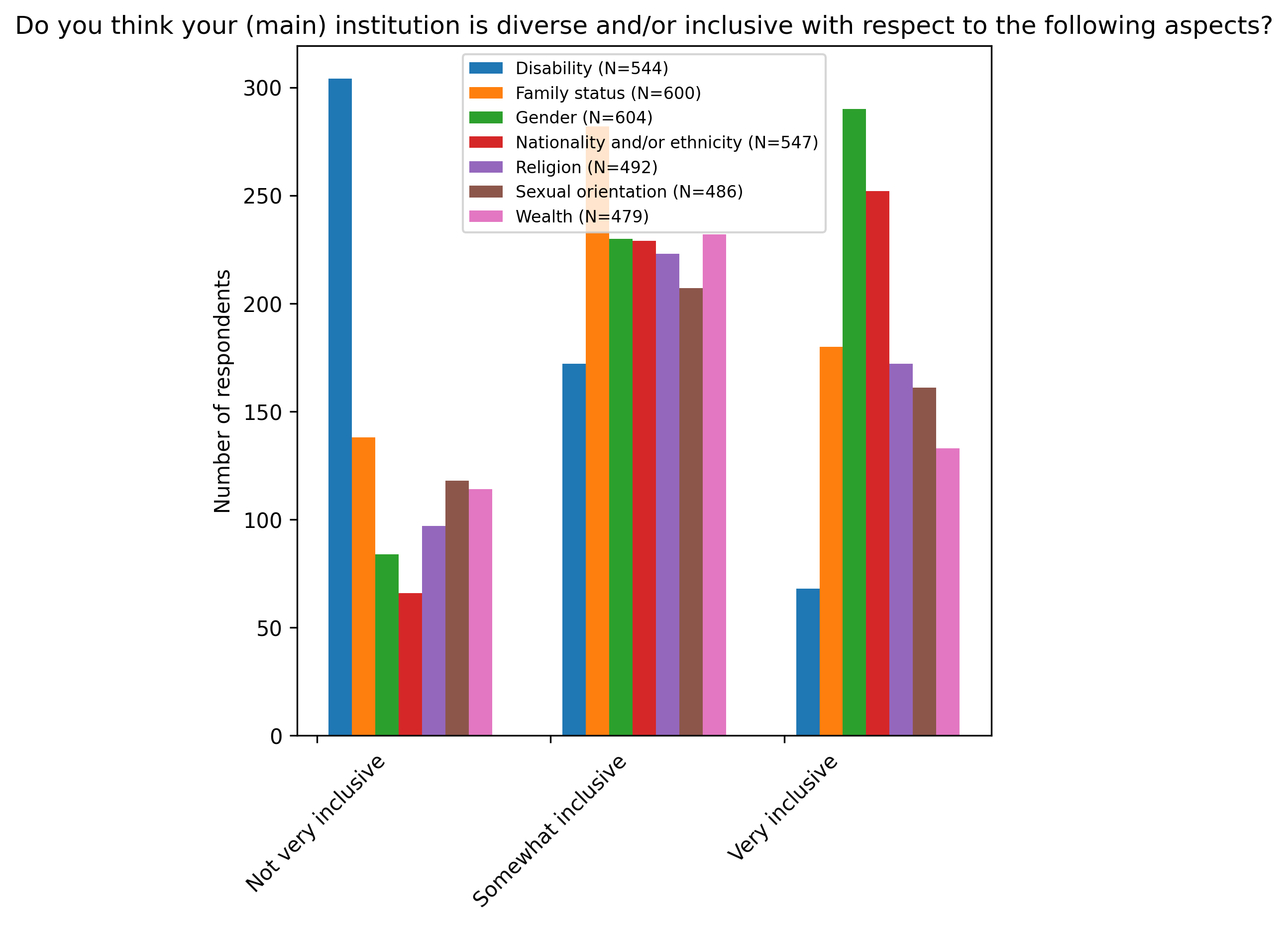}
    \caption{Survey perceptions of inclusion at home institutes across various characteristics.} % \color{blue}(Needs error bars.)\color{black}
    \label{fig:inclusion}
\end{figure}

\begin{figure}[H]
    \centering
    \includegraphics[width=0.9\linewidth]{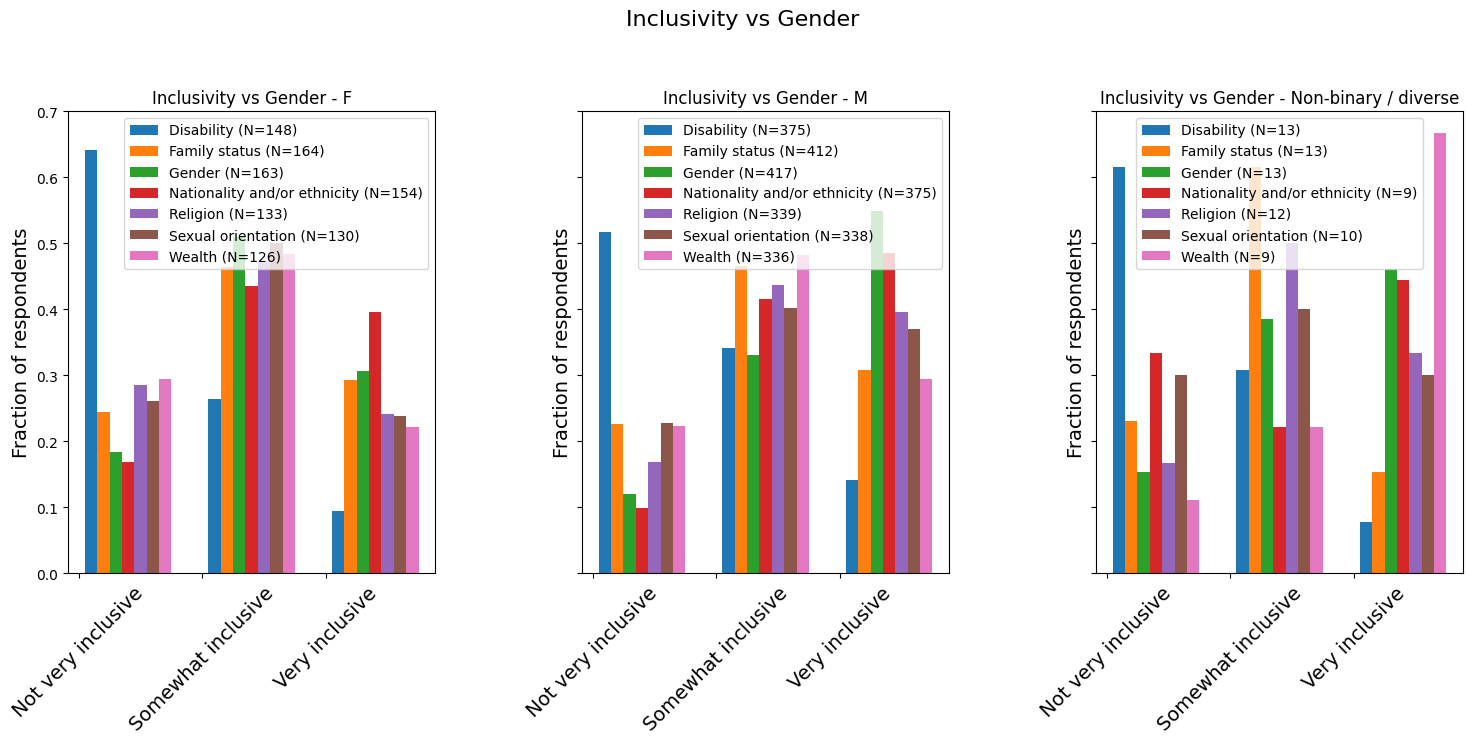}
    \caption{Survey perceptions of inclusion at home institutes across various characteristics comparing different genders. } % \color{blue}(Plots require larger labels. Needs error bars.)\color{black}}
    \label{fig:inclusion-gender}
\end{figure}

\begin{figure}[H]
    \centering
    \includegraphics[width=0.8\linewidth]{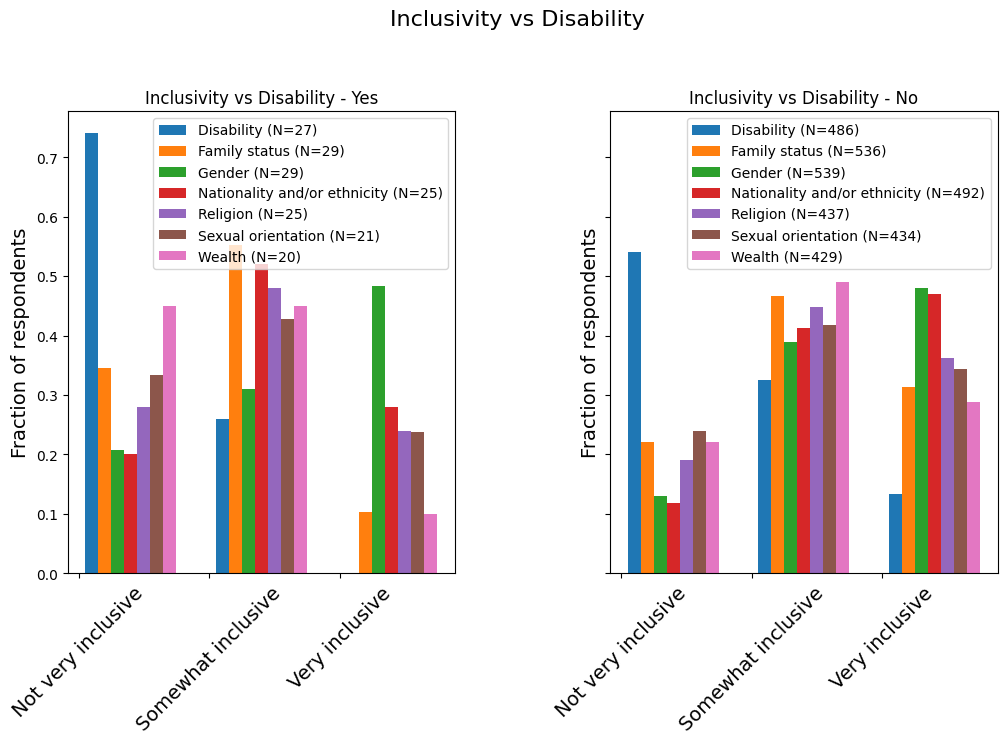}
    \caption{Survey perceptions of inclusion at home institutes across various characteristics comparing different abilities. } % \color{blue}(Needs error bars.)\color{black}}
    \label{fig:inclusion-disability}
\end{figure}

\begin{figure}[H]
    \centering
    \includegraphics[width=0.6\linewidth]{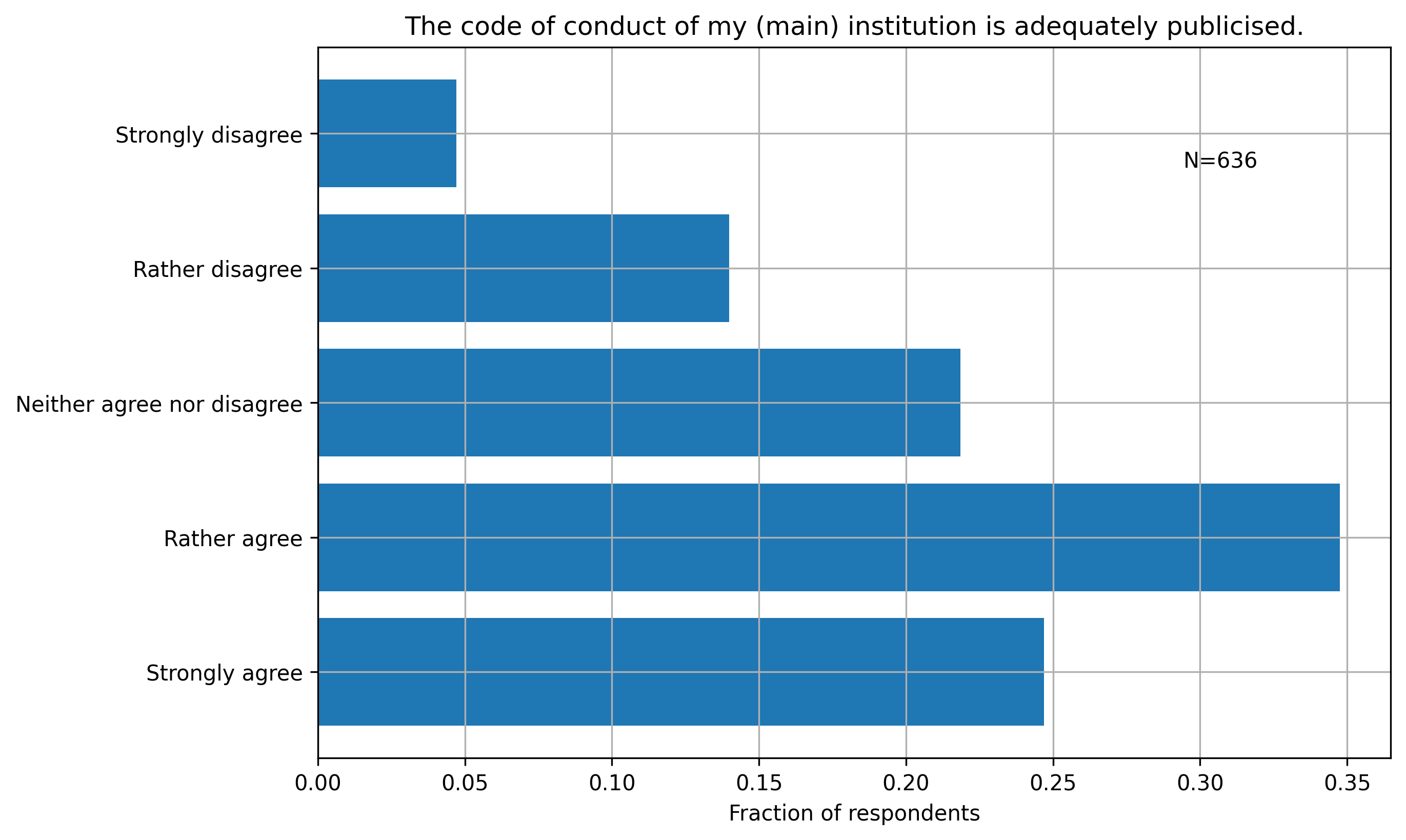}
    \caption{Survey feedback on the visibility of the code of conduct in home institutions. } % \color{blue}(Needs error bars.)\color{black}}
    \label{fig:code-of-conduct_publicised}
\end{figure}

\begin{figure}[H]
    \centering
    \includegraphics[width=0.6\linewidth]{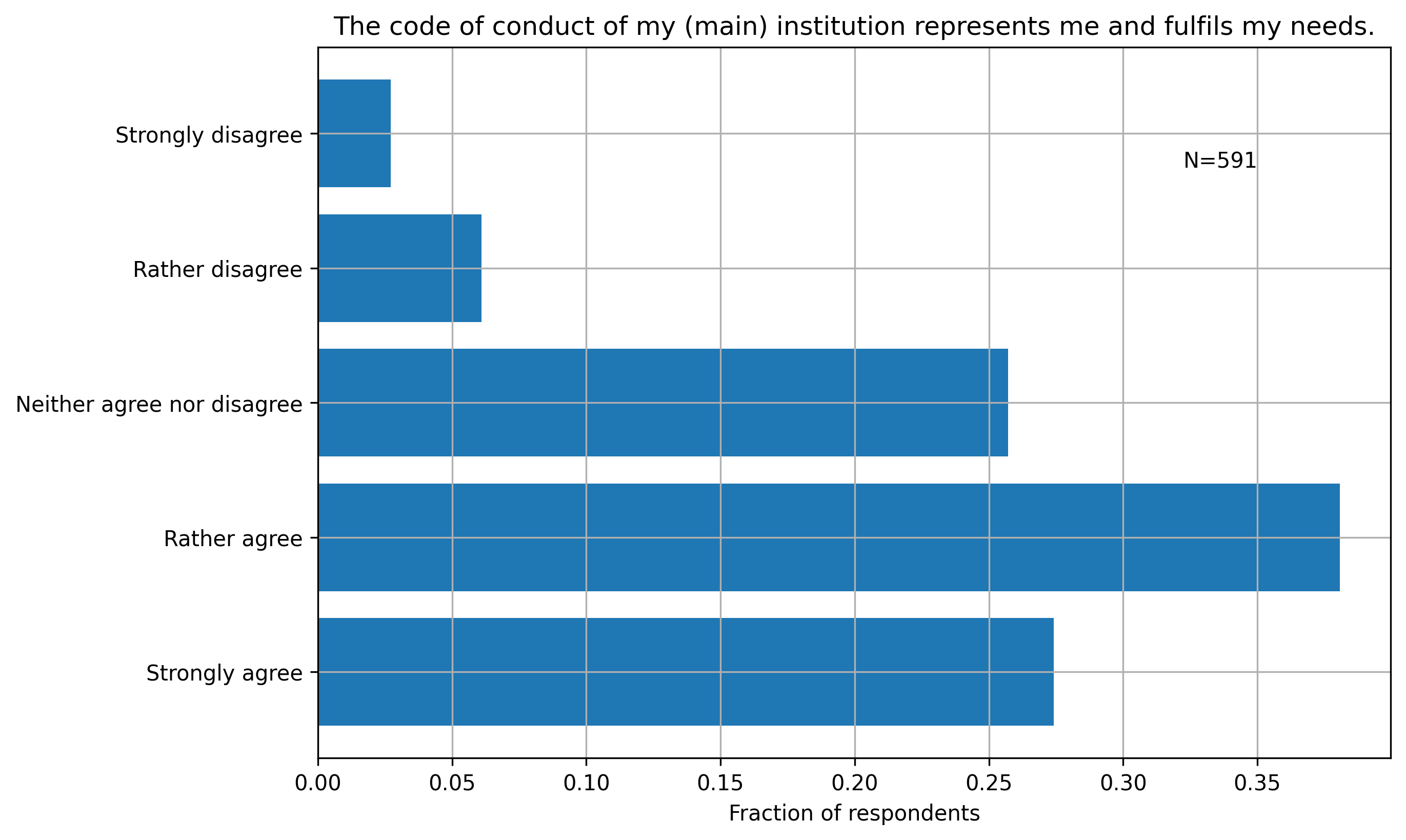}
    \caption{Survey responses on representation in the code of conduct in home institutions. } % \color{blue}(Needs error bars.)\color{black}}
    \label{fig:code-of-conduct_representation}
\end{figure}

\begin{figure}[H]
    \centering
    \begin{subfigure}{.49\textwidth}
	 	 	\centering
	 	 	\includegraphics[width=\linewidth]{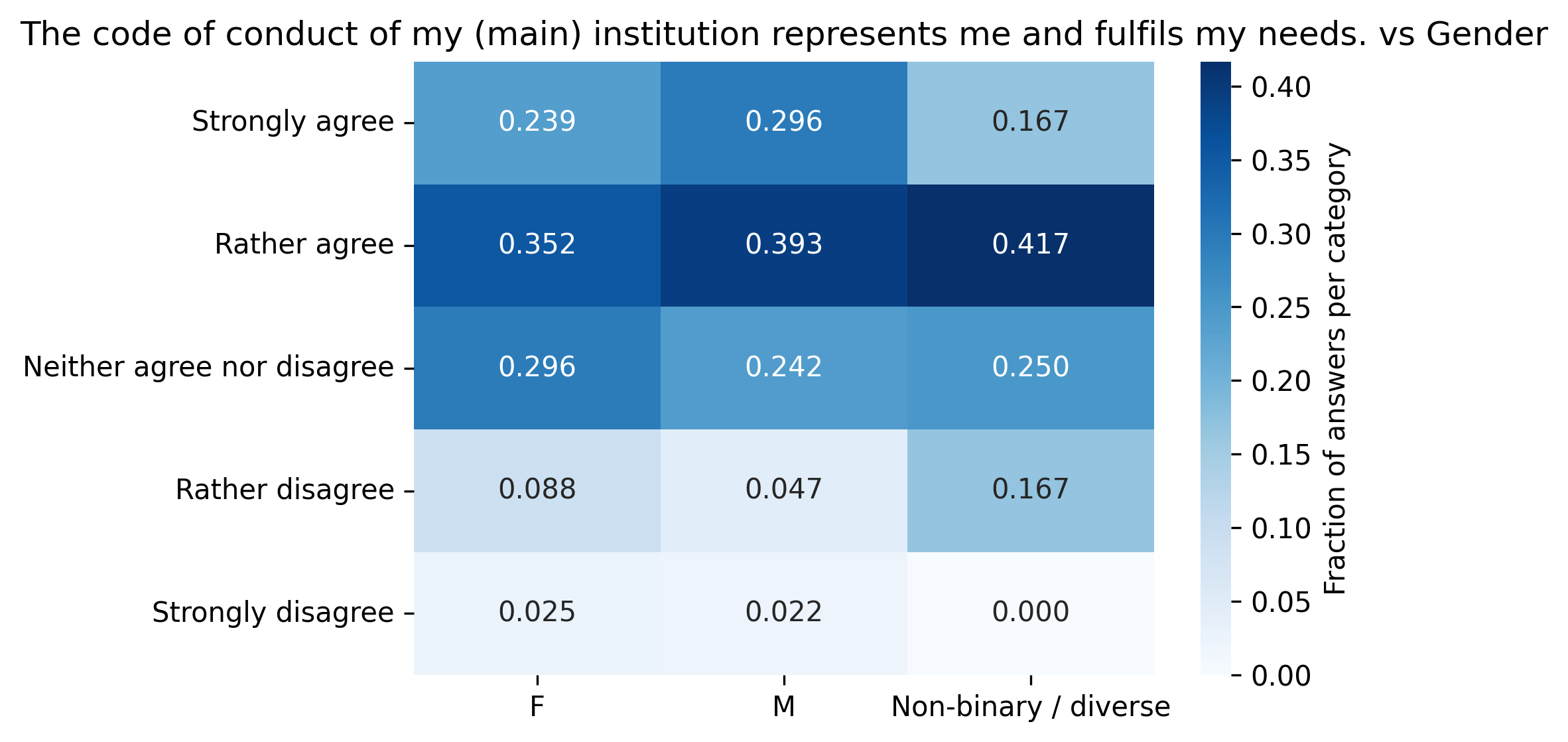}
            \label{fig:CC_rep_gender_frac}
            \caption{}
	  \end{subfigure} 
      \hfill
    \begin{subfigure}{.49\textwidth}
	 	 	\centering
	 	 	\includegraphics[width=\linewidth]{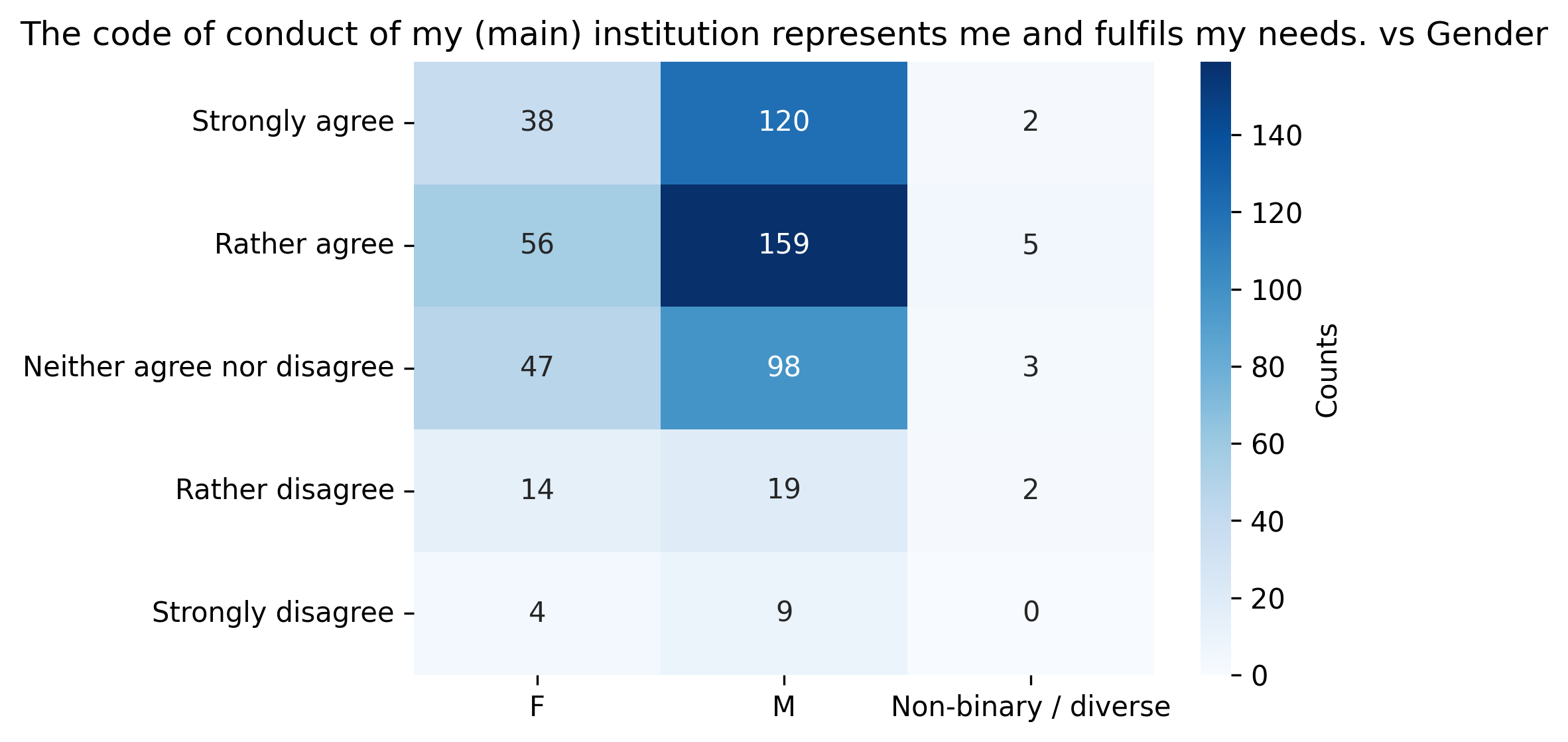}
            \label{fig:CC_rep_gender}
            \caption{}
	 	 \end{subfigure}
    \caption{Survey responses on representation in the code of conduct in home institutions, comparing different genders a) by fraction and b) by count.}
    \label{fig:code-of-conduct_representation-gender}
\end{figure}

\begin{figure}[H]
    \centering
    \begin{subfigure}{.49\textwidth}
	 	 	\centering
	 	 	\includegraphics[width=\linewidth]{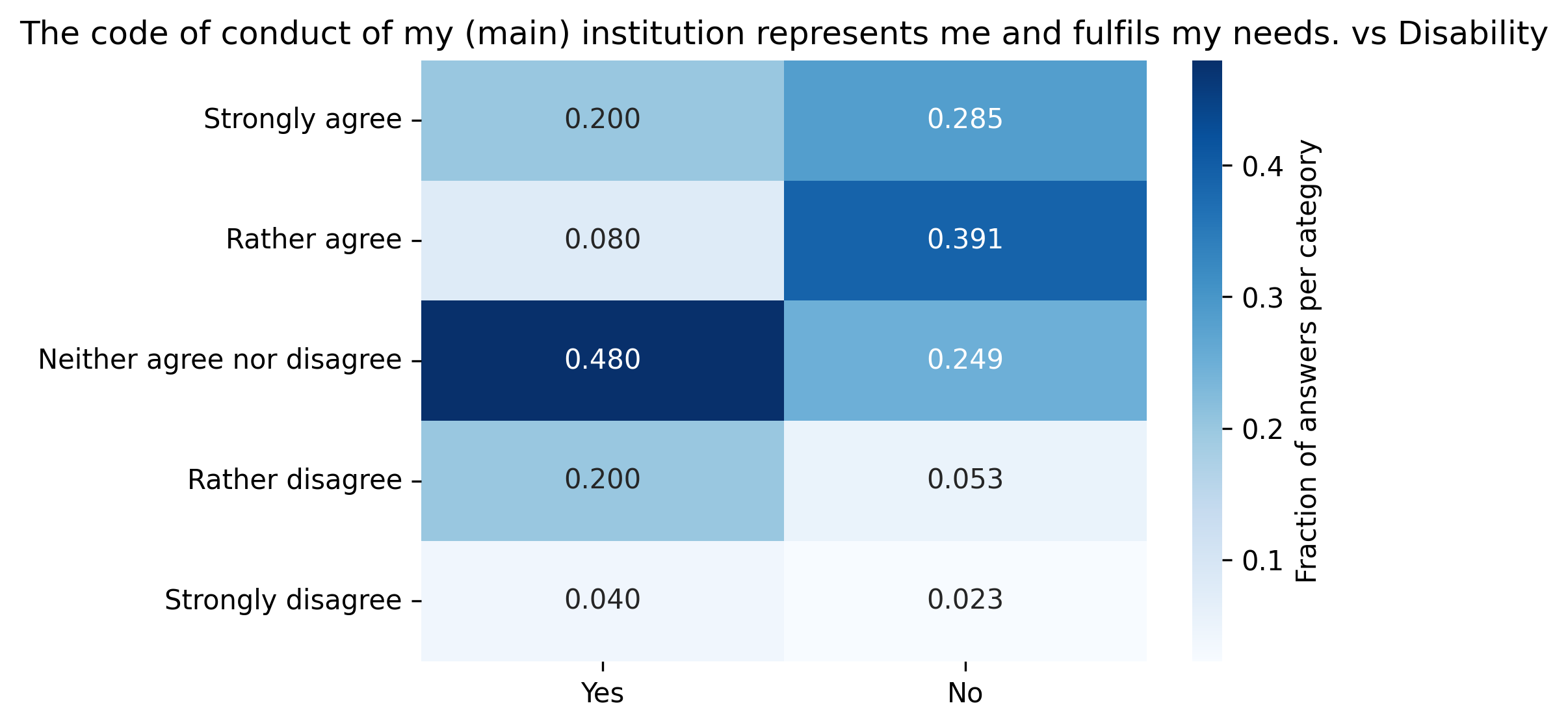}
            \label{fig:CC_rep_disability_frac}
            \caption{}
	  \end{subfigure} 
      \hfill
    \begin{subfigure}{.49\textwidth}
	 	 	\centering
	 	 	\includegraphics[width=\linewidth]{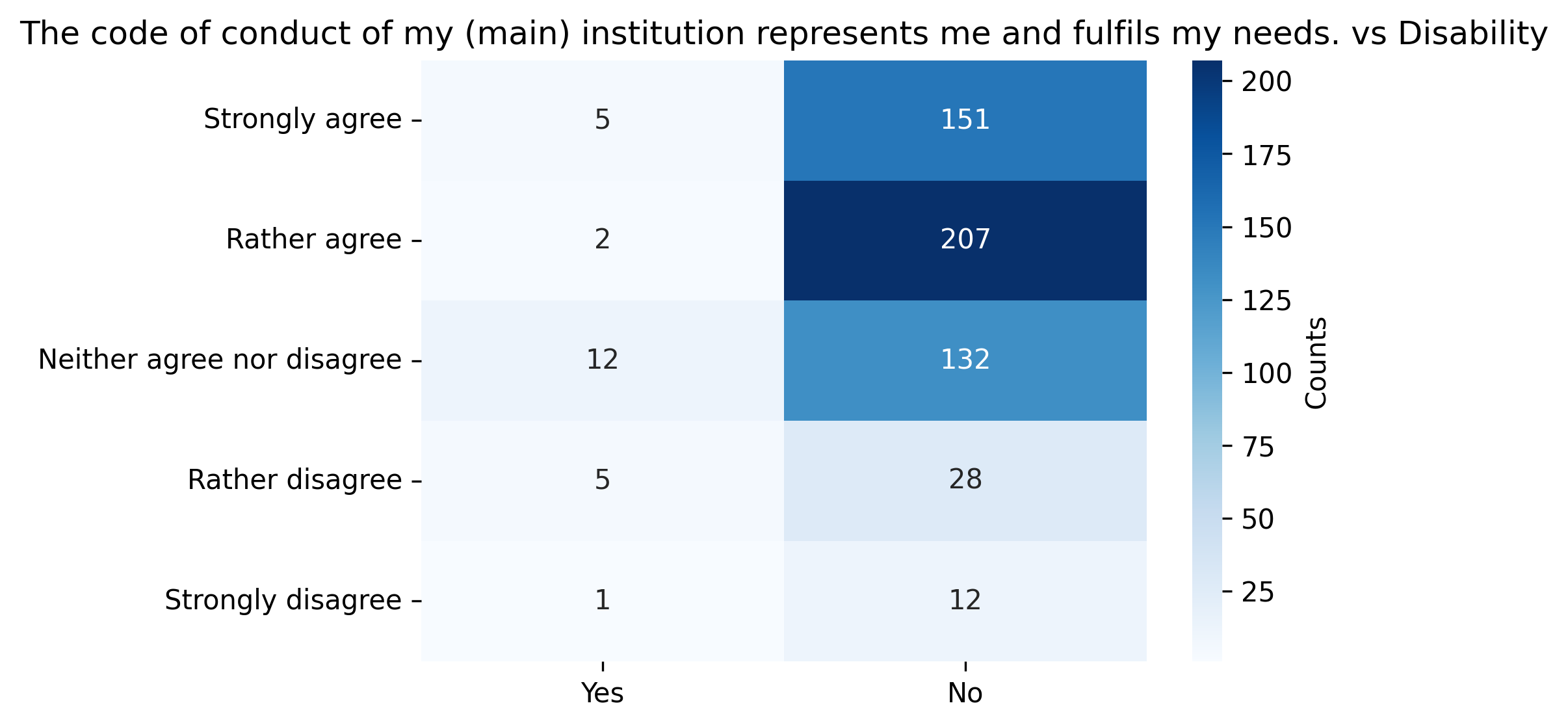}
            \label{fig:CC_rep_disability}
            \caption{}
	 	 \end{subfigure}
    \caption{Correlations between whether survey respondents consider themselves to have a disability or chronic condition and how well-represented they feel by the code of conduct in home institutions a) by fraction and b) by count. }
    \label{fig:code-of-conduct_representation-disability}
\end{figure}

\begin{figure}[H]
    \centering
    \includegraphics[width=0.75\linewidth]{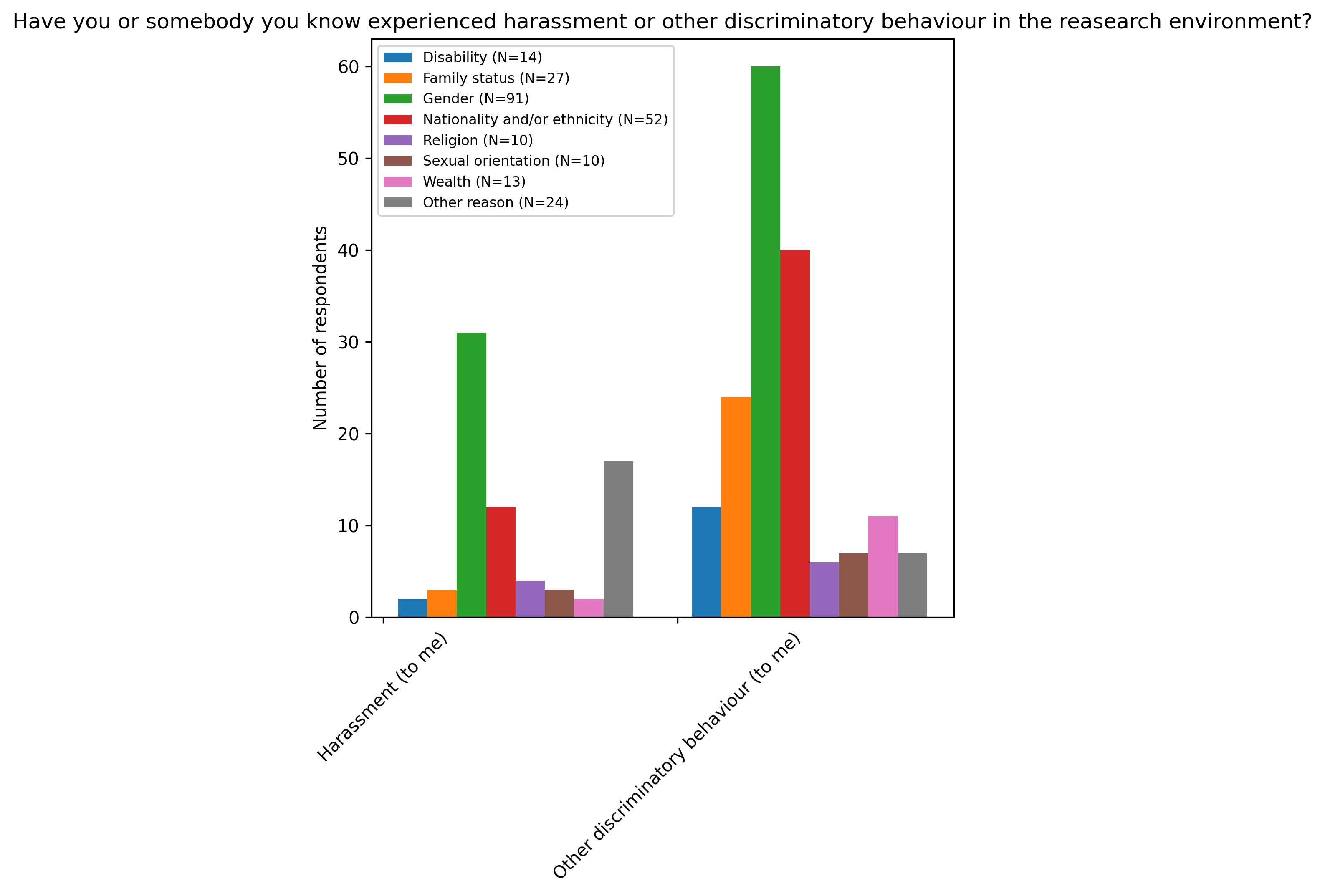}
    \caption{Survey responses on experiences of harassment and discrimination on the basis of various characteristics. } % \color{blue}(Needs error bars.)\color{black}}
    \label{fig:harassment-self}
\end{figure}

\begin{figure}[H]
    \centering
    \includegraphics[width=0.75\linewidth]{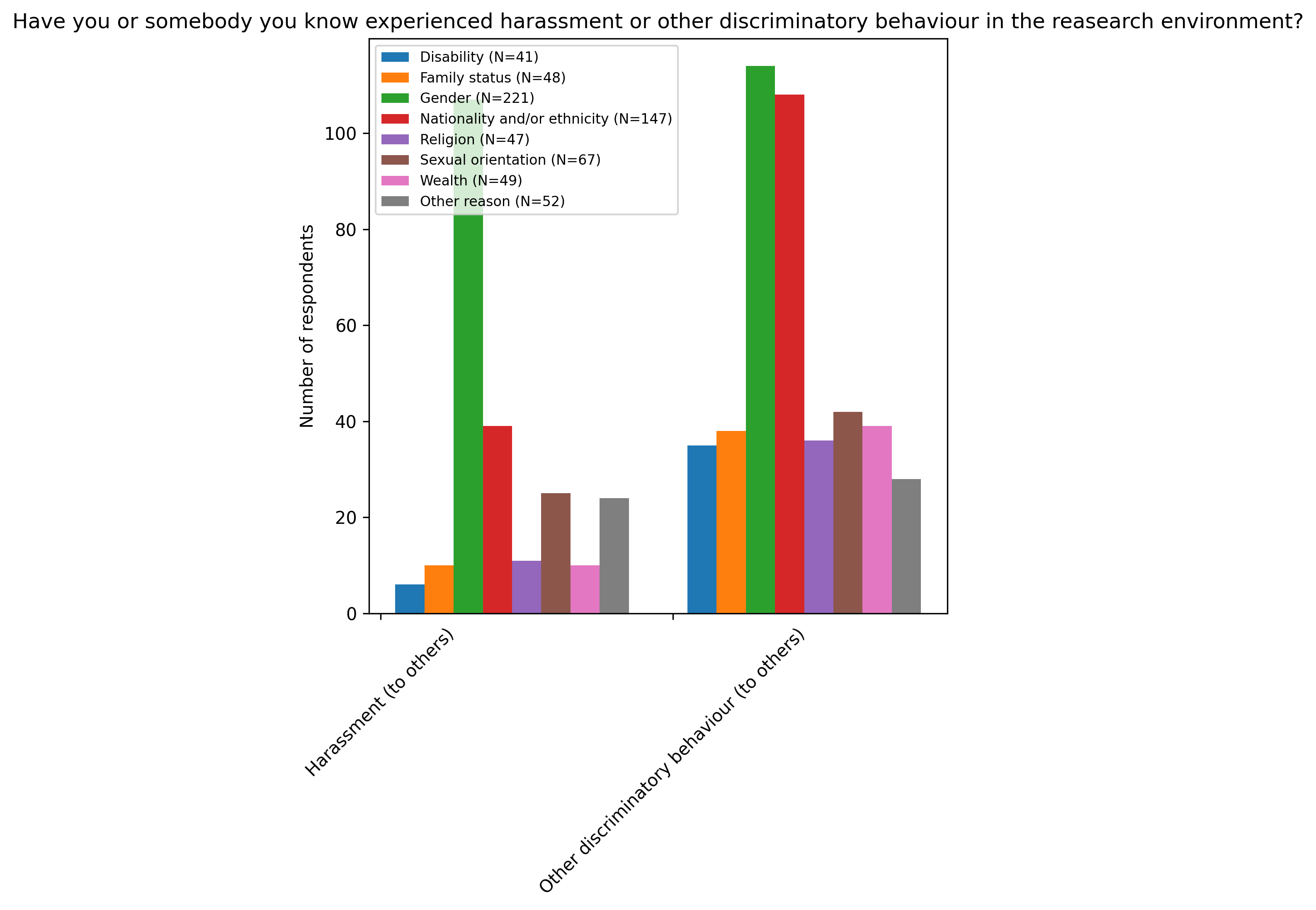}
    \caption{Survey responses on experiences of harassment and discrimination of another person on the basis of various characteristics. } % \color{blue}(Needs error bars.)\color{black}}
    \label{fig:harassment-others}
\end{figure}

\begin{figure}[H]
    \centering
    \includegraphics[width=\linewidth]{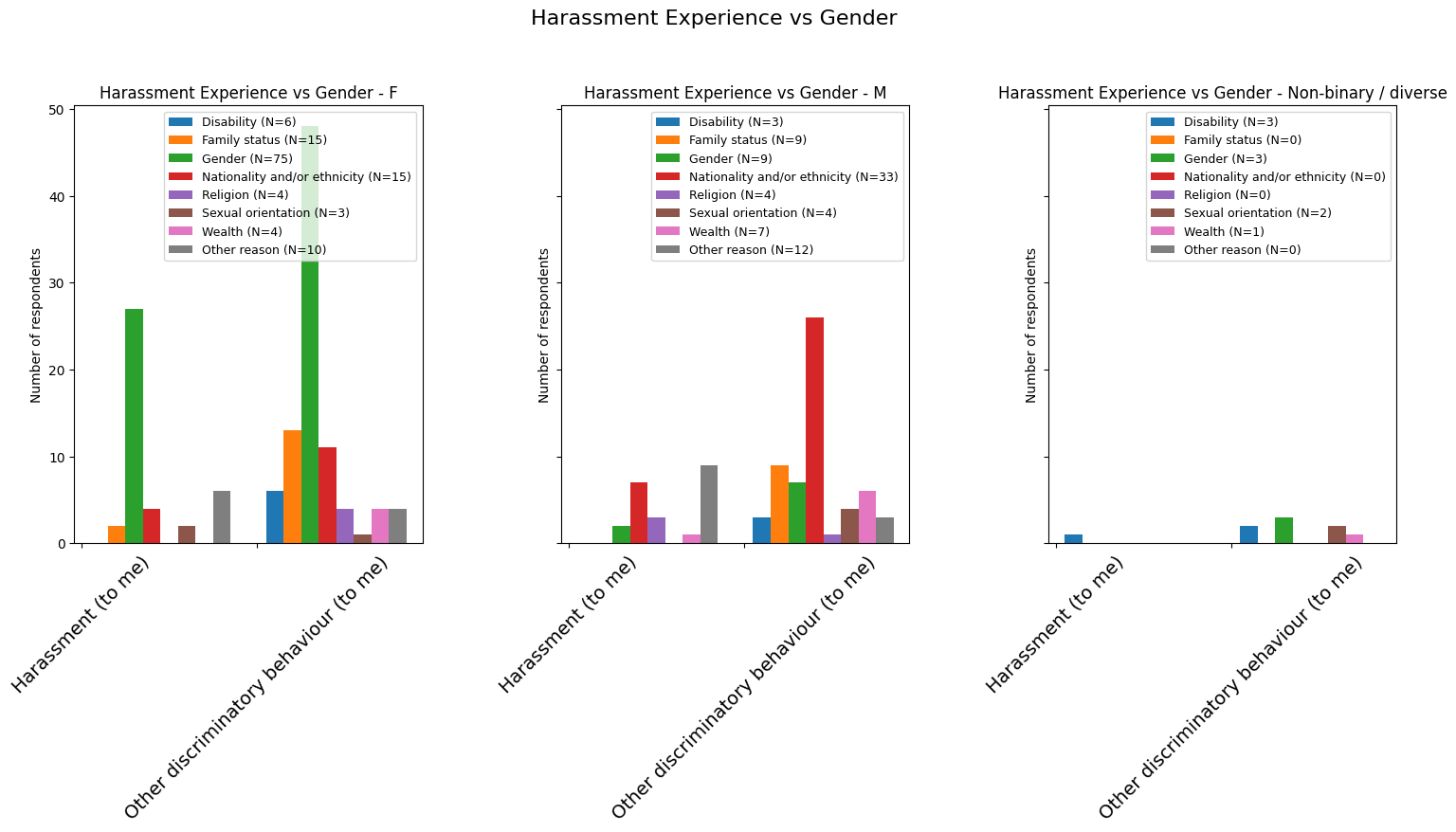}
    \caption{Survey responses on experiences of harassment and discrimination on the basis of various characteristics, comparing different genders. } % \color{blue}(Needs error bars.)\color{black}}
    \label{fig:harassment-self-gender}
\end{figure}

\begin{figure}[H]
    \centering
    \includegraphics[width=\linewidth]{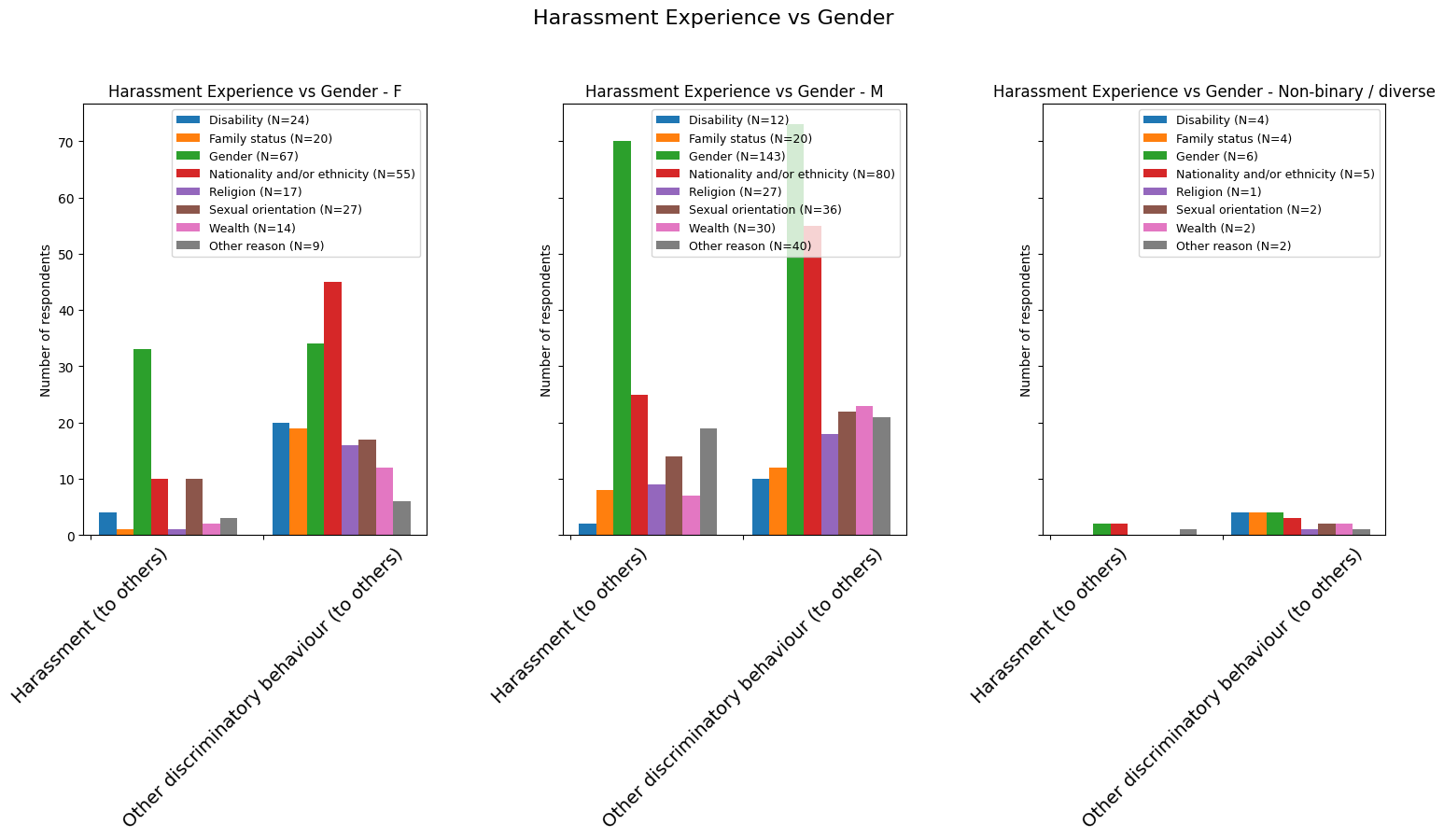}
    \caption{Survey responses on experiences of harassment and discrimination of another person on the basis of various characteristics, comparing different genders. } % \color{blue}(Needs error bars.)\color{black}}
    \label{fig:harassment-others-gender}
\end{figure}

\begin{figure}[H]
    \centering
    \includegraphics[width=0.8\linewidth]{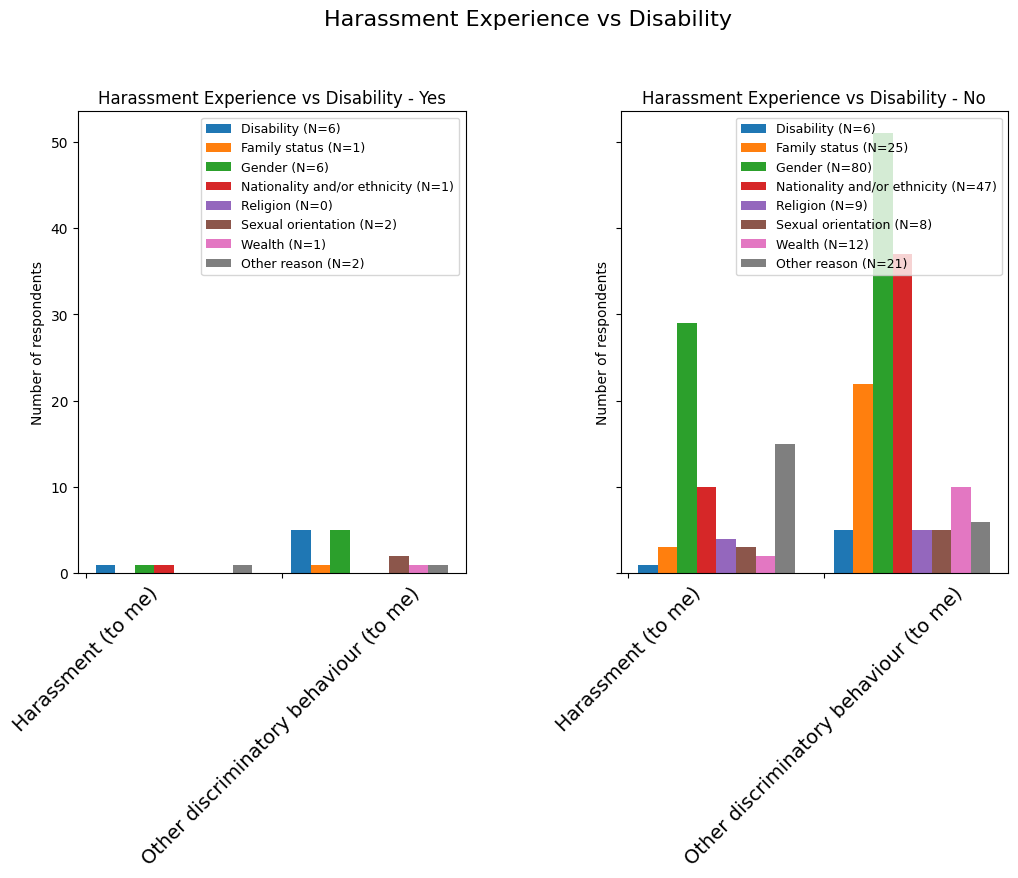}
    \caption{Survey responses on experiences of harassment and discrimination on the basis of various characteristics, comparing different abilities. } % \color{blue}(Needs error bars.)\color{black}}
    \label{fig:harassment-self-disability}
\end{figure}

\begin{figure}[H]
    \centering
    \includegraphics[width=0.8\linewidth]{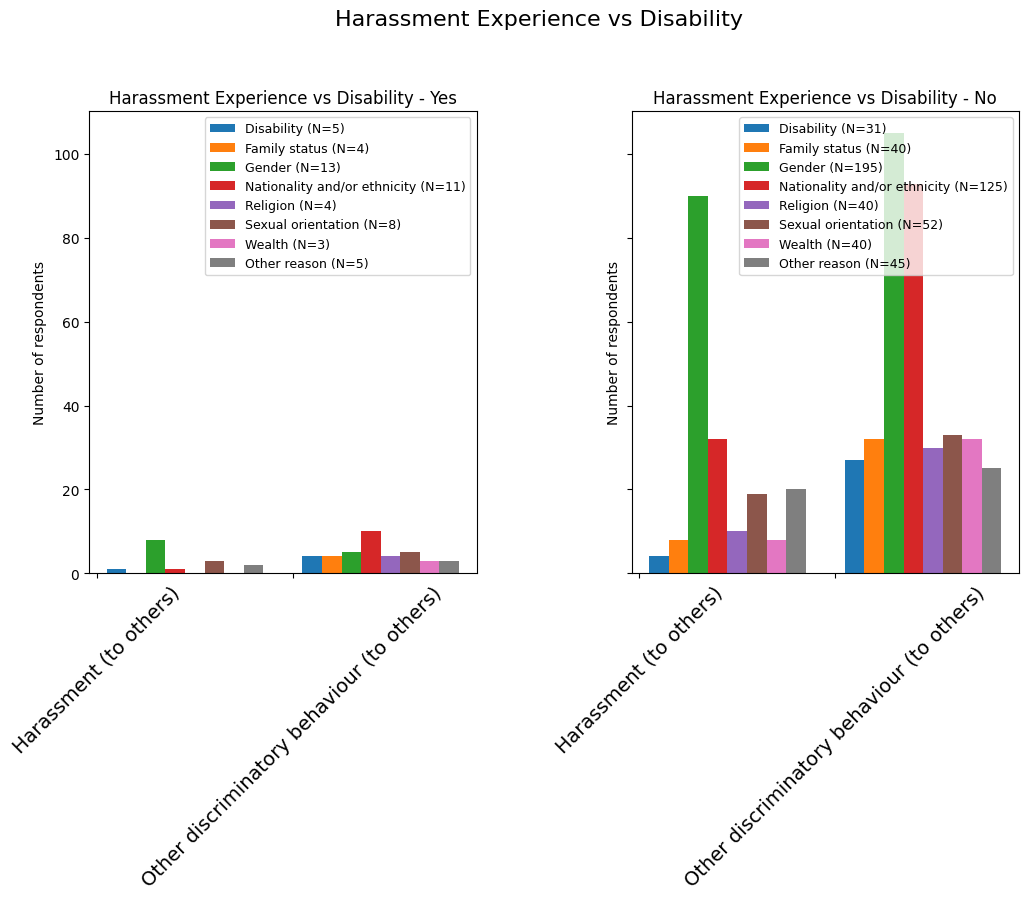}
    \caption{Survey responses on experiences of harassment and discrimination of another person on the basis of various characteristics, comparing different abilities. } % \color{blue}(Needs error bars.)\color{black}}
    \label{fig:harassment-others-disability}
\end{figure}

\begin{figure}[H]
    \centering
    \includegraphics[width=0.8\linewidth]{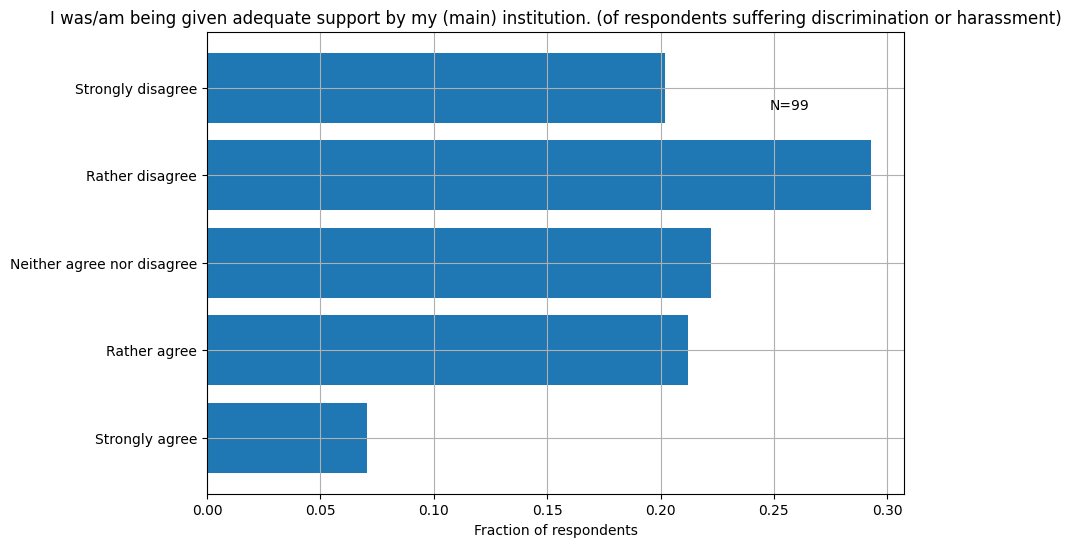}
    \caption{Survey responses on support received on respondents that have experienced harassment or discrimination. } % \color{blue}(Needs error bars.)\color{black}}
    \label{fig:support_harassment}
\end{figure}

% Until here or plots are reference on the text

\begin{figure}[H]
    \centering
    \begin{subfigure}{.49\textwidth}
        \centering
        \includegraphics[width=\linewidth]{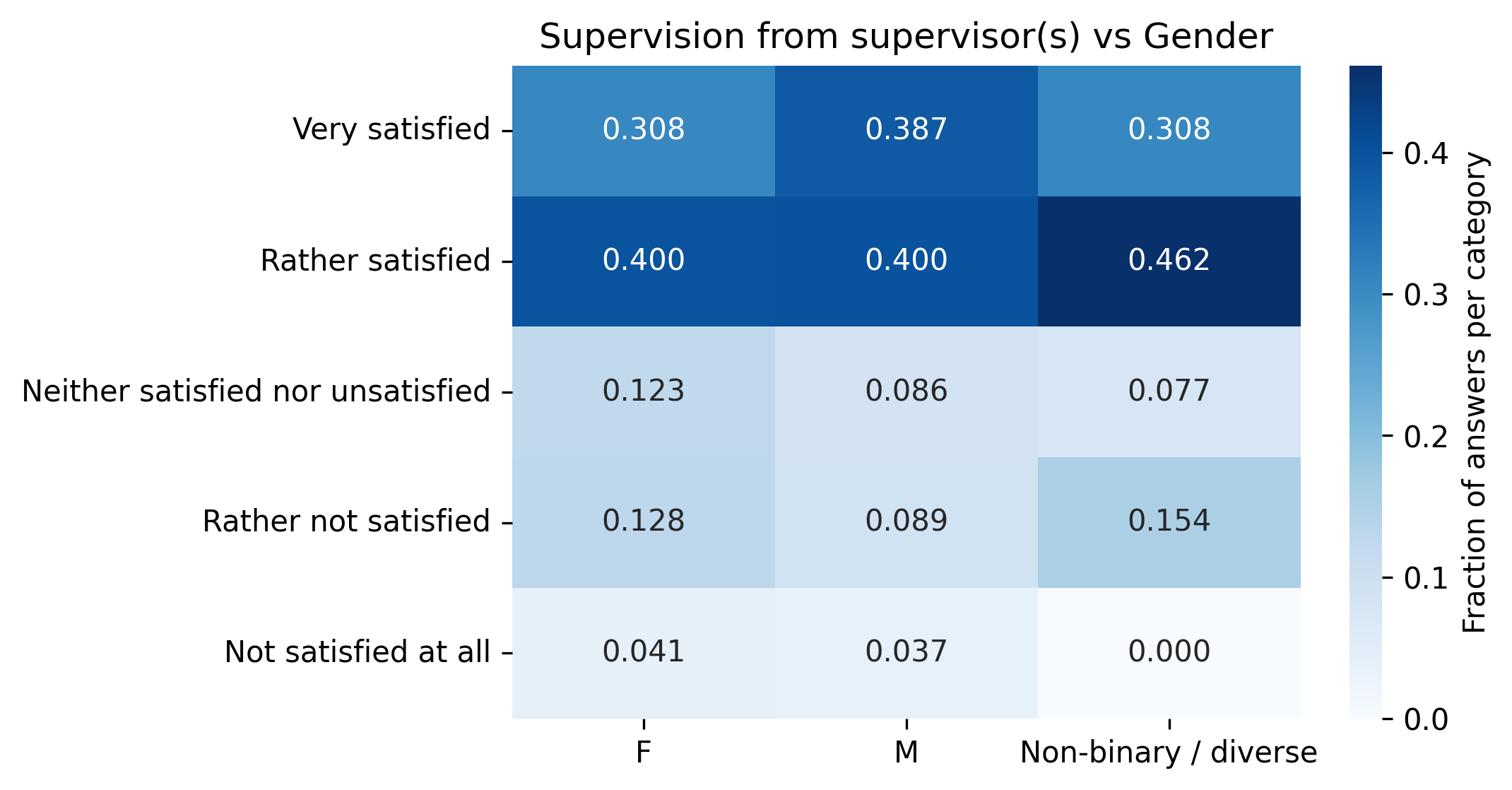}
        \label{fig:supervisor-gender-frac}
        \caption{}
    \end{subfigure}
    \hfill
    \begin{subfigure}{.49\textwidth}
        \centering
        \includegraphics[width=\linewidth]{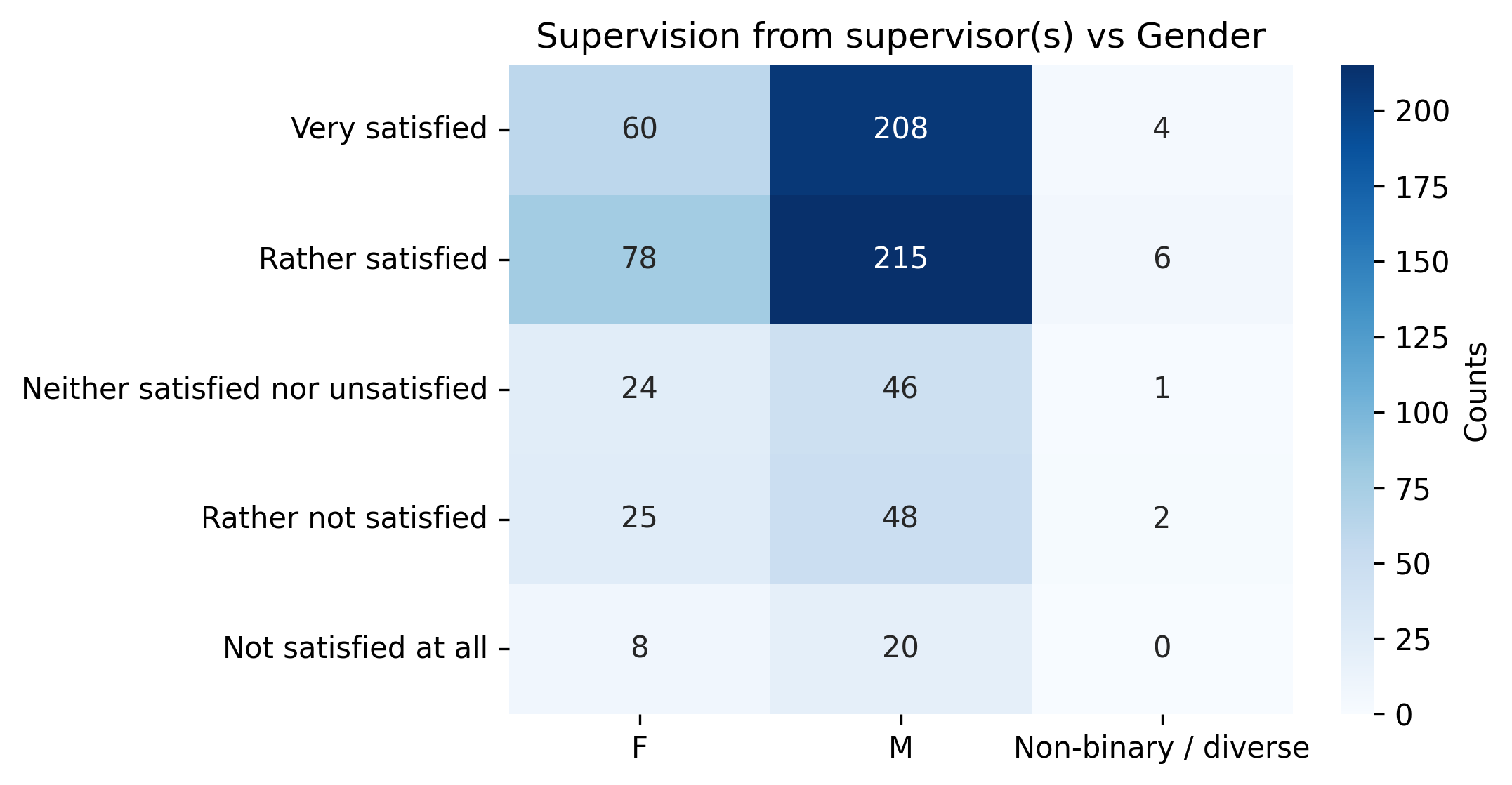}
        \label{fig:supervisor-gender-abs}
        \caption{}
    \end{subfigure}
    \centering
    \begin{subfigure}{.49\textwidth}
        \centering
        \includegraphics[width=\linewidth]{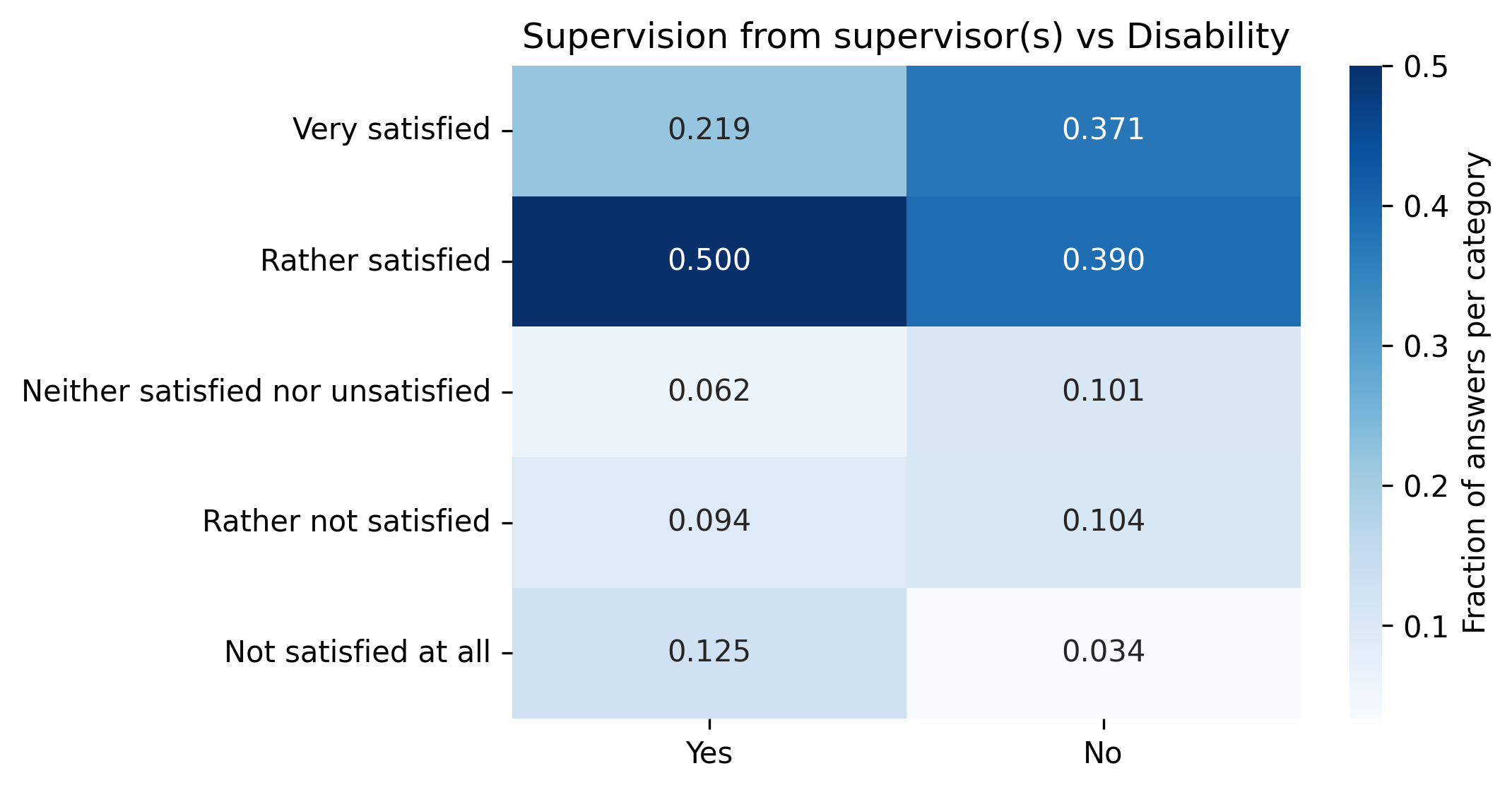}
        \label{fig:supervisor-disability-frac}
        \caption{}
    \end{subfigure}
    \hfill
    \begin{subfigure}{.49\textwidth}
        \centering
        \includegraphics[width=\linewidth]{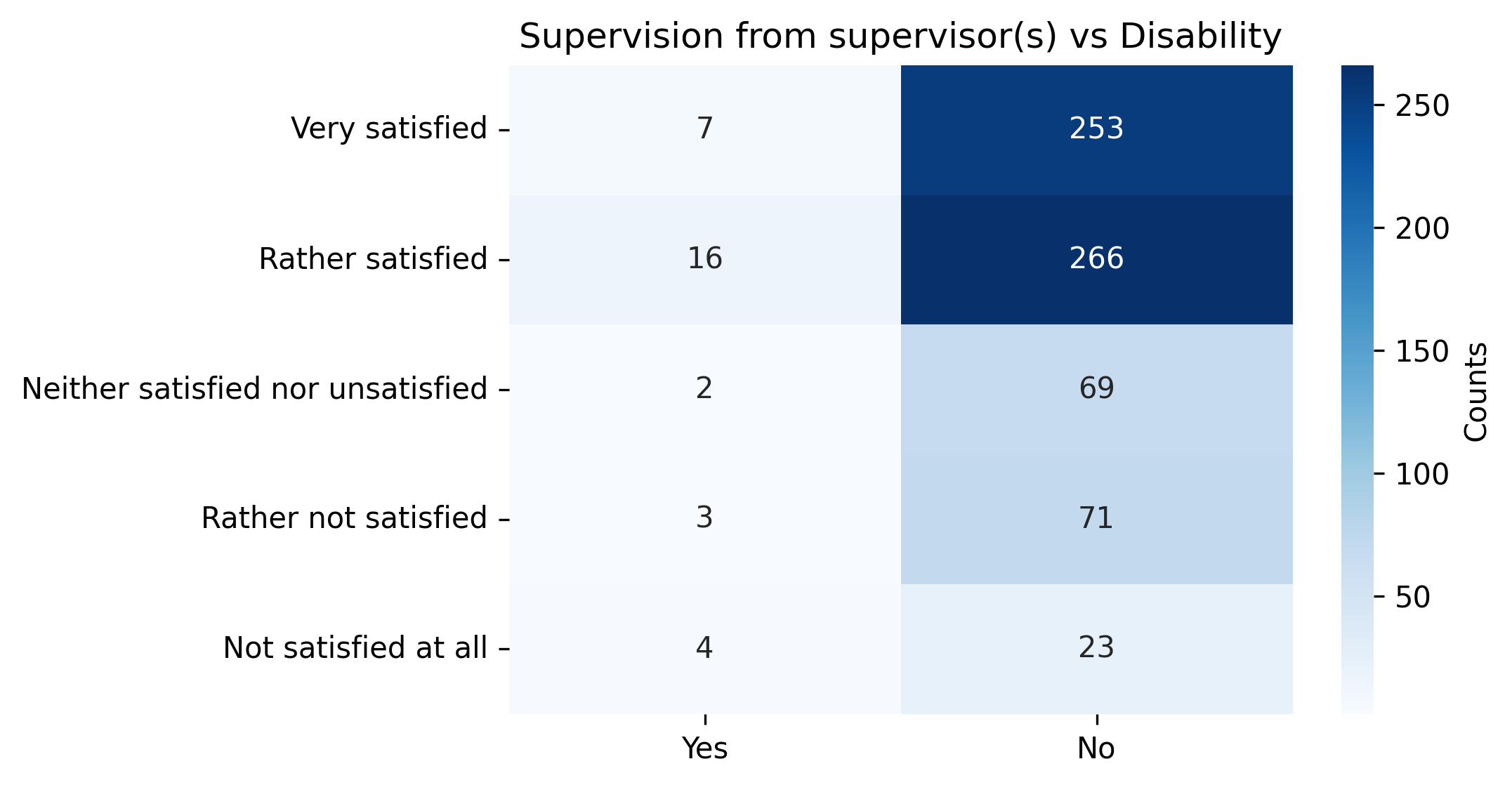}
        \label{fig:supervisor-disability-abs}
        \caption{}
    \end{subfigure}
    \caption{Survey responses on supervisor satisfaction correlated with gender (top row) and whether respondents consider themselves to have a disability or chronic condition (bottom row) presented as fractions (left) and counts (right).}
    \label{fig:supervision_gender_disability}
\end{figure}

\begin{figure}[H]
    \centering
    \includegraphics[width=0.8\linewidth]{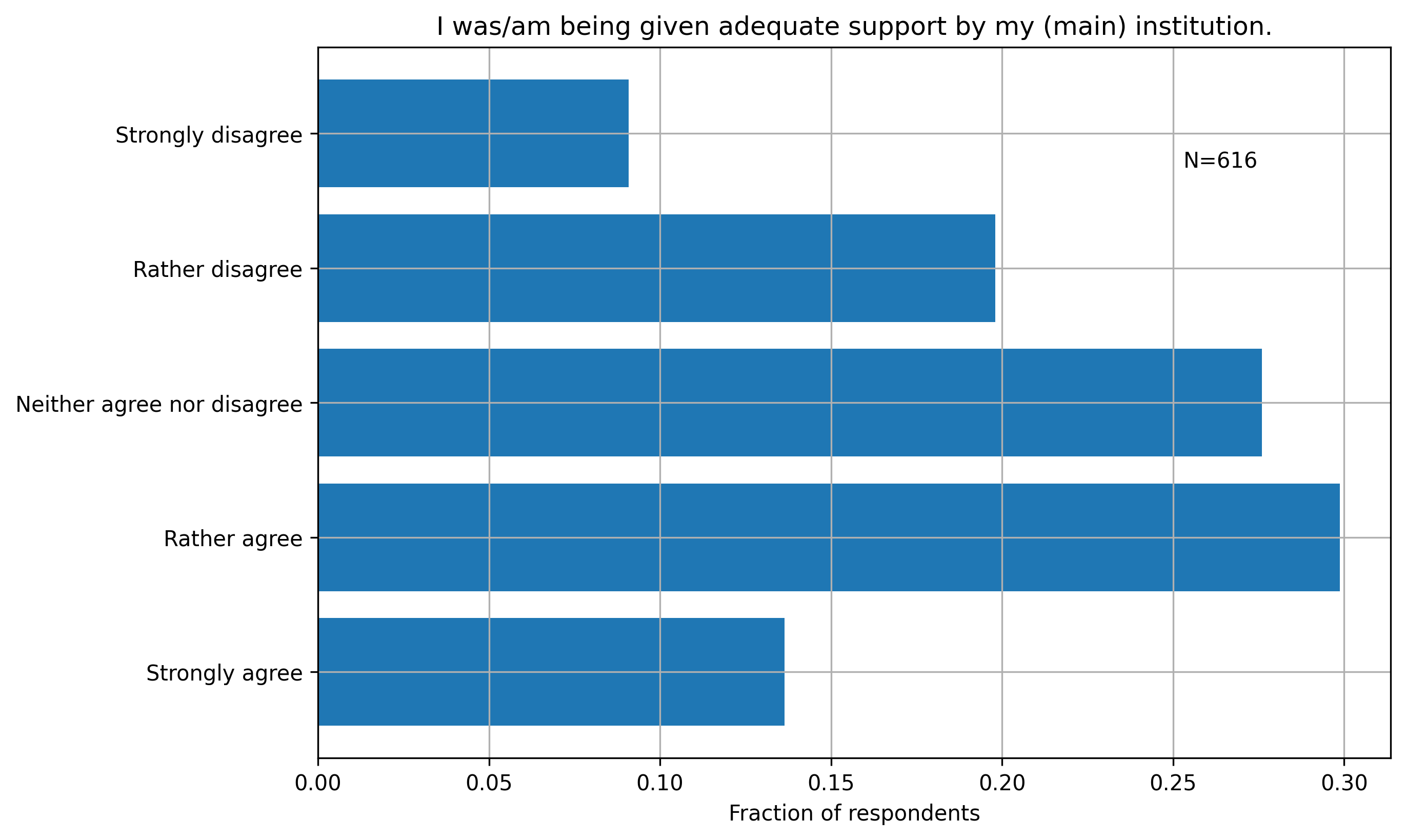}
    \caption{Survey results on adequacy of support during research career. } % \color{blue}(Needs error bars.)\color{black}}
    \label{fig:support}
\end{figure}

\begin{figure}[H]
    \centering
    \includegraphics[width=0.9\linewidth]{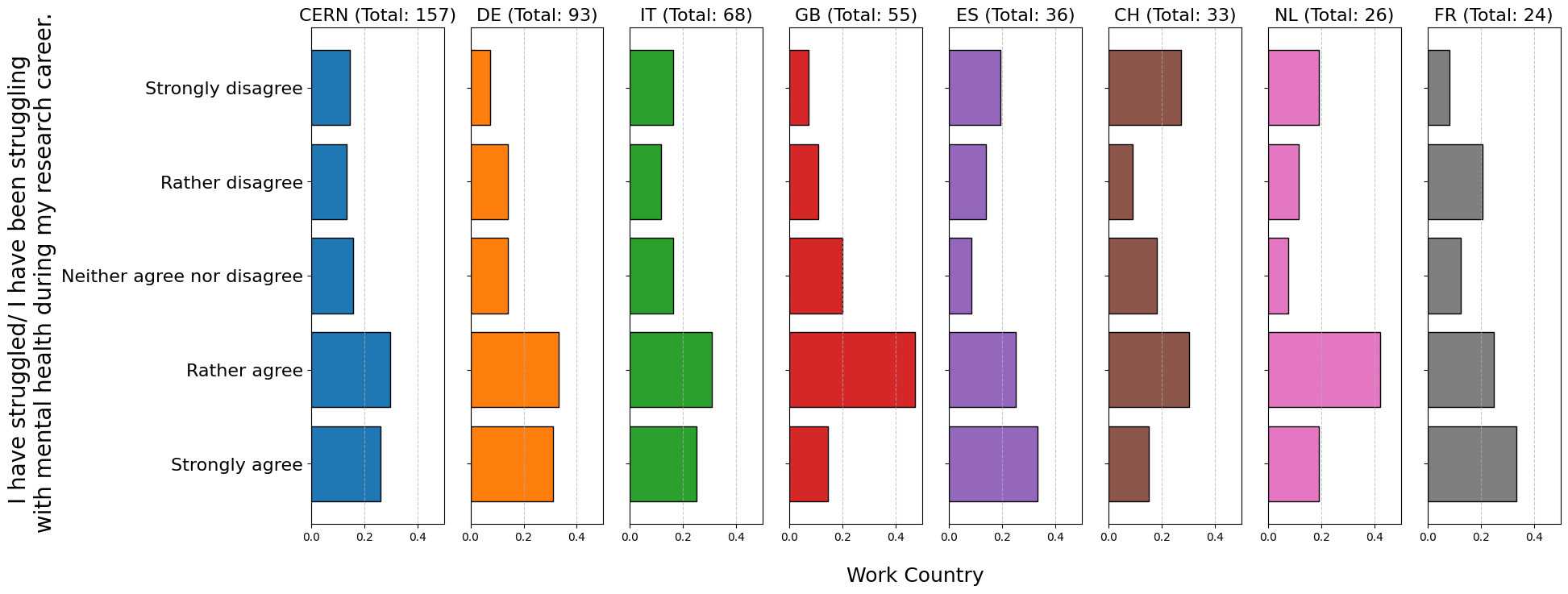}
    \caption{Survey responses on mental health struggles during research career correlated with country of work. } % \color{blue}(Plot requires larger axes and tick labels. Needs error bars.)\color{black}}
    \label{fig:enter-label}
\end{figure}

\begin{figure}[H]
    \centering
    \includegraphics[width=\linewidth]{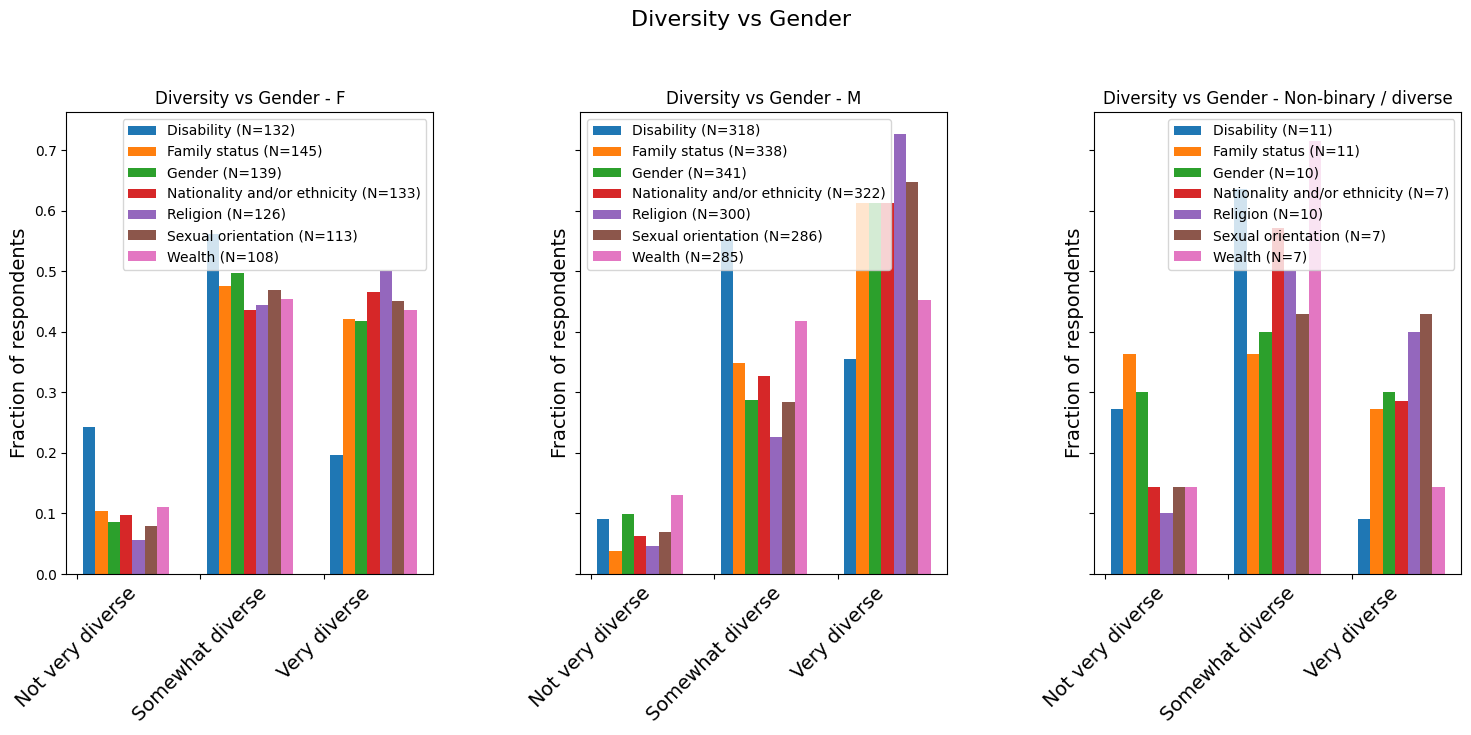}
    \caption{Survey perceptions of diversity at home institutes across various characteristics comparing different genders. } % \color{blue}(Maybe have ``Diversity" as the headings?Needs error bars)\color{black}}
    \label{fig:diverse-gender}
\end{figure}

\begin{figure}[H]
    \centering
    \includegraphics[width=0.8\linewidth]{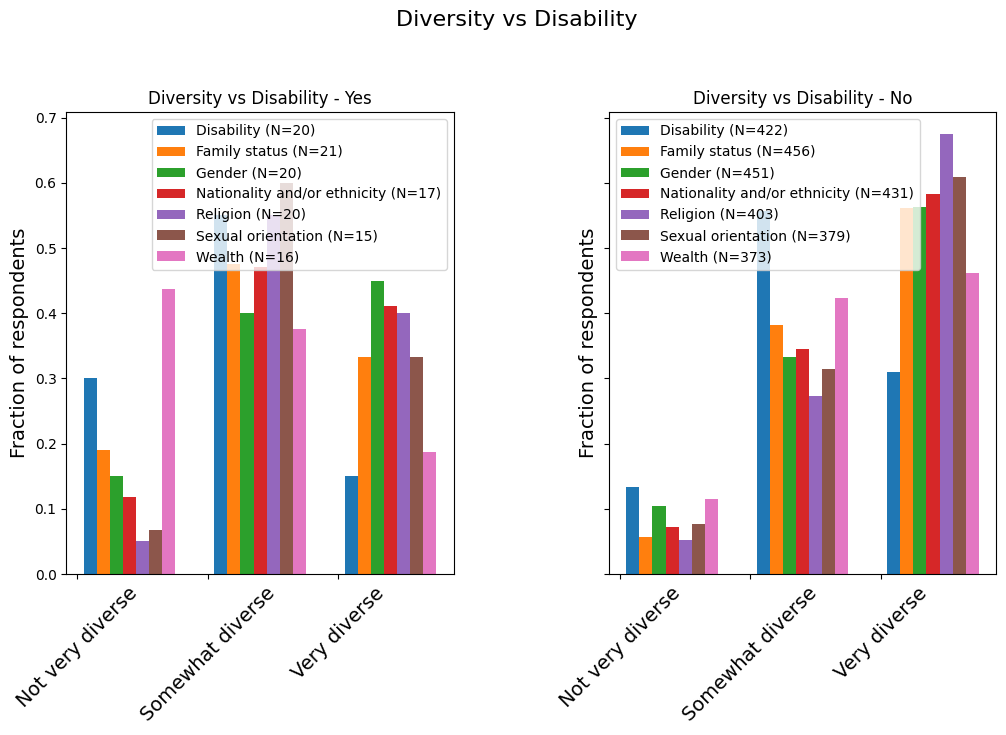}
    \caption{Survey perceptions of diversity at home institutes across various characteristics comparing different abilities.} % \color{blue}(Maybe have ``Diversity" as the headings? Needs error bars)\color{black}}
    \label{fig:diverse-disability}
\end{figure}

\subsection{Communicating the importance of particle physics}

\begin{table}[H]
    \centering
    \begin{tabular}{l c c}
        Question & Yes & No \\
        \toprule
        Are you motivated to engage with the public on topics of future project in particle physics? &  & \\
        - \textit{all respondents} & 614 & 105\\
        -\textit{ respondents for whom it is important to have a future flagship project }& 488 & 65\\
        -\textit{ respondents who plan to stay in the field after their current position }& 546 & 68\\
        \midrule
        Do you feel adequately prepared (trained) to engage with the public on topics of particle physics? &  & \\
         - \textit{all respondents} & 440 & 303\\
        \midrule
        Are you participating in outreach activities? &  & \\
        - \textit{all respondents} & 442 & 309\\
        -\textit{ respondents who are motivated for outreach }& 387 & 208\\
        -\textit{ respondents who are motivated for outreach and plan to stay in the field }& 349 & 178\\
        \midrule
        Do you feel valued adequately for doing outreach work? &  & \\
        - \textit{respondents who chose to answer the supplementary questions and are active in outreach } & 135 & 96\\
        \midrule
        Have you received any training on science communication or outreach? (teaching included) &  & \\
        - \textit{respondents who chose to answer the supplementary questions} & 167 & 208\\
        \midrule
        Did you find any of the training you received useful? &  & \\
        - \textit{respondents who chose to answer the supplementary questions and received training} & 135 & 24\\
        \bottomrule
    \end{tabular}
    \caption{Survey responses on engagement and training in science communication and outreach across different groups of respondents.}
    \label{tab:comm_wg_raw_data}
\end{table}

\begin{figure}[H]
    \centering
    \begin{subfigure}{.49\textwidth}
        \centering
        \includegraphics[width=\linewidth]{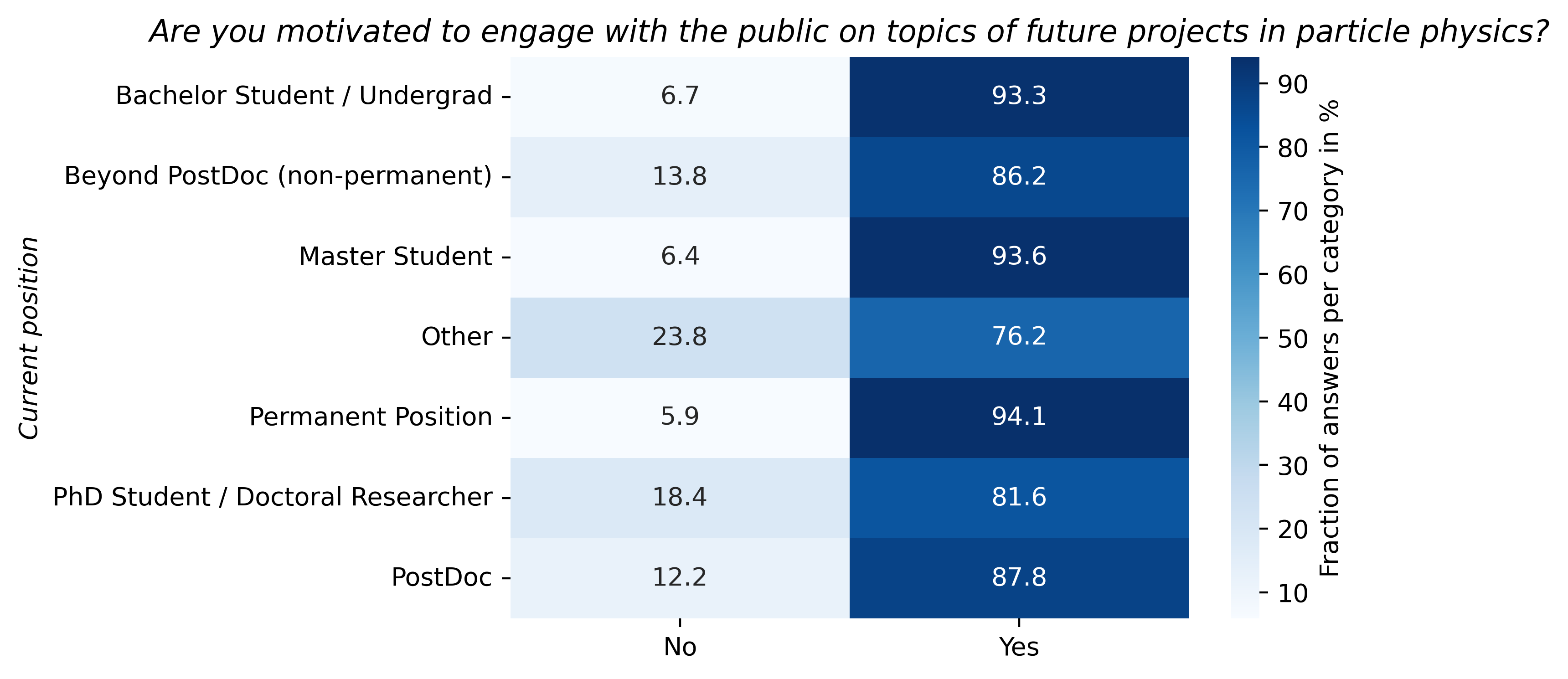}
        \label{fig:comm_mot_vs_pos_rel}
        \caption{}
    \end{subfigure}
    \hfill
    \begin{subfigure}{.49\textwidth}
        \centering
        \includegraphics[width=\linewidth]{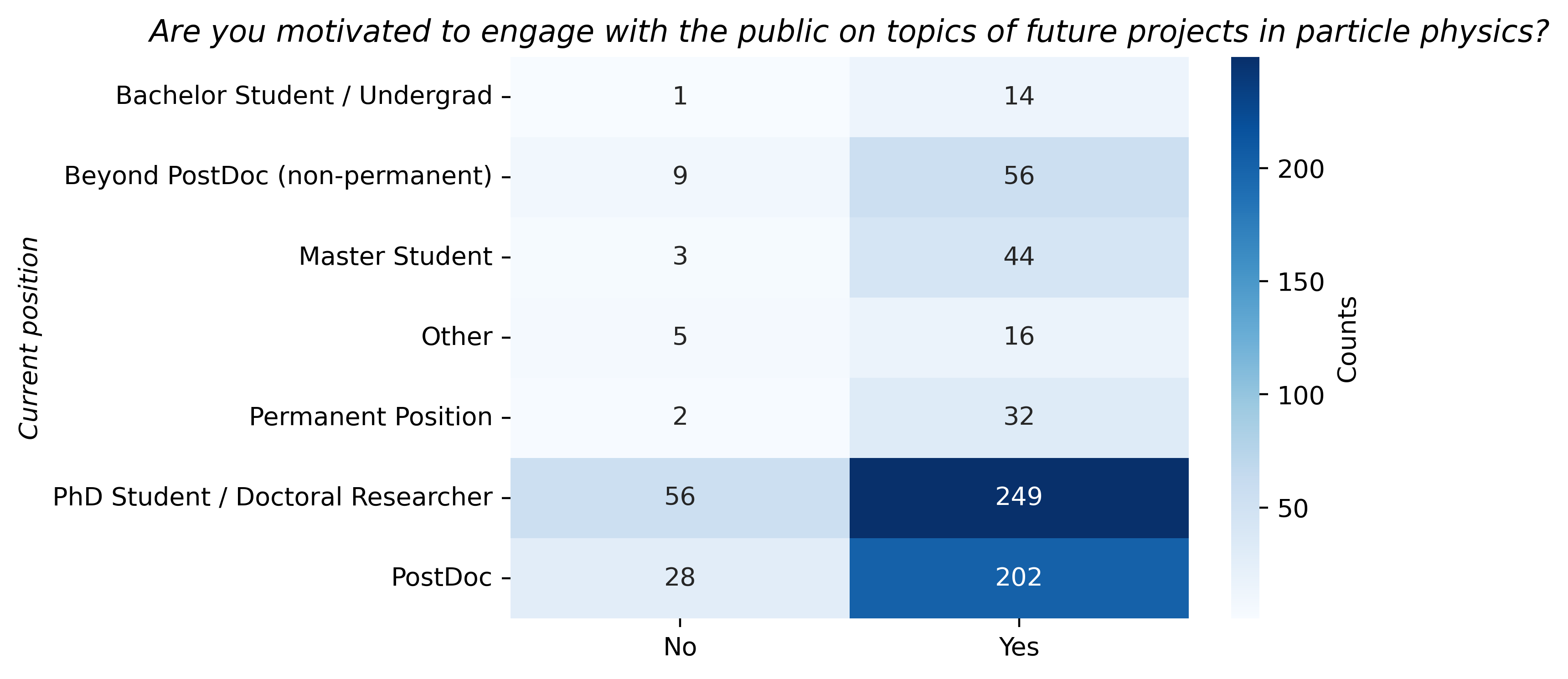}
        \label{fig:comm_mot_vs_pos_abs}
        \caption{}
    \end{subfigure}
    \centering
    \begin{subfigure}{.49\textwidth}
        \centering
        \includegraphics[width=\linewidth]{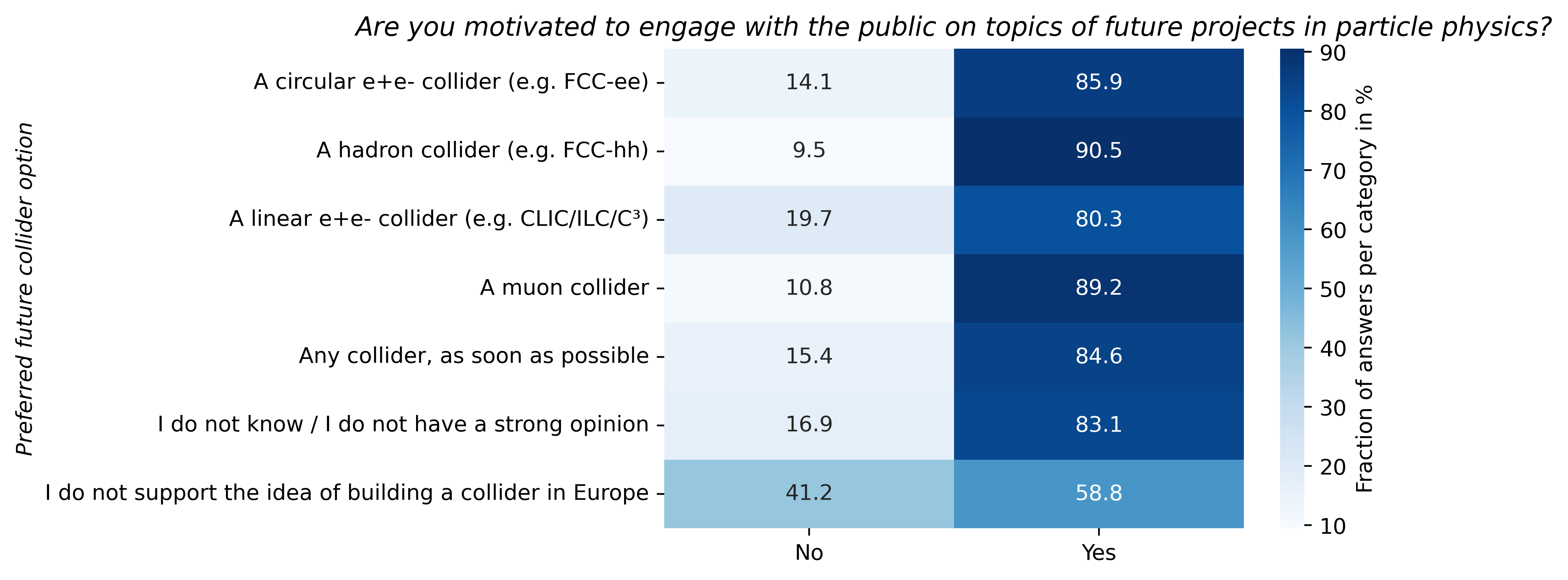}
        \label{fig:comm_mot_vs_future_rel}
        \caption{}
    \end{subfigure}
    \hfill
    \begin{subfigure}{.49\textwidth}
        \centering
        \includegraphics[width=\linewidth]{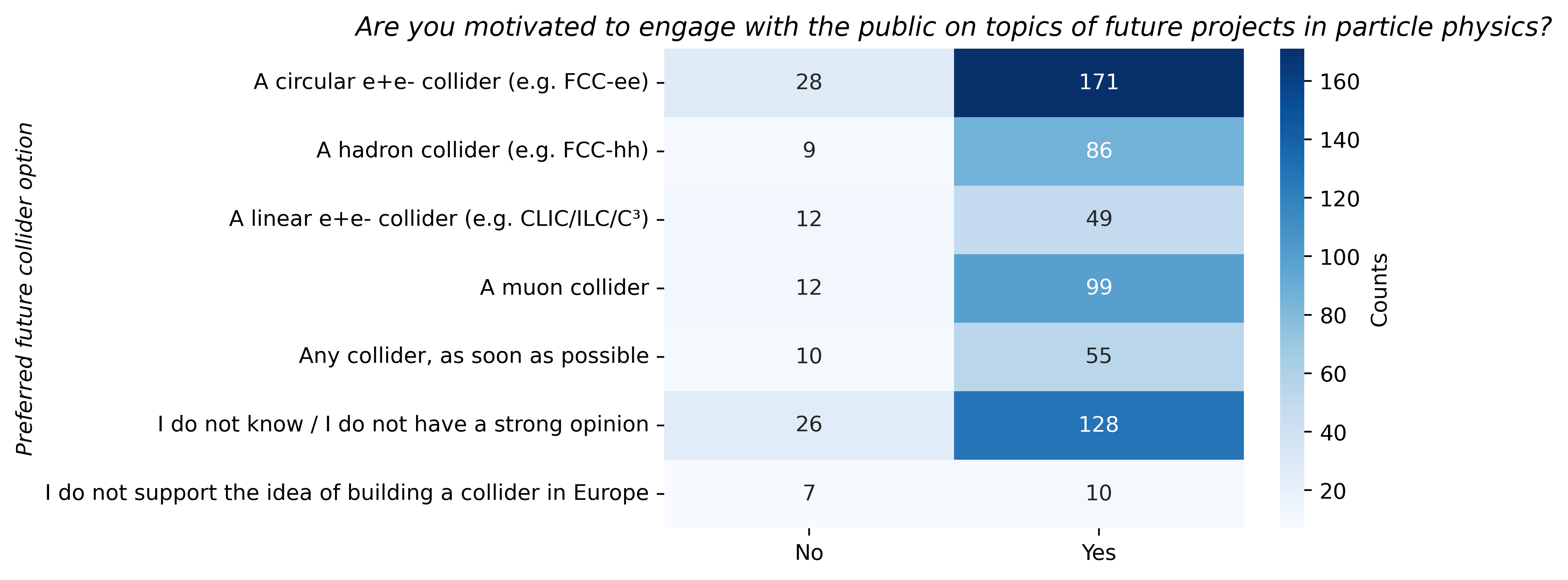}
        \label{ffig:comm_mot_vs_future_rel}
        \caption{}
    \end{subfigure}
    \caption{Survey responses on motivation to engage with the public on future projects in particle physics correlated with career status (top row) and preferred future collider project (bottom row) presented as fractions (left) and absolute counts (right). } % \color{blue}(Plots require larger labels)\color{black}}
    \label{fig:comm_motivation_splits}
\end{figure}

\begin{figure}[H]
    \raggedright
    \hspace{5mm}
    \includegraphics[width=0.7\textwidth]{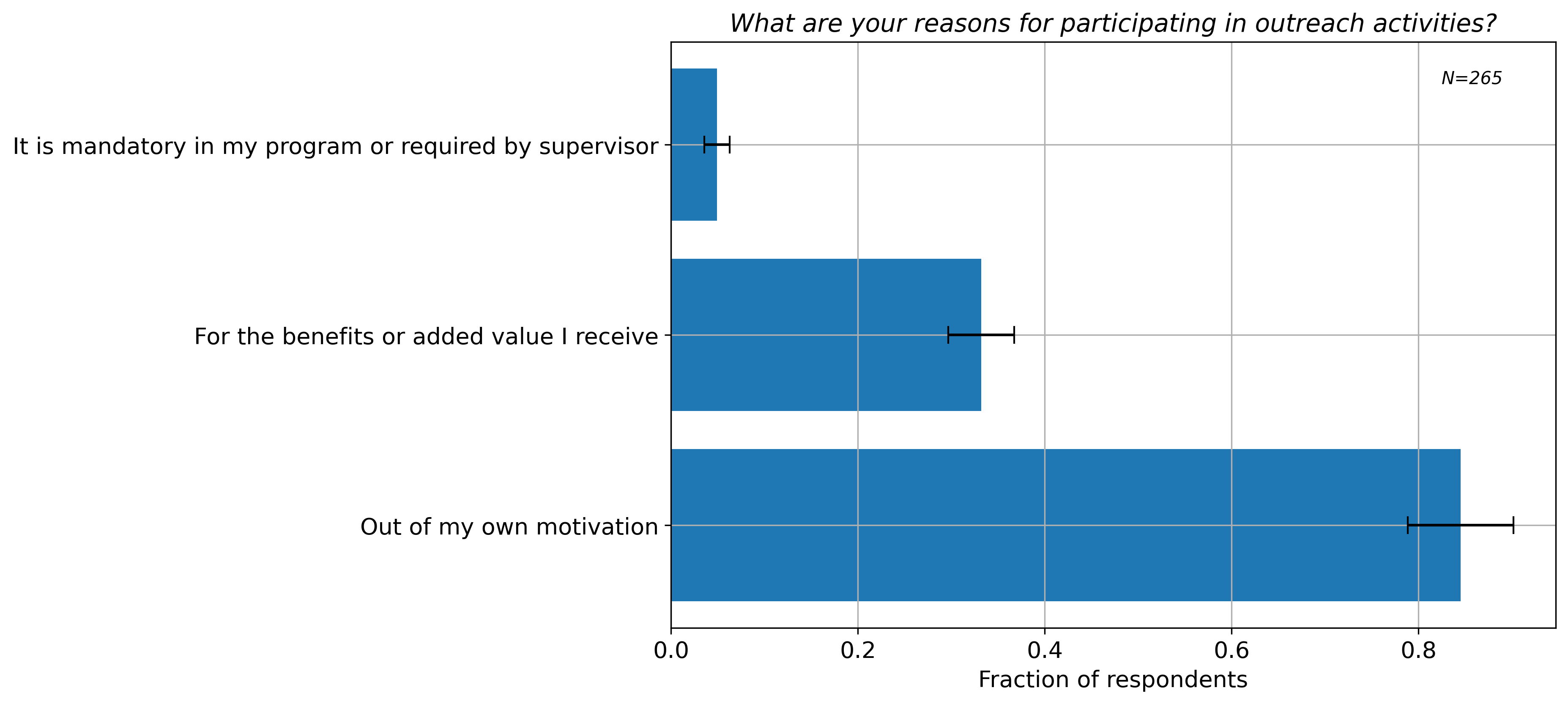}
    \caption{Reasons behind their active involvement in outreach for the subset of respondents who chose to answer the supplementary questions and reported being active in outreach.}
    \label{fig:comm_reas_participation}
\end{figure}

\begin{figure}[!htb]
    \raggedright
    \hspace{0pt}
    \includegraphics[width=0.78\textwidth]{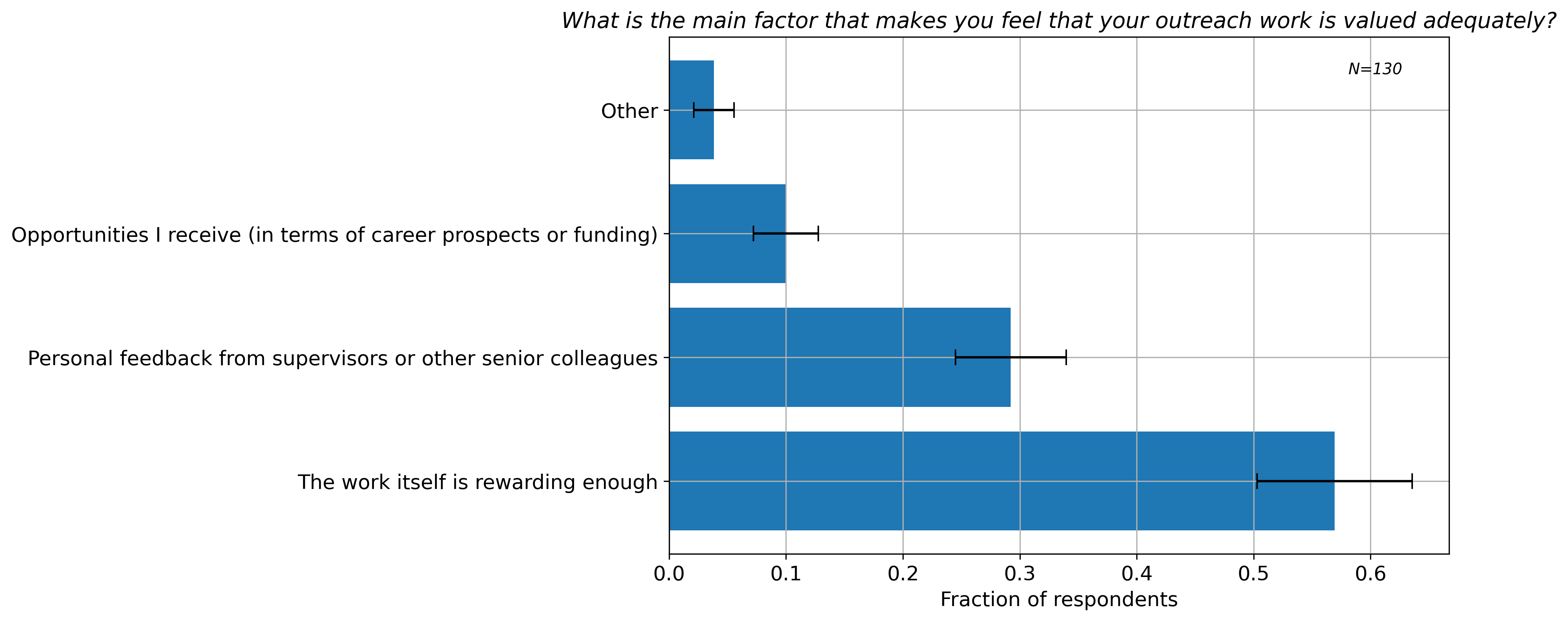}
    \caption{Reasons behind feeling adequately valued for their outreach work for the subset of respondents who chose to answer the supplementary questions and reported being active in outreach.}
    \label{fig:comm_reas_valued}
\end{figure}

\begin{figure}[H]
    \centering
    \begin{subfigure}{.49\textwidth}
        \centering
        \includegraphics[width=\linewidth]{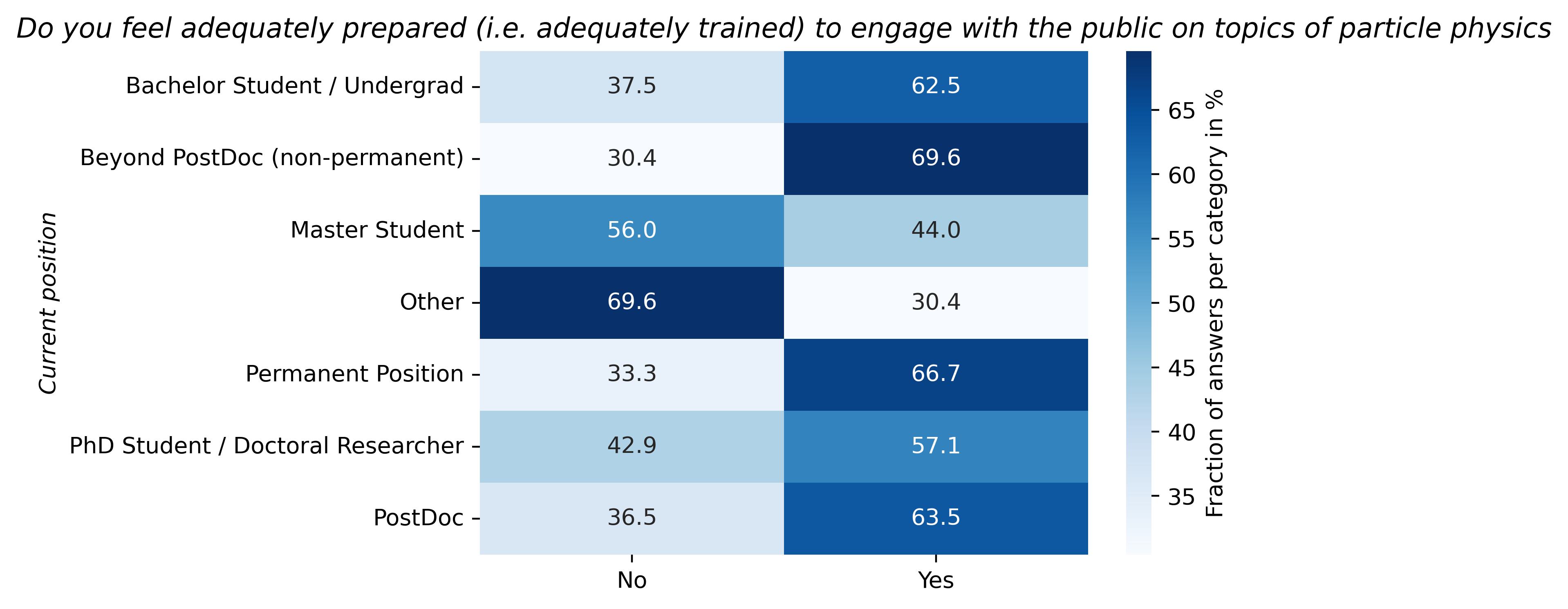}
        \label{fig:comm_prep_vs_pos_rel}
        \caption{}
    \end{subfigure}
    \hfill
    \begin{subfigure}{.49\textwidth}
        \centering
        \includegraphics[width=\linewidth]{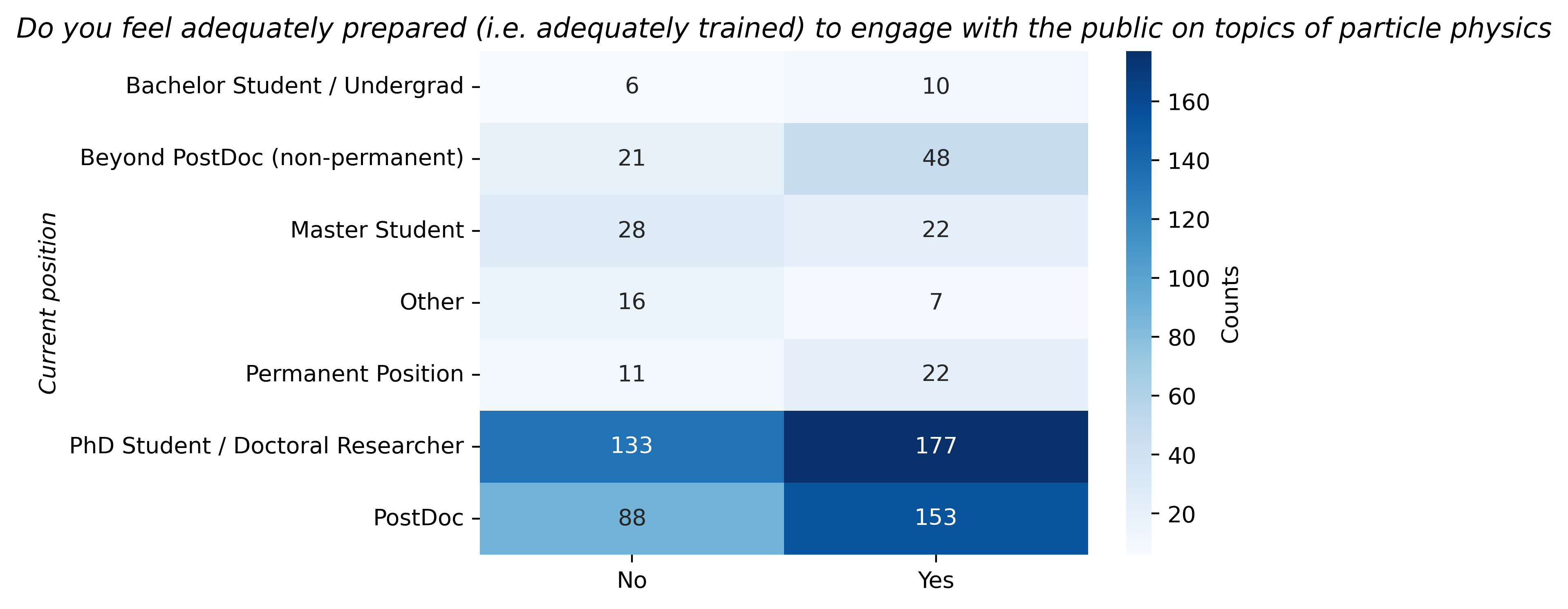}
        \label{fig:comm_prep_vs_pos_abs}
        \caption{}
    \end{subfigure}
    \centering
    \begin{subfigure}{.49\textwidth}
        \centering
        \includegraphics[width=\linewidth]{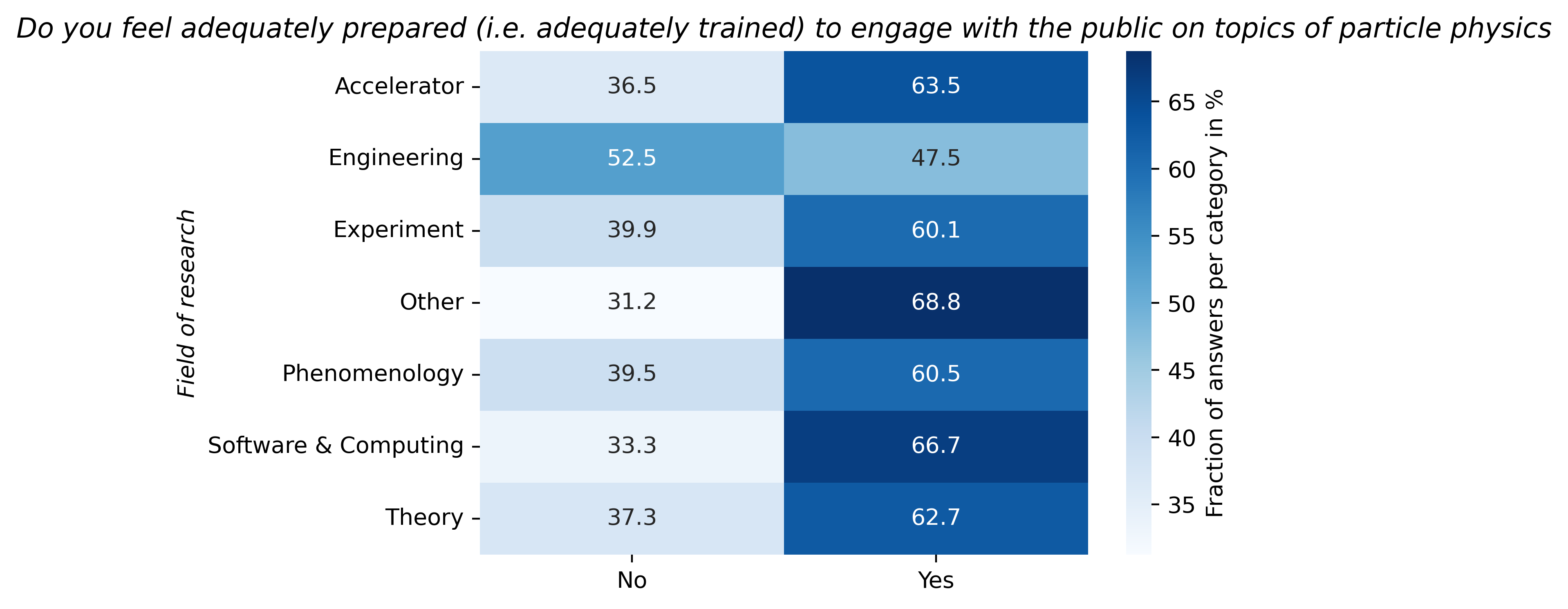}
        \label{fig:comm_prep_vs_field_rel}
        \caption{}
    \end{subfigure}
    \hfill
    \begin{subfigure}{.49\textwidth}
        \centering
        \includegraphics[width=\linewidth]{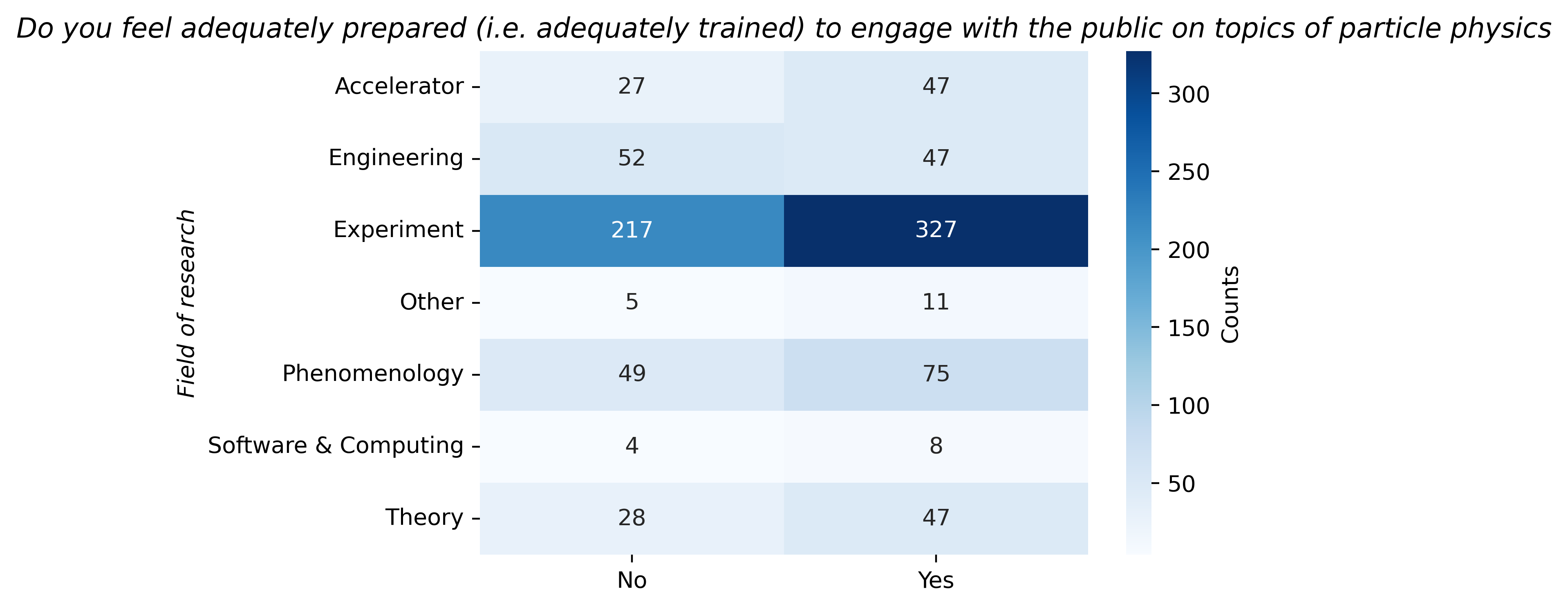}
        \label{ffig:comm_prep_vs_field_abs}
        \caption{}
    \end{subfigure}
    \caption{Survey responses on preparedness to engage with the public on topics of particle physics correlated with career status (top row) and field of research(bottom row) presented as fractions (left) and absolute counts (right).} % \color{blue}(Plots require larger labels.)\color{black}}
    \label{fig:comm_preparedness_splits}
\end{figure}

\begin{figure}[!htb]
    \raggedright
    \includegraphics[width=0.78\linewidth]{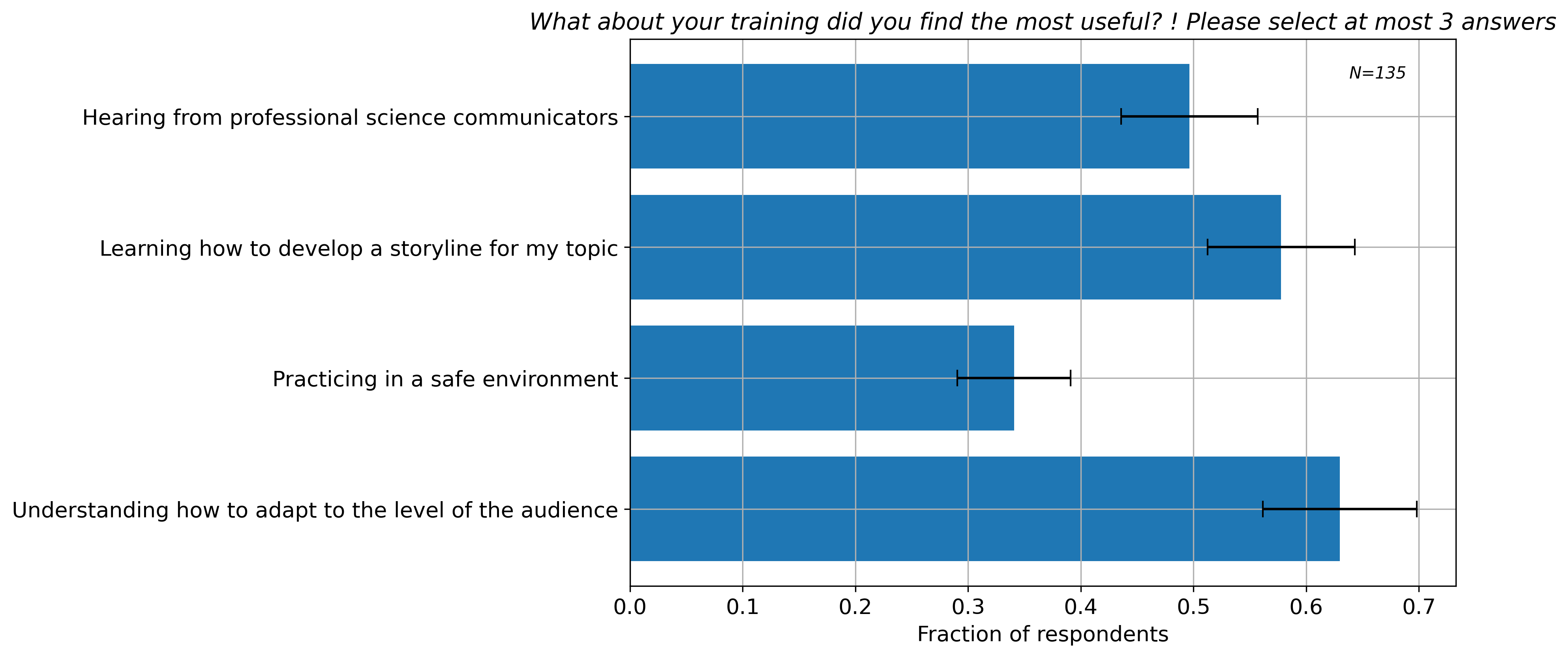}
    \caption{Fraction of respondents reporting what aspects of the science communication training they received they found the most useful.}
    \label{fig:comm_training_useful}
\end{figure}

\begin{figure}[H]
    \hspace{0pt} 
    \hspace{5mm}
    \includegraphics[width=0.7\linewidth]{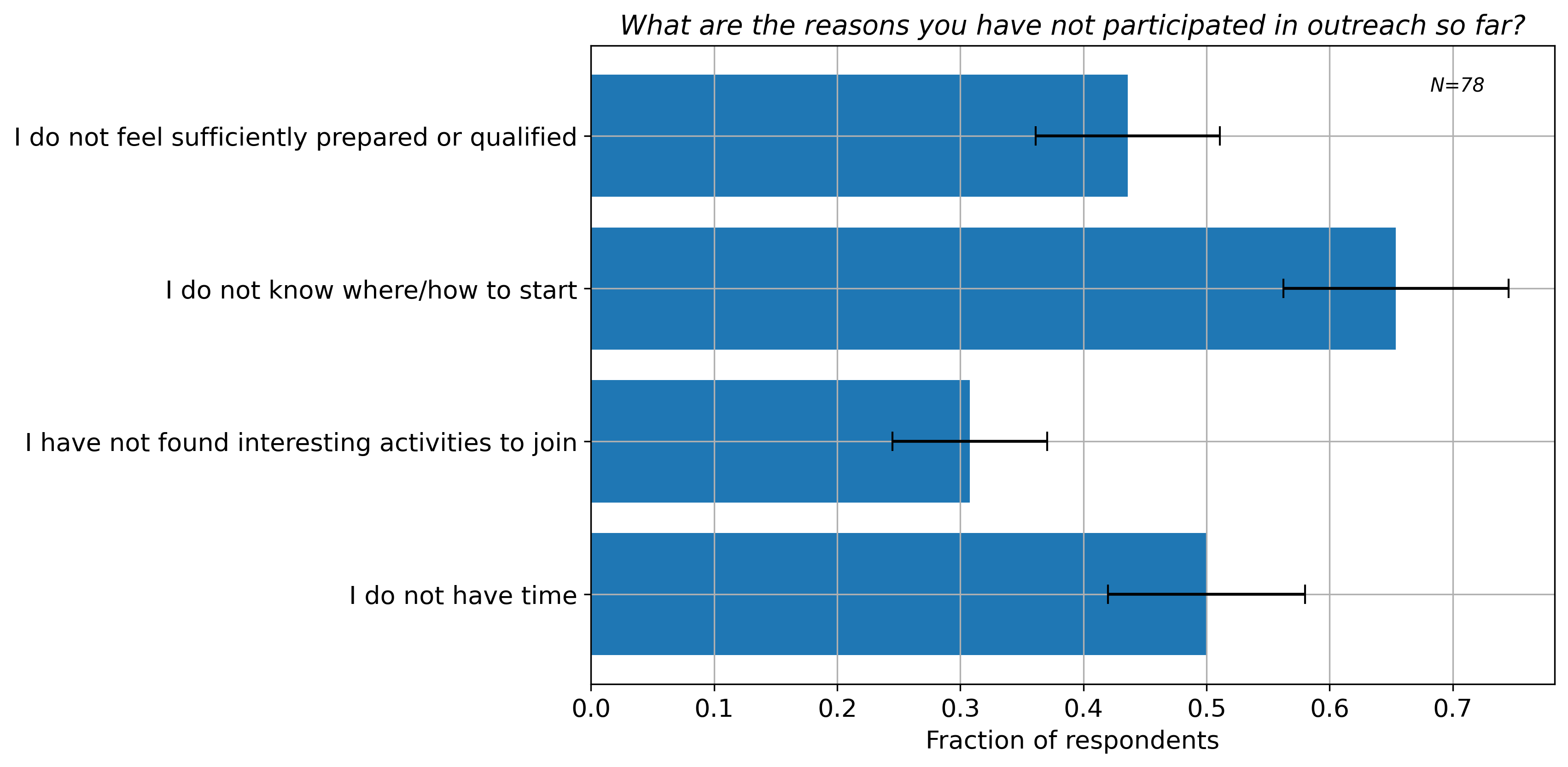}
    \caption{Fraction of survey participants expressing the motivation of no involvement in outreach activities. The sample size is small, as this was an additional question.}
    \label{fig:comm_reason_no_inv}
\end{figure}

\begin{figure}[H]
    \hspace{25mm}
\includegraphics[width=0.6\linewidth]{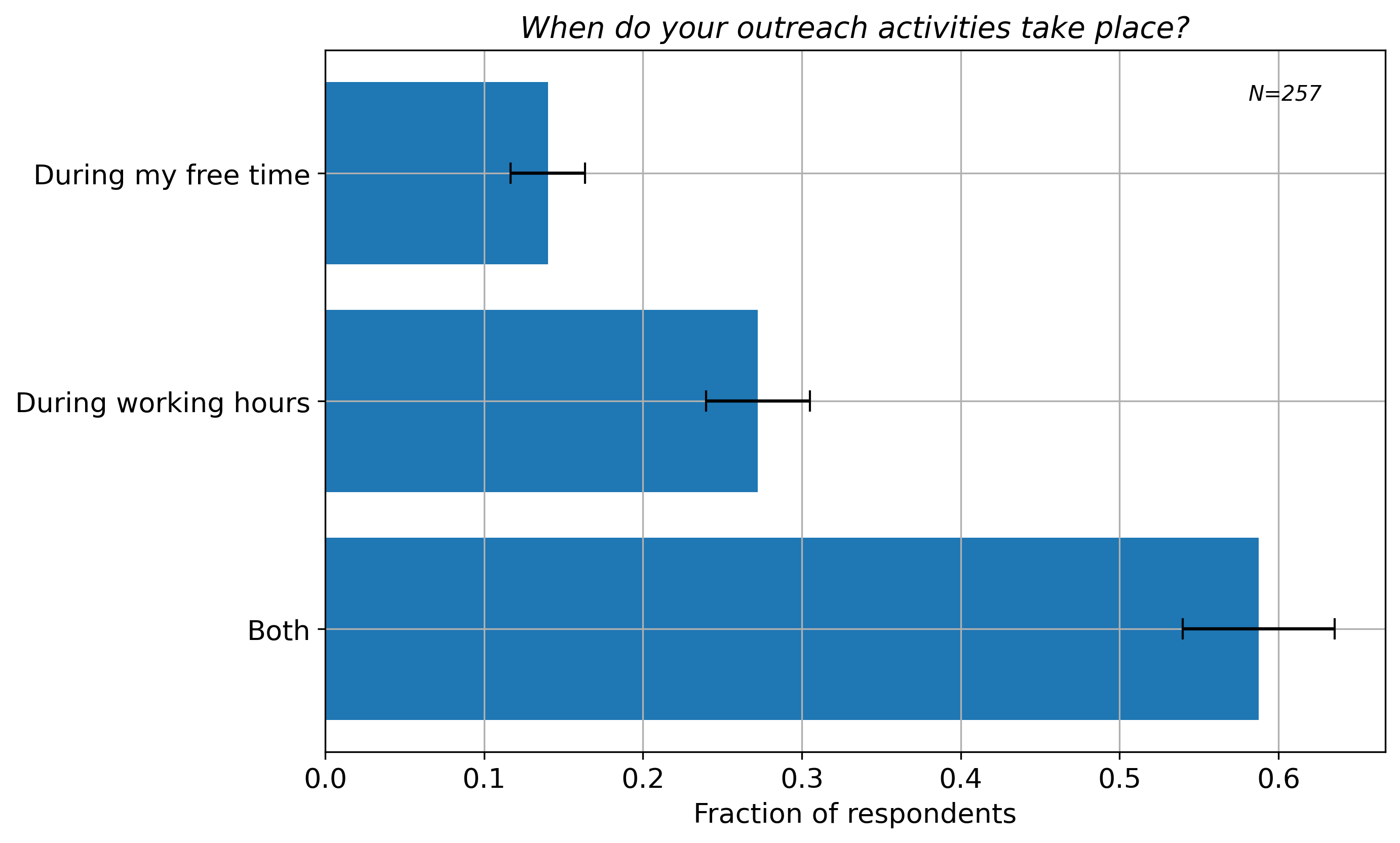}
    \caption{Fraction of survey participants expressing which time slot they allocate to pursue outreach activities. This was an additional question.}
    \label{fig:comm_time}
\end{figure}

\subsection{Future colliders}
\begin{figure}[H]
    \centering
    \includegraphics[width=0.95\linewidth]{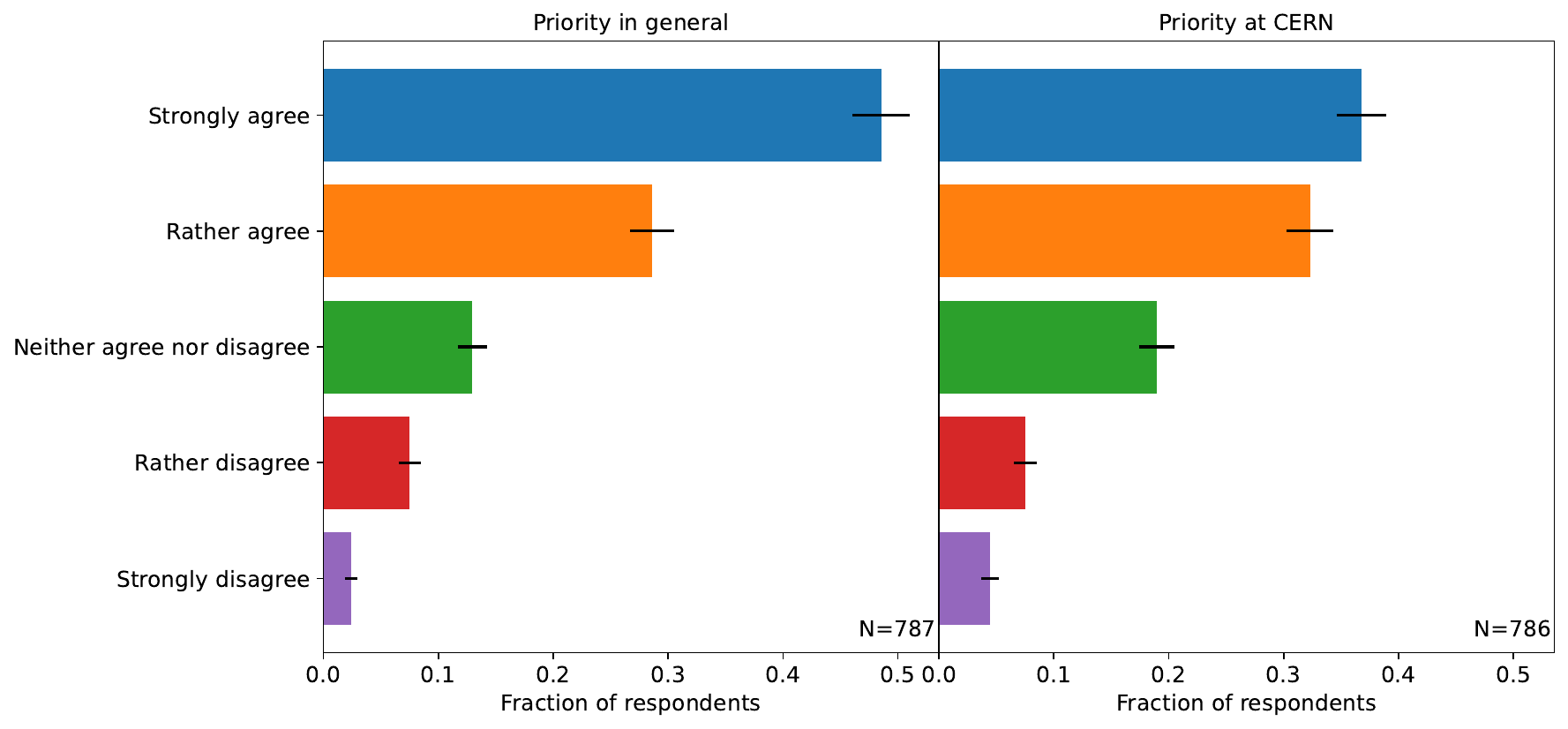}
    \caption{Distribution of responses to the question on the priority of a next flagship project in general (left) and whether it should be built at CERN (right).} % \color{blue}(Maybe add the word ``Priority" in the headings, i.e., Priority in general, and Priority at CERN?)\color{black}}
    \label{fig:fc-priority-next-flagship}
\end{figure}

\begin{figure}[H]
    \centering
    \includegraphics[width=0.95\linewidth]{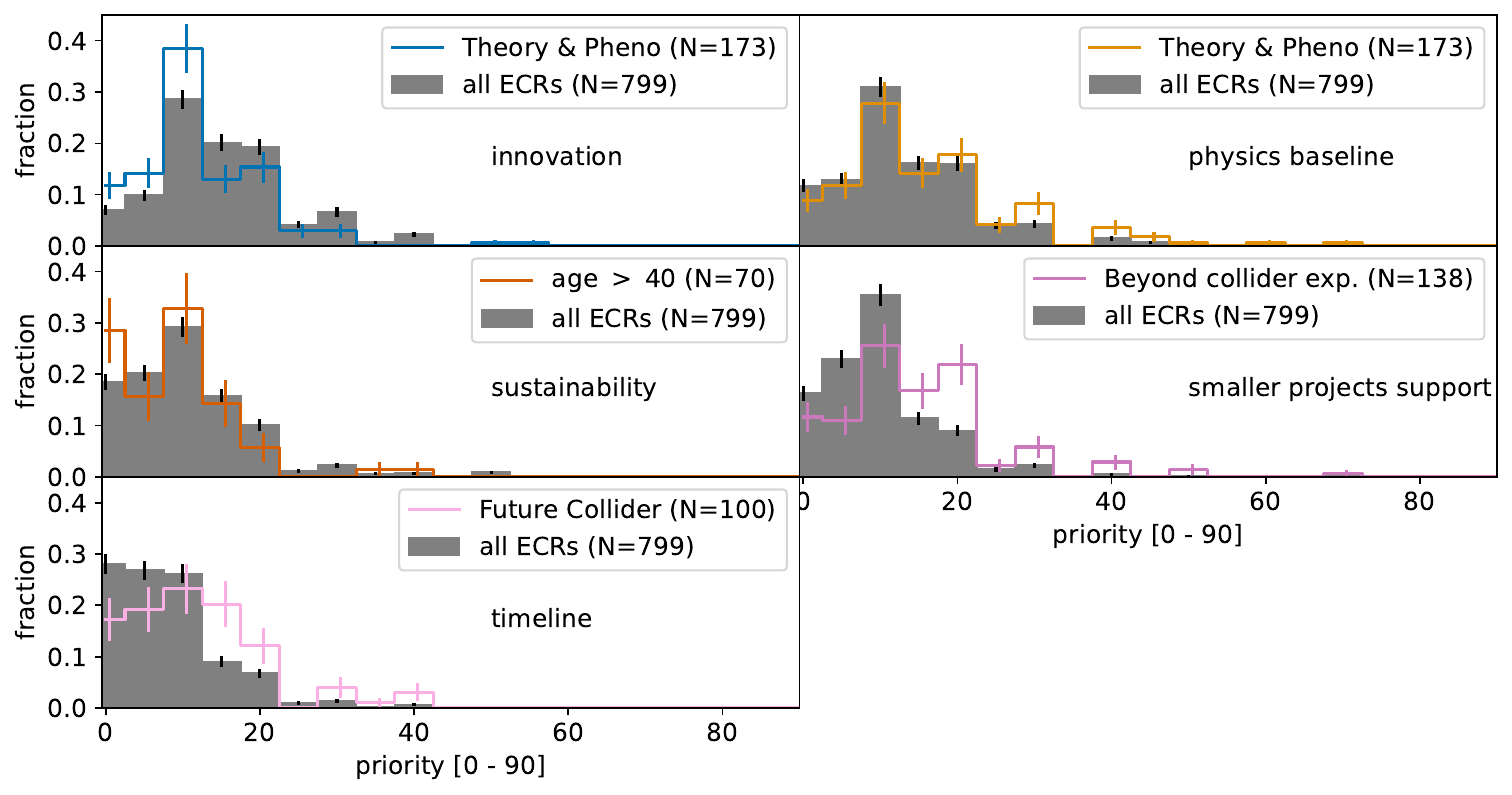}
    \caption{Comparison of the normalised distributions for selected criteria and subgroups of ECRs (coloured) with the distributions from all ECRs. Note that the selection of \emph{age $>$ 40} is done on the set of all answers and includes non-ECR and non-ECFA participants.}
    \label{fig:fc_priorities_highlights}
\end{figure}

\begin{figure}[H]
    \centering
    \includegraphics[width=0.75\linewidth]{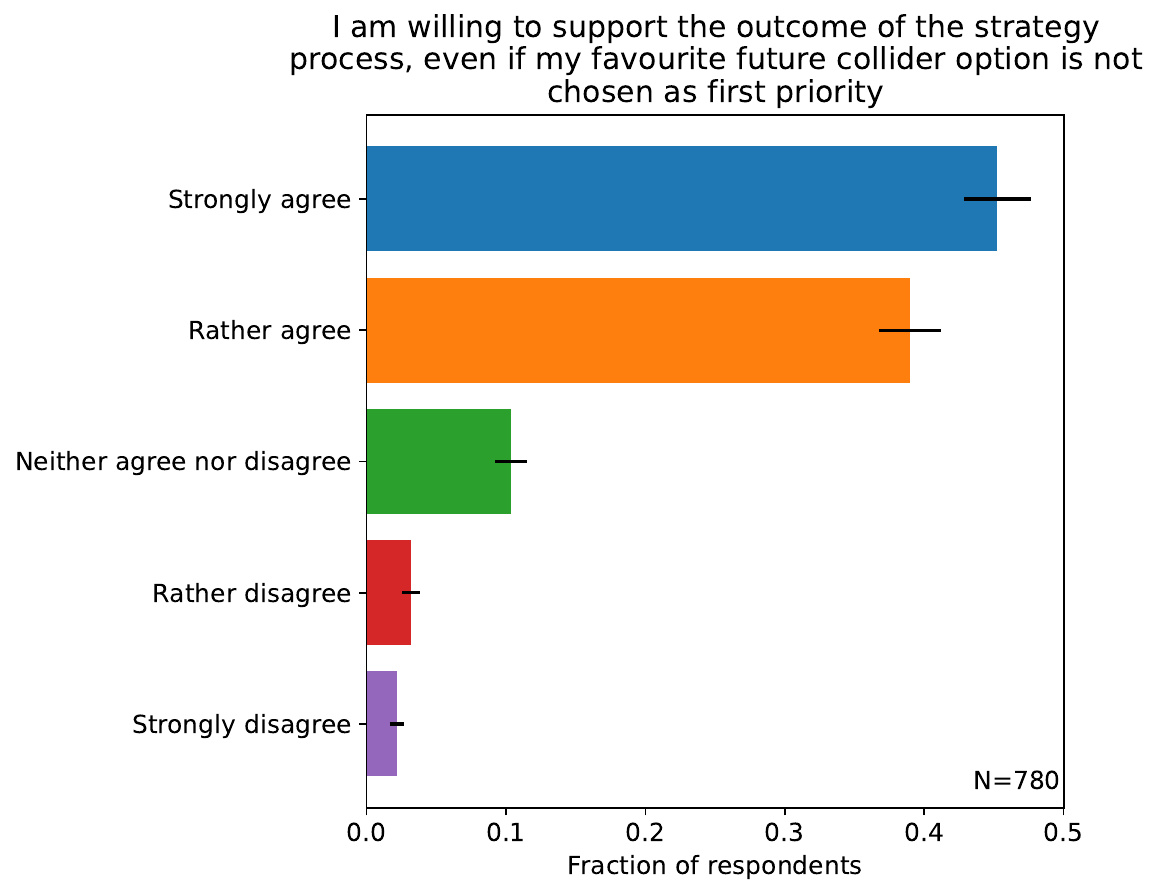}
    \caption{Distribution of responses on the question of whether ECRs would support the strategy outcome even if their favourite project is not the first priority.}
    \label{fig:fc-support-non-favorite}
\end{figure}

\begin{figure}[H]
    \centering
    \includegraphics[width=0.63\linewidth]{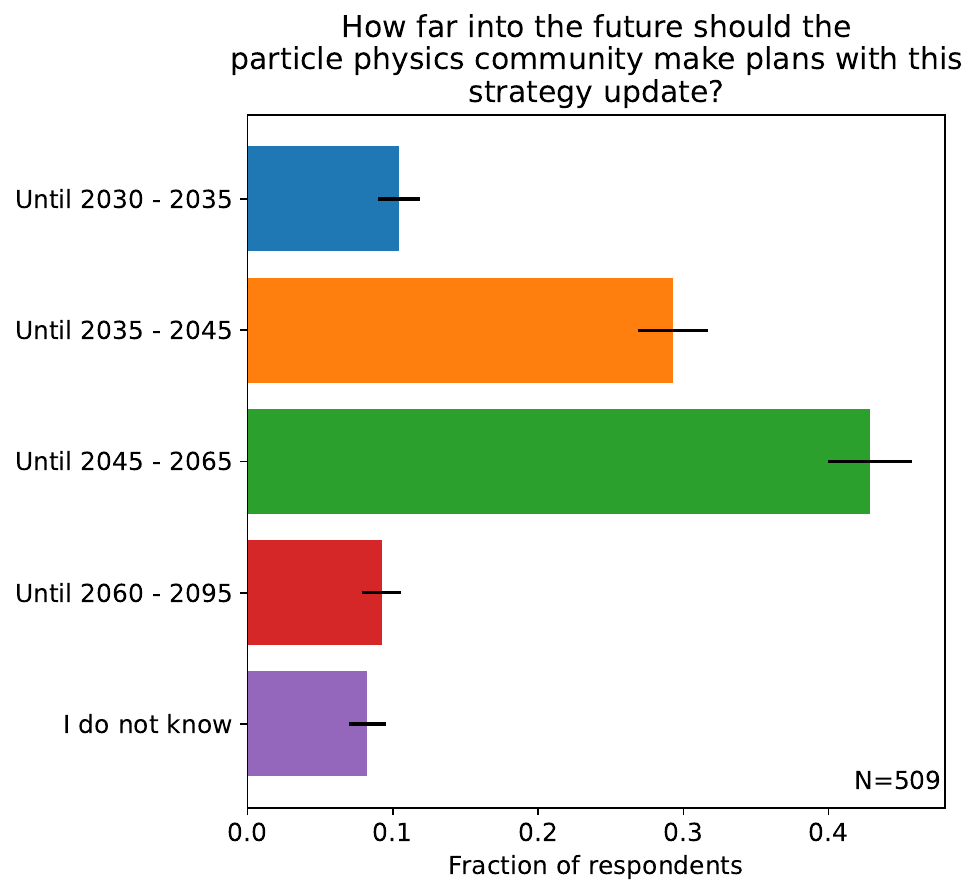}
    \caption{Proportional responses to a question on what timescale should the decision in this EPPSU process concern. } 
    \label{fig:fc-strategy-reach}
\end{figure}
\begin{figure}[H]
    \centering
    \includegraphics[width=\linewidth]{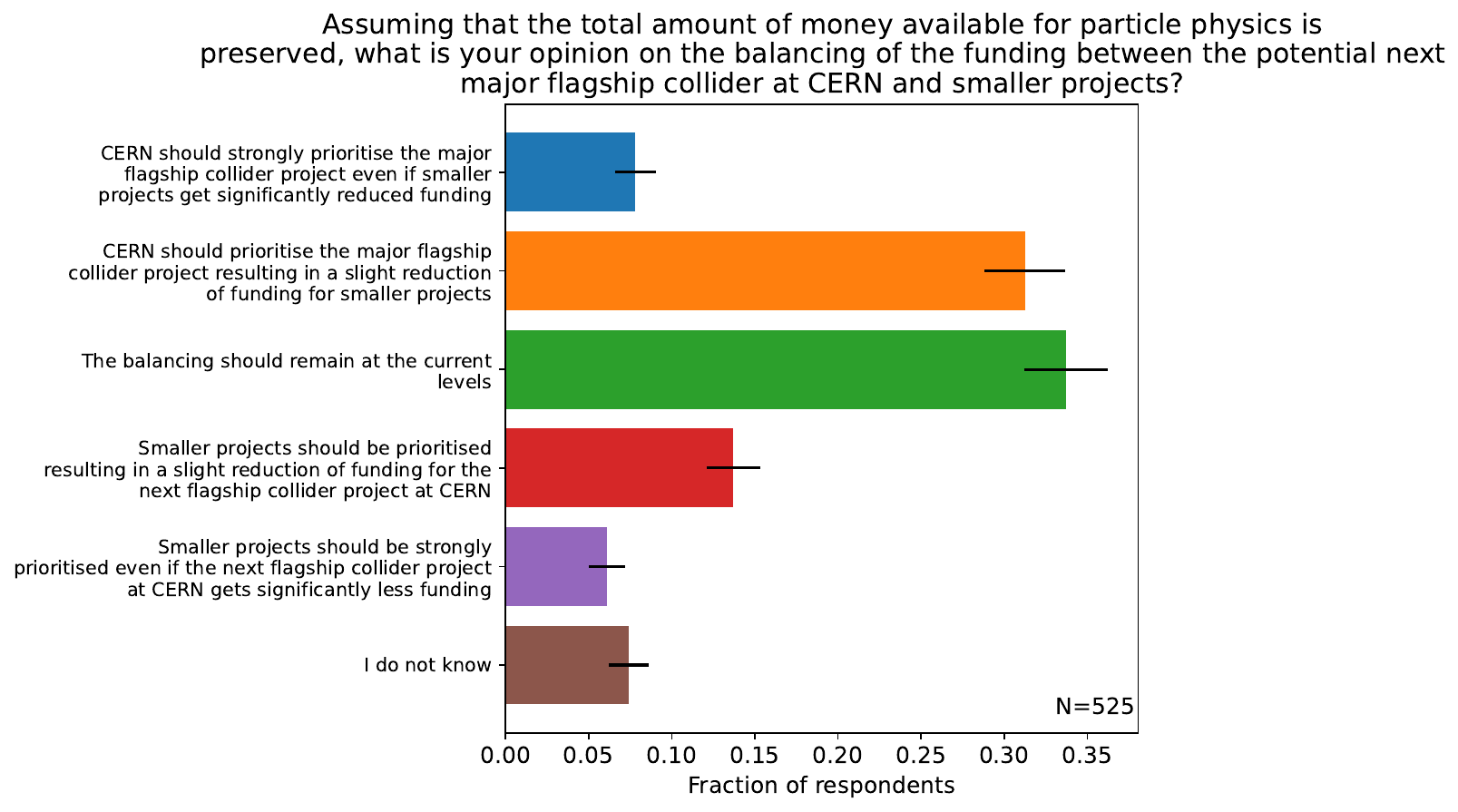}
    \caption{Proportional responses to a question whether funding share between the next flagship collider and smaller projects should remain at current levels or increase either way.  } % \color{blue}(Text is wrongly split over line break.)\color{black} } 
    \label{fig:fc-funding-split}
\end{figure}

\begin{figure}[H]
    \centering
    \includegraphics[width=0.5\linewidth]{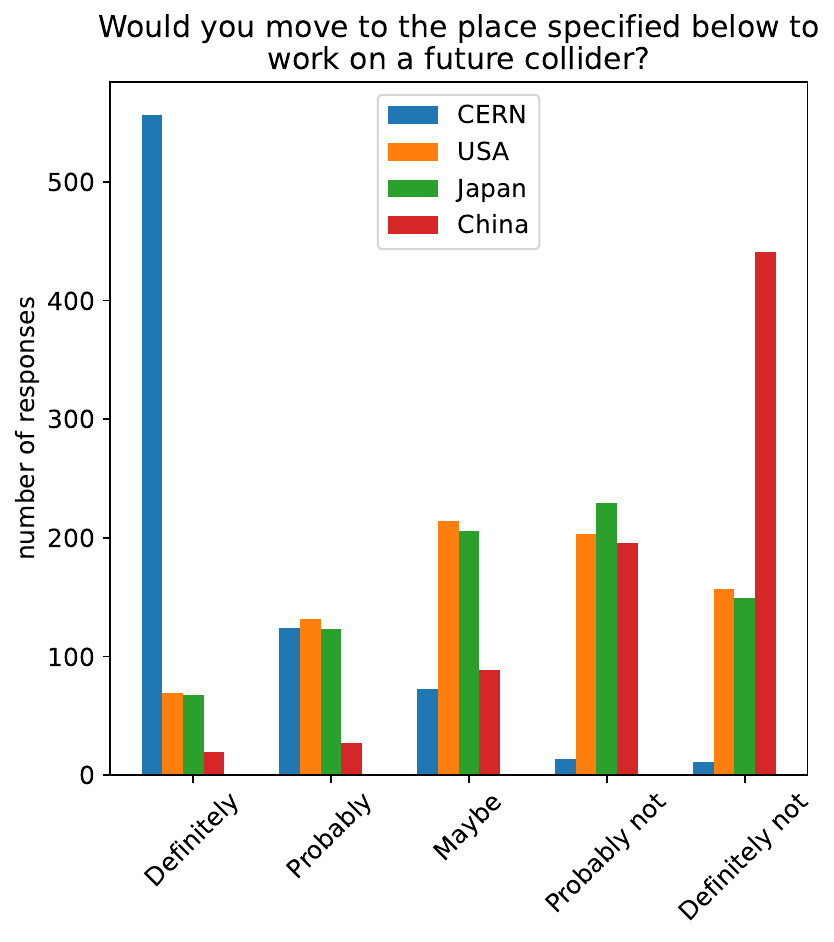}
    \caption{Distribution of responses to the question whether participants would move to the specified place to work on a future collider project. } % \color{blue}(Needs error bars)\color{black} }
    \label{fig:fc-mobility}
\end{figure}

\begin{figure}[h]
    	\centering
    	\centering
	 	 \begin{subfigure}{.45\textwidth}
	 	 	\centering
	 	 	\includegraphics[width=\linewidth]{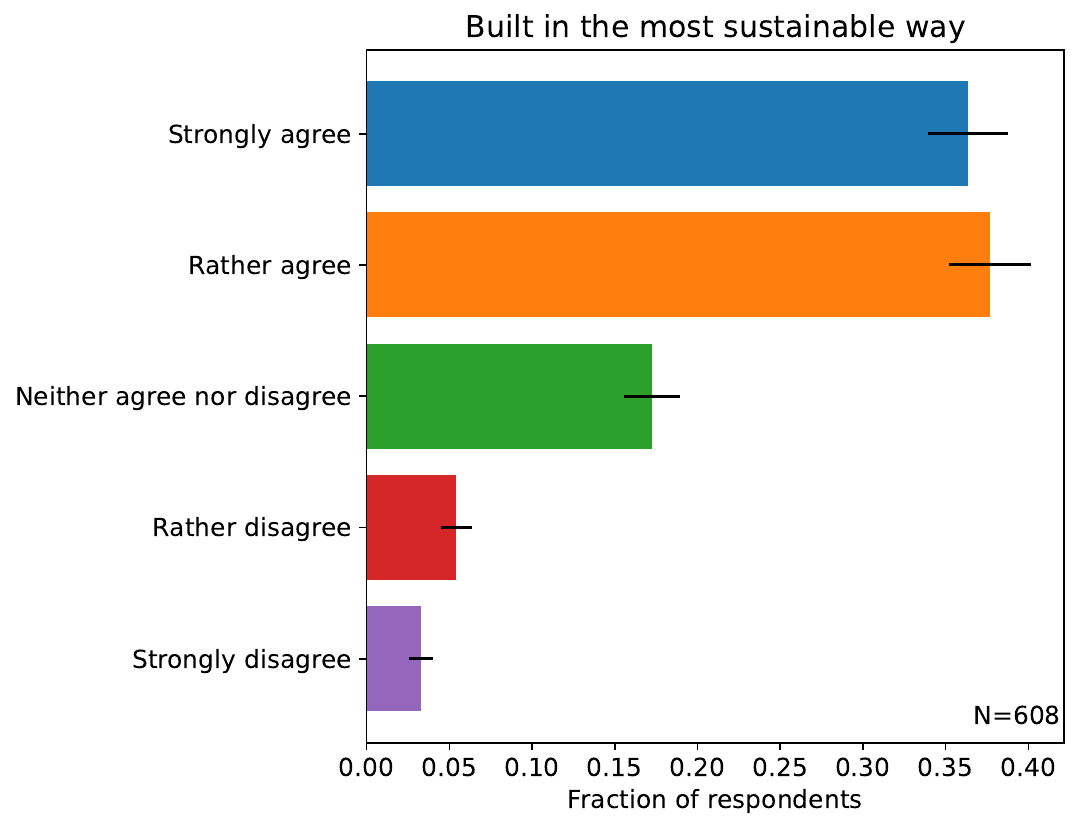}
                    \caption{}
	 	 \end{subfigure}%
	 	 \begin{subfigure}{.48\textwidth}
	 	 	\centering
	 	 	\includegraphics[width=\linewidth]{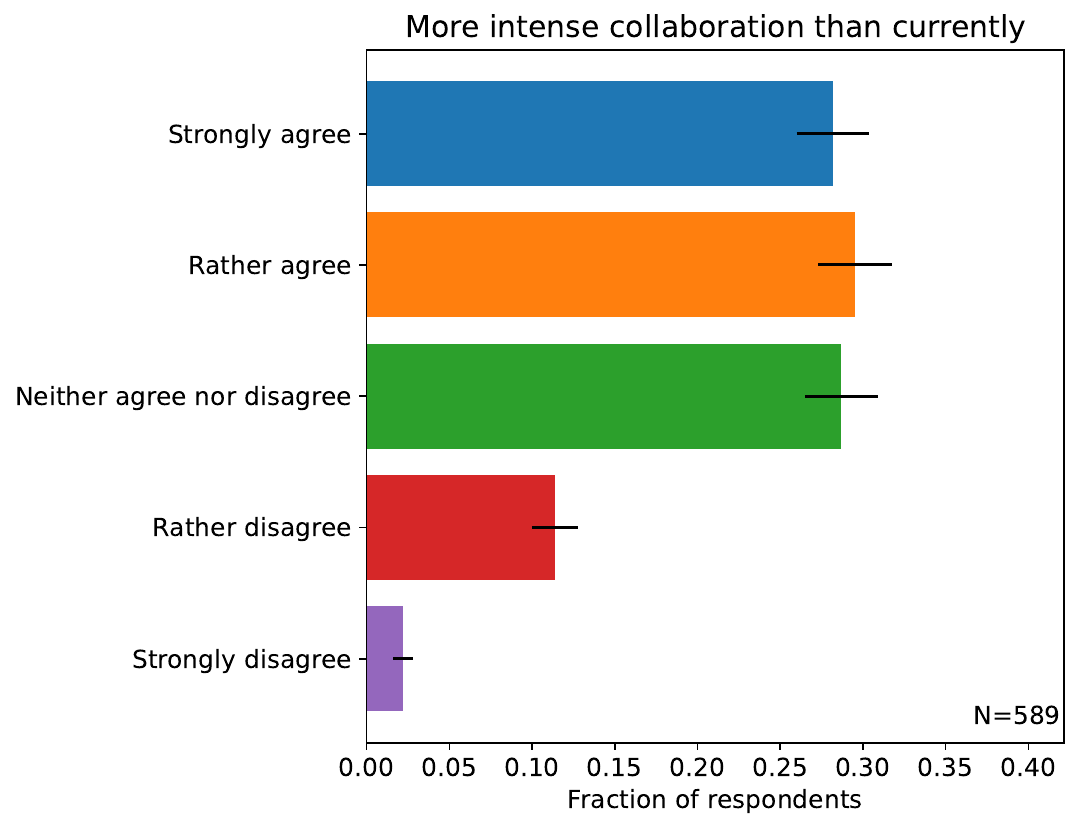}
                    \caption{}
	 	 \end{subfigure}
              \begin{subfigure}{.5\textwidth}
	 	 	\centering
	 	 	\includegraphics[width=\linewidth]{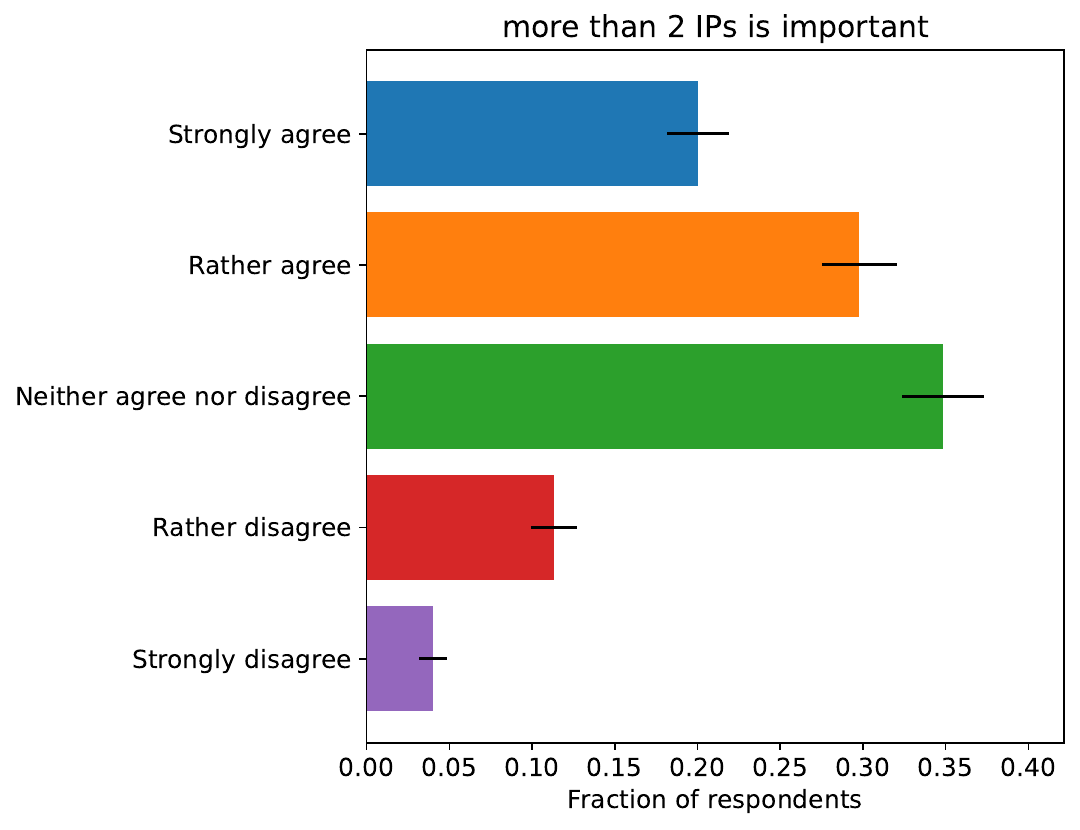}
                    \caption{}
	 	 \end{subfigure}
\caption{Proportional responses to a set of questions on specific features of the future collider. Survey participants were asked if they agree that: the collider should be built and run in the most sustainable way (a), collaboration with CERN non-member states should be intensified (b) and having more than two interaction points, and thus detectors, is important to them (c).  } % \color{blue}(Plot titles smaller than the pie chart labels, better to have them same size.)\color{black}} 
    	\label{fig:fc-specific-prios}
\end{figure}

\subsection{Future experiments beyond colliders}

\begin{figure}[H]
    \centering
    \includegraphics[width=1.0\linewidth]{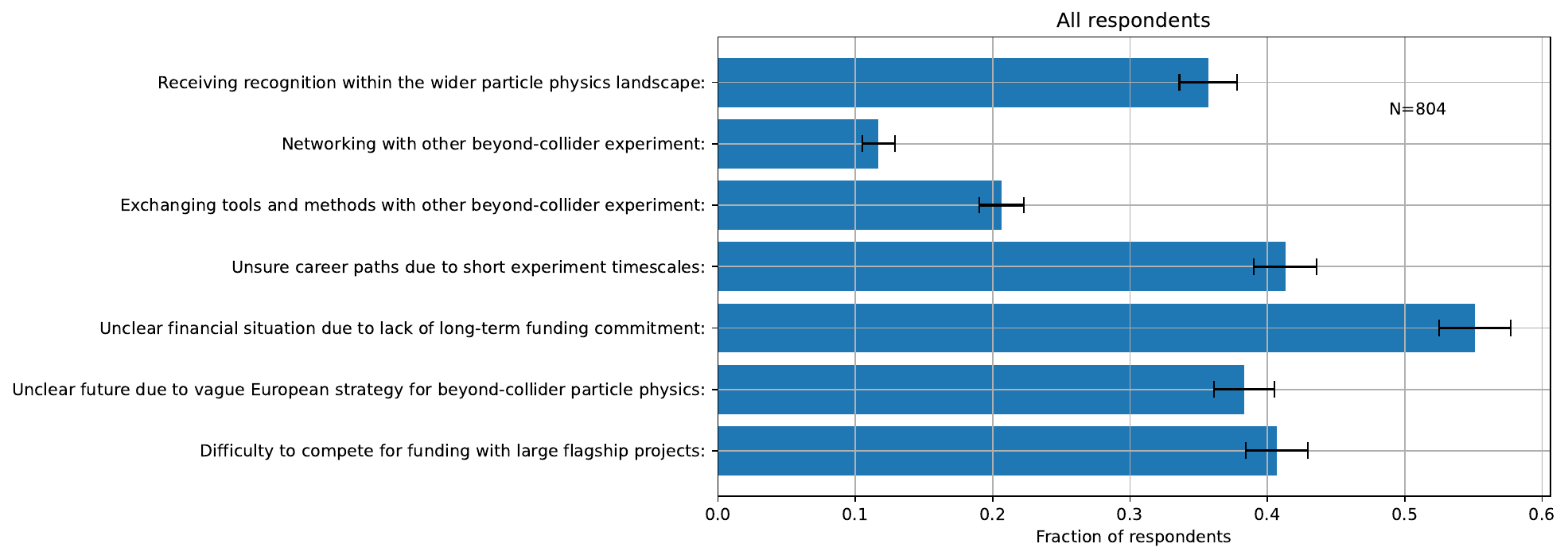}
    \includegraphics[width=1.0\linewidth]{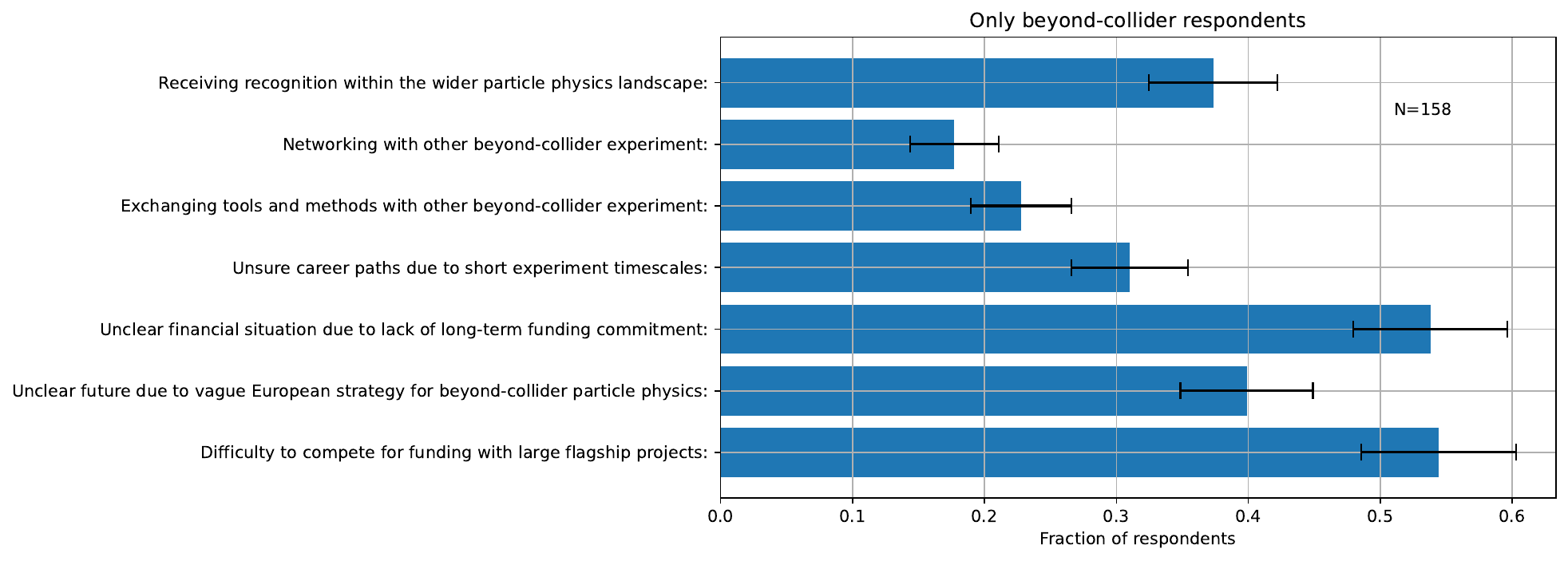}
    \caption{Fraction of respondents who chose the answers (up to a maximum of three) that applied the most to the question: ``Maintaining a strong programme for beyond-collider particle physics in Europe has many benefits, such as exploring parameter space that is not accessible at colliders, producing broad
researcher expertise, and strengthening the collaboration with neighbouring disciplines. The domain also
faces a range of challenges. In your view, what are the main challenges for beyond-collider particle
physics experiments in Europe?". The figure above includes all respondents who fit the ECFA ECR definition, and the lower figure the subset who are active in beyond-collider experiments. The error bars represent simply the square-root of the number of responses.} % \color{blue}(Plots require larger labels.)\color{black}}
    \label{fig:bc_question1}
\end{figure}

\begin{figure}[H]
    \centering
    \includegraphics[width=1.0\linewidth]{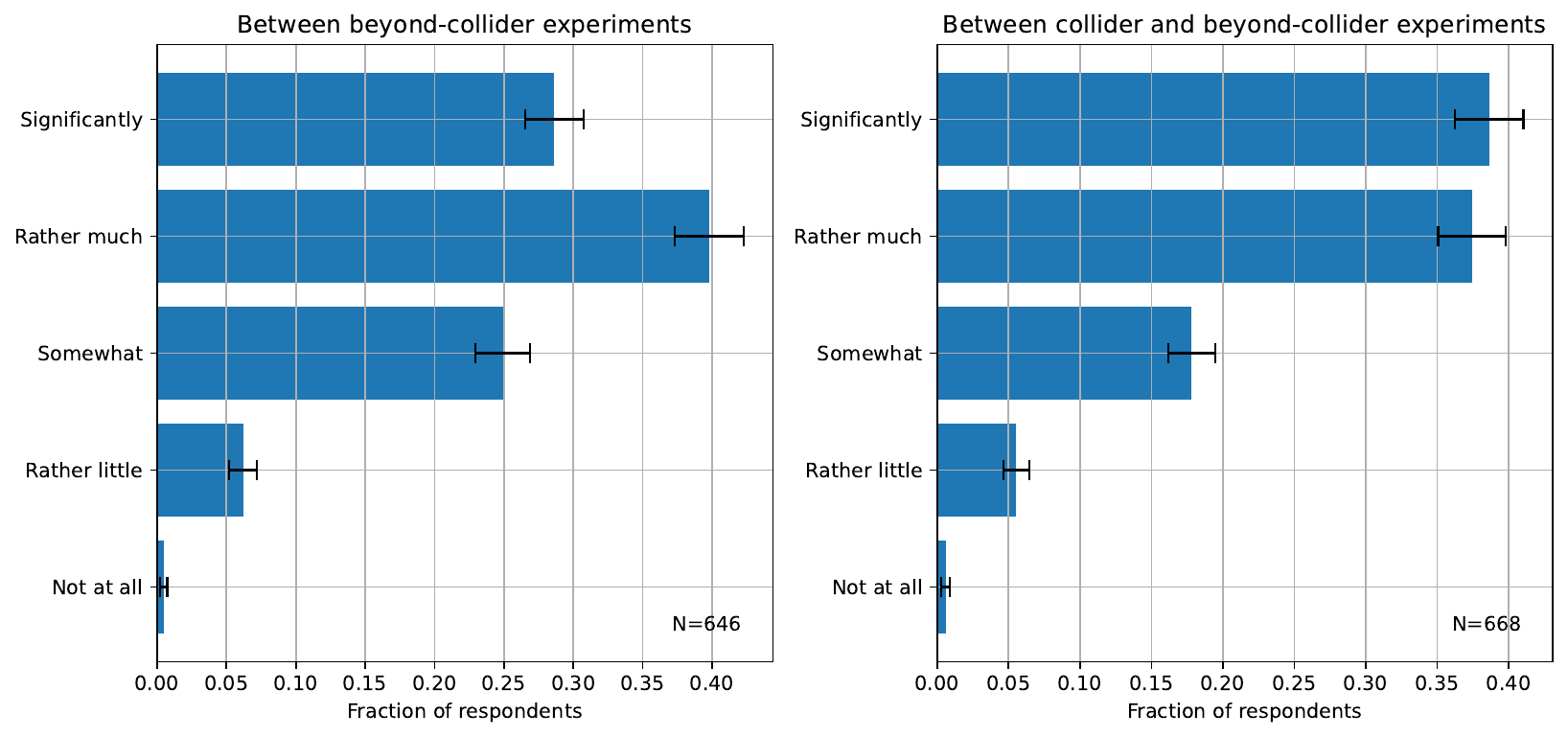}
    \caption{Proportional responses to the question: ``In your view, what possible actions would benefit beyond-collider particle physics experiments in Europe?" considering the sub-questions ``Easier career transitions between beyond-collider experiments" (left) and ``Easier career transitions between collider and beyond-collider experiments" (right).}
    \label{fig:bc_career_paths}
\end{figure}

\begin{figure}[H]
    \centering
    \includegraphics[width=1.0\linewidth]{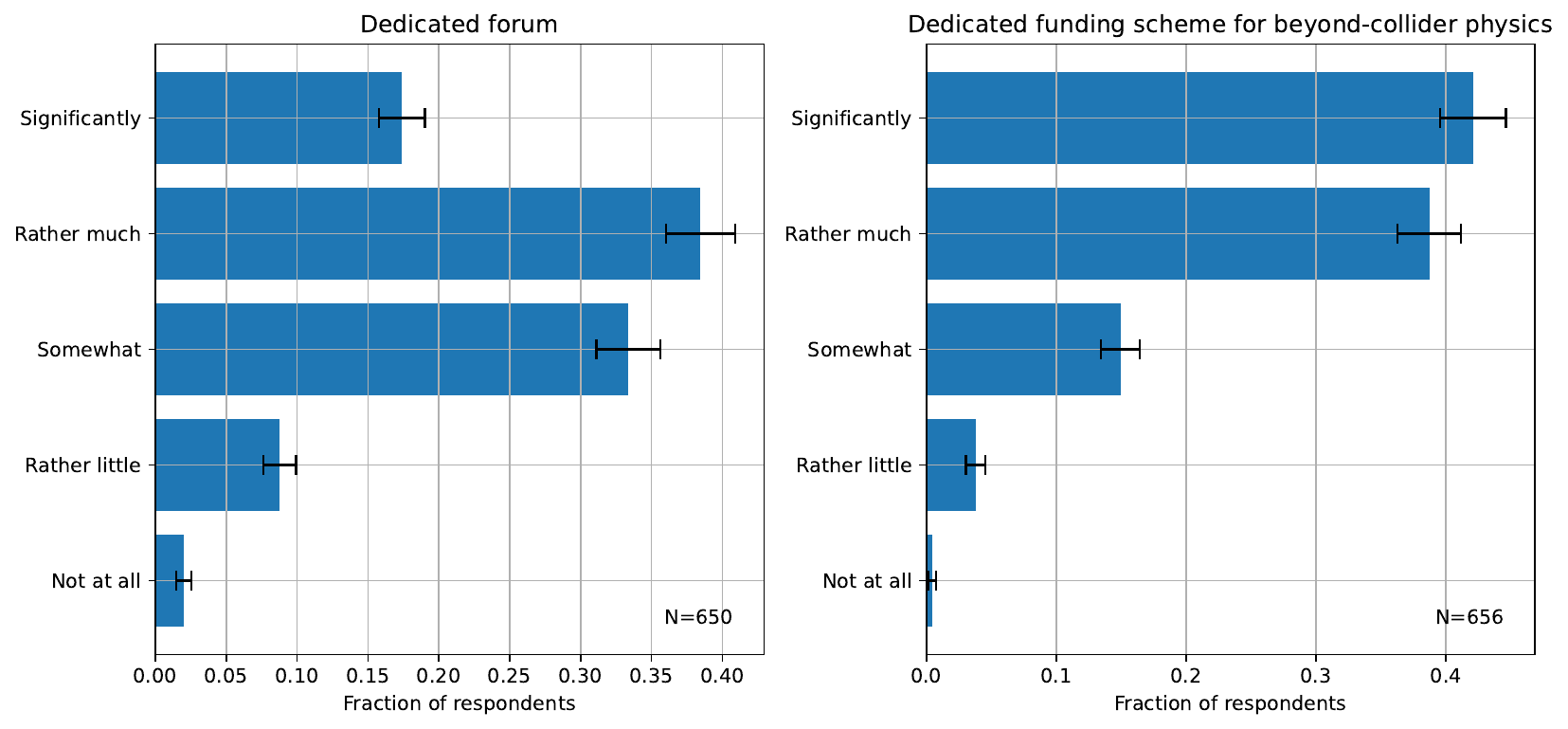}
    \caption{Proportional responses to the question: ``In your view, what possible actions would benefit beyond-collider particle physics experiments in Europe?" considering the sub-questions ``A dedicated beyond-collider particle physics forum for enhanced networking, community formation, and career opportunity sharing" (left) and ``A dedicated funding scheme for beyond-collider particle physics" (right).}
    \label{fig:bc_forum_funding_scheme}
\end{figure}

\begin{figure}[H]
    \centering
    \includegraphics[width=1.\linewidth]{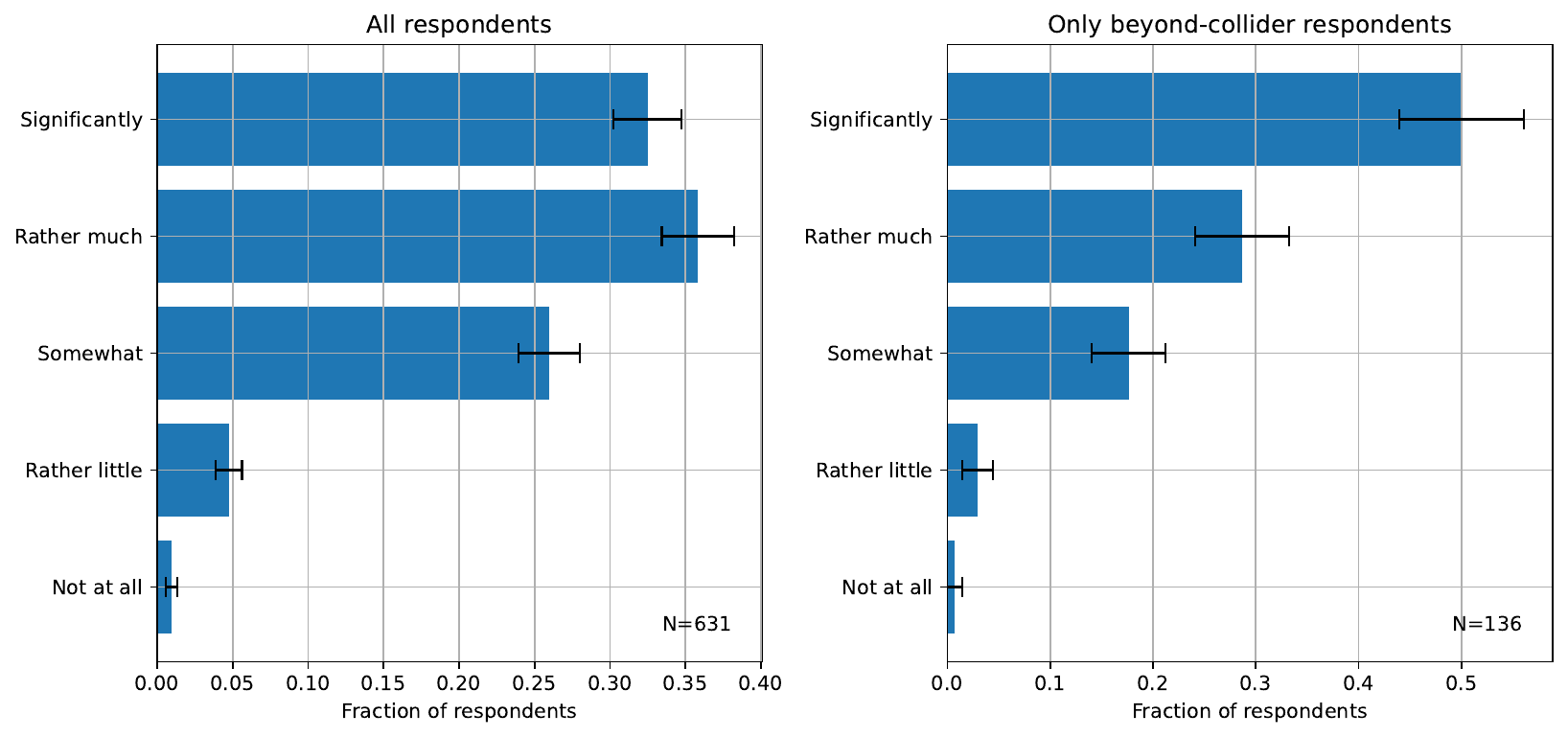}
    \caption{Proportional responses to the question: ``In your view, what possible actions would benefit beyond-collider particle physics experiments in Europe?"  considering the sub-question ``Concrete recommendations for beyond-collider physics in the EPPSU" for all respondents who fit the ECFA ECR definition (left) and the subset who are active in beyond-collider experiments (right).}
    \label{fig:bc_recommendations}
\end{figure}

\begin{figure}[H]
    \centering
    \includegraphics[width=1.\linewidth]{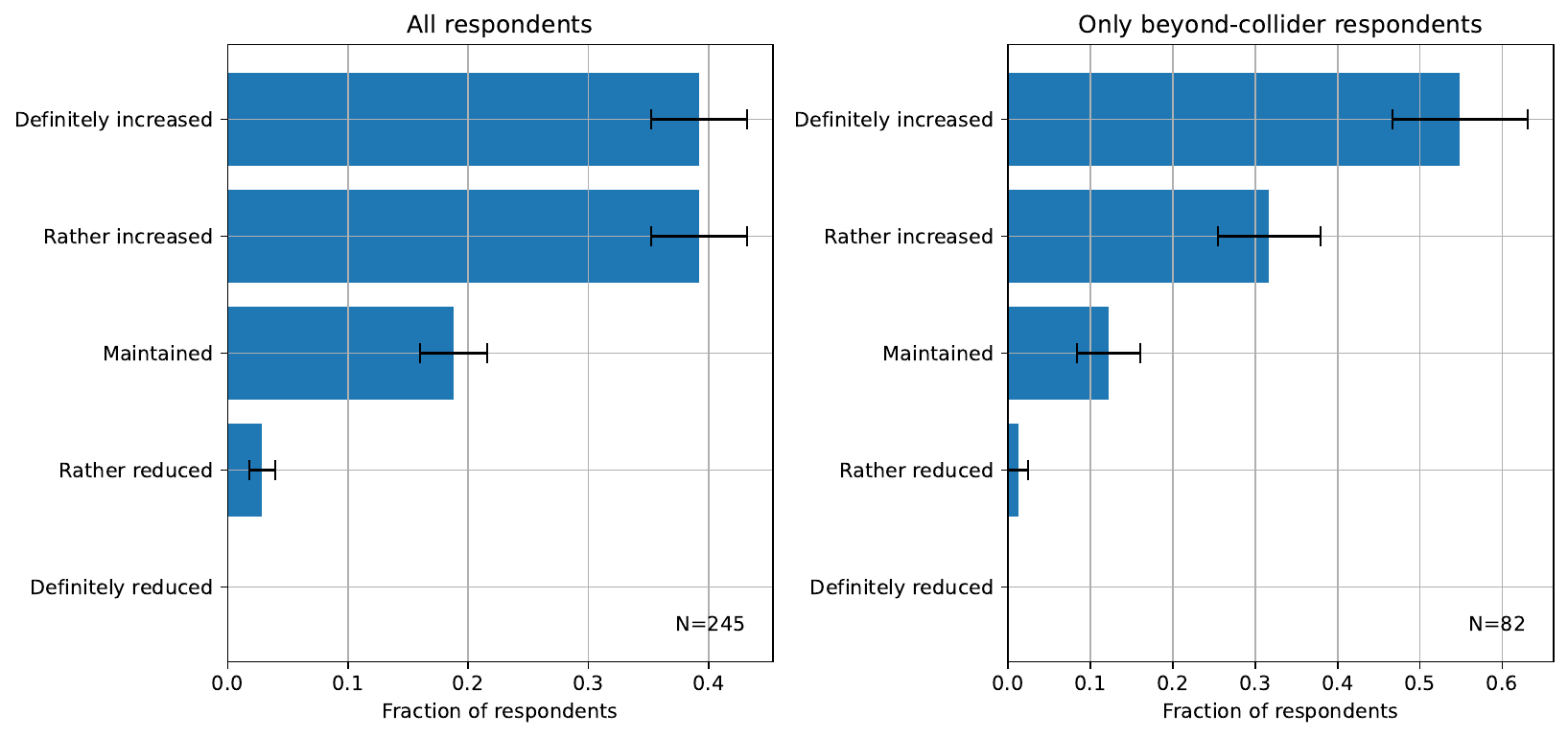}
    \caption{Proportional responses to the question ``To what extent should CERN participate in beyond-collider physics and in its different subfields, e.g. fixed target, nuclear physics, astroparticle physics, etc.? Compared to the current level, the activity should be:" considering the sub-question ``Beyond-collider physics generally" for all respondents who fit the ECFA ECR definition (left) and the subset who are active in beyond-collider experiments (right).}
    \label{fig:bc_q3_beyond_colliders_generally}
\end{figure}

\begin{figure}[H]
    \centering
    \includegraphics[width=1.\linewidth]{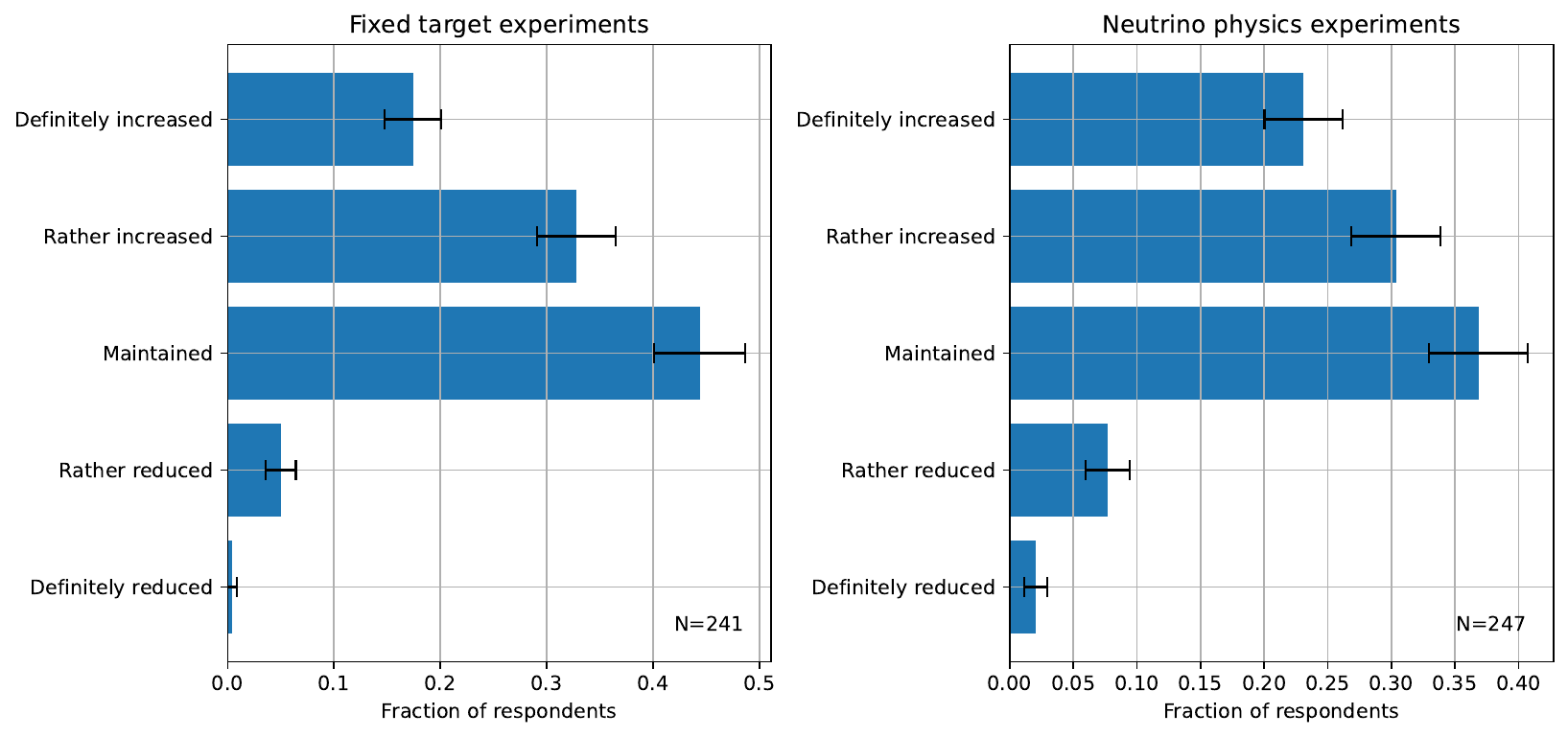}
    \includegraphics[width=1.\linewidth]{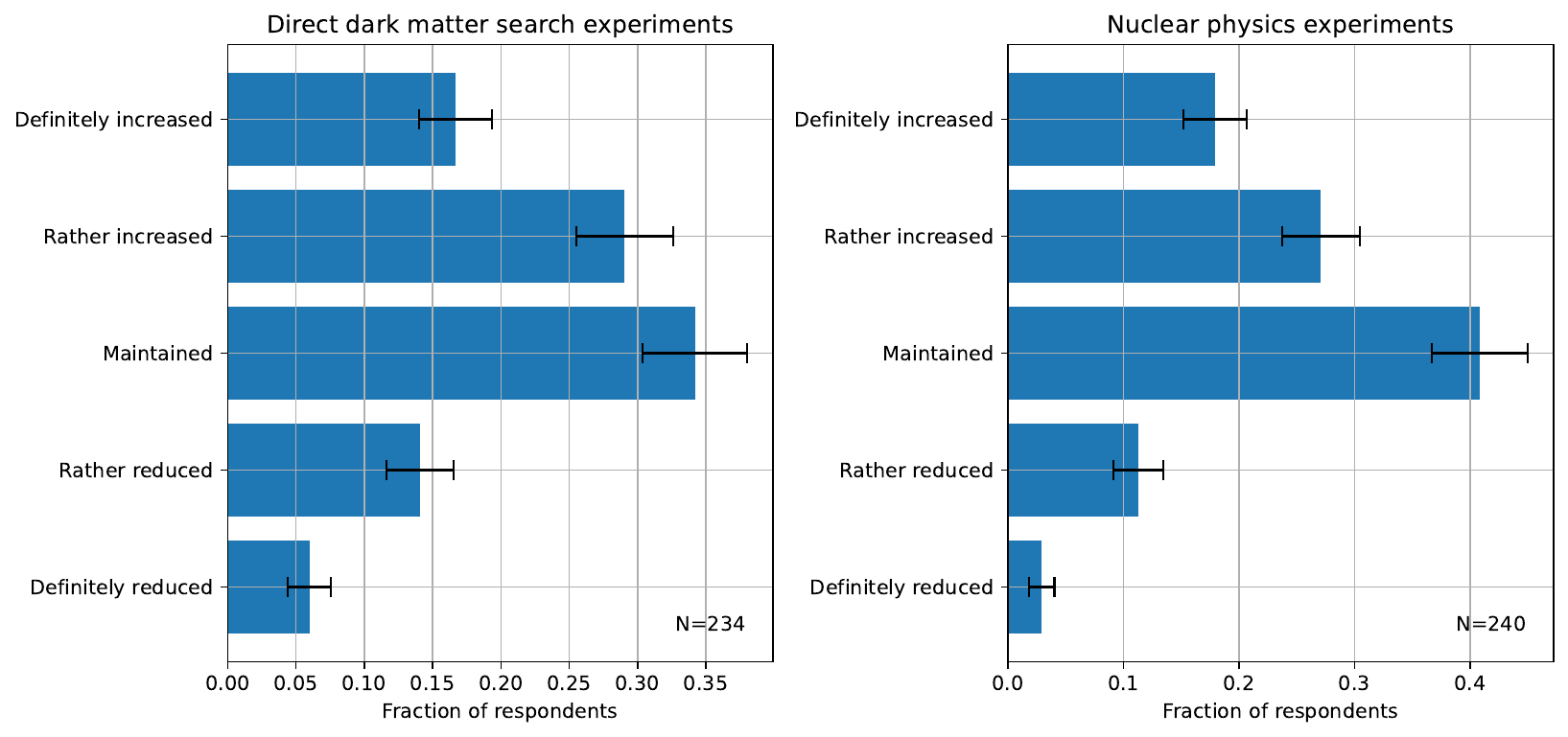}
    \caption{Proportional responses to the question ``To what extent should CERN participate in beyond-collider physics and in its different subfields, e.g. fixed
target, nuclear physics, astroparticle physics, etc.? Compared to the current level, the activity should be:" for the different sub- and adjacent fields, fixed target experiments, neutrino physics, direct dark matter search experiments, and nuclear physics. }
    \label{fig:bc_q3_subfields_part1}
\end{figure}

\begin{figure}[H]
    \centering
    \includegraphics[width=1.\linewidth]{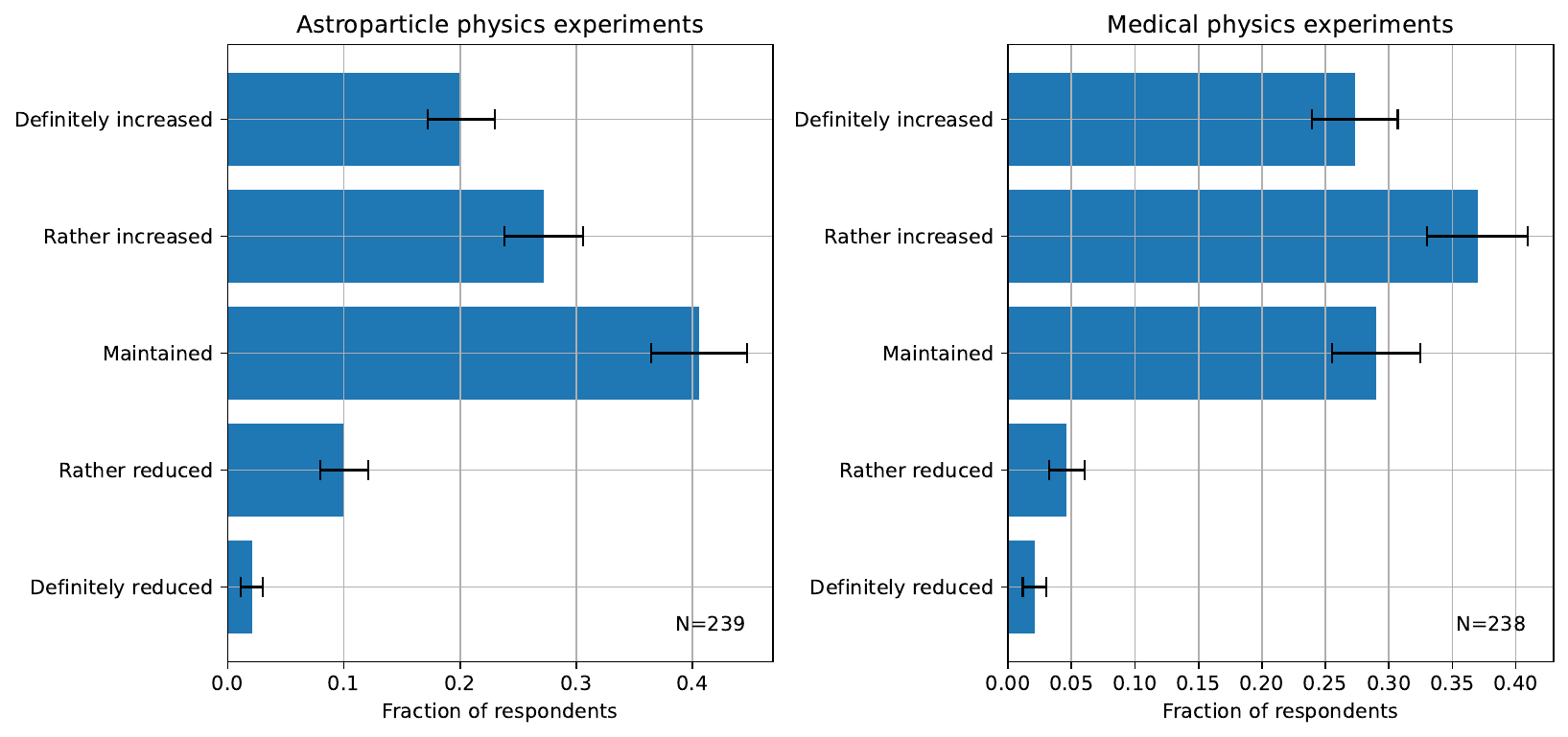}
    \includegraphics[width=0.55\linewidth]{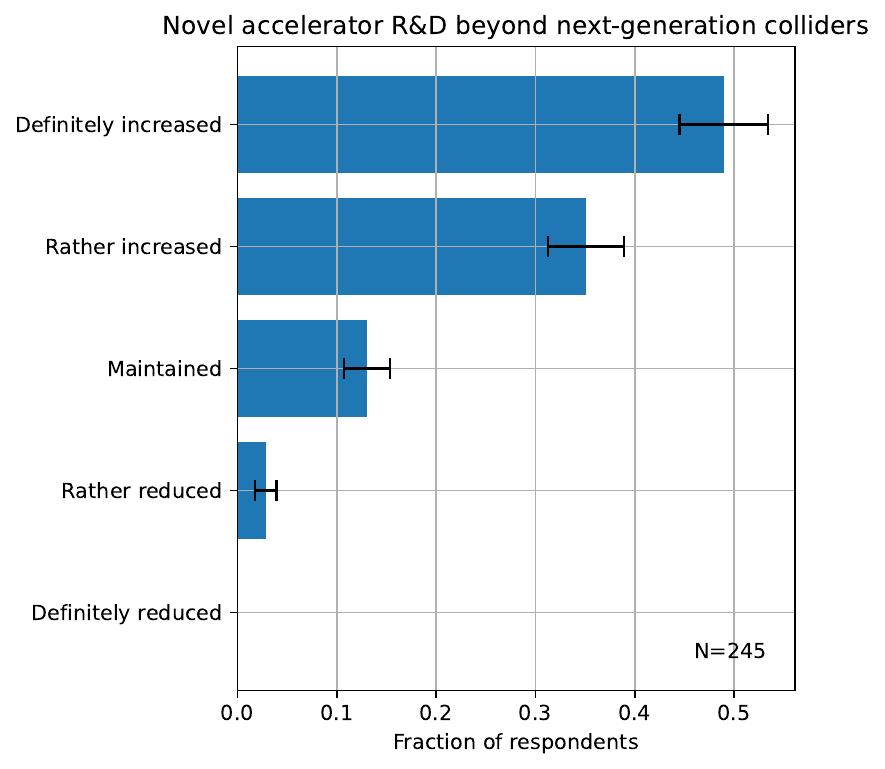}
    \caption{Proportional responses to the question ``To what extent should CERN participate in beyond-collider physics and in its different subfields, e.g. fixed
target, nuclear physics, astroparticle physics, etc.? Compared to the current level, the activity should be:" for the different sub- and adjacent fields, astroparticle physics, medical physics, and novel accelerator R\&D beyond next-generation colliders. }
    \label{fig:bc_q3_subfields_part2}
\end{figure}

\section{Glossary}
\renewcommand{\glossarysection}[2][]{}
\printglossaries

\end{document}